\documentclass[a4paper]{article}
\usepackage{etex}
\usepackage{a4wide}
\usepackage{graphicx}
\usepackage{amsbsy,amssymb,amsgen,amsfonts}
\usepackage{amsmath,amsfonts,color,latexsym}
\usepackage[pdftex,breaklinks=true,bookmarks=true,colorlinks=true,linkcolor=blue,citecolor=blue,urlcolor=blue]{hyperref}
\usepackage[authoryear,round,longnamesfirst]{natbib}
\shortcites{amoura:10,caporaloni:75,eames:11,gerashenko:08,glowinski:99,lsvezzo:10,mordant:02,qureshi:08,qureshi:07,tencate:04,calzavarini:09,villalba:12,tencate:04,kidanemariam:13}
\newcommand{\ben} {\begin{equation}}
\newcommand{\een} {\end{equation}}
\newcommand{\be} [1] {\begin{equation} \label{#1}}
\newcommand{\ee} {\end{equation}}
\newcommand{\bse} [1] {\begin{subequations} \label{#1}}
\newcommand{\ese} {\end{subequations}}
\newcommand{\ban} {\begin{eqnarray*} }
\newcommand{\ean} {\end{eqnarray*} }
\newcommand{\bea} {\begin{eqnarray}}
\newcommand{\eea} {\end{eqnarray}}

\newcommand{\solid}{---$\!$---$\!$--}

\newcommand{\dashed}{\hbox{{--}\,{--}\,{--}\,{--}}}

\newsavebox{\mybox}
\sbox{\mybox}{\dashed}

\newcommand{\revision}[2]{#2}
\newcommand{\revisions}[1]{}

\newcommand{\figpap}[2]{#1#2}
\usepackage{sectsty}
\allsectionsfont{\sffamily}
\let\LaTeXmaketitle\maketitle
\renewcommand{\maketitle}{{\sf\LaTeXmaketitle}}

\begin{document}
%
  \title{
    The motion of a single heavy sphere in ambient fluid:\\
    a benchmark for interface-resolved particulate flow 
    simulations with significant relative velocities
  }


\author{Markus Uhlmann$^\ast$ and 
  Jan Du\v{s}ek$^\dagger$
  \\[1ex]
  \it\small
  $^\ast$Institute for Hydromechanics, Karlsruhe Institute of
  Technology (KIT), 76131 Karlsruhe, Germany\footnote{%
    \tt markus.uhlmann@kit.edu}\\
  \it\small
  $^\dagger$Institut de M\'ecanique des Fluides et des Solides, 
  Universit\'e de Strasbourg, 67000 Strasbourg, France\footnote{%
    \tt dusek@unistra.fr}
}
\date{
}%
\maketitle
\begin{abstract}
  %
  Detailed data describing the motion of a rigid sphere settling in
  unperturbed fluid is generated by means of highly-accurate
  spectral/spectral-element simulations with the purpose of serving as 
  a future benchmark case. 
  A single solid-to-fluid density ratio of 1.5 is
  chosen, while the value of the Galileo number is varied from 144 to
  250 such as to cover the four basic regimes of particle motion
  (steady vertical, steady oblique, oscillating oblique,
  chaotic). 
  \revision{}{%
    This corresponds to a range of the particle Reynolds number from
    185 to 365. 
  }
  In addition to the particle velocity data, 
  extracts of the fluid velocity field are provided, as well as the
  pressure distribution on the sphere's surface.  
  Furthermore, the same solid-fluid system is simulated with a
  particular non-boundary-conforming approach, i.e.\ the immersed
  boundary method proposed by \cite{uhlmann:04}, using various
  spatial resolutions. 
  It is shown that the current benchmark case allows to adjust the
  resolution requirements for a given error tolerance in each flow
  regime. 
\end{abstract}

%
%
\section{Introduction}\label{sec-intro}
The gravity-induced settling or rising of a spherical rigid body in a
viscous fluid exhibits a rich set of dynamical features, involving
a variety of patterns of motion from steady vertical to fully chaotic
in different regions of the parameter space. 
Many aspects of the flow physics have been discussed in a recent
review by \cite{ern:12}. 
When considering spheres settling in a priori ambient surroundings 
all deviations from a straight vertical path as well as all
unsteadiness originate from the characteristics of the fluid motion
in the near-field around the immersed object and in its wake. 
Therefore, the analysis of the motion of settling/rising objects
really implies an investigation of the features of particle wakes. 

Beyond their relevance to particle trajectories, wakes generated by
moving particles are of significance in the context of
particle-induced turbulence generation and modification. 
One question which is often posed in particulate
flow systems pertains to the amount of turbulence enhancement or
attenuation due to the addition of particles to a given fluid flow. 
Elucidating the physics of wakes shed by single (and multiple)
mobile particles is expected to contribute to a better understanding
of the technologically important problem of turbulence-particle
interaction to which a considerable effort has been devoted 
\citep{balachandar:10}. 

As a complement to modern experimental techniques, it has now become
feasible to simulate numerically the flow around a reasonably large
amount of moving immersed objects based upon the Navier-Stokes
equations \citep[e.g.][]{tencate:04,uhlmann:08a,lucci:10,lucci:11,villalba:12,gao:13}. 
For reasons of computational efficiency, most of the simulations of
this kind employ numerical techniques which do not rely on
geometry-conforming grids, thereby avoiding the necessity for
repeated remeshing and complex data structures. 
Instead, the general idea of these methods is to allow for the
treatment of a single medium throughout the domain occupied by both
the fluid and the solid, while imposing locally the constraint of
rigid body motion through some kind of manipulation of the
Navier-Stokes equations. While the general concept of these
non-conforming methods as well as their efficiency has now been widely
established, it is felt that rigorous resolution criteria have not yet
been determined in all situations. 

Typically, finite-size particle flow simulation codes are validated
with respect to a sub-set of the following test cases:
\begin{enumerate}
\item Flow around a fixed sphere with uniform, steady inflow versus 
  standard drag correlations \cite[such as][]{clift:78} or 
  high-fidelity numerical data \citep[e.g.][]{johnson:99,bouchet:06}. 
\item Gravitational settling of a single heavy sphere versus reference
  data, e.g.\ by \cite{mordant:00}. 
\item ``Drafting-kissing-tumbling'': gravitational settling of a pair
  of cylinders (in two space dimensions) or spheres initially trailing
  each other, for which no rigorous reference data exists to our
  knowledge. 
\item Rotation of a single fixed cylinder
  (in two space-dimensions) or ellipsoid in Couette flow
  versus reference data from experiments \citep{zettner:01}, 
  analytical solutions \citep{jeffery:22} 
  or numerical reference data \citep[e.g.][]{ding:00}. 
\item Lateral migration of a single neutrally-buoyant particle in
  laminar Hagen-Poiseuille flow versus analytical results
  \citep{asmolov:99} and experimental data \citep{matas:04}. 
  For this case numerical reference data is available
  \citep[e.g.][]{yang:05b}. 
  The computationally less demanding case of two-dimensional flow
  around migrating circular disks in plane channel flow has been
  studied numerically by \cite{inamuro:00}, \cite{pan:02b} and
  \cite{joseph:02}.  
  %
  %
  %
\end{enumerate}
%
\begin{figure}
  \figpap{
    \centering
    \begin{minipage}{.4\linewidth}
      \includegraphics[width=\linewidth,clip=true,
      viewport=90 398 350 685]
      {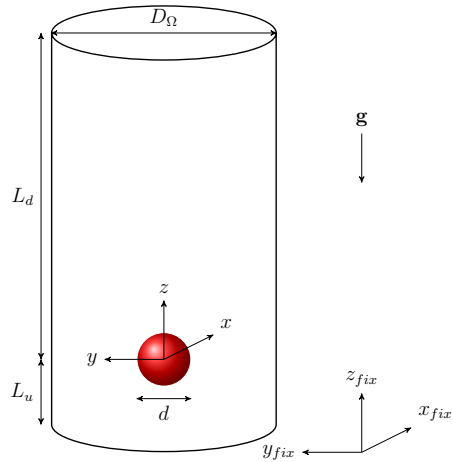}
    \end{minipage}
  }{
    \caption{%
      The geometry of the problem and the computational domain as
      employed in the reference method described in \S~\ref{sec-ref}. 
      \protect\label{fig-ref-schematic}
    }
  }
\end{figure}
\begin{figure}
  \figpap{
  \centering
  \begin{minipage}{3ex}
    $\displaystyle\frac{\rho_p}{\rho_f}$
  \end{minipage}
  \begin{minipage}{.5\linewidth}
    \includegraphics[width=\linewidth]
    {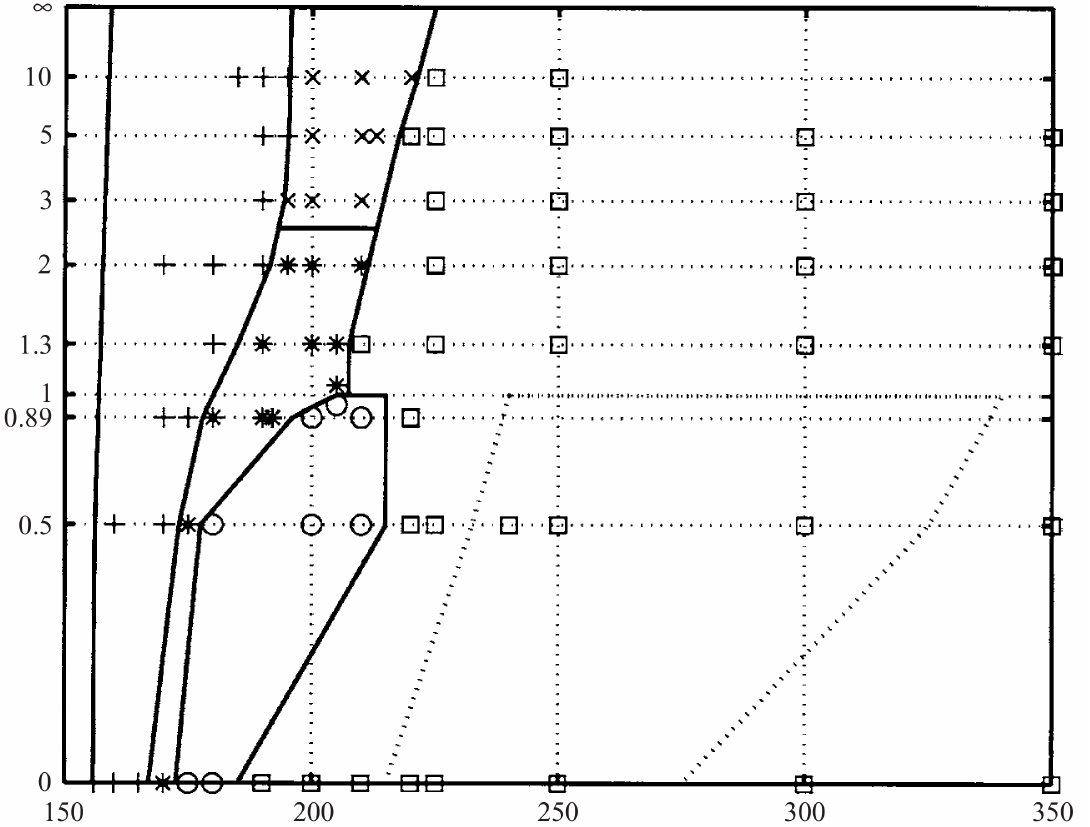}
    \\
    \centerline{$G$}
  \end{minipage}
  }{
  \caption{%
    Diagram of flow regimes in the $(G,\rho_p/\rho_f)$ parameter
    plane \citep{jenny:04}. 
    To the left of the leftmost line the wake is axisymmetric and the
    particle trajectory is steady and vertical. 
    The different symbols correspond to simulations classified as: 
    $+$,~steady oblique path; 
    $\ast$,~oscillating oblique path with low frequency; 
    $\times$,~oscillating oblique path with high frequency;
    $\circ$,~periodic zigzagging trajectory; 
    $\Box$,~three-dimensional chaotic trajectory. 
    The dotted line delimits the region of coexistence of a chaotic
    and a periodic zigzagging state. 
    The parameter values considered in the present work all lie on a
    horizontal line at $\rho_p/\rho_f=1.5$. 
    \protect\label{fig-jenny-et-al-diagram}
  }
  }
\end{figure}
Concerning flows with significant relative velocities between the
solid and the fluid phase (e.g.\ due to buoyancy effects as in
sedimentation systems),  
our personal experience has shown that the above array of validation
tests might not be 
representative of all relevant flow features. 
In particular, the subtle dynamics of
particle motion due to differences in wake characteristics in the
various regions of the parameter space may not be sufficiently
captured by a numerical code at a given resolution although it might 
perform reasonably well in the above cases. 
Therefore, the purpose of the present work is to provide a further 
benchmark configuration serving as a test of simulation tools for
fully-resolved fluid-particle motion.

The case of a single settling 
unconfined 
sphere in the absence of solid
boundaries appears an attractive configuration in this context.
%
On one hand, high-fidelity data can be generated
by means of relatively efficient reference simulations with spectral
accuracy \citep{jenny:04b}. 
%
%
In the reference method, the mesh is translated with the immersed object
which avoids remeshing \citep{mougin:02}.
On the other hand, as mentioned above, the settling process of a
single sphere covers all the essential dynamics involved in general
sedimentation problems, including very subtle effects of wake-induced
non-trivial trajectories, while excluding additional complexity due to
inter-particle collisions. 
It is as such a challenging and rigorous test case for any
non-geometry-conforming numerical simulation method. 
At the same time the benchmark simulations need not be 
excessively demanding, 
since the size of the computational domain can be kept relatively small.
Furthermore, the initial state and the boundary conditions of the
problem are simple and well-defined. 

\begin{figure}
  \figpap{
  \centering
  \begin{minipage}{.5\linewidth}
    \includegraphics[width=\linewidth,clip=true,
    viewport=90 360 390 685]
    {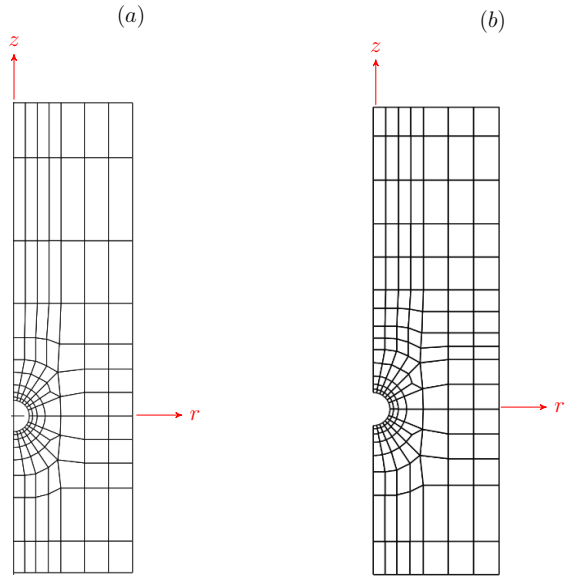}
  \end{minipage}
  }{
  \caption{%
    The spectral element meshes used in the axial/radial plane of the
    coordinate system attached to the particle in cases with lateral
    domain size $D_\Omega=7.54$: 
    $(a)$ mesh with 134 elements, used for simulations with
    $G=\{144,178.46\}$;  
    $(b)$ mesh with 169 elements, used for $G=\{190,250\}$. 
    \protect\label{fig-sem-grid}
  }
  }
\end{figure}
For this purpose we have generated detailed data for the flow field and the
rigid body motion in the case of a single heavy sphere settling in
quiescent surroundings, using a highly accurate
spectral/spectral-element method. The simulations are similar to those
performed by and described in \cite{jenny:04}. 
However, in the present work the computational domain was purposefully
kept small, thereby requiring new simulations. 
Furthermore, in the present paper we aim at reporting a complete set
of data \citep[contrary to the previous publication of][]{jenny:04} 
for the purpose of validating alternative numerical methods. 

In parallel, we report results from computations of the same flow
configuration obtained by means of a non-geometry-conforming code 
based upon an immersed boundary method 
\citep[IBM, ][]{uhlmann:04}. 
We have
performed refinement tests from which the required small-scale
resolution can be deduced in each flow regime. 

The outline of the paper is the following. In
\S~\ref{sec-ref} the flow geometry, boundary conditions
and the numerical method used to generate the reference data is
described, before we proceed to present the benchmark data. 
In \S~\ref{sec-ibm} we further illustrate the validation procedure by
describing simulations performed with 
an immersed boundary method; 
the numerical approach is 
first summarized (\S~\ref{sec-ibm-numa}) and then the results are
compared to the reference data (\S~\ref{sec-ibm-results}). 
The paper closes with a summary and discussion in
\S~\ref{sec-conclusion}. 

%
\section{Reference case}\label{sec-ref}

\subsection{ Flow configuration and governing equations}
\label{sec-reference-config}
\begin{figure}
  \figpap{
  \centering
  \begin{minipage}{.7\linewidth}
    \includegraphics[width=\linewidth,clip=true,
    viewport=90 430 550 685]
    {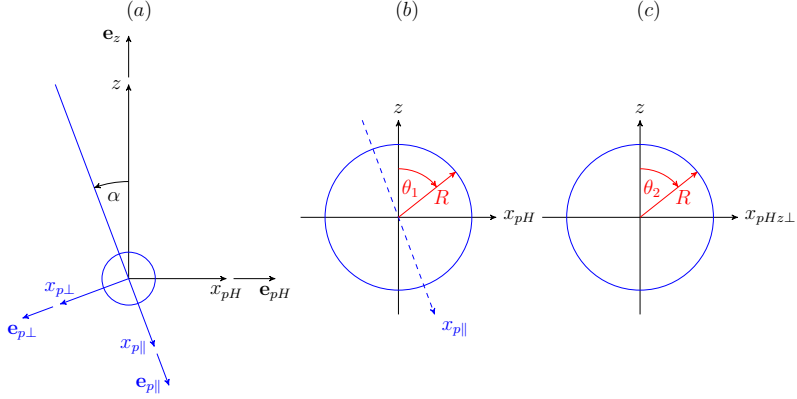}
  \end{minipage}
  }{
  \caption{%
    $(a)$ 
    Sketch of the notation concerning directional unit vectors and
    coordinates in the plane spanned by the vertical axis
    $\mathbf{e}_z$ and the direction of the particle motion
    $\mathbf{e}_{p\parallel}$, as defined in
    \S~\ref{sec-ref-results-notation}. 
    Note that the direction $\mathbf{e}_{pHz\perp}$
    is perpendicular to the plane of the sketch which corresponds to
    the plane defined by the trajectory and the vertical
    direction. The axes 
    $\mathbf{e}_{pH}$, $\mathbf{e}_{pHz\perp}$ and $\mathbf{e}_z$ 
    (as well as $\mathbf{e}_{p\perp}$, $\mathbf{e}_{pHz\perp}$,
    $\mathbf{e}_{p\parallel}$) 
    form a right-handed coordinate system.
    $(b,c)$ Definition of two great circles on the sphere (located in
    two planes which are perpendicular to each other), as
    used for the presentation of the pressure coefficient below. These  
    graphs define the angles $\theta_1$ and $\theta_2$ along the great
    circles. 
    \protect\label{fig-ref-notation-1}
  }
  }
\end{figure}
We are considering the motion of a spherical solid body with diameter
$d$ immersed in a fluid under the action of a gravitational field. 
Figure~\ref{fig-ref-schematic} illustrates the geometry of the problem
as well as the definition of the different coordinate systems which
will be used in the following. 
The first set of Cartesian coordinates $(x,y,z)$ describes a position with
respect to the center of the sphere. 
Secondly, the Cartesian coordinates with respect to a fixed origin
are denoted as $(x_{fix},y_{fix},z_{fix})$. 
The directions of the axes in both of these Cartesian coordinate
systems are the same, with the $z$ and $z_{fix}$ axes pointing into the
direction opposite to gravity. 
The position of the sphere in the fixed coordinate system is
henceforth denoted as $\mathbf{x}_{sphere}$. 
Alternatively, we use a cylindrical coordinate system 
(the origin of which is attached to the center of the
particle), 
with the coordinates denoted as $(z,r,\theta)$, $r$ being the radial
coordinate and $\theta$ the azimuthal angle in the horizontal plane.

The equations for the flow of a viscous incompressible fluid can be
written as
\begin{subequations}\label{equ-navier-stokes-1}
  \begin{eqnarray}\label{equ-navier-stokes-1-mom}
    \partial_t\mathbf{u}
    +\left(\left[\mathbf{u}-\mathbf{u}_p\right]\cdot\nabla\right)\mathbf{u} 
    +\nabla p
    &=&
    \frac{1}{G} \nabla^2 \mathbf{u}
    \,,
    \\\label{equ-navier-stokes-1-cont}
    \nabla\cdot\mathbf{u}&=&0
    \,.
  \end{eqnarray}
\end{subequations}
In (\ref{equ-navier-stokes-1}) the fluid velocity vector with respect
to 
\revision{the inertial frame}{%
the fixed frame} 
is denoted by $\mathbf{u}$, 
$\mathbf{u}_p$ is the sphere's translational velocity vector 
\revision{in the inertial frame}{%
in the fixed frame} 
(with components $u_p,v_p,w_p$),  
and $p$ is the hydrodynamic pressure without the hydrostatic part.  
The equations given in (\ref{equ-navier-stokes-1}) have been made
dimensionless by means of the reference scales $\ell_{ref}=d$, 
$u_{ref}=(a_{ref}\ell_{ref})^{1/2}$, $t_{ref}=\ell_{ref}/u_{ref}$ and
$p_{ref}=\rho_f u_{ref}^2$ for length, velocity, time
and pressure, respectively.
In doing so, the characteristic acceleration
$a_{ref}=|\rho_p/\rho_f-1|g$ has been used, where $\rho_f$ is the
fluid density, $\rho_p$ the sphere's density and $g$ the magnitude of
the vector of gravitational acceleration, i.e.\ $g=|\mathbf{g}|$. 
The dimensionless parameter $G$ appearing in the Navier-Stokes
equations (\ref{equ-navier-stokes-1}) under this choice of reference
scales is the Galileo number defined as: 
\begin{equation}\label{equ-def-galileo}
  G=\frac{\sqrt{\left|\frac{\rho_p}{\rho_f}-1\right|\,g\,d^3}}{\nu}
  \,.
\end{equation}
Note that the Galileo number is equivalent to a Reynolds number
defined with the sphere diameter as the length scale and the
gravitational velocity $u_{ref}=(|\rho_p/\rho_f-1|gd)^{1/2}$ as the
velocity scale.  
\revision{}{%
  At a point on the sphere surface ${\cal S}$, the no-slip boundary
  condition accounting for the sphere translation and rotation reads:  
  \begin{equation}
    \label{bcsphere}
    \mathbf{u}|_{\cal S} = \mathbf{u}_p \,+\, \boldsymbol{\omega}_p \times
    \mathbf{r}_{\cal S}
    \,,
  \end{equation}
  where $\boldsymbol{\omega}_p$ is the angular velocity vector
  describing the rotation of the sphere with respect to its center and
  $\mathbf{r}_{\cal S}$ the position vector at the sphere surface with
  respect to the center.}
The motion of the immersed solid sphere is described by the following
equations  
\begin{subequations}\label{equ-newton-1}
  \begin{eqnarray}\label{equ-newton-1-lin}
    \frac{\rho_p}{\rho_f}\frac{\mbox{d}\mathbf{u}_p}{\mbox{d}t}
    &=&
    %
    \frac{6}{\pi}\oint_{\cal S}
    \left(
      \boldsymbol{\tau}\cdot\mathbf{n}-p\mathbf{n}
    \right)
    \,\mbox{d}S
    -\mathbf{i}
    \,,
    \\\label{equ-newton-1-ang}
    \frac{\rho_p}{\rho_f}\frac{\mbox{d}\boldsymbol{\omega}_p}{\mbox{d}t}
    &=&
    \frac{60}{\pi}\oint_{\cal S}
    \mathbf{r}_S\times
    \left(
      \boldsymbol{\tau}\cdot\mathbf{n}
    \right)
    \,\mbox{d}S
    \,,
  \end{eqnarray}
\end{subequations}
where the same reference quantities as in (\ref{equ-navier-stokes-1})
have been used. 
\revision{In (\ref{equ-newton-1}) 
  $\boldsymbol{\omega}_p$ is the sphere's angular velocity vector (with
  components $\omega_{px},\omega_{py},\omega_{pz}$),  
  $\boldsymbol{\tau}$ is the viscous stress tensor whose components are
  given by $\tau_{ij}=(\partial u_i/\partial x_j+\partial u_j/\partial
  x_i)/G$, $\mathbf{n}$ denotes the outward pointing unit vector normal
  to the surface  ${\cal S}$ of the sphere, 
  $\mathbf{i}$ is the unit vector pointing in the vertical direction, 
  and $\mathbf{r}_S$ is the position vector of a point on the surface
  ${\cal S}$ with respect to the sphere's center.}{%
  In (\ref{equ-newton-1}) the angular velocity vector
  $\boldsymbol{\omega}_p$ has components
  $\omega_{px},\omega_{py},\omega_{pz}$,   
  $\boldsymbol{\tau}$ is the viscous stress tensor whose components are
  given by $\tau_{ij}=(\partial u_i/\partial x_j+\partial u_j/\partial
  x_i)/G$, $\mathbf{n}$ denotes the outward pointing unit vector normal
  to the surface ${\cal S}$ of the sphere and 
  $\mathbf{i}$ is the unit vector pointing in the vertical direction.}

The coupled system of field equations for the fluid flow
(\ref{equ-navier-stokes-1}) and ordinary differential equations for
the sphere motion (\ref{equ-newton-1}) is fully characterized by two 
non-dimensional numbers, namely the Galileo number $G$ (as defined in
\ref{equ-def-galileo}) and the density ratio $\rho_p/\rho_f$. 
Note that for steady motion of the sphere the value of the density
ratio is not relevant any more and the problem is fully determined by
the value of the Galileo number. 
%
Figure~\ref{fig-jenny-et-al-diagram} gives an overview of the features
exhibited by the motion of an immersed sphere in the parameter space
spanned by the values of $G$ and $\rho_p/\rho_f$ \citep{jenny:04}. 
It will be further discussed in \S~\ref{sec-reference-general-results}
below. 

\subsection{Numerical method}\label{sec-reference-numa}
\begin{table}
  \centering
  \setlength{\tabcolsep}{5pt}
  \begin{tabular}{*{4}{c}}
    case&
    $\rho_p/\rho_f$&
    $G$&
    $D_{\Omega}$
    \\[1ex]
    AS&
    $1.5$&
    $144$&
    $5.34$
    \\
    AL&
    $1.5$&
    $144$&
    $7.54$
    \\[1ex]
    BS&
    $1.5$&
    $178.46$&
    $5.34$
    \\
    BL&
    $1.5$&
    $178.46$&
    $7.54$
    \\[1ex]
    CS&
    $1.5$&
    $190$&
    $5.34$
    \\
    CL&
    $1.5$&
    $190$&
    $7.54$
    \\[1ex]
    DS&
    $1.5$&
    $250$&
    $5.34$
    \\
    DL&
    $1.5$&
    $250$&
    $7.54$
    \\
  \end{tabular}
  \caption{
    Parameter points in the $(G,\rho_p/\rho_f)$ plane and diameter
    $D_{\Omega}$ of the cylindrical domain in the present reference
    simulations.  
    In the descriptive name of each case (first column), ``S'' refers
    to the smaller domain with $D_{\Omega}=5.34$ and ``L'' to
    $D_{\Omega}=7.54$. 
  }    
  \label{tab-parameters-ref}
\end{table}
The numerical method used as the basis for the present development has
been described in \cite{ghidersa00} and \cite{jenny:04b}. The spatial
discretization takes advantage of the axisymmetry of the computational
domain for expanding the variables into a rapidly converging azimuthal
Fourier series. The so obtained azimuthal Fourier modes are functions
of only the radial distance $r$ and of the axial projection $z$.  They
obey a set of two-dimensional equations coupled via the advective
terms. The discretization in the radial--axial plane $(r,z)$ uses the
spectral element decomposition \citep{patera84}. The time
discretization is chosen in view of solving high Reynolds number
flows. In this case the adopted time splitting approach, used already
in \cite{patera84}, is both accurate and efficient. The non-linear
terms are treated explicitly (in our case we use the third order
Adams-Bashforth method), which un-couples linear two-dimensional
Stokes-like problems  in individual azimuthal subspaces numbered by
the azimuthal wavenumber $m$. The latter are solved by splitting the
pressure--velocity coupling into a Poisson pressure equation and a
Helmholtz equation for the velocity. In the literature
\citep[e.g.][]{karniadakis91}, the splitting is considered before the
discretization. 
\cite{kotouc08} have noted
that, if the whole augmented matrix of the Stokes--like problem is
created, the matrix obtained by multiplying the discretized divergence
by the discretized gradient is not exactly the same as that of the
diffusion operator. The so obtained improvement of accuracy was
combined with a considerable reduction of computational costs achieved
by replacing the iterative (conjugate gradient) pressure solver by a
direct method. 

At the inflow (bottom) cylinder basis the velocity is set equal to
zero to simulate an asymptotically quiescent fluid. At the outflow
(top) cylinder basis and at its side a no stress Neumann boundary
condition is imposed on the velocity field and a zero pressure is set.    

In the present simulations we have employed grids with different
numbers of spectral elements: for the small domain size, the
radial/axial plane was tesselated with 129 elements; 
for the tesselation of the larger domain size 134 elements were used
at the lower Galileo number values ($G=144$ and $G=178$), 
while 169 elements were used at $G=190$ and $G=250$. 
Figure~\ref{fig-sem-grid} shows these grids.
In all cases 6 collocation points in each of the two spatial
directions internal to each element were used. 
Furthermore, the azimuthal Fourier expansion was truncated above
mode~7.  
Finally, the time step has been adjusted such that the CFL number
takes a value of 0.25.
Extensive validation and grid convergence studies 
\citep{jenny:04b,bouchet:06} have demonstrated the adequacy of this
spatial and temporal resolution in the present parameter range.
%

\subsection{Overview of sphere dynamics and flow regimes}
\label{sec-reference-general-results}
%
The state diagram in figure~\ref{fig-jenny-et-al-diagram} features 5
different symbols corresponding to the following classification. 
%
\begin{table}
  \centering
  \setlength{\tabcolsep}{5pt}
  \begin{tabular}{*{4}{c}}
    case&
    $u_{pV}$&
    $L_r$
    &
    \revision{}{%
      $Re_\parallel$
    }
    \\[1ex]
    AS&
    $-1.292$& 
    $1.385$ 
      &
    \revision{}{%
      $186.05$
    }
    \\
    AL&
    $-1.285$& 
    $1.383$ 
    &
    \revision{}{%
      $185.08$
    }
  \end{tabular}
  \caption{
    Reference results for flow cases in the steady axi-symmetrical
    flow regime ($G=144$). 
  }    
  \label{tab-results-ref-A}
\end{table}
For Galileo numbers below a value of approximately 155
\citep[at all density ratios, cf.\ also][]{fabre2012}, 
steady vertical particle motion with full
axisymmetry in the horizontal plane is obtained. 
When increasing the Galileo number beyond the  
threshold of that primary 
bifurcation,  
%
the axisymmetry of the wake is broken, and a regime with steady oblique
particle motion exists. 
Further increasing the Galileo number for a given density ratio, 
a Hopf bifurcation occurs, leading to oscillating oblique paths; 
the diagram in figure~\ref{fig-jenny-et-al-diagram} actually shows two
such oscillating oblique regimes, distiguished by the value of the
oscillation frequency, roughly occurring for $\rho_p/\rho_f$ above or
below a value of 2.5. 
%
%
\revision{Below $\rho_p/\rho_f=1$ (for rising spheres),}{%
   For density ratios smaller than unity (i.e.\ rising spheres)} 
the oblique oscillating state was found to give way to a `zig-zagging' 
state (marked by open circles in
figure~\ref{fig-jenny-et-al-diagram}), i.e.\ to a rise along a
periodic and wavy trajectory remaining vertical in the mean. The frequency was
shown to be about three times smaller than that of the low frequency
oblique oscillating state. 
For $\rho_p/\rho_f>1$ (for falling spheres)
the oblique oscillating state becomes directly chaotic when 
further increasing the Galileo number. 
Conversely, for rising spheres intermittent chaos was shown to arise
from the zig-zagging state. 
The chaotic states are all marked by the same symbol in
figure~\ref{fig-jenny-et-al-diagram}, although there
are significant qualitative differences between highly intermittent
states close to the right limit (upper limit in terms of the Galileo 
number) of the stability of the periodic zig-zagging state and much
less ordered states at high Galileo numbers and high density
ratios. 
In view of subsequent experimental observations
\citep{Veldhuis2007,horowitz2010}, the region delimited (roughly) by
the dotted line in figure~\ref{fig-jenny-et-al-diagram} is especially
noteworthy. It corresponds to the region of bi-stability between
chaotic and periodic states. The periodic states are, again, vertical
in the mean (zig-zagging), however, their frequency is significantly
higher than that of the states marked by open circles. The frequency
is close to that evidenced in \cite{Veldhuis2007} and
\cite{horowitz2010} for experimental zig-zagging trajectories. The
low-frequency zig-zagging state has never been observed
experimentally, very likely because of its 
weak stability 
\citep{jenny:04}.   
%

\subsection{Reference data}
\label{sec-ref-results}
\begin{figure}
  \figpap{
  \centering
  \begin{minipage}{.13\linewidth}
    \centerline{$(a)$}
    \includegraphics[width=\linewidth,clip=true,
    viewport=1150 827 1400 1750]
    {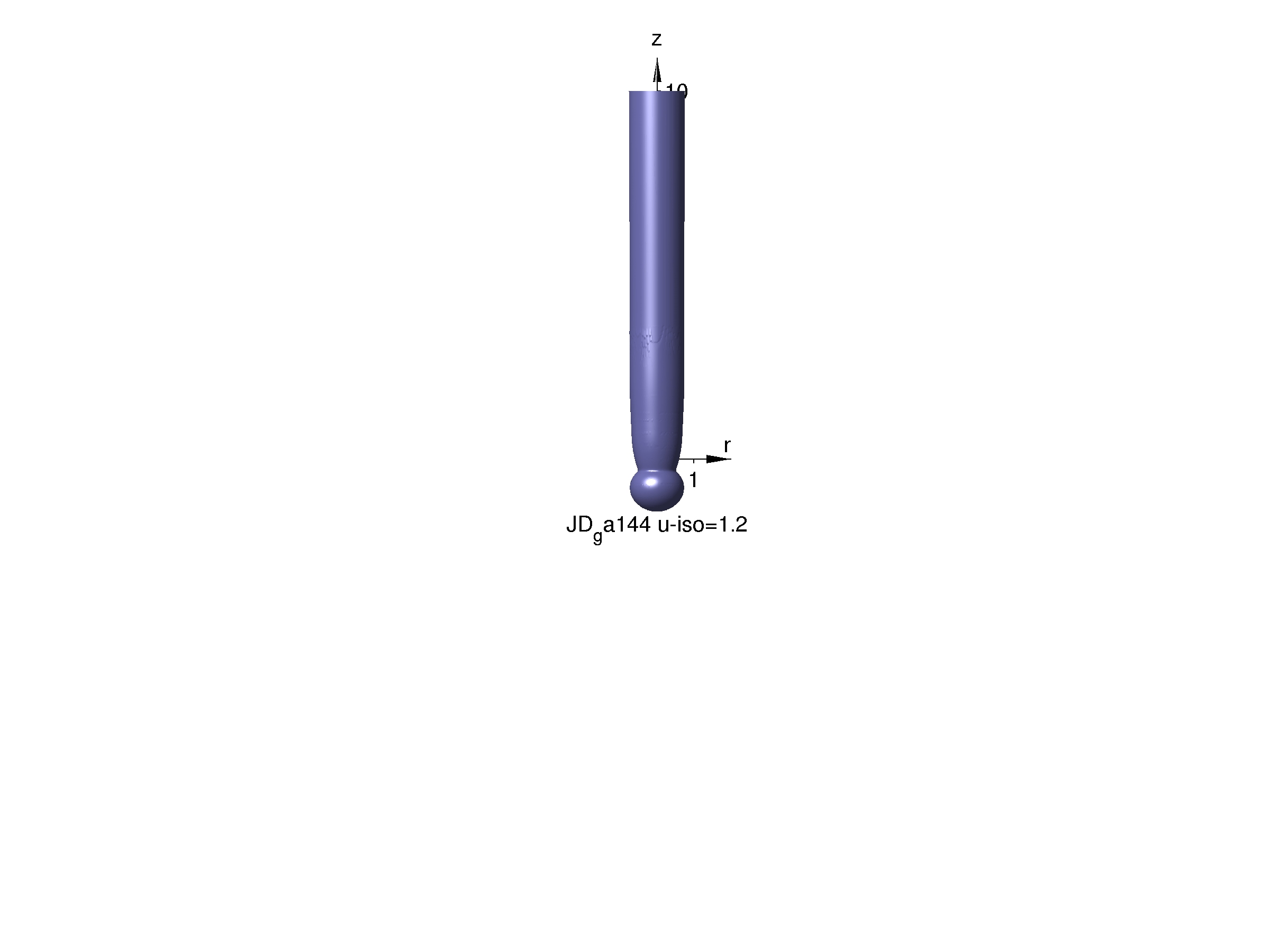}
  \end{minipage}
  \begin{minipage}{.13\linewidth}
    \centerline{$(b)$}
    \includegraphics[width=\linewidth,clip=true,
    viewport=1150 827 1400 1750]
    {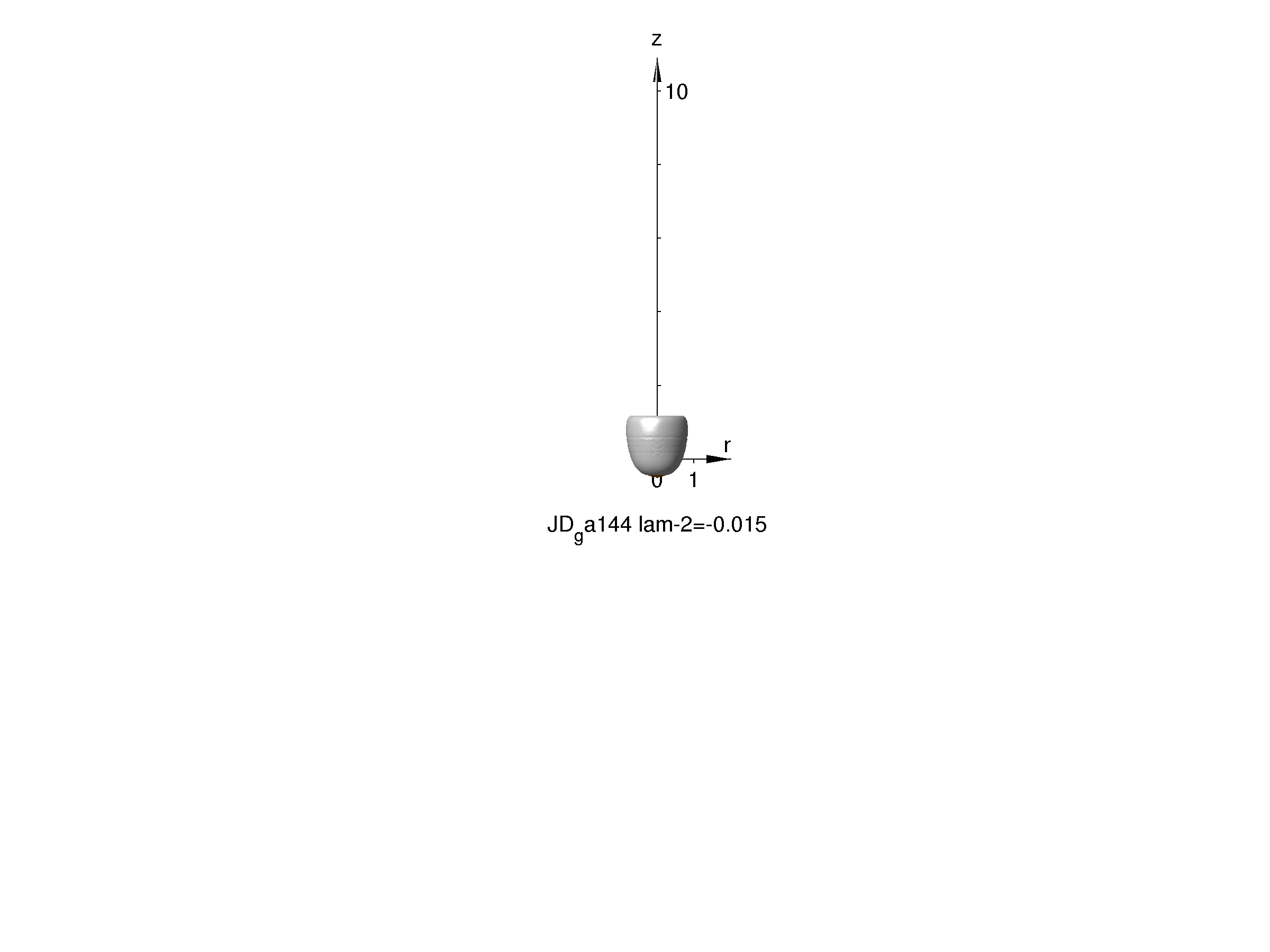}
  \end{minipage}
  \begin{minipage}{1ex}
    $z$
  \end{minipage}
  \begin{minipage}{.18\linewidth}
    \centerline{$(c)$}
    \includegraphics[width=\linewidth]
    {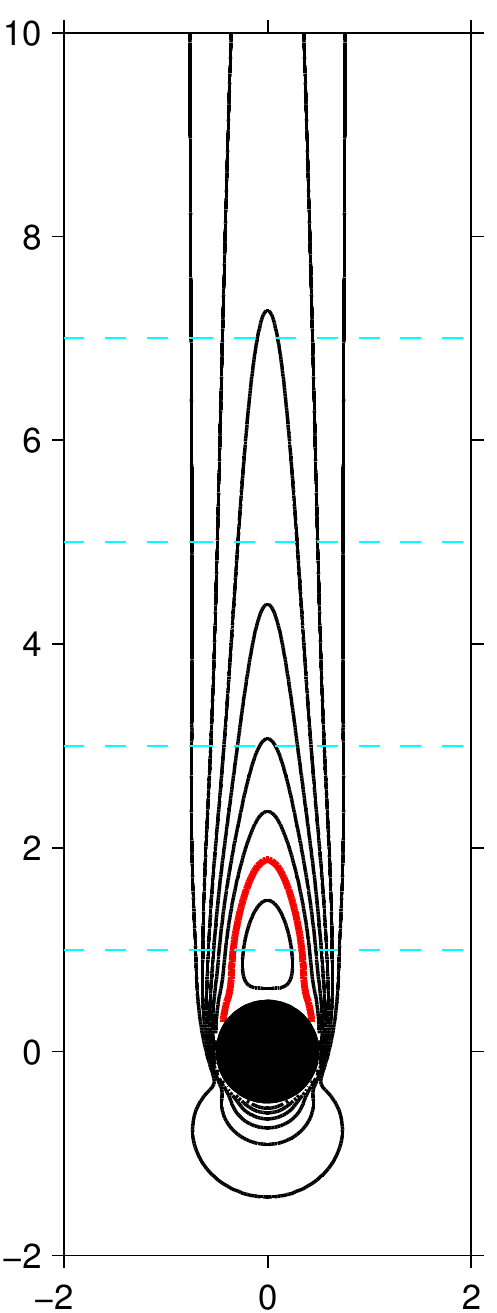}
    \centerline{$r$}
  \end{minipage}
  \revision{%
  \begin{minipage}{3.5ex}
    $\frac{u_{r\,\parallel}}{u_{r\parallel\infty}}$
  \end{minipage}
  \begin{minipage}{.45\linewidth}
    \centerline{$(d)$}
    \includegraphics[width=\linewidth]
    {figure_5d.pdf}
    \centerline{$z$}
  \end{minipage}
  }{}
  }{
  \caption{%
    Reference results for case AL ($G=144$).  
    $(a)$ The surface where the vertical relative velocity 
    $u_{r\parallel}=1.2$ (corresponding to $0.93|u_{pV}|$).  
    $(b)$ The surface where $\lambda_2=-0.015$
    (cf.\ definition in the text).
    $(c)$ Contours of the vertical relative velocity
    $u_{r\parallel}$ in the
    vertical/radial plane passing through the particle center. 
    Contours are shown for values (-0.2:0.2:1.2);  
    the red line marks the extent of the recirculation region (i.e.\ 
    $u_{r\parallel}=0$).     
    \revision{The cyan colored dashed lines indicate the
      positions where velocity profiles are presented in
      figure~\ref{fig-results-ref-cross-profiles-a1l} below.}{%
      The cyan colored dashed lines indicate the
      positions of the velocity profiles which are given in the
      supplementary data and which are compared to the IBM results in
      figure~\ref{fig-results-ibm-cross-profiles-a1c}.%
    } 
    \revision{%
    $(d)$ 
    Profile of the quantity $u_{r\parallel}$ on the vertical axis
    through the particle center. 
    }{}
    \protect\label{fig-results-ref-contour-a1l}
  }
  }
\end{figure}
With the purpose of providing data for 
validation and benchmarking of numerical simulation codes, we have
selected a set of parameter points which are representative of the
different regimes of sphere motion. 
With the aim of keeping the data-set tractable, we have considered 
a single density ratio which was chosen as $\rho_p/\rho_f=1.5$. 
This value corresponds to particles with a moderately higher density
than the fluid (e.g.\ polyester in water). 
%
%
Concerning the Galileo number, four values were considered such that
each case corresponds to one of the observed regimes of motion 
of falling spheres:
\begin{itemize}
\item $G=144$: steady vertical fall, axi-symmetric wake \revision{}{(case A)};
\item $G=178$: steady oblique fall, double-threaded wake \revision{}{(case B)};
\item $G=190$: oscillating oblique fall \revision{}{(case C)};
\item $G=250$: chaotic motion \revision{}{(case D)}. 
\end{itemize}
Therefore, the chosen parameter points sample a cross-section of
the parameter map shown in figure~\ref{fig-jenny-et-al-diagram} at
$\rho_p/\rho_f=1.5$.
With increasing $G$ these parameter points represent a sequence of
flow cases with increasing physical complexity and -- due to the onset
of unsteadiness and further bifurcations -- of increasing demand from
the point of view of a numerical method.  

In the present work we have strived to use a relatively small
computational domain in order to maintain the computational effort in  
subsequent studies, where successive refinement will be performed, 
manageable. As a consequence, we have chosen two values for the
horizontal diameter of the cylindrical domain, 
$D_\Omega=\{5.34,7.54\}$, cf.\ sketch in
figure~\ref{fig-ref-schematic}, while maintaining the vertical length
fixed at 
$L_x=15$ (with $L_u=5$ and $L_d=10$ the vertical length
upstream and downstream of the particle center, respectively). 
All reference simulations were run in both, the wider and the smaller
cylindrical domains. 
As a side effect, the sensitivity of the results with respect to the
domain size can be gauged. 

Please refer to table~\ref{tab-parameters-ref} for a summary of the
parameter values which have been simulated in the present work. 
\revision{%
  The table also shows the abbreviations which will be used in the
  following when referring to the respective flow cases.}{%
  The table also shows the two-letter abbreviations which will be used
  in the   following when referring to the respective flow cases: the
  first letter denotes the flow regime (from A to D), the second one
  designates the lateral domain size (S or L, i.e.\ ``small'' or
  ``large'').
}
\begin{figure}
  \figpap{
    \centering
  \begin{minipage}{1ex}
    $y$
  \end{minipage}
  \begin{minipage}{.26\linewidth}
      \centerline{$(a)$}
    \includegraphics[width=\linewidth]
    {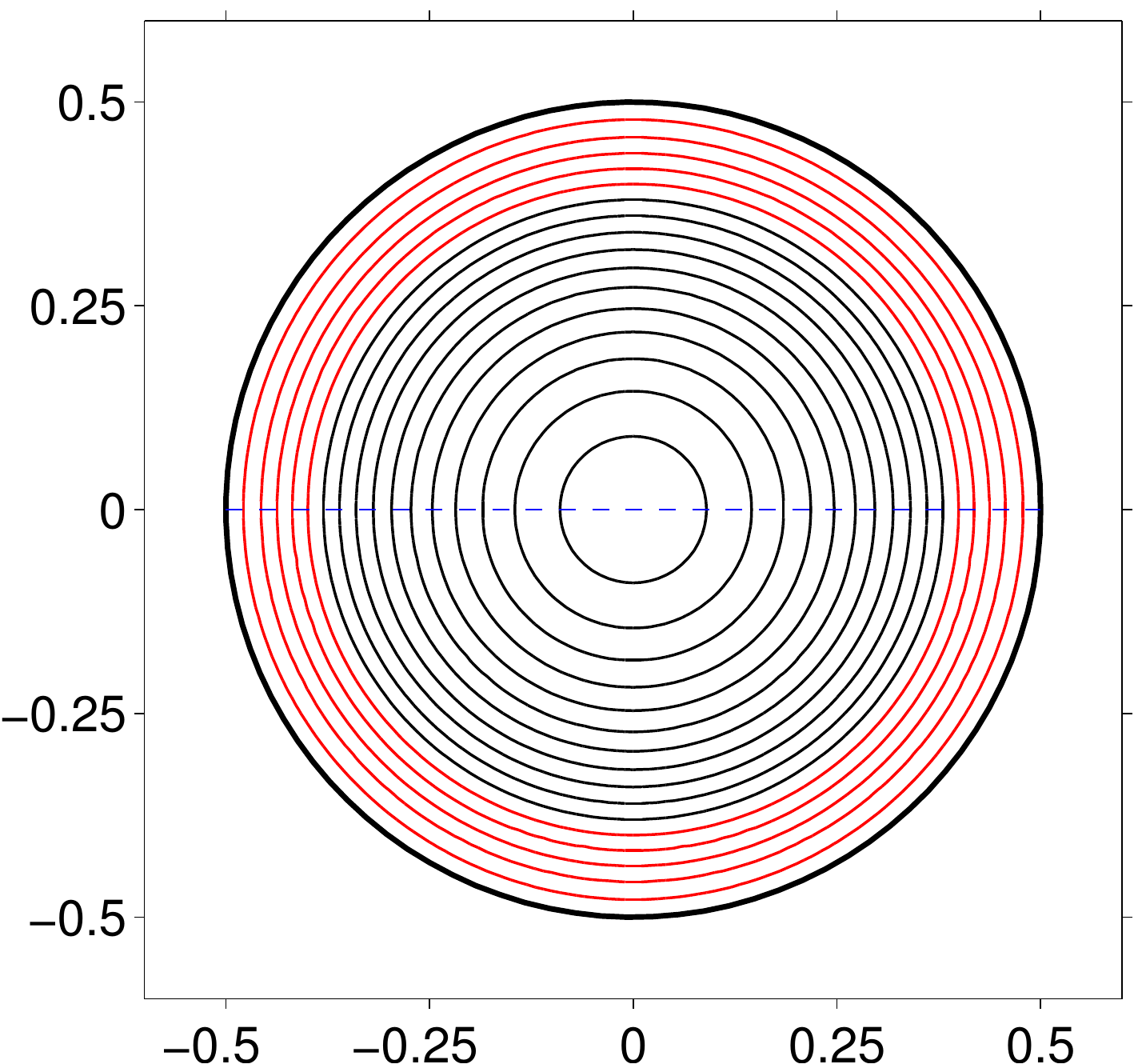}
    \centerline{$x$}
  \end{minipage}
  \begin{minipage}{.26\linewidth}
    \centerline{$(b)$}
    \includegraphics[width=\linewidth]
    {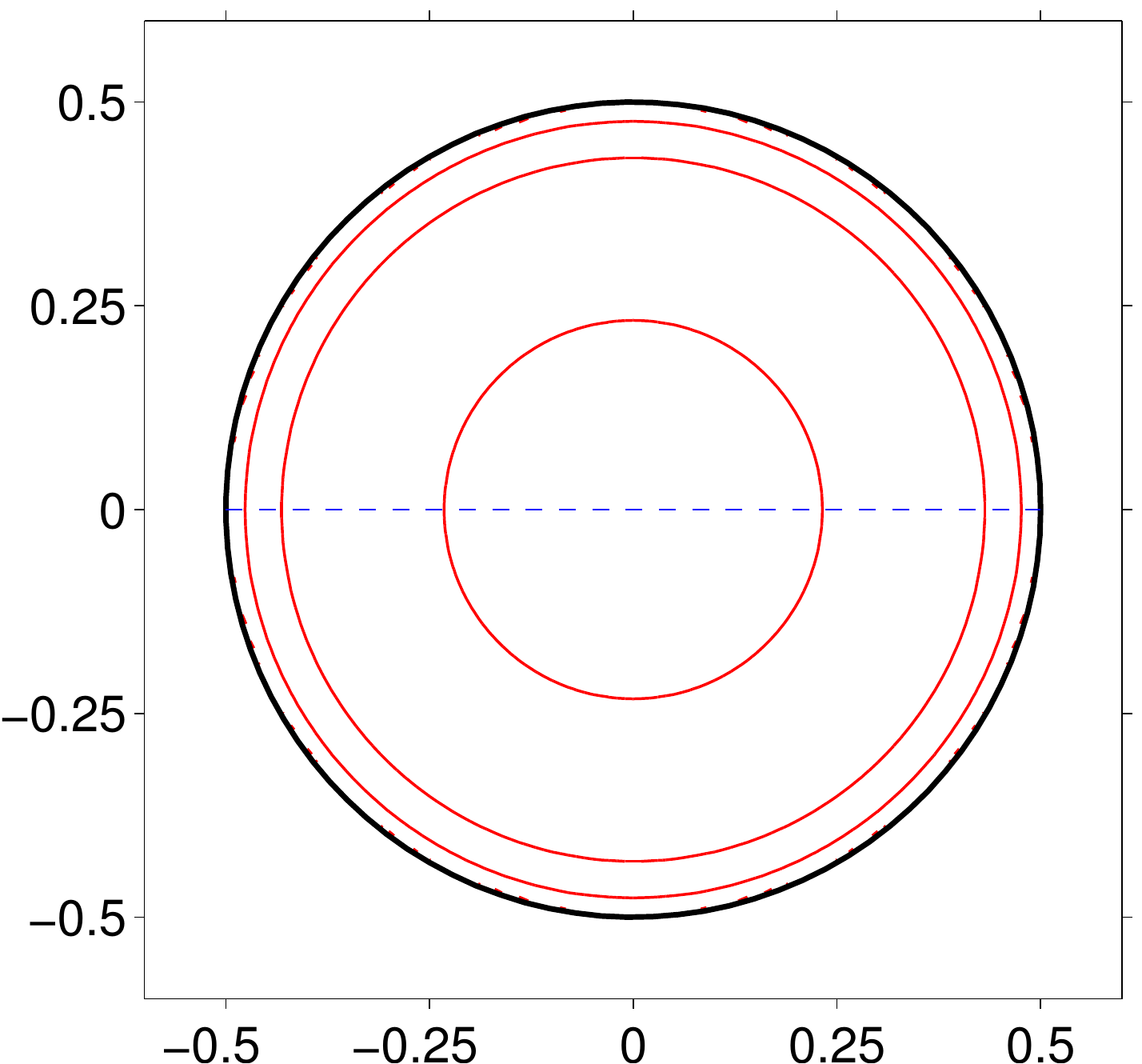}
    \centerline{$x$}
  \end{minipage}
  \revision{%
    \hfill
    \begin{minipage}{1.5ex}
      $c_p$
    \end{minipage}
    \begin{minipage}{.45\linewidth}
      \centerline{$(c)$}
      \includegraphics[width=\linewidth]
      {figure_7c.pdf}
      \centerline{$\theta$}
    \end{minipage}
  }{}
  }{
  \caption{%
    Reference results for case AL showing the pressure coefficient $c_p$. 
    $(a)$ Contours of $c_p$ on the surface of the sphere, projected
    upon the horizontal plane through the sphere's center
    (upstream-facing side). Contours are shown for values
    (-0.5:0.1:1.1), with negative values being indicated by red color.  
    $(b)$ The same as $(a)$, but for the downstream-facing side of the
    sphere. 
    \revision{%
      $(c)$ 
      Profile of $c_p$ along a great circle on the sphere, in a vertical
      plane (along blue dashed lines in graphs $a$ and $b$). 
      The upstream stagnation point corresponds to a value of the angle  
      $\theta=\pm\pi$. Please refer to figure~\ref{fig-ref-notation-1}
      for the definition of the angle $\theta$. 
    }{}
    %
    %
    \protect\label{fig-results-ref-press-a1l}
  }
  }
\end{figure}
\subsubsection{Geometric definitions and notation}
\label{sec-ref-results-notation}
Let us first fix the notation used in the subsequent presentation of
flow and particle data.
The particle velocity relative to the ambient fluid velocity
$\mathbf{u}_\infty$ is defined as
\begin{equation}\label{equ-res-ref-def-upr}
  \mathbf{u}_{pr}=\mathbf{u}_p-\mathbf{u}_\infty
  \,,
\end{equation}
with Cartesian components
$\mathbf{u}_{pr}=(u_{pr},v_{pr},w_{pr})^T$. Note that throughout 
\S~\ref{sec-ref-results} we have 
$\mathbf{u}_\infty=0$.    
\revision{}{%
  The Reynolds number based upon the magnitude of the relative
  particle velocity is simply obtained as follows
\begin{equation}\label{equ-res-ref-def-reynolds-par}
  Re_\parallel
  =
  \frac{|\mathbf{u}_{pr}|\,u_{ref}\,d}{\nu}
  =
  |\mathbf{u}_{pr}|\,G
  \,. 
\end{equation}
Its values will be listed in the tables below for convenience.} 

%
\begin{table}
  \centering
  \setlength{\tabcolsep}{5pt}
  \begin{tabular}{*{6}{c}}
    case&
    $u_{pV}$&
    $u_{pH}$&
    $\omega_{pH}$&
    $L_r$
    &
    \revision{}{%
      $Re_\parallel$
    }
    \\[1ex]
    BS&   
    $-1.363$& 
    $0.1270$& 
    $0.0136$& 
    $1.631$ 
    &
    \revision{}{%
      $244.29$
    }
    \\
    BL&
    $-1.356$& 
    $0.1245$& 
    $0.0137$& 
    $1.629$
    &
    \revision{}{%
      $243.01$
    }
  \end{tabular}
  \caption{
    Reference results for flow cases in the steady oblique 
    flow regime ($G=178.46$). 
  }    
  \label{tab-results-ref-B}
\end{table}
\begin{figure}
  \figpap{
  \begin{minipage}{.22\linewidth}
    \centerline{$(a)$}
    \includegraphics[width=\linewidth,clip=true,
    viewport=1050 827 1500 1750]
    {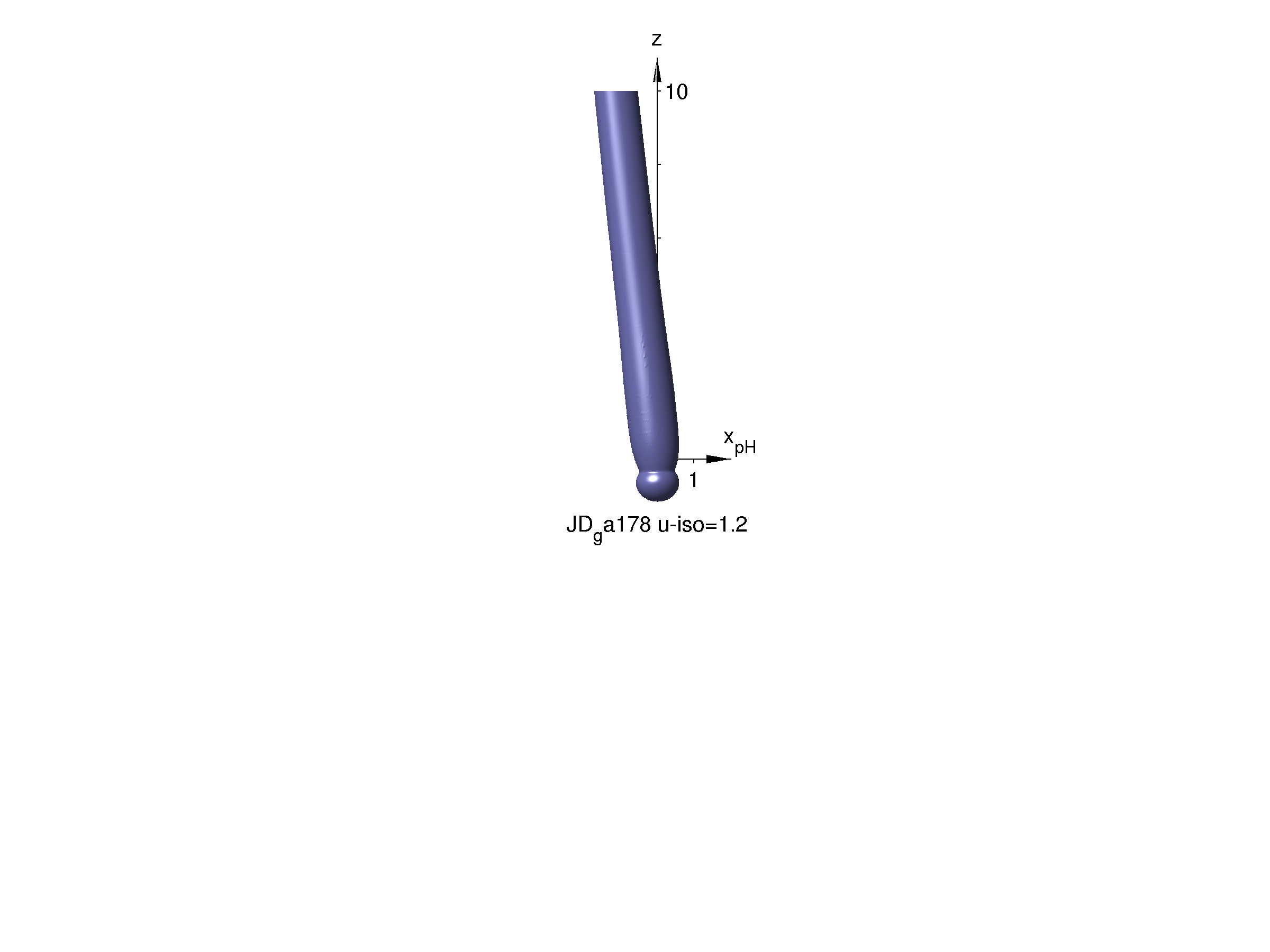}
  \end{minipage}
  \hfill
  \begin{minipage}{.22\linewidth}
    \centerline{$(b)$}
    \includegraphics[width=\linewidth,clip=true,
    viewport=1050 827 1500 1750]
    {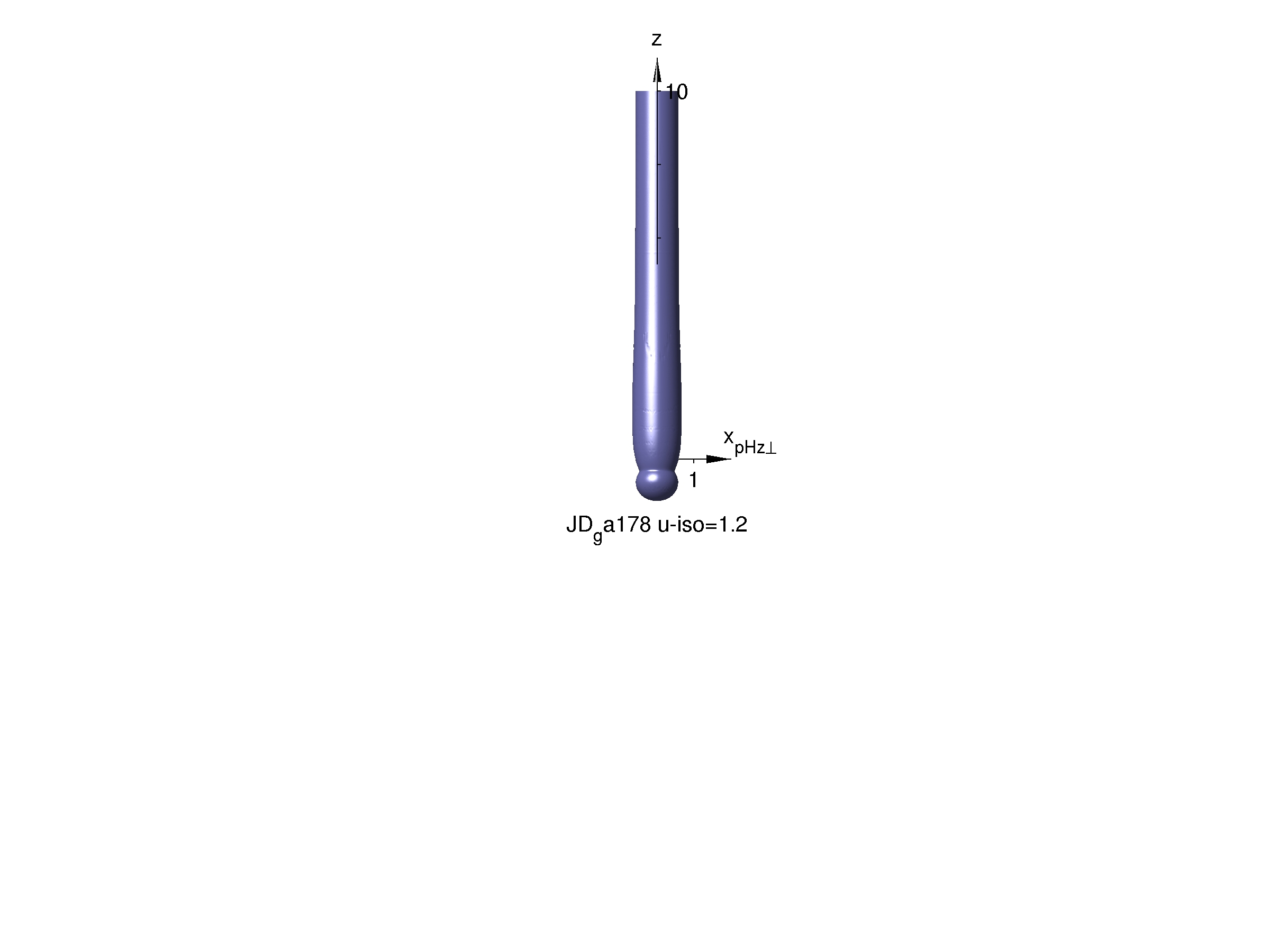}
  \end{minipage}
  \hfill
  \begin{minipage}{.22\linewidth}
    \centerline{$(c)$}
    \includegraphics[width=\linewidth,clip=true,
    viewport=1050 827 1500 1750]
    {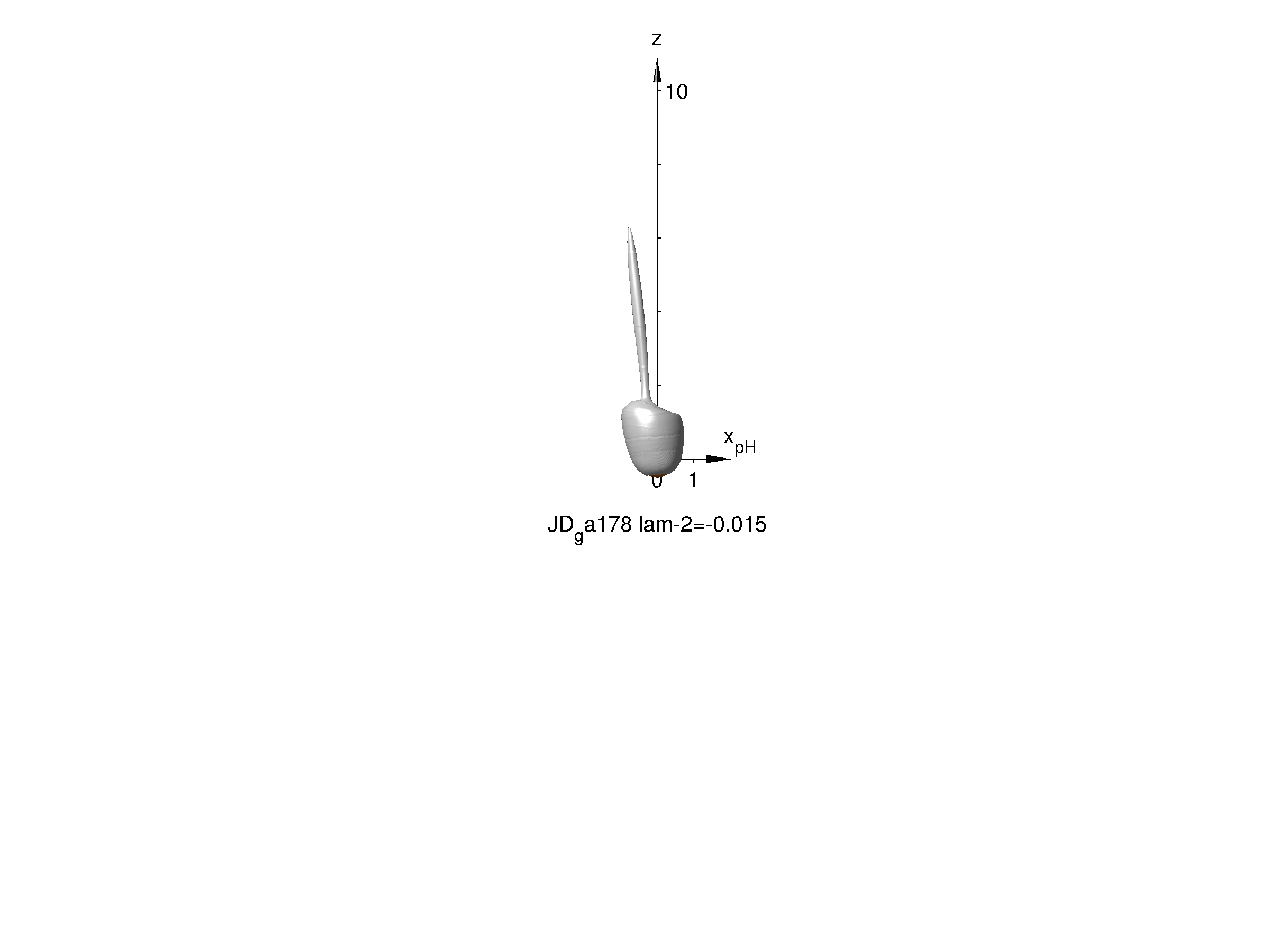}
  \end{minipage}
  \hfill
  \begin{minipage}{.22\linewidth}
    \centerline{$(d)$}
    \includegraphics[width=\linewidth,clip=true,
    viewport=1050 827 1500 1750]
    {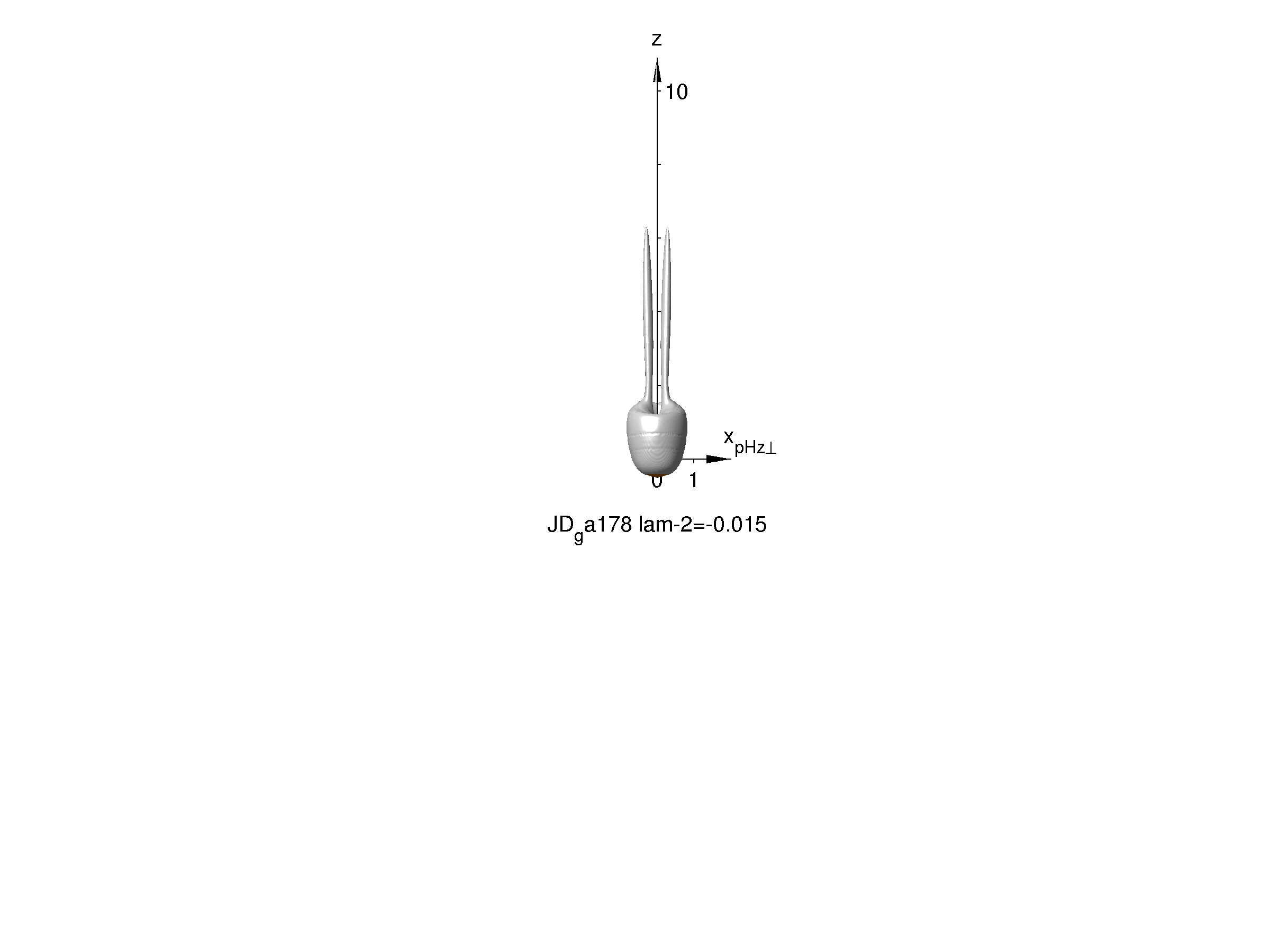}
  \end{minipage}
  }{
  \caption{%
    Reference results for case BL ($G=178.46$). 
    Graphs $(a)$ and $(b)$ show the surface where
    $u_{r\parallel}=1.2$ (corresponding to $0.88|\mathbf{u}_{pr}|$), 
    Graphs $(c)$ and $(d)$ show the surface where $\lambda_2=-0.015$
    (cf.\ definition in the text). 
    In $(a)$ and $(c)$ the view is directed along $\mathbf{e}_{pHz\perp}$;  
    in $(b)$ and $(d)$ it is directed along $\mathbf{e}_{pH}$.
    \protect\label{fig-results-ref-3d-iso-b1l}
  }
  }
\end{figure}
The magnitude of the particle velocity (relative to the ambient) in
the horizontal plane is denoted as $u_{pH}$, defined through
\begin{equation}\label{equ-res-ref-def-uph}
  u_{pH}=\sqrt{u_{pr}^2+v_{pr}^2}
  \,.
\end{equation}
The vertical component of the particle velocity relative to the
ambient fluid velocity 
is given by 
\begin{equation}\label{equ-res-ref-def-upv}
  u_{pV}=w_{pr}
  \,.
\end{equation}
\begin{figure}
  \figpap{
    \centering
  \begin{minipage}{1ex}
    $z$
  \end{minipage}
  \begin{minipage}{.22\linewidth}
    \centerline{$(a)$}
    \includegraphics[width=\linewidth]
    {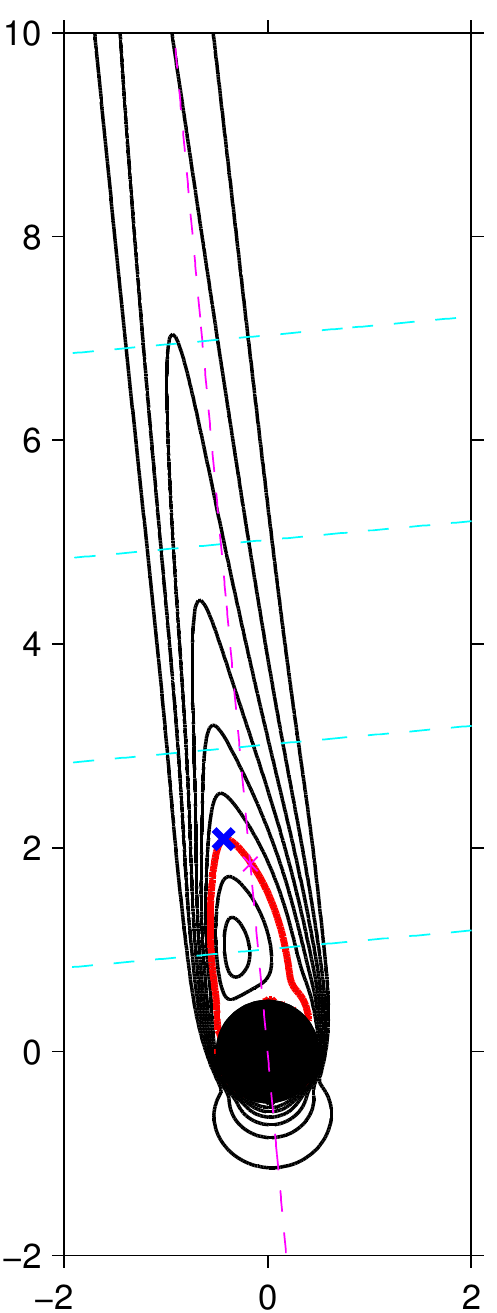}
    \centerline{${x}_{pH}$}
  \end{minipage}
  \begin{minipage}{2ex}
    $x_{p\parallel}$
  \end{minipage}
  \begin{minipage}{.22\linewidth}
    \centerline{$(b)$}
    \includegraphics[width=\linewidth]
    {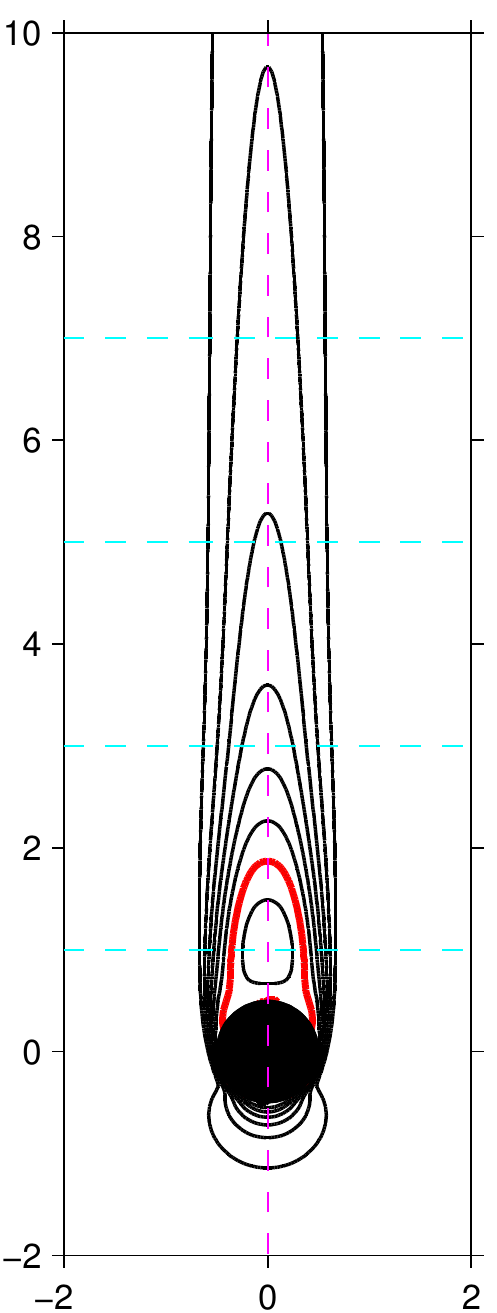}
    \centerline{${x}_{pHz\perp}$}
  \end{minipage}
  \revision{%
  \begin{minipage}{3.5ex}
    $\frac{u_{r\,\parallel}}{u_{r\parallel\infty}}$
  \end{minipage}
  \begin{minipage}{.45\linewidth}
    \centerline{$(c)$}
    \includegraphics[width=\linewidth]
    {figure_9c.pdf}
    \centerline{$x_{p\,\parallel}$}
  \end{minipage}}{}
  }{
  \caption{%
    Reference results for case BL ($G=178.46$). 
    $(a)$ Contours of the projected relative velocity
    $u_{r\,\parallel}$ in the plane which is spanned by the vertical
    direction $\mathbf{e}_z$ and the direction of the particle motion
    $\mathbf{e}_{p\parallel}$, passing through the sphere's center. 
    Contours are shown for values (-0.4:0.2:1.2); the red line marks
    the extent of the recirculation region (i.e.\ $u_{r\,\parallel}=0$). 
    The blue cross marks the location which defines the recirculation
    length $L_r$. 
    The magenta colored dashed line indicates the direction of the particle
    motion given by the unit vector $\mathbf{e}_{p\,\parallel}$.
    %
    $(b)$ Same as $(a)$, but in the plane spanned by
    $\mathbf{e}_{pHz\perp}$ and $\mathbf{e}_{p\parallel}$, 
    passing through the sphere's center. 
    \revision{The cyan colored dashed lines in $(a)$ and $(b)$ indicate the
    positions 
    where profiles are presented in
    figure~\ref{fig-results-ref-cross-profiles-b1l} below.}{%
    The cyan colored dashed lines in $(a)$ and $(b)$ indicate the
    positions of the velocity profiles which are given in the
      supplementary data and which are compared to the IBM results in
      figure~\ref{fig-results-ibm-cross-profiles-b1c-24}.%
  }    
    \revision{$(c)$ 
    Profile of the same quantity $u_{r\,\parallel}$ along the axis
    passing through the particle center and following the direction of
    the particle motion (along magenta dashed line in $a$ and $b$).}{}
    \protect\label{fig-results-ref-contour-b1l}
  }
  }
\end{figure}
The horizontal ($\omega_{pH}$) and vertical components ($\omega_{pV}$)
of the angular particle velocity are similarly defined as
\begin{subequations}\label{equ-res-ref-def-omp}
  \begin{eqnarray}\label{equ-res-ref-def-omph}
    \omega_{pH}&=&\sqrt{\omega_{px}^2+\omega_{py}^2}
    \,,
    \\\label{equ-res-ref-def-ompv}
    \omega_{pV}&=&\omega_{pz}
    \,.
  \end{eqnarray}
\end{subequations}
The directional unit vector $\mathbf{e}_{pH}$ of the particle motion
in the horizontal plane (in cases of non-vertical motion,
$u_{pH}>0$) is given by  
\begin{equation}\label{equ-res-ref-def-eph}
  \mathbf{e}_{pH}=\left(u_{pr},v_{pr},0\right)^T/u_{pH}
  \,.
\end{equation}
The component of the position vector in the direction
$\mathbf{e}_{pH}$ (measured from the sphere's center) will be denoted
by $x_{pH}$;  
the direction perpendicular to $\mathbf{e}_{pH}$ in the horizontal
plane (i.e.\ also perpendicular to $\mathbf{e}_z$) is referred to as  
\begin{equation}\label{equ-res-ref-def-ephperp}
  \mathbf{e}_{pHz\perp}=
  \left(-v_{pr},u_{pr},0\right)^T/u_{pH}
  \,,
\end{equation}
(with associated coordinate $x_{pHz\perp}$) in the following.  
In the case of purely vertical motion (corresponding to purely
axi-symmetric flow), 
the definitions (\ref{equ-res-ref-def-eph}) and
(\ref{equ-res-ref-def-ephperp}) do not apply;  
cylindrical coordinates $(z,r,\theta)$ will be chosen instead, as
defined in \S~\ref{sec-reference-config}. 

\begin{figure}
  \figpap{
    \centering
  \begin{minipage}{3ex}
    $x_{pH}$
  \end{minipage}
  \begin{minipage}{.26\linewidth}
    \centerline{$(a)$}
    \includegraphics[width=\linewidth]
    {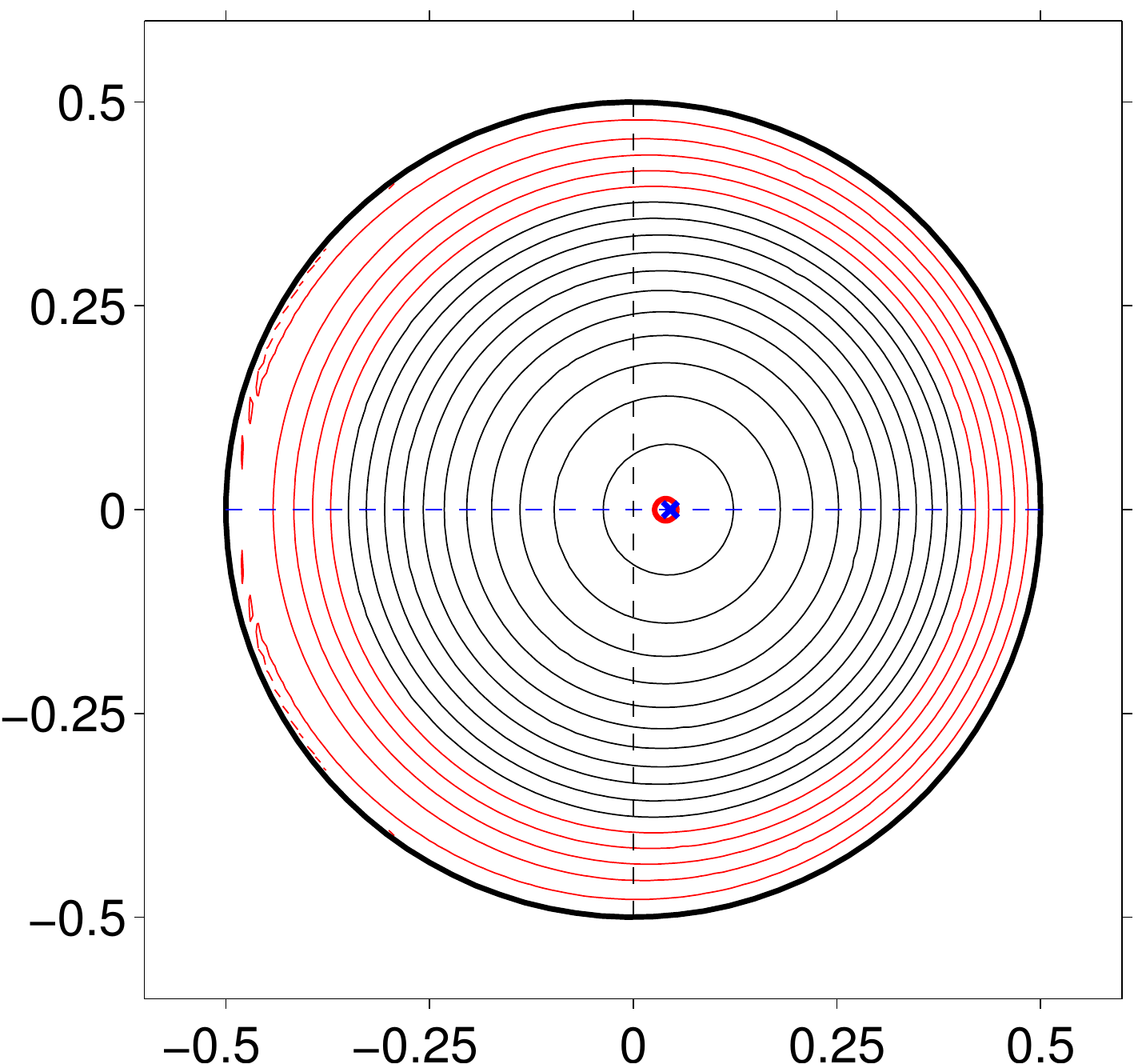}
    \centerline{$x_{pHz\perp}$}
  \end{minipage}
  \begin{minipage}{.26\linewidth}
    \centerline{$(b)$}
    \includegraphics[width=\linewidth]
    {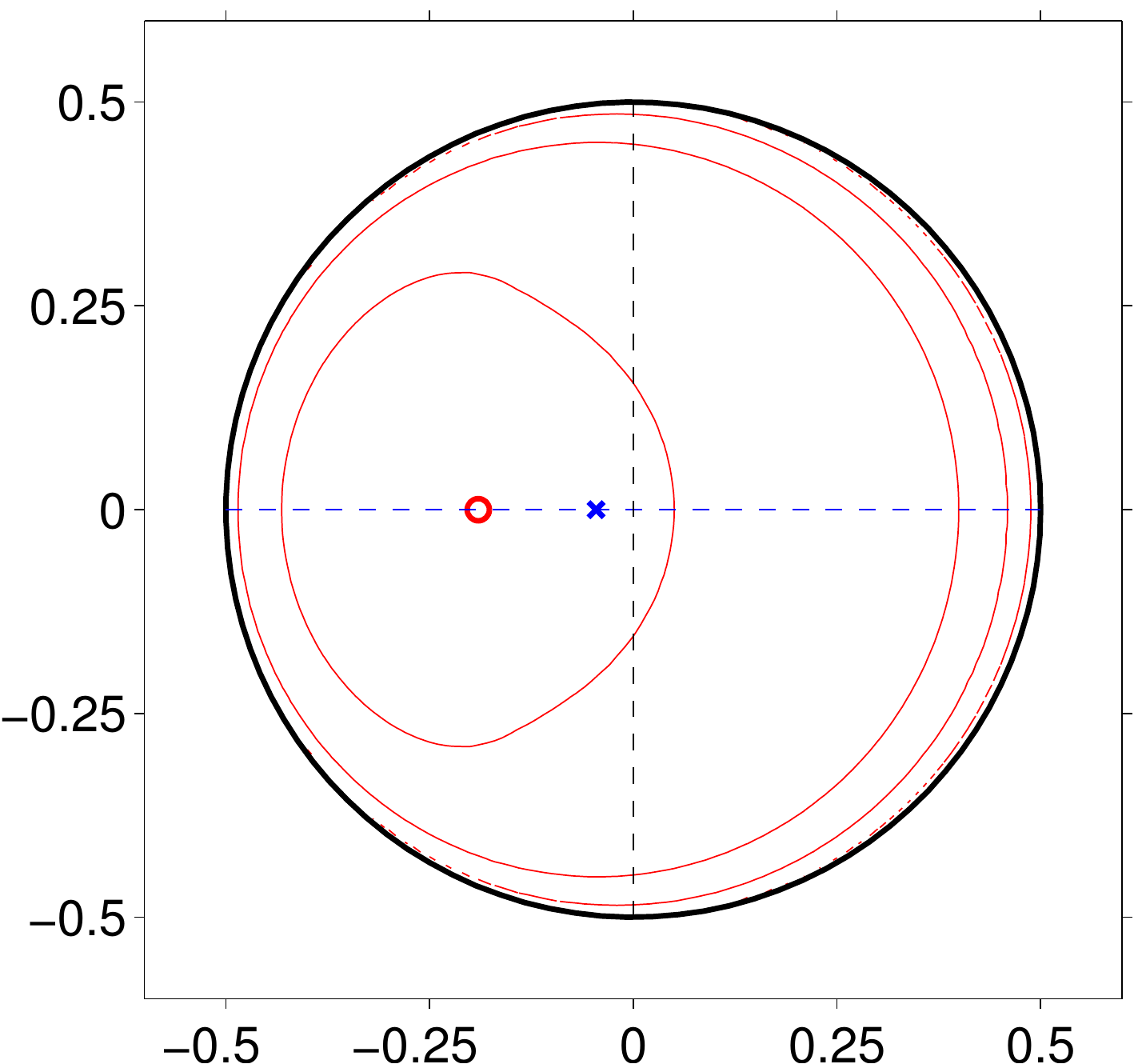}
    \centerline{$x_{pHz\perp}$}
  \end{minipage}
  }{
  \caption{%
    Reference results for case BL showing the pressure coefficient
    $c_p$.  
    $(a)$ Contours of $c_p$ on the surface of the sphere, projected
    upon the horizontal plane through the sphere's center 
    (upstream-facing side). Contours are shown for values
    (-0.5:0.1:1.1), with negative values being indicated by red
    color. 
    The blue cross marks the point where the axis of motion, 
    $\mathbf{e}_{p\parallel}$, crosses the sphere's surface. 
    The red circle marks the location of the local pressure maximum on
    the depicted hemisphere. 
    $(b)$ The same as $(a)$, but for the downstream-facing side of the 
    sphere. 
    \revision{%
    $(c)$ 
    Profile of $c_p$ along two great circles on the sphere, in a plane
    given by the vertical direction $\mathbf{e}_z$ and
    $\mathbf{e}_{pH}$ ($\mathbf{e}_{pHz\perp}$) shown in black (blue)
    color,  
    corresponding to the black (blue) dashed lines in graphs $(a,b)$. 
    The upstream stagnation point corresponds to a value of the angle  
    $\theta\approx\pm\pi$.}{}
    \protect\label{fig-results-ref-press-b1l}
  }
  }
\end{figure}
The fluid velocity field expressed in the coordinate system attached
to the particles (i.e.\ the fluid velocity relative to the
particle motion) is defined as
\begin{equation}\label{equ-res-ref-def-urel}
  \mathbf{u}_{r}(\mathbf{x},t)=
    \mathbf{u}(\mathbf{x},t)-\mathbf{u}_p(t) 
  \,,
\end{equation}
with Cartesian components $\mathbf{u}_{r}=(u_r,v_r,w_r)^T$. 
The unit vector pointing in the direction of the particle motion
relative to the ambient is defined as 
\begin{equation}\label{equ-res-ref-def-eppar}
  \mathbf{e}_{p\,\parallel}=\mathbf{u}_{pr}/||\mathbf{u}_{pr}||
  \,,
\end{equation}
and the distance from the sphere's center along this direction is
denoted by $x_{p\,\parallel}$. 
The direction which is perpendicular to both $\mathbf{e}_{pHz\perp}$
and $\mathbf{e}_{p\,\parallel}$  
is given by
\begin{equation}\label{equ-res-ref-def-epperpzH}
  \mathbf{e}_{p\,\perp}=
  \mathbf{e}_{pHz\perp}
  \times
  \mathbf{e}_{p\,\parallel}
  \,,
\end{equation}
again with corresponding coordinate $x_{p\,\perp}$.
A sketch of these geometrical definitions in the plane given by the
vertical coordinate direction and the particle velocity vector is
shown in figure~\ref{fig-ref-notation-1}. 

The relative fluid velocity projected upon the direction 
opposite to
the particle velocity vector (relative to the ambient) is given by 
\begin{equation}\label{equ-res-ref-def-urpar}
  u_{r\,\parallel}=
  \mathbf{u}_r\cdot\left(-\mathbf{e}_{p\,\parallel}\right)
  \,,
\end{equation}
(note the negative sign), and the components in the two remaining
coordinate directions of the frame
($\mathbf{e}_{p\perp}$,$\mathbf{e}_{pHz\perp}$,$\mathbf{e}_{p\parallel}$)
are denoted as:
\begin{subequations}\label{equ-?}
  \begin{eqnarray}
    u_{r\,\perp}&=&
    \mathbf{u}_r\cdot\mathbf{e}_{p\perp}
    \,,
    \\
    u_{r\,Hz\perp}&=&
    \mathbf{u}_r\cdot\mathbf{e}_{pHz\perp}
    \,.
    \end{eqnarray}
\end{subequations}
The quantity $u_{r\,\parallel}$ is used for the definition of the
sphere wake recirculation length which is determined as follows. 
Let us define a curve ${\cal C}$ as the connection of locations where
the projected relative velocity $u_{r\,\parallel}$ changes sign 
in a plane passing through the sphere center and
which is parallel to both the vertical direction $\mathbf{e}_z$ and
the direction of the sphere's translational velocity
$\mathbf{e}_{p\,\parallel}$.  
Then the recirculation length $L_r$ is measured as the largest
distance between the sphere's surface and any point on the curve ${\cal
  C}$.  
%
%
A graphical impression of our definitions of the recirculation length
can be gathered from figure~\ref{fig-results-ref-contour-b1l} which
will be discussed in \S~\ref{sec-ref-results-steady-oblique}.  
%
%
With the present notation, the pressure coefficient can be defined as
follows:
\begin{equation}\label{equ-def-pressure-coeff-cp}
  c_p=
  \frac{p-p_\infty}{|\mathbf{u}_{pr}|^2/2}
  \,,
\end{equation}
where $p_\infty$ is the pressure of the ambient fluid, 
and the relative particle velocity $\mathbf{u}_{pr}$ is defined
in (\ref{equ-res-ref-def-upr}). Note that the fluid density is absent
in the denominator of (\ref{equ-def-pressure-coeff-cp}) due to the
choice of reference quantities (cf.\ \S~\ref{sec-reference-config}). 
%
\subsubsection{Steady axi-symmetric regime}
\label{sec-ref-results-axisymm}
\begin{table}
  \centering
  \setlength{\tabcolsep}{5pt}
  \begin{tabular}{*{11}{c}}
    case&
    $\overline{u}_{pV}$&
    $\overline{u}_{pH}$&
    $\overline{\omega}_{pH}$&
    $u_{pV}^\prime$&
    $u_{pH}^\prime$&
    $\omega_{pH}^\prime$&
    $f$
    &
    $\overline{L}_r$&
    ${L}_r^\prime$
    &
    \revision{}{%
      $Re_\parallel$
    }
    \\[1ex]
    CS&
    $-1.383$& 
    $0.137$& 
    $0.013$& 
    $0.010$& 
    $0.040$& 
    $0.010$& 
    $0.068$  
    &
    --&
    --
    &
    \revision{}{%
      $264.06$
    }
    \\
    CL&
    $-1.376$& 
    $0.136$& 
    $0.012$& 
    $0.008$& 
    $0.033$& 
    $0.008$& 
    $0.071$&  
    $1.681$& 
    $0.069$  
    &
    \revision{}{%
      $262.71$
    }
    %
  \end{tabular}
  \caption{
    Reference results for flow cases in the oscillating oblique 
    flow regime ($G=190$). 
    Overbars stand for mean values and primes for fluctuation
    amplitudes, as defined in
    (\ref{equ-res-ref-def-mean}-\ref{equ-res-ref-def-ampli}).  
    \revision{}{%
      The Reynolds number $Re_\parallel$ (defined in
      \ref{equ-res-ref-def-reynolds-par}) is computed from the mean
      velocity values. 
    }
  }    
  \label{tab-results-ref-C}
\end{table}
\begin{figure}[b]
  \figpap{
  \centering
  \begin{minipage}{4ex}
    $u_{pV}$
  \end{minipage}
  \begin{minipage}{.37\linewidth}
    \centerline{$(a)$}
    \includegraphics[width=\linewidth]
    {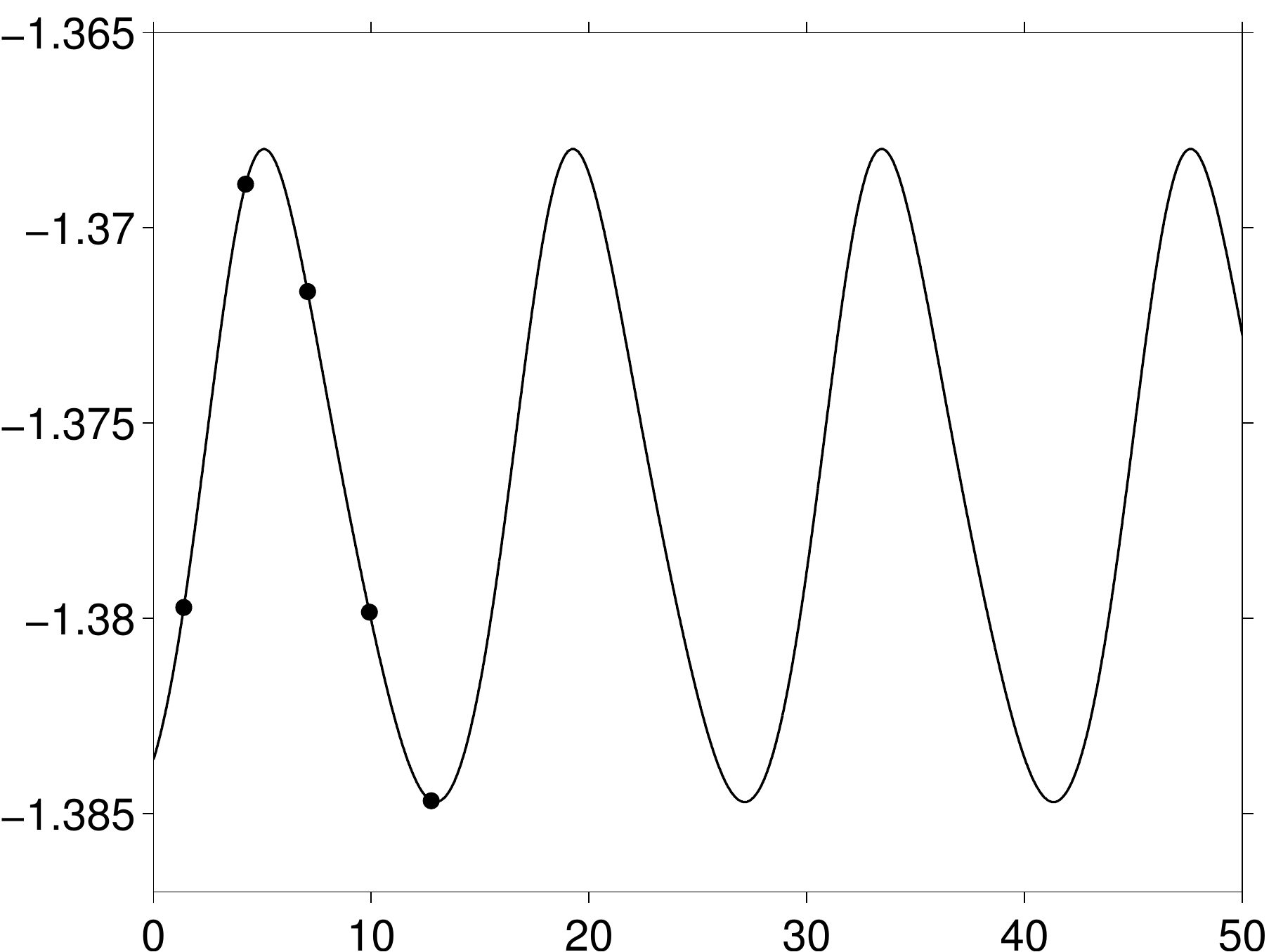}
    \\
    \centerline{$t$}
  \end{minipage}
  \revision{%
    \\[1ex]}{\hspace*{5ex}}
  \begin{minipage}{4ex}
    $u_{pH}$
  \end{minipage}
  \begin{minipage}{.37\linewidth}
    \centerline{$(b)$}
    \includegraphics[width=\linewidth]
    {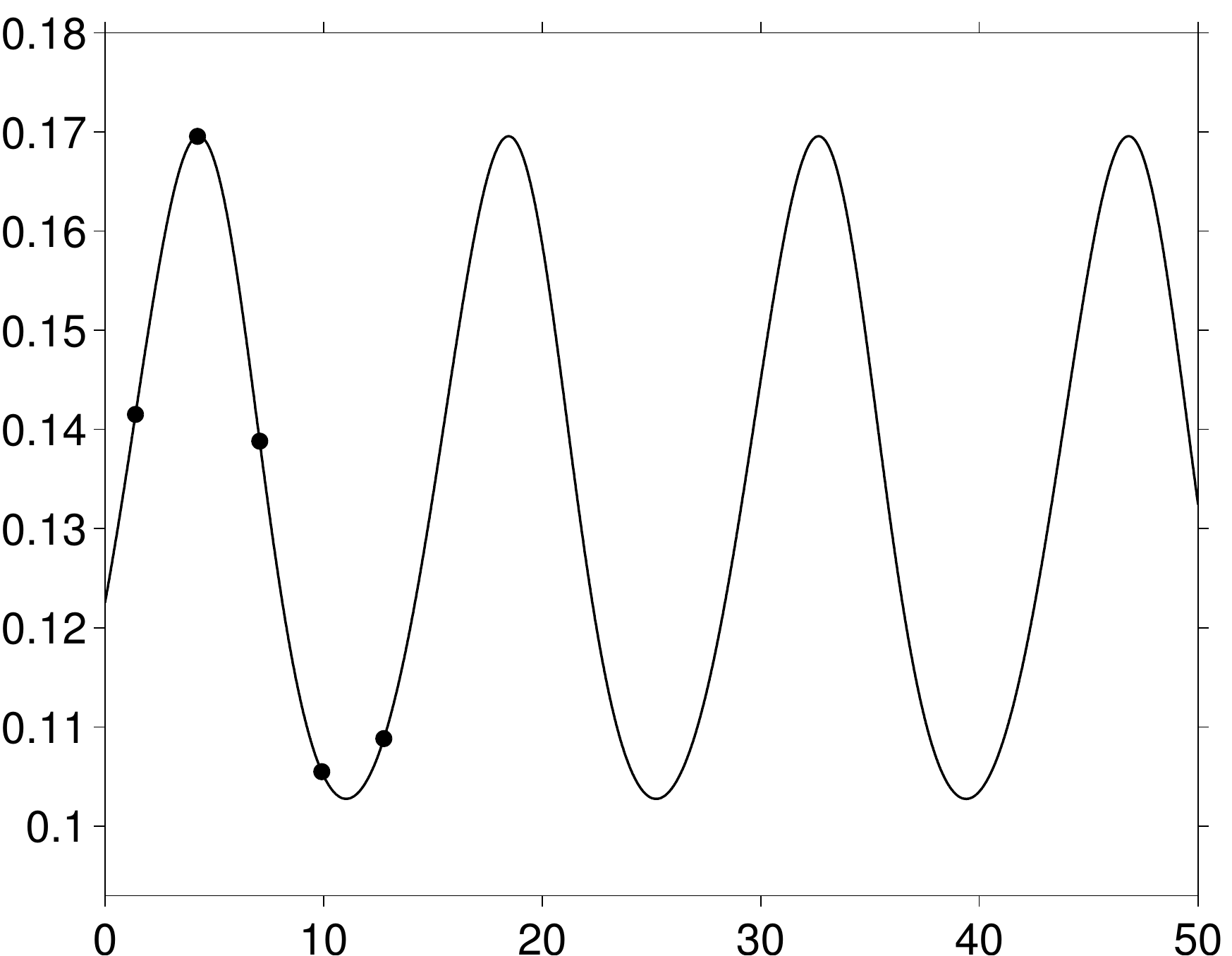}
    \\
    \centerline{$t$}
  \end{minipage}
  \\[1ex]
  \revision{%
  \begin{minipage}{4ex}
    $\omega_{pH}$
  \end{minipage}
  \begin{minipage}{.37\linewidth}
    \centerline{$(c)$}
    \includegraphics[width=\linewidth]
    {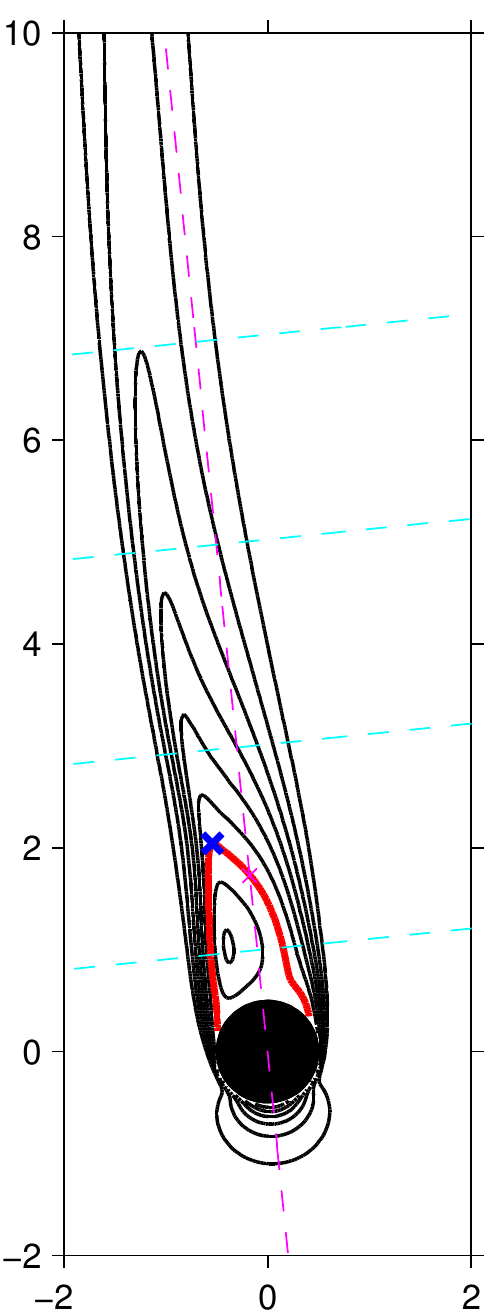}
    \\
    \centerline{$t$}
  \end{minipage}
  \\
  }{}
  }{
  \caption{%
    Reference results for case CL ($G=190$), exhibiting time-periodic
    dynamics.  
    The graphs show the temporal evolution of: 
    $(a)$ the vertical particle velocity component; 
    $(b)$ the horizontal particle velocity component%
    \revision{; 
      $(c)$ the horizontal angular particle velocity component.}{.}
    The symbols indicate the time instants corresponding to the flow
    fields shown in figures~\ref{fig-results-ref-3d-iso-lambda2-c1l} and
    \ref{fig-results-ref-contour-c1l}. 
    \protect\label{fig-results-ref-history-c1l}
  }
  }
\end{figure}
When the Galileo number is set to $G=144$ the particle wake under
fully established conditions is axi-symmetric, the particle motion is
steady and it follows a straight vertical path. Therefore, the angular
particle velocity is identically zero, 
and its translational velocity only has one 
non-zero 
component, $u_{pV}$. 
Table~\ref{tab-results-ref-A} lists the asymptotic, steady-state
values of $u_{pV}$  obtained for the two domain sizes which have been
simulated. It can be seen that the difference is small
(approximately $0.5$\%), which implies that a variation of the domain
size in the present range ($D_\Omega=5.34$ to $7.54$) has an almost
negligible influence on the particle motion.   

Figure~\ref{fig-results-ref-contour-a1l} gives a visual impression of
the flow field around the particle in steady-state motion for case
AL. 
\revision{}{%
  The graph in figure~\ref{fig-results-ref-contour-a1l}$(a)$ shows an
  iso-surface of the vertical component of the flow velocity relative
  to the particle motion. 
}
It can be seen that after approximately one diameter downstream of the
rear stagnation point, the wake is almost aligned with the vertical
direction. 
Also included in the figure is an iso-surface plot of the
second largest eigenvalue of the tensor
$\mathbf{S}^2+\boldsymbol{\Omega}^2$ (where $\mathbf{S}$ and
$\boldsymbol{\Omega}$ are the symmetrical and anti-symmetrical parts
of the velocity gradient tensor, respectively), henceforth denoted as
$\lambda_2$. 
It has been proposed by \cite{jeong:95} that vortical structures can
be identified with regions where the quantity $\lambda_2$
takes negative values. 
Figure~\ref{fig-results-ref-contour-a1l}$(b)$ shows that vortical
motion in the present case is concentrated in a thin torus-shaped
region enclosing the sphere.  

Contours of the fluid velocity relative to the particle motion,
$u_{r\parallel}$ (cf.\ definition in equ.\
\ref{equ-res-ref-def-urpar}), are shown in 
figure~\ref{fig-results-ref-contour-a1l}$(c)$. The extent of the
region where $u_{r\parallel}<0$ is marked in red therein. The length
of this recirculation region is given in
table~\ref{tab-results-ref-A}. Again, it can be seen that the
influence of the lateral domain size is almost negligible (less than
$0.2$\%). 
\revision{
  The plot in figure~\ref{fig-results-ref-contour-a1l}$(d)$
  shows the vertical component of the relative fluid velocity along
  the vertical 
  axis through the sphere's center. The curve exhibits the expected
  rapid deceleration when approaching the front stagnation point,
  negative values in the recirculation region, and a slower recovery
  (proportional to the inverse of the distance from the downstream
  stagnation point) further downstream along the axis.

  In order to allow for use of the present results as reference data in
  future studies, we present a number of profiles of velocity components
  in figure~\ref{fig-results-ref-cross-profiles-a1l}. 
  In the cylindrical coordinate system attached to the particle center,
  only the axial component $u_{r\parallel}$ and the radial component
  (henceforth denoted as $u_{rad}$) are non-zero. The figure shows
  profiles of these two quantities taken along the radial direction at
  four different axial locations: $x_{p\parallel}=\{-1,-3,-5,-7\}$
  (along the dashed lines marked in
  figure~\ref{fig-results-ref-contour-a1l}).}{%
  The vertical component of the relative fluid velocity along
  the vertical axis through the sphere's center (i.e.\ along a
  vertical cut through figure~\ref{fig-results-ref-contour-a1l}$c$)  
  exhibits the expected
  rapid deceleration when approaching the front stagnation point,
  takes negative values in the recirculation region, and has a slower
  recovery (proportional to the inverse of the distance from the
  downstream stagnation point) further downstream along the axis. 
  This information is included in the supplementary data and it is
  compared to the IBM results in
  figure~\ref{fig-results-ibm-cross-profiles-a1c} below. 
  %
  Furthermore, radial profiles of the two non-zero velocity components
  in the cylindrical coordinate system attached to the particle center
  (the axial component $u_{r\parallel}$ and the radial component,
  henceforth denoted as $u_{rad}$) 
  at four different axial locations: $x_{p\parallel}=\{-1,-3,-5,-7\}$
  (along the dashed lines marked in
  figure~\ref{fig-results-ref-contour-a1l}$c$)
  are provided in the supplementary data.} 

\begin{figure}
  \figpap{
  \begin{minipage}{.195\linewidth}
    \centerline{$(a)$}
    \includegraphics[width=\linewidth,clip=true,
    viewport=1050 830 1500 1750]
    {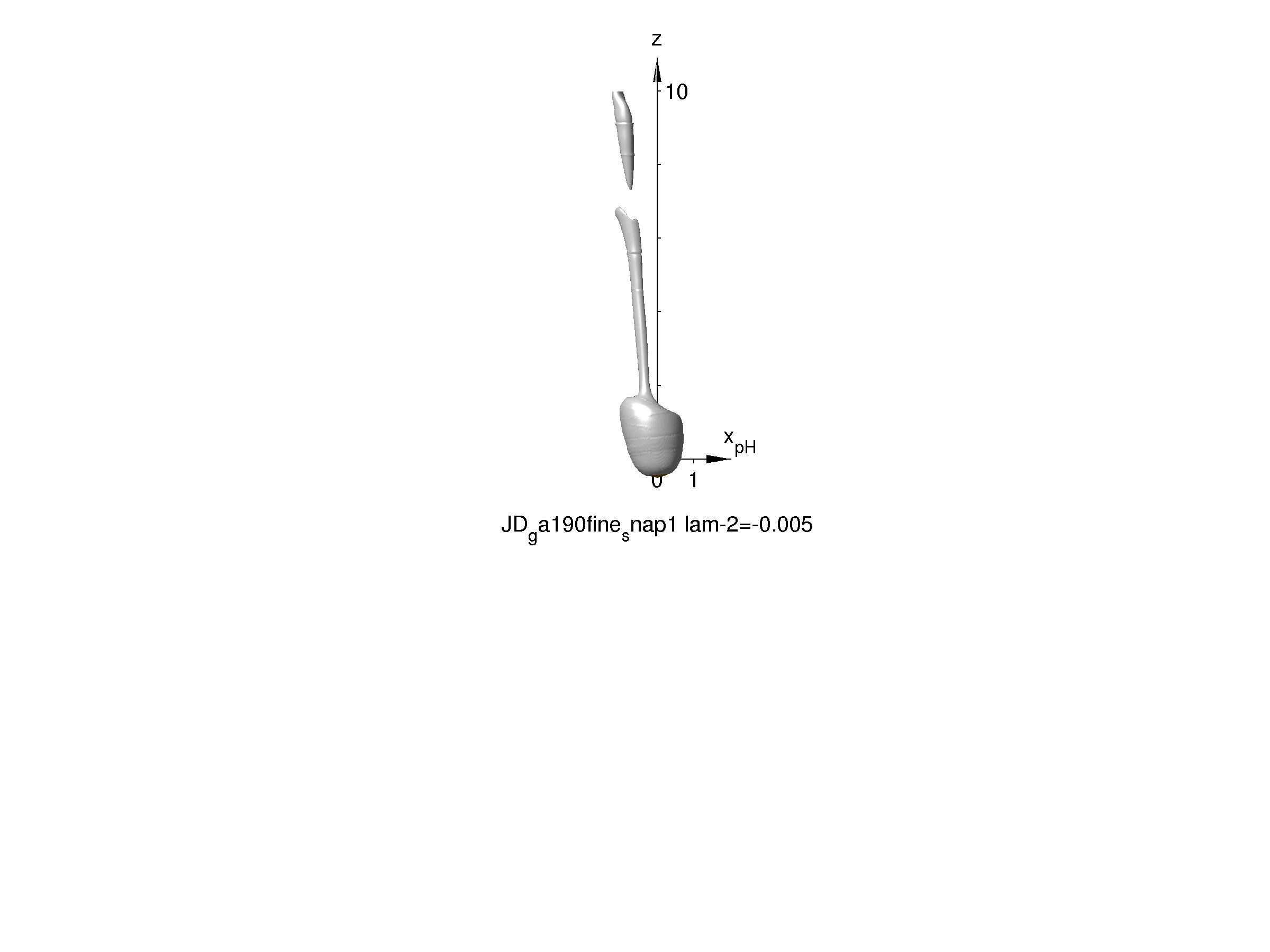}
  \end{minipage}
  \begin{minipage}{.195\linewidth}
    \centerline{$(b)$}
    \includegraphics[width=\linewidth,clip=true,
    viewport=1050 830 1500 1750]
    {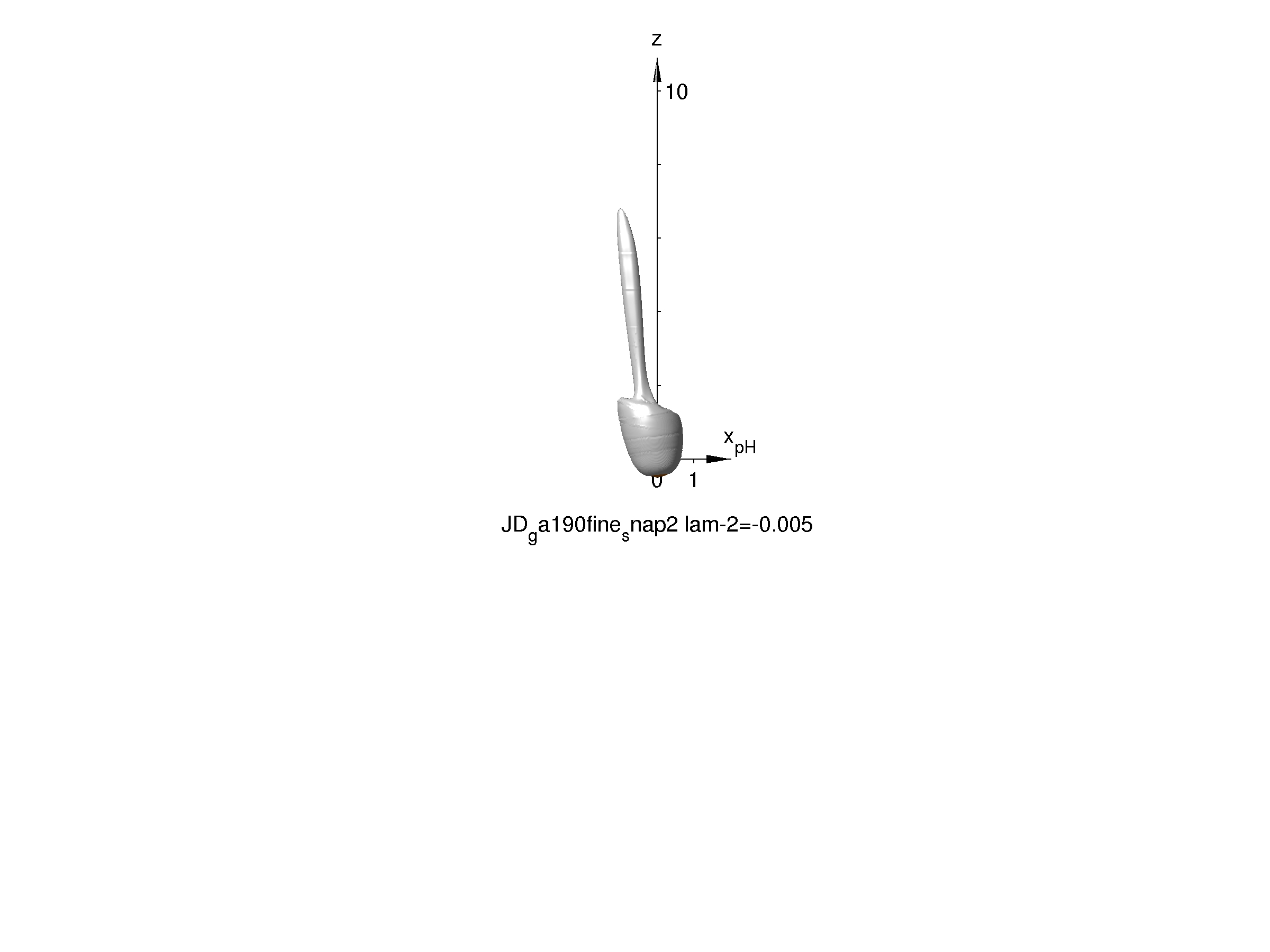}
  \end{minipage}
  \begin{minipage}{.195\linewidth}
    \centerline{$(c)$}
    \includegraphics[width=\linewidth,clip=true,
    viewport=1050 830 1500 1750]
    {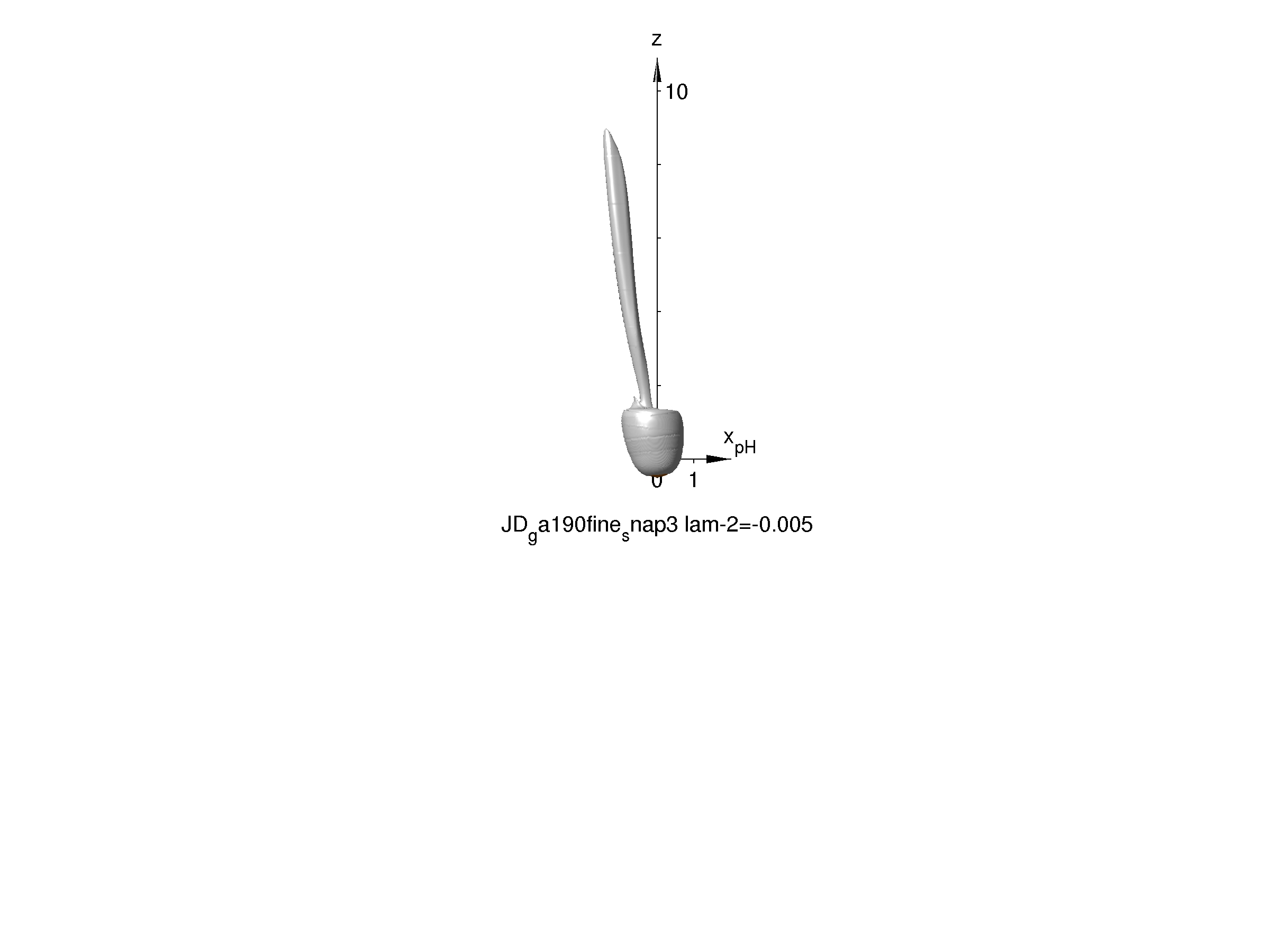}
  \end{minipage}
  \begin{minipage}{.195\linewidth}
    \centerline{$(d)$}
    \includegraphics[width=\linewidth,clip=true,
    viewport=1050 830 1500 1750]
    {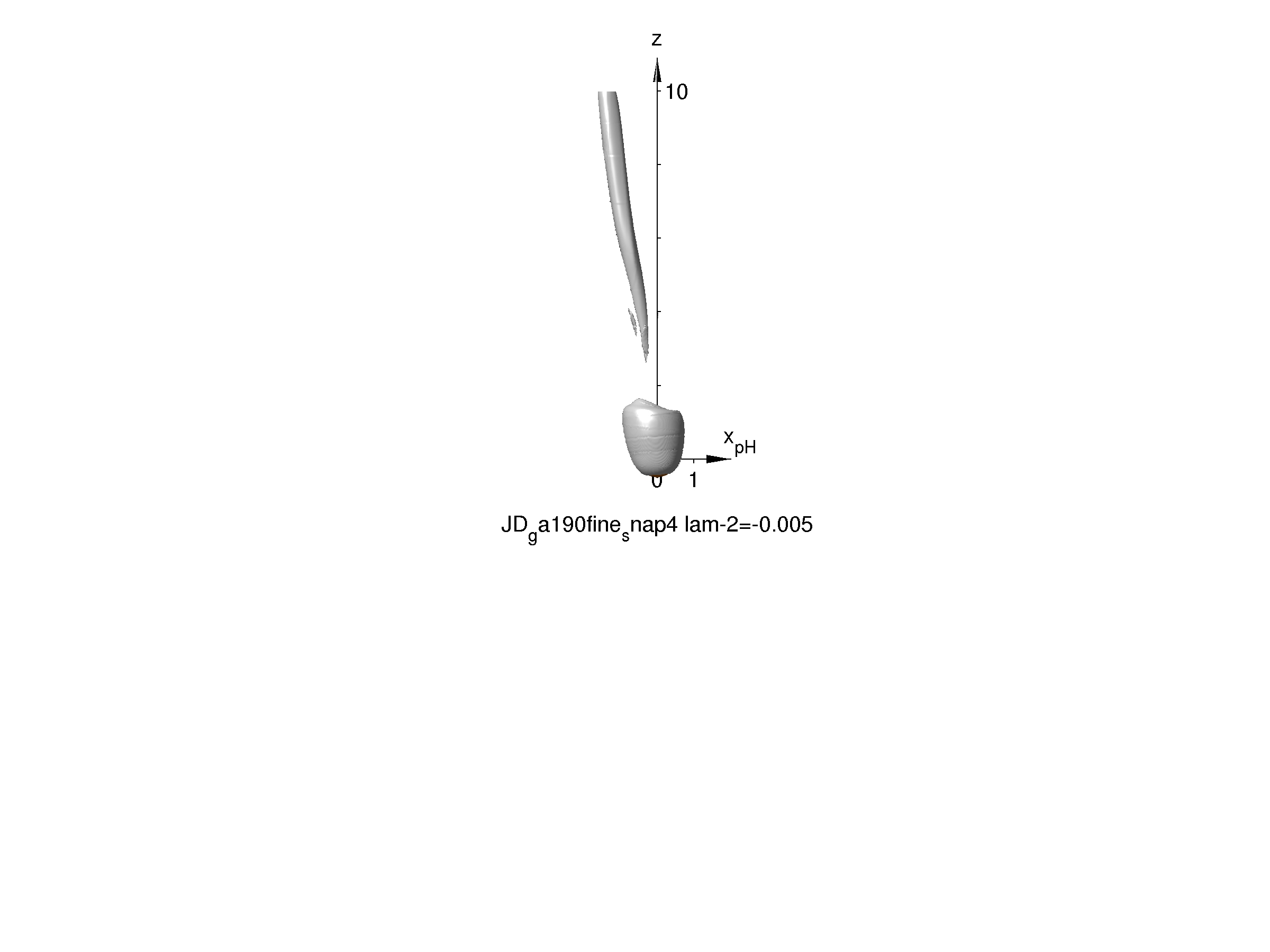}
  \end{minipage}
  \begin{minipage}{.195\linewidth}
    \centerline{$(e)$}
    \includegraphics[width=\linewidth,clip=true,
    viewport=1050 830 1500 1750]
    {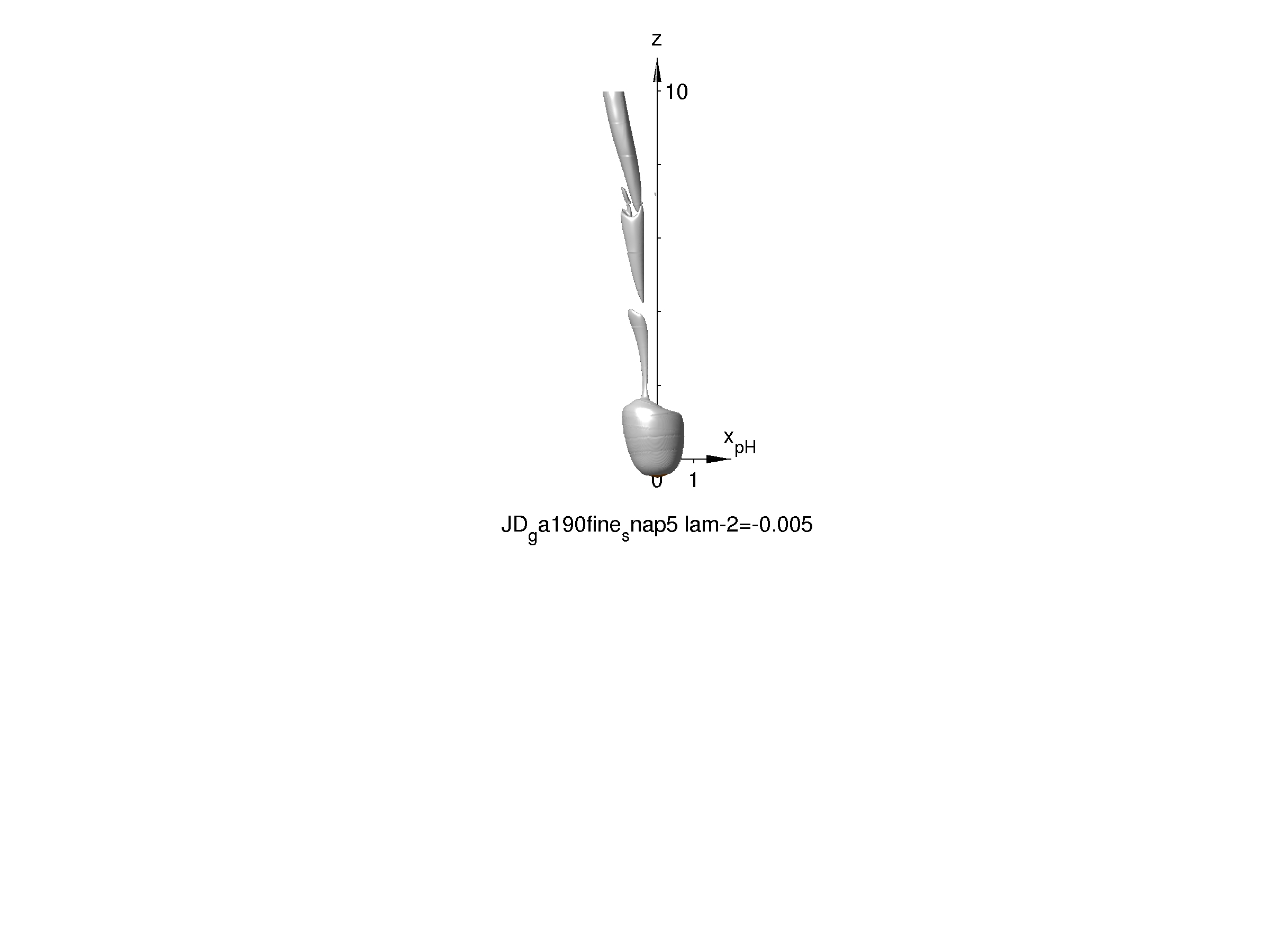}
  \end{minipage}
  \\
  \begin{minipage}{.195\linewidth}
    \centerline{$(f)$}
    \includegraphics[width=\linewidth,clip=true,
    viewport=1050 830 1500 1750]
    {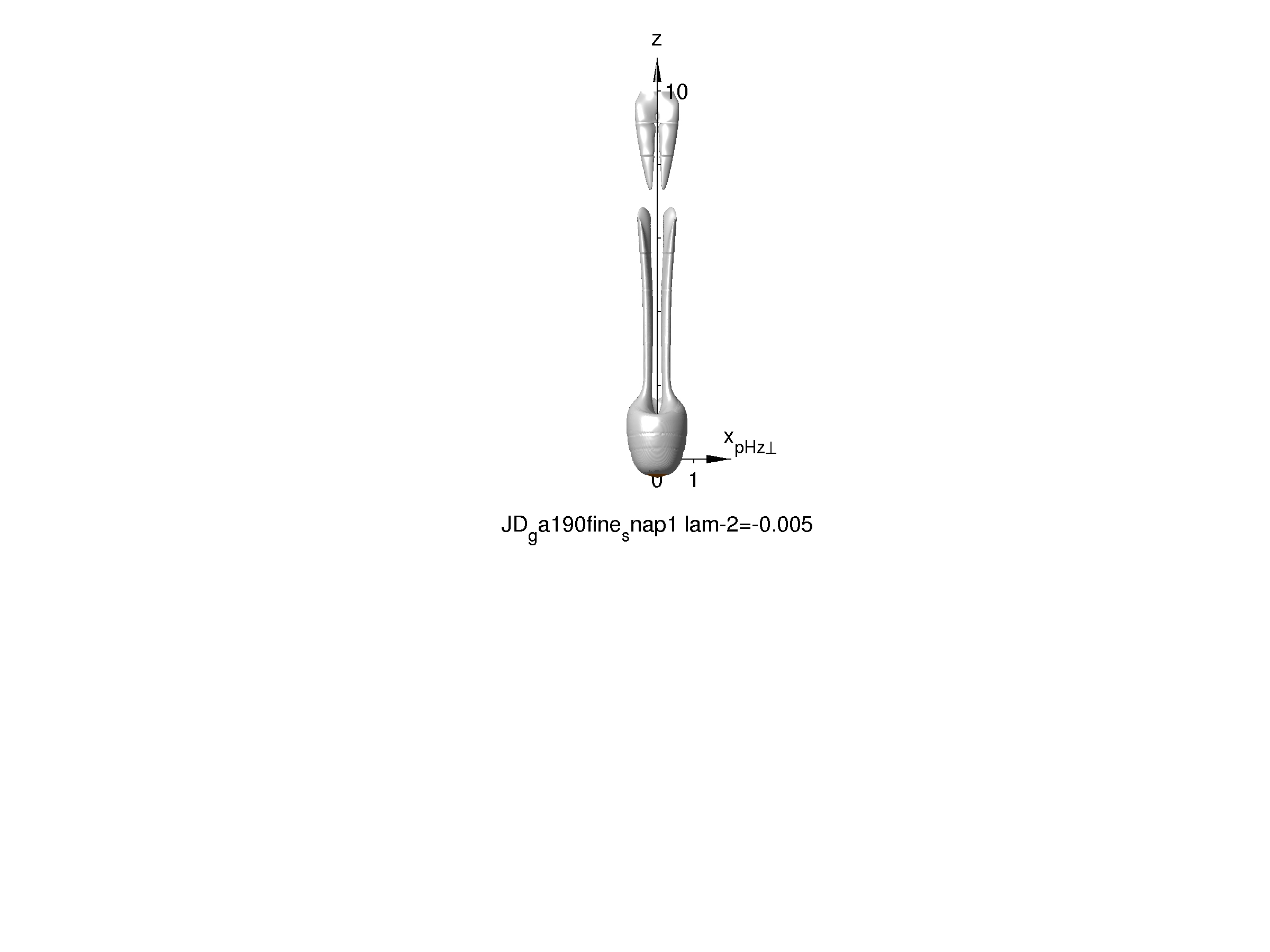}
  \end{minipage}
  \begin{minipage}{.195\linewidth}
    \centerline{$(g)$}
    \includegraphics[width=\linewidth,clip=true,
    viewport=1050 830 1500 1750]
    {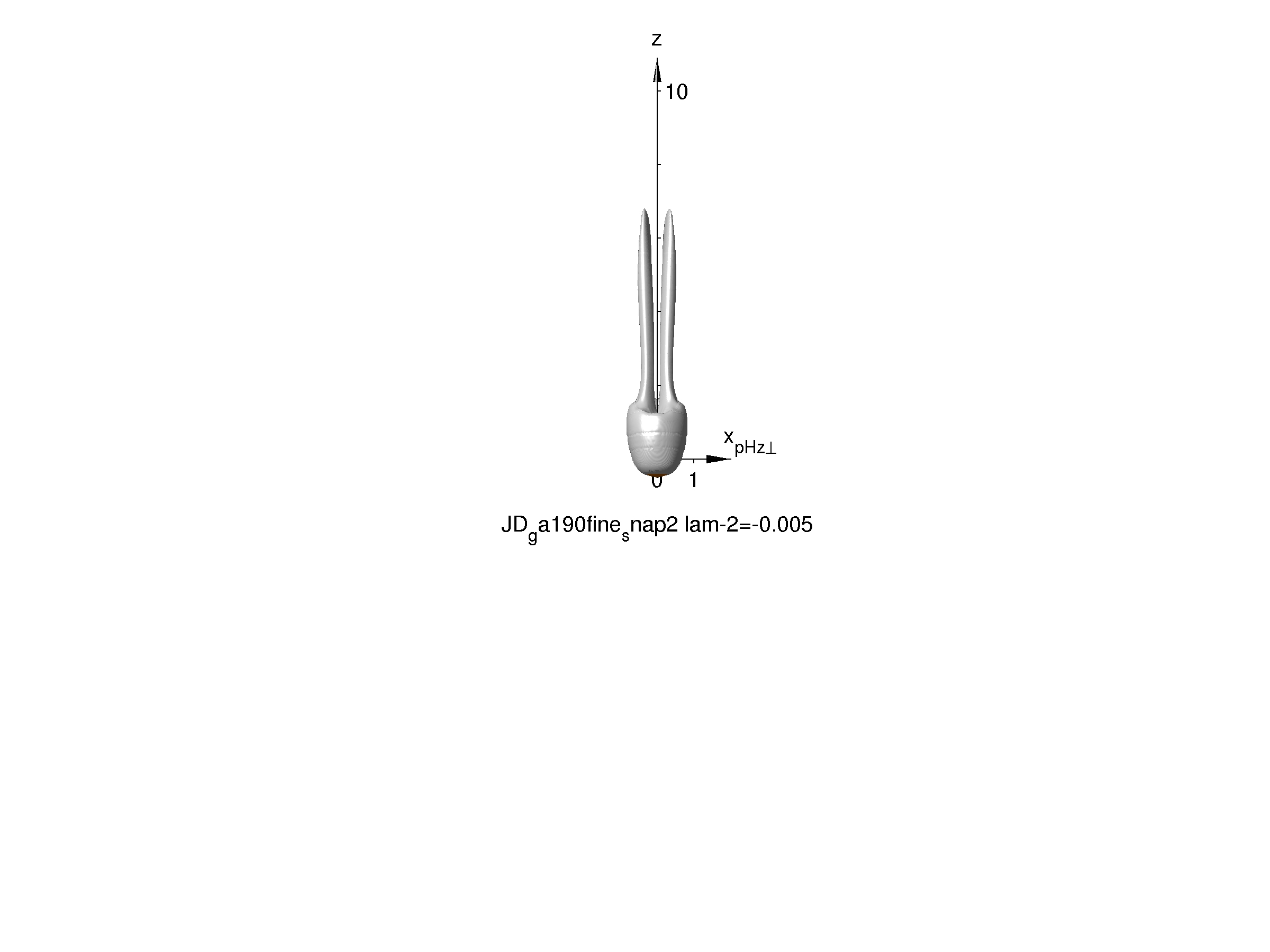}
  \end{minipage}
  \begin{minipage}{.195\linewidth}
    \centerline{$(h)$}
    \includegraphics[width=\linewidth,clip=true,
    viewport=1050 830 1500 1750]
    {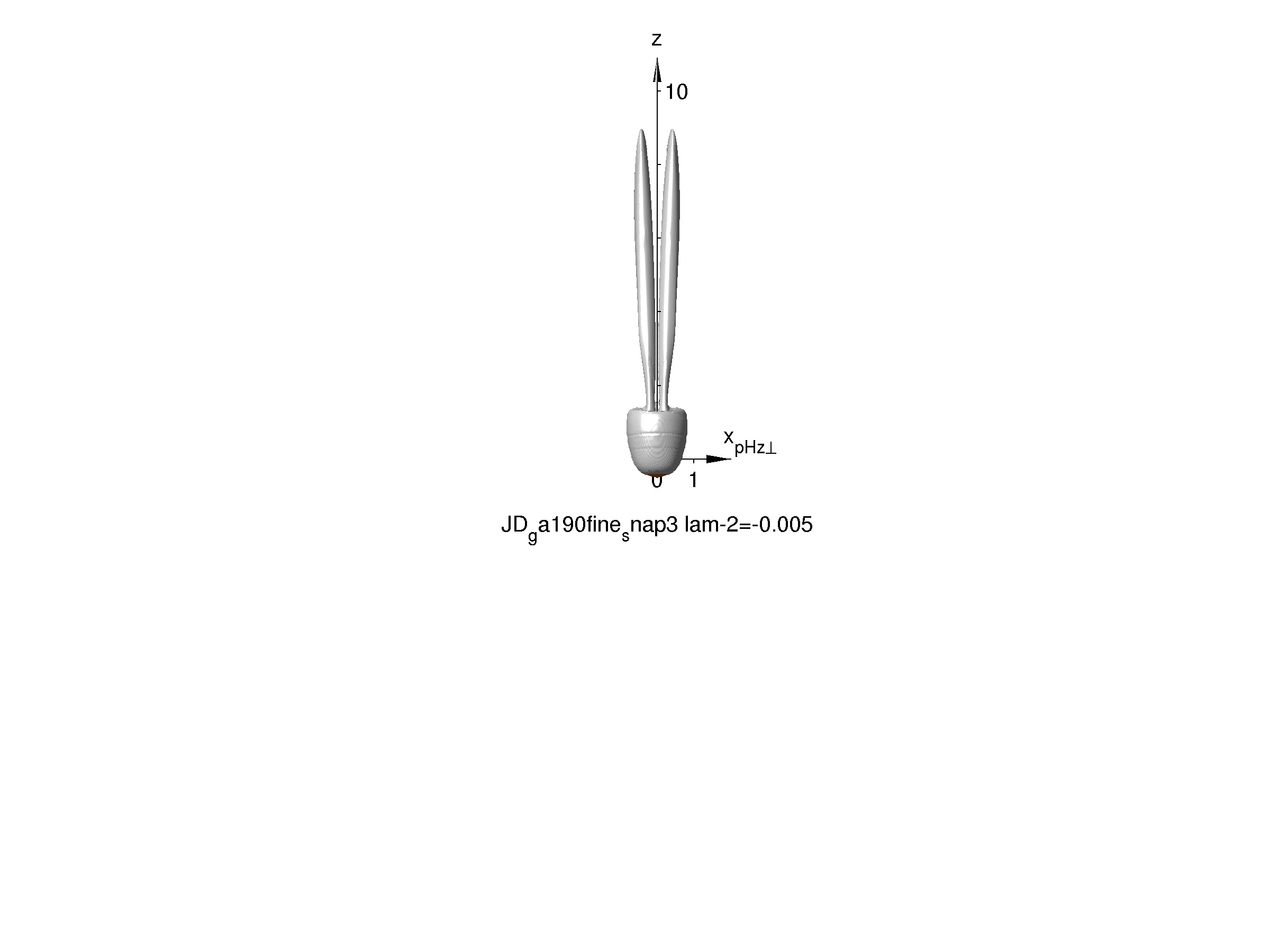}
  \end{minipage}
  \begin{minipage}{.195\linewidth}
    \centerline{$(i)$}
    \includegraphics[width=\linewidth,clip=true,
    viewport=1050 830 1500 1750]
    {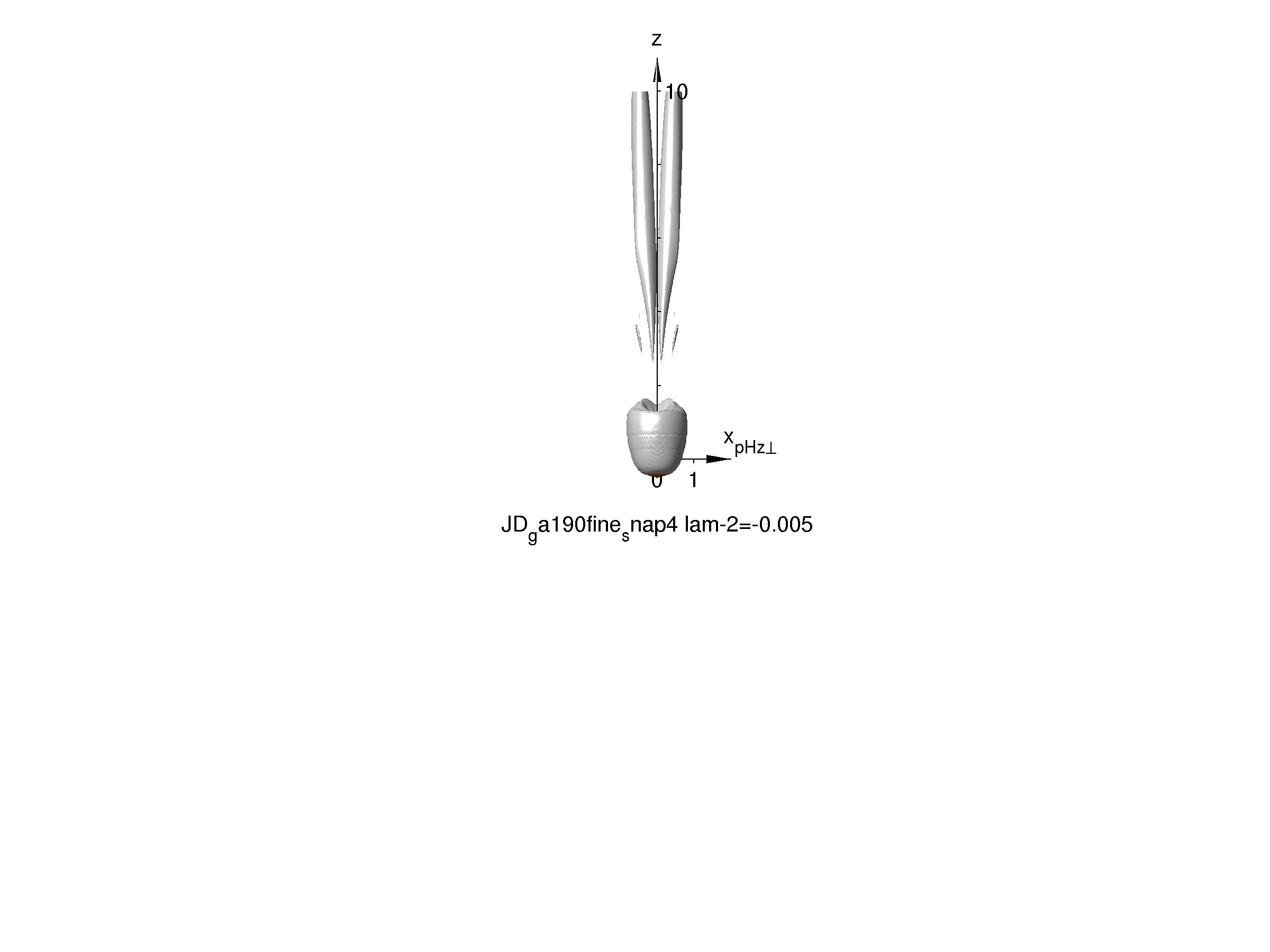}
  \end{minipage}
  \begin{minipage}{.195\linewidth}
    \centerline{$(j)$}
    \includegraphics[width=\linewidth,clip=true,
    viewport=1050 830 1500 1750]
    {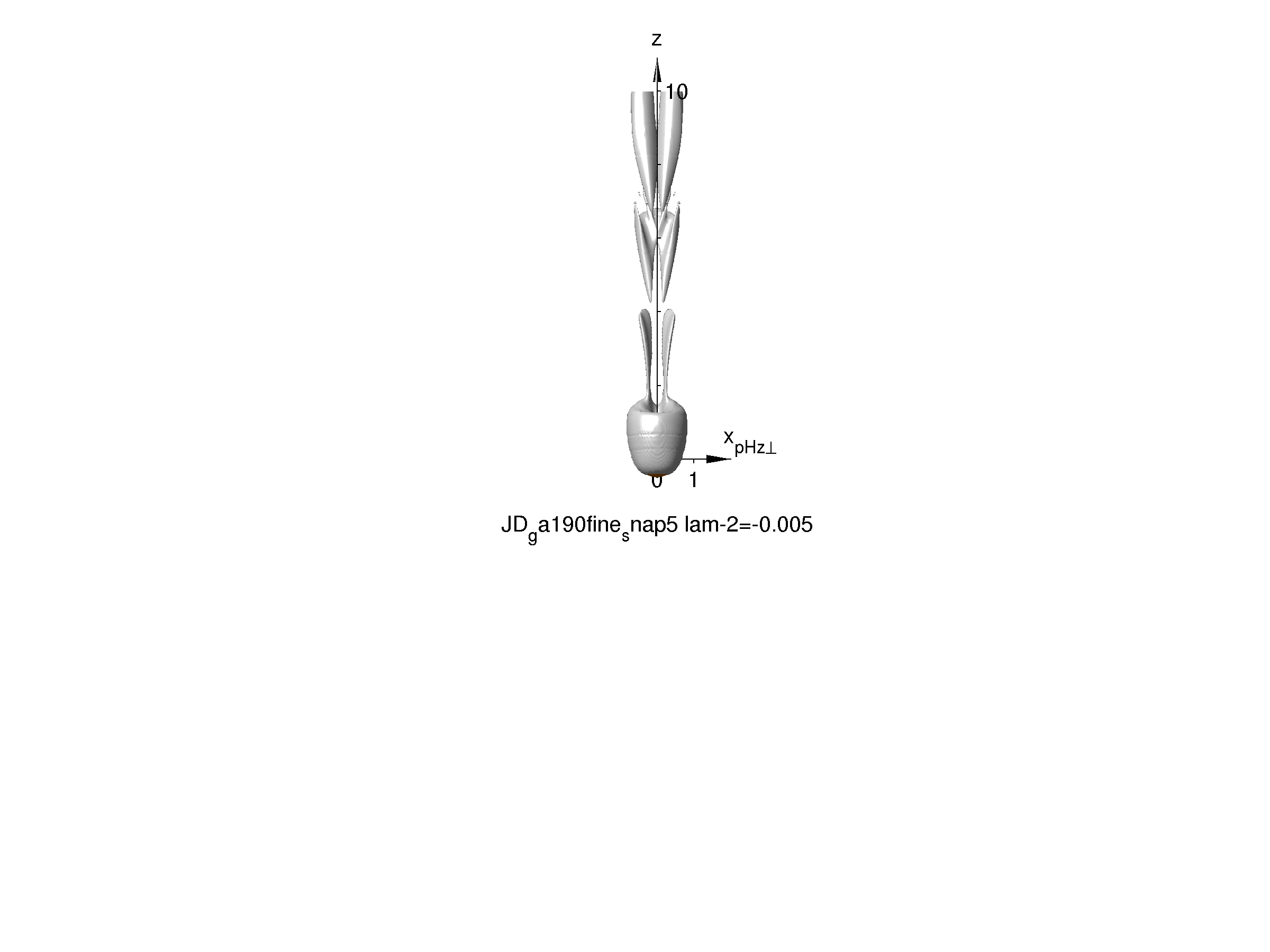}
  \end{minipage}\\
  }{
  \caption{%
    \revision{%
      As figure~\ref{fig-results-ref-3d-iso-u-c1l}, but showing the
      surface where $\lambda_2=-0.005$ (cf.\ definition in the
      text).}{%
      Reference results for case CL ($G=190$, time-periodic dynamics),
      showing instantaneous flow fields, with time increasing from
      left to right by increments of $T/5$ ($T$ being the period). 
      The surface where $\lambda_2=-0.005$ (cf.\ definition in the
      text) is visualized. 
      In $(a$-$e)$ the view is directed along $\mathbf{e}_{pHz\perp}$;  
      in $(f$-$j)$ it is directed along $\mathbf{e}_{pH}$.}
    \protect\label{fig-results-ref-3d-iso-lambda2-c1l}
  }
  }
\end{figure}
\revision{
Finally,  in figure~\ref{fig-results-ref-press-a1l} we report data for
the pressure on the sphere's surface. First, contours of the surface 
pressure projected upon a horizontal plane (on the upstream and
downstream sides) are presented in
figure~\ref{fig-results-ref-press-a1l}$(a,b)$. From the different
density of contour-lines along the radial direction it can be inferred 
that the pressure gradients are much larger on the upstream side as
compared to the downstream side. 
The values of the pressure coefficient $c_p$ (cf.\ definition in
\ref{equ-def-pressure-coeff-cp}) along a great circle are shown in 
figure~\ref{fig-results-ref-press-a1l}$(c)$. Note the geometrical
definition in figure~\ref{fig-ref-notation-1}$(b)$ which is such that
the upstream stagnation point is located at an angle $\theta=\pm\pi$. 
Here the incomplete pressure recovery in the recirculation zone is
clearly visible.}{%
Finally, in figure~\ref{fig-results-ref-press-a1l} we report data for
the pressure on the sphere's surface. Contours of the 
pressure coefficient $c_p$ (cf.\ definition in
\ref{equ-def-pressure-coeff-cp}) projected upon a horizontal plane (on
the upstream and downstream sides of the sphere) are presented.
From the different
density of contour-lines along the radial direction it can be inferred 
that the pressure gradients are much larger on the upstream side as
compared to the downstream side, clearly showing the incomplete
pressure recovery in the recirculation zone. 
The values of the pressure coefficient along a great circle are 
included in the supplementary material; they will be compared to the
IBM data in figure~\ref{fig-results-ibm-press-a1c} below. 
}
\subsubsection{Steady oblique regime}
\label{sec-ref-results-steady-oblique}
\begin{figure}
  \figpap{
  \begin{minipage}{2.5ex}
    $z$
  \end{minipage}
  \begin{minipage}{.188\linewidth}
    \centerline{$(a)$}
    \includegraphics[width=\linewidth]
    {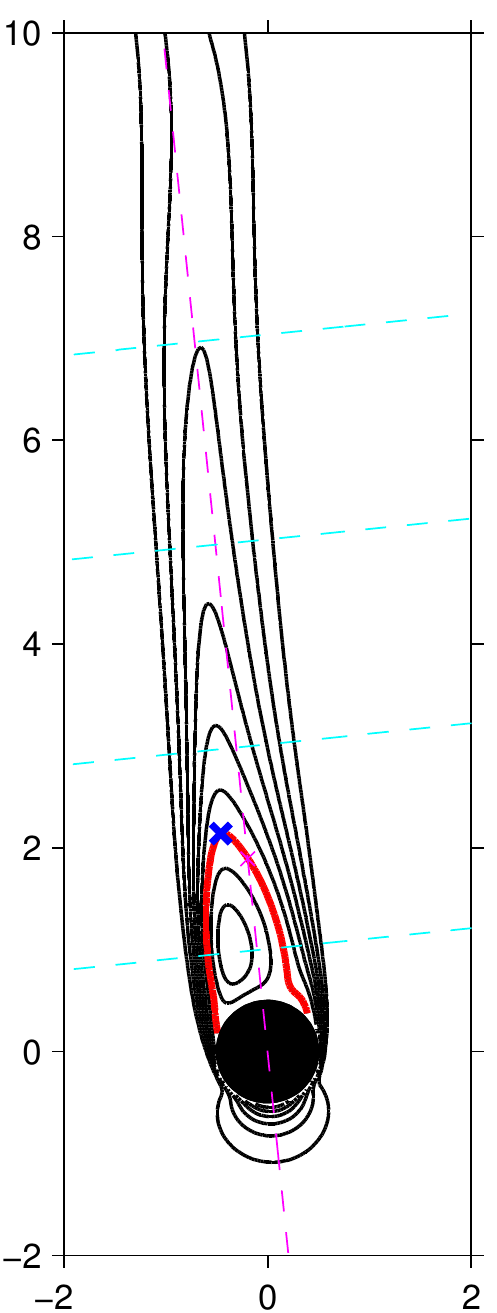}
    \centerline{${x}_{pH}$}
  \end{minipage}
  \begin{minipage}{.188\linewidth}
    \centerline{$(b)$}
    \includegraphics[width=\linewidth]
    {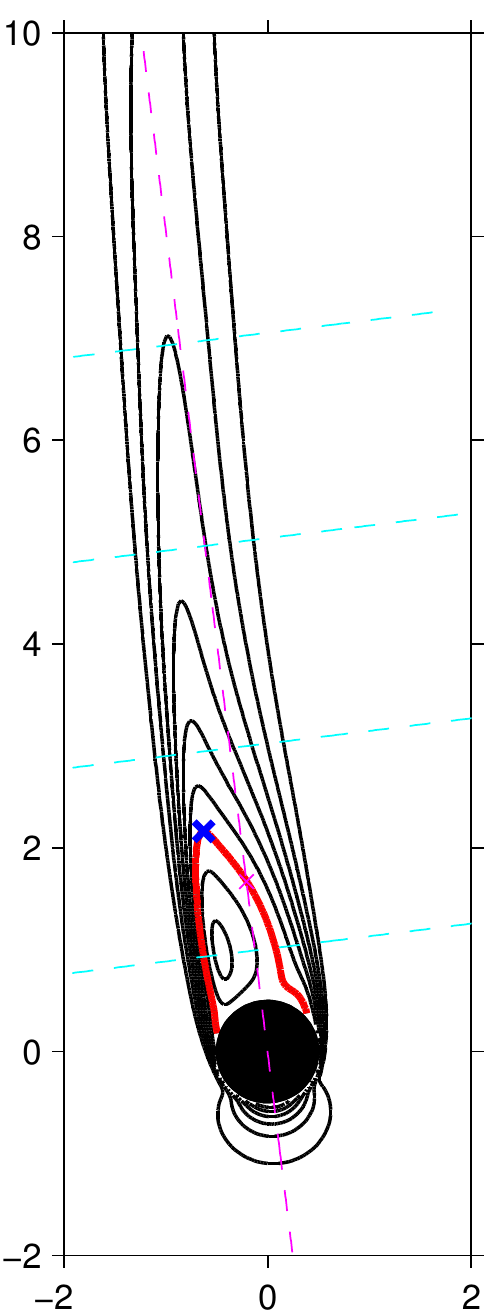}
    \centerline{${x}_{pH}$}
  \end{minipage}
  \begin{minipage}{.188\linewidth}
    \centerline{$(c)$}
    \includegraphics[width=\linewidth]
    {figure_12c.pdf}
    \centerline{${x}_{pH}$}
  \end{minipage}
  \begin{minipage}{.188\linewidth}
    \centerline{$(d)$}
    \includegraphics[width=\linewidth]
    {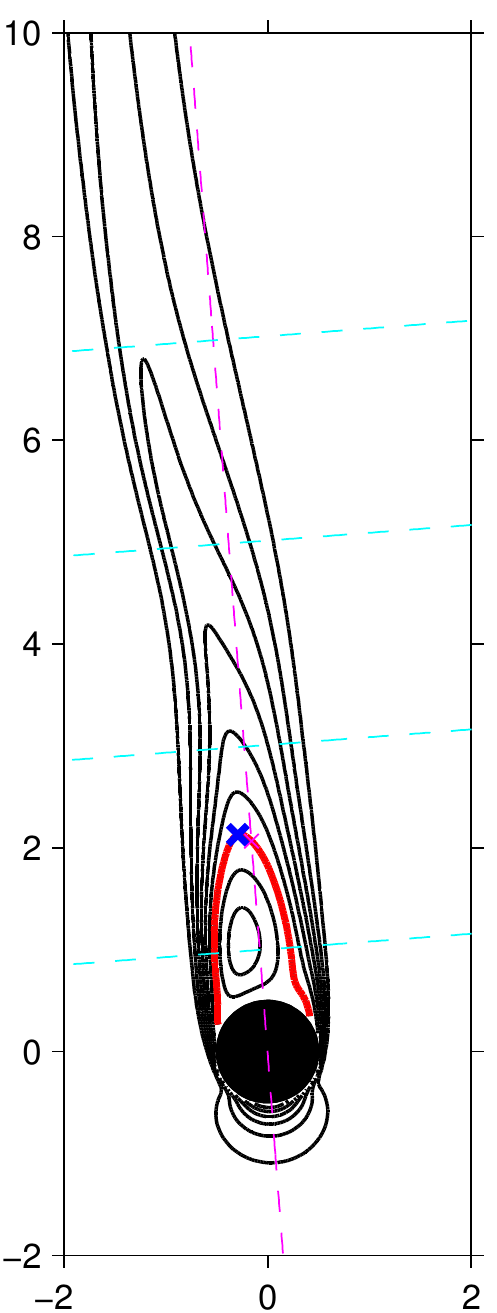}
    \centerline{${x}_{pH}$}
  \end{minipage}
  \begin{minipage}{.188\linewidth}
    \centerline{$(e)$}
    \includegraphics[width=\linewidth]
    {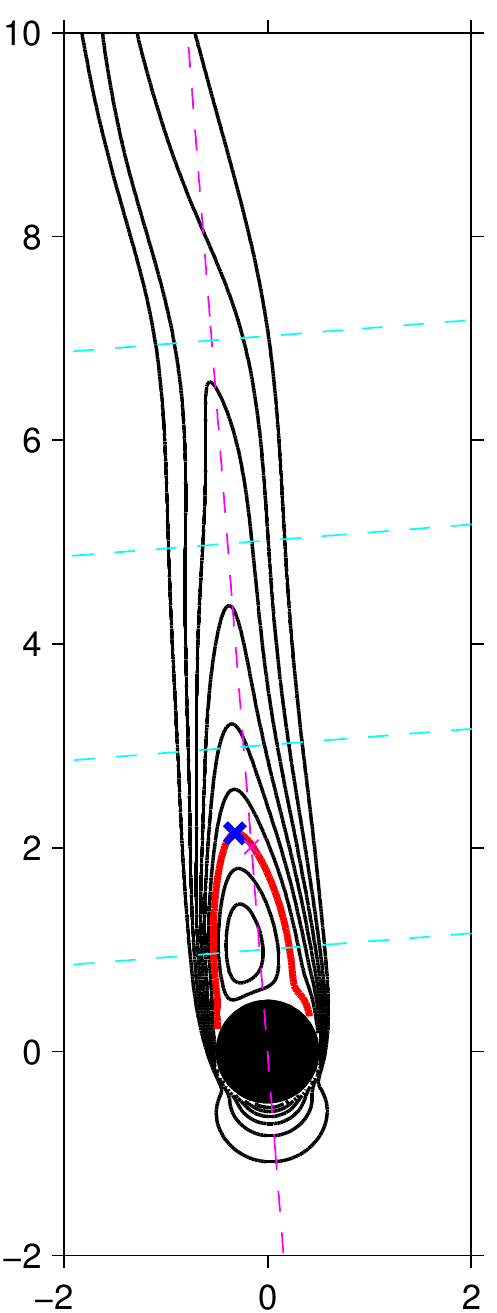}
    \centerline{${x}_{pH}$}
  \end{minipage}
  \\
  \begin{minipage}{2.5ex}
    $x_{p\parallel}$
  \end{minipage}
  \begin{minipage}{.188\linewidth}
    \centerline{$(f)$}
    \includegraphics[width=\linewidth]
    {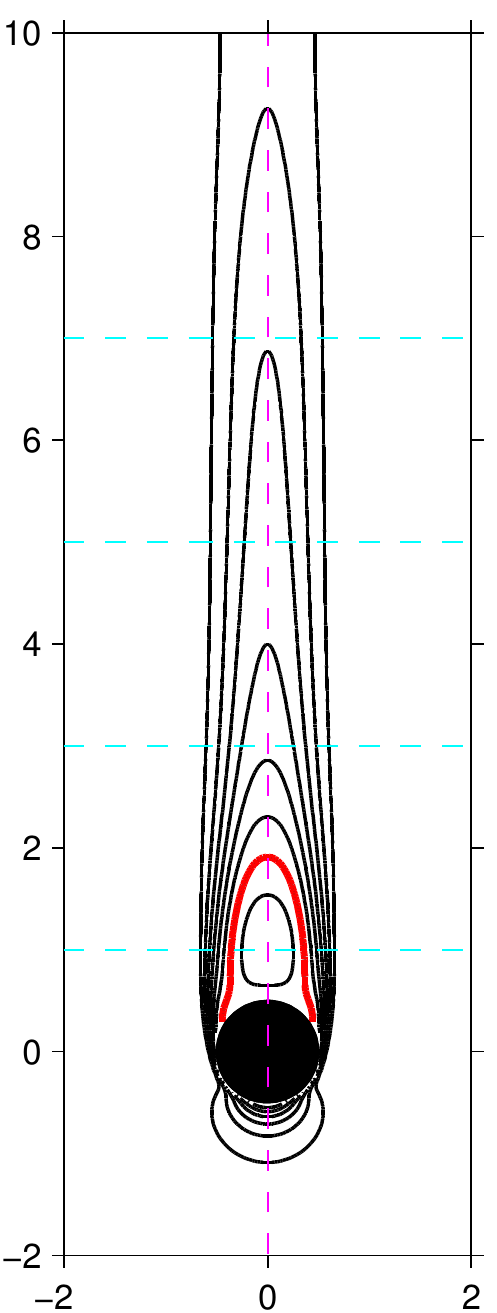}
    \centerline{${x}_{pHz\perp}$}
  \end{minipage}
  \begin{minipage}{.188\linewidth}
    \centerline{$(g)$}
    \includegraphics[width=\linewidth]
    {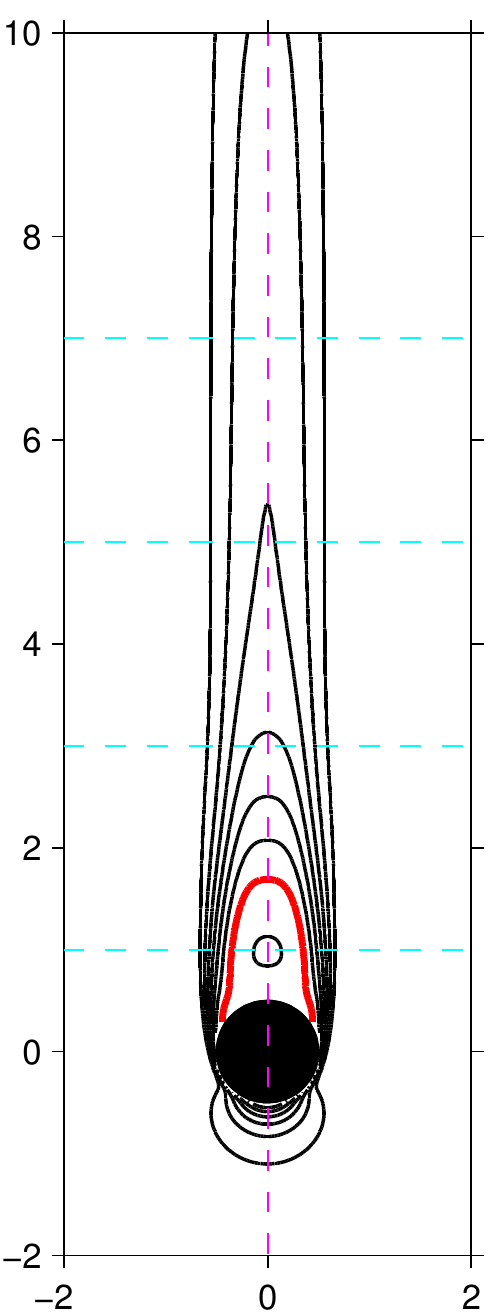}
    \centerline{${x}_{pHz\perp}$}
  \end{minipage}
  \begin{minipage}{.188\linewidth}
    \centerline{$(h)$}
    \includegraphics[width=\linewidth]
    {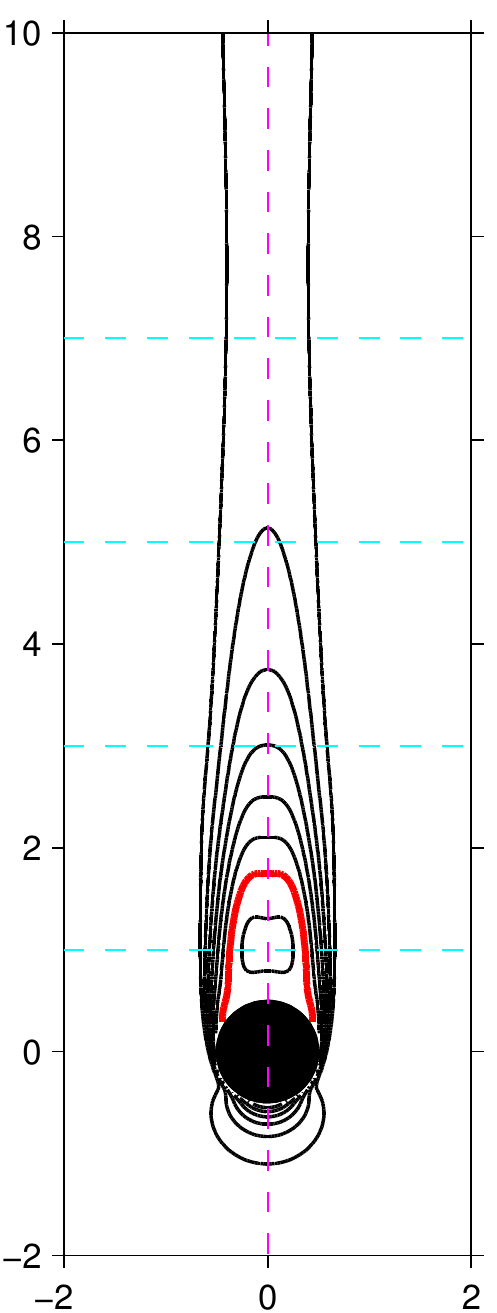}
    \centerline{${x}_{pHz\perp}$}
  \end{minipage}
  \begin{minipage}{.188\linewidth}
    \centerline{$(i)$}
    \includegraphics[width=\linewidth]
    {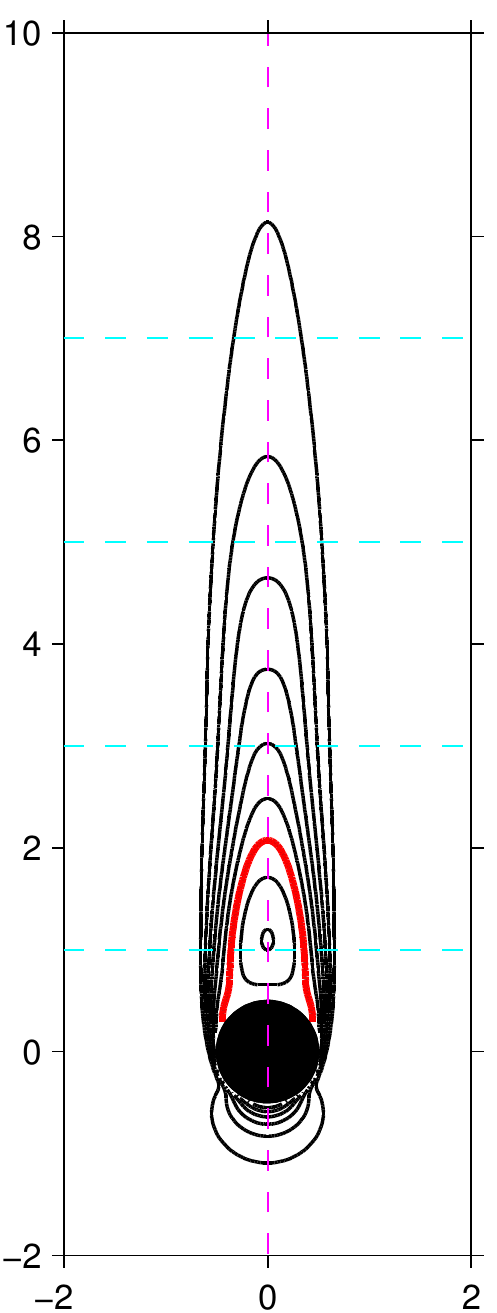}
    \centerline{${x}_{pHz\perp}$}
  \end{minipage}
  \begin{minipage}{.188\linewidth}
    \centerline{$(j)$}
    \includegraphics[width=\linewidth]
    {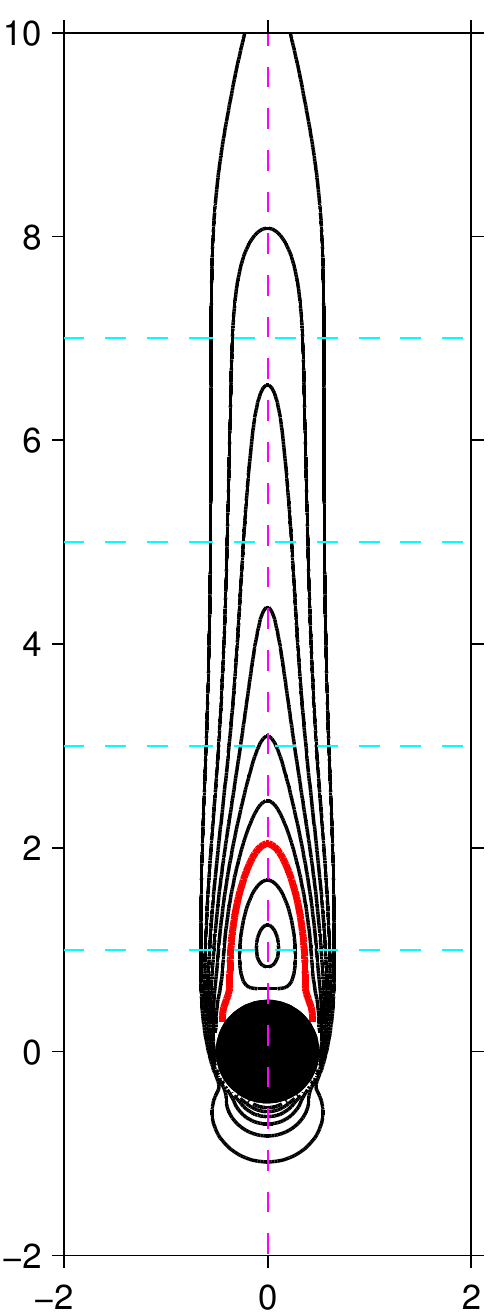}
    \centerline{${x}_{pHz\perp}$}
  \end{minipage}
  }{
  \caption{%
    \revision{%
      As figure~\ref{fig-results-ref-3d-iso-u-c1l}, but showing 
      contours of the projected relative velocity
      $u_{r\,\parallel}$.}{%
      As figure~\ref{fig-results-ref-3d-iso-lambda2-c1l}, but showing 
      contours of the projected relative velocity
      $u_{r\,\parallel}$.}
    In $(a$-$e)$ a plane spanned by the vertical
    direction $\mathbf{e}_z$ and the direction of the particle motion
    $\mathbf{e}_{p\parallel}$, passing through the sphere's center, is
    chosen. 
    In $(f$-$j)$ the plane spanned by
    $\mathbf{e}_{pHz\perp}$ and $\mathbf{e}_{p\parallel}$, 
    passing through the sphere's center, is chosen.
    The red line marks the instantaneous extent of the recirculation
    region (i.e.\ $u_{r\,\parallel}=0$). 
    The magenta colored dashed line indicates the direction of the particle
    motion given by the unit vector $\mathbf{e}_{p\,\parallel}$.
    %
    \protect\label{fig-results-ref-contour-c1l}
  }
  }
\end{figure}
In the second regime, here simulated with a Galileo value of
$G=178.46$, the particle wake (in the fully developed state) is no
longer axi-symmetric, but still steady. Therefore, the particle motion
is along a non-vertical straight path. Consequently, in addition to
the vertical particle velocity component, $u_{pV}$, the horizontal one
($u_{pH}$, along unit vector $\mathbf{e}_{pH}$) is non-zero;
furthermore, the particle 
rotates with angular velocity 
$\omega_{pH}$ around the horizontal axis
$\mathbf{e}_{pHz\perp}$ (i.e.\ perpendicular to $\mathbf{e}_{pH}$).  
Table~\ref{tab-results-ref-B} lists the numerical values obtained in
our simulations for both domain sizes. Again, it is found that the
influence of the domain size is very small. 
An interesting
parameter is the angle of particle motion with respect to the
vertical, whose tangent is given by the ratio of horizontal to
vertical amplitude, viz.
\begin{equation}\label{equ-res-ref-steady-oblique-def-angle}
  \tan(\alpha)=\frac{u_{pH}}{|u_{pV}|}
  \,.
\end{equation}
\revision{%
  The values for the angle $\alpha$ are $5.323$ and $5.225$ for case BS
  and BL, respectively (in degrees).}{%
  The values for the angle $\alpha$ are $5.323$ and $5.225$ degrees
  for case BS and BL, respectively.}

Figure~\ref{fig-results-ref-3d-iso-b1l} provides an impression of the
flow field around the sphere in 
\revision{case BL.}{%
  case BL, showing iso-surfaces of the relative velocity
  $u_{r\parallel}$ and of $\lambda_2$ as seen from two different angles. 
}
In particular, the
iso-surface plot of $\lambda_2$ reveals the double-threaded structure
of the wake, with both threads lying slightly off-center, therefore
generating a horizontal force component upon the sphere. 
This observation is confirmed in
figure~\ref{fig-results-ref-contour-b1l}$(a,b)$, where contours of the  
relative velocity 
\revision{(projected upon the sphere's motion vector)}{%
  (projected upon the direction opposite to the sphere's velocity
  relative to the ambient fluid) 
}
are shown in two perpendicular planes. The maximum extent of the
recirculation region is found off center (i.e.\ the recirculation
length $L_r$ is larger than $L_{r\parallel}$, cf.\
table~\ref{tab-results-ref-B}). 
\revision{For future reference,
  figure~\ref{fig-results-ref-contour-b1l}$(c)$ shows the relative
  fluid velocity $u_{r\parallel}$ along the axis defined by the
  particle velocity (i.e.\ along $\mathbf{e}_{p\parallel}$ through the
  sphere's center).}{%
  The relative fluid velocity $u_{r\parallel}$ along the axis defined
  by the   particle velocity (i.e.\ along $\mathbf{e}_{p\parallel}$
  through the sphere's center) is included in the supplementary
  data-set; it will also be used in the comparison with the IBM data
  below (cf.\ figure~\ref{fig-results-ibm-wake-deficit-steady-oblique}
  and \ref{fig-results-ibm-dthalf-wake-deficit-steady-oblique}).} 
%

\begin{figure}
  \figpap{
  \centering
  \begin{minipage}{3ex}
    $w_{pr}$
  \end{minipage}
  \begin{minipage}{.45\linewidth}
    \centerline{$(a)$}
    \includegraphics[width=\linewidth]
    {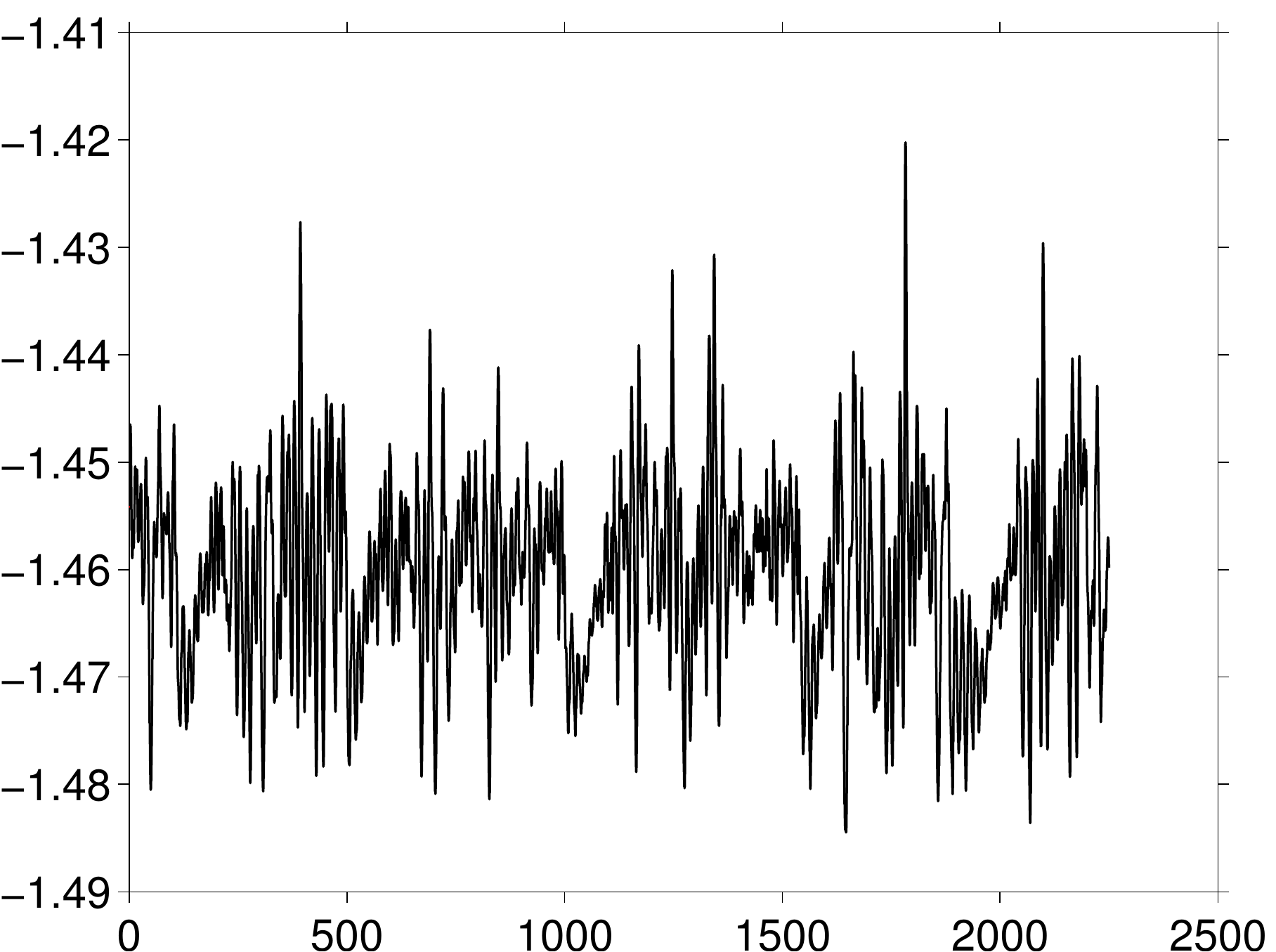}
      \\
    \centerline{$t$}
  \end{minipage}
  \hfill
  \begin{minipage}{3ex}
    $u_{pr}$
  \end{minipage}
  \begin{minipage}{.35\linewidth}
    \centerline{$(b)$}
    \includegraphics[width=\linewidth]
    {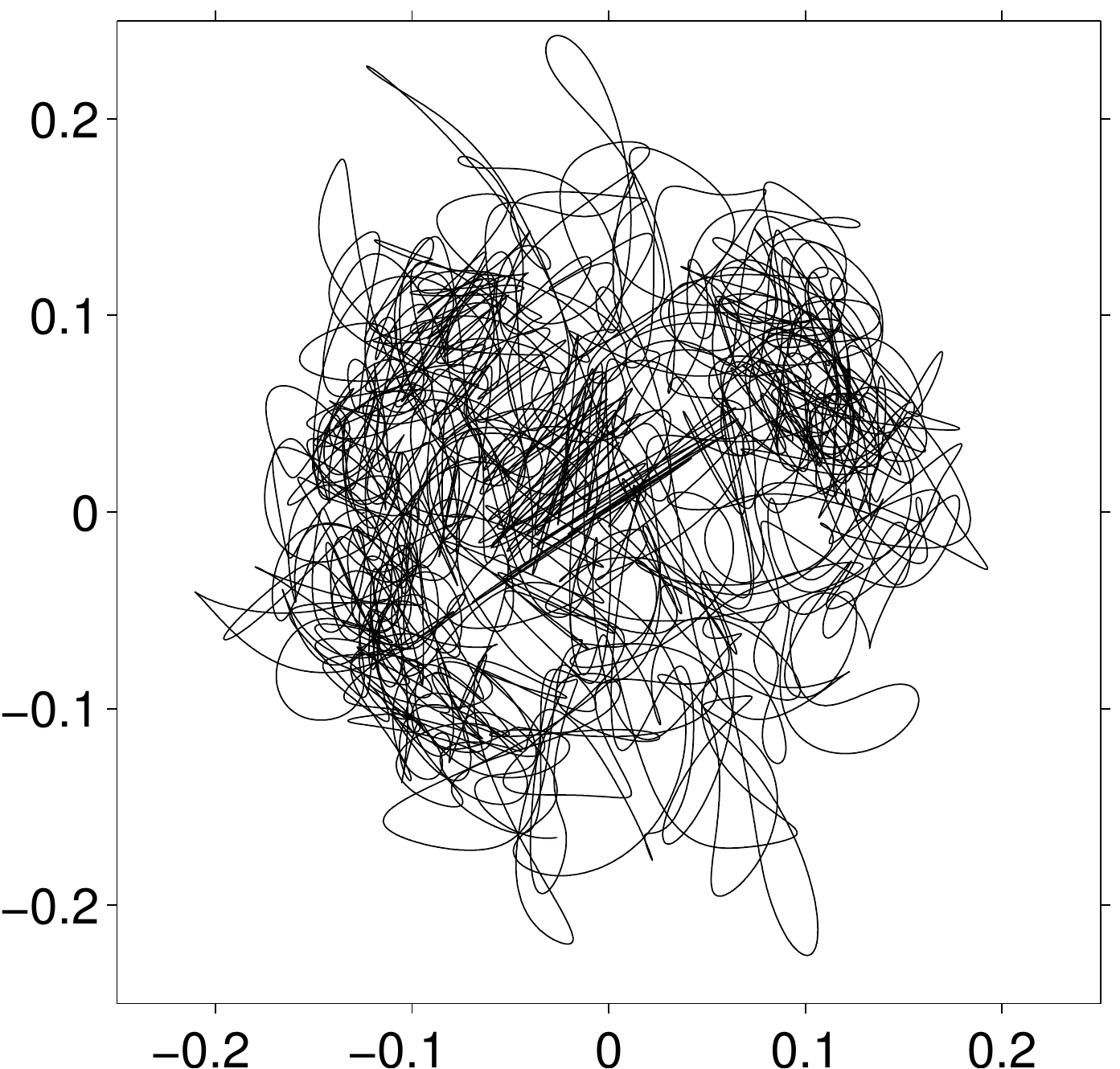}
    \\
    \centerline{$v_{pr}$}
  \end{minipage}
  }{
  \caption{%
    Reference results for case DL ($G=250$), exhibiting chaotic
    dynamics. 
    $(a)$ Temporal evolution of the vertical particle velocity
    component (measured relative to the ambient fluid velocity). 
    $(b)$ Phase-space plot in the two-dimensional space spanned by the
    two horizontal particle velocity components. 
    \protect\label{fig-results-ref-history-d1l}
  }
  }
\end{figure}
%
\revision{
  For the purpose of providing benchmark data, relative velocity
  profiles along the directions $\mathbf{e}_{pHz\perp}$ (at
  $x_{p\perp}=0$) and along $\mathbf{e}_{p\perp}$ (at $x_{pHz\perp}=0$)
  are shown in figure~\ref{fig-results-ref-cross-profiles-b1l} for
  various distances $x_{p\parallel}=\{-1,-3,-5,-7\}$ downstream of the
  rear stagnation point.  
  From the plots in
  figure~\ref{fig-results-ref-cross-profiles-b1l}$(a,c)$ it can be
  clearly seen how the symmetry is broken, i.e.\ the wake is more
  intense for $x_{pHz\perp}>0$. 
  Contrarily, the symmetry along the direction $\mathbf{e}_{p\perp}$ (at
  $x_{pHz\perp}=0$, crossing the axis of the sphere's motion) is
  retained, as can be seen from the plots in
  figure~\ref{fig-results-ref-cross-profiles-b1l}$(b,d,f)$. 
  Note that the velocity component $u_{rHz\perp}$ along
  $\mathbf{e}_{pHz\perp}$, cf.\
  figure~\ref{fig-results-ref-cross-profiles-b1l}$(e)$, is identically
  zero due to the retained symmetry.}{%
  Note that relative velocity
  profiles along the directions $\mathbf{e}_{pHz\perp}$ (at
  $x_{p\perp}=0$) and along $\mathbf{e}_{p\perp}$ (at $x_{pHz\perp}=0$)
  at 
  various distances $x_{p\parallel}=\{-1,-3,-5,-7\}$ downstream of the
  rear stagnation point 
  are included in the supplementary data-set. This information is used
  in the comparison with the IBM data in
  figure~\ref{fig-results-ibm-cross-profiles-b1c-24} below.} 
%

The surface pressure data for case BL is visualized in
figure~\ref{fig-results-ref-press-b1l}. 
The contours projected upon the horizontal plane are roughly
similar to the axi-symmetric case (cf.\
figure~\ref{fig-results-ref-press-a1l}). 
However, the pressure maximum on the upstream side is now
shifted towards a small positive value of $x_{pHz\perp}$
(approximately coinciding with the point where the axis of motion
crosses the sphere's surface, cf.\ blue cross in
figure~\ref{fig-results-ref-press-b1l}$a$). 
Contrarily, on the downstream side -- due to the non-axi-symmetric wake
-- the local pressure maximum considerably moves off center (red
circle in figure~\ref{fig-results-ref-press-b1l}$b$), much more
than the inclination of the sphere's axis of motion (blue cross in
figure~\ref{fig-results-ref-press-b1l}$b$).  
\subsubsection{Oscillating oblique regime}
\label{sec-ref-results-oscill-oblique}
At a value of the Galileo number of $G=190$ the asymptotic state of
the particle motion is characterized by a periodic temporal
evolution. 
Particle motion still takes place in a single plane given by the
(time-independent) vectors $\mathbf{e}_z$ and $\mathbf{e}_{pH}$ (as in
the steady oblique regime of \S~\ref{sec-ref-results-steady-oblique}), 
but the quantities themselves are time-dependent. 
Therefore, in this regime there are three non-zero, time-dependent
components of the translational and angular particle velocity, i.e.\
$u_{pH}(t)$, $u_{pV}(t)$ and $\omega_{pH}(t)$.   
As can be seen from figure~\ref{fig-results-ref-history-c1l}, 
the signals of these quantities are similar to a single harmonic, but
a closer analysis reveals that they are in fact anharmonic. 
Instead of providing a complete fit to a multicomponent sine base
(which appears to converge only slowly), we provide simple measures of
the oscillating signals as follows. 
The mean and the amplitude of a particle-related quantity $\phi(t)$ are
defined from the maxima and minima as 
\begin{equation}\label{equ-res-ref-def-mean}
  \overline{\phi}\equiv
  \frac{\phi_{max}+\phi_{min}}{2}
  \,,
\end{equation}
and 
\begin{equation}\label{equ-res-ref-def-ampli}
  \phi^\prime\equiv
  \frac{\phi_{max}-\phi_{min}}{2}
  \,,
\end{equation}
respectively.
The oscillation period $T$ is determined from a count of the
zero-crossings of the fluctuation values $\phi-\overline{\phi}$ over a
time sufficiently larger than (i.e.\ several multiples of) the
period. The oscillation frequency $f$ is then simply obtained as the 
inverse of $T$.

Table~\ref{tab-results-ref-C} lists the numerical values describing
the oscillating signals. 
First, it is once more observed that the
lateral domain size does not play a significant role, as the
difference between choosing $D_\Omega=5.34$ and $7.54$ is below 
0.5\% (0.2\%) of the value $u_{pV}$ in case CS for the translational
(angular) velocity components. The oscillation frequency $f$
differs by approximately 4\% between cases CS and CL.   
%
%
Secondly, it is found that the mean values for translational and angular
particle velocities are similar to the steady-state values obtained at
$G=178.46$. 
Note that the oscillation frequency is indeed small. The observed
period $T$ corresponds roughly to the time during which the sphere has
covered a vertical distance of 20 diameters. 

\revision{%
  Figures~\ref{fig-results-ref-3d-iso-u-c1l} and
  \ref{fig-results-ref-3d-iso-lambda2-c1l} show visualizations of the
  flow field (in terms of iso-surfaces of the relative velocity projected
  upon the sphere's axis of motion and of $\lambda_2$, respectively) for
  five snapshots equally distributed over one oscillation cycle. 
  It can be seen that the shape of the wake significantly varies over
  the oscillation period. In particular, it is found that the wake
  region sways back and forth in the plane of particle motion (i.e.\
  along the direction pointed by $\mathbf{e}_{p\perp}$), with the
  displacement length increasing at increasing distance downstream of
  the sphere. 
  The vortical structure in the near-field of the sphere
  (figure~\ref{fig-results-ref-3d-iso-lambda2-c1l}) still principally
  exhibits a double-threaded character, much alike case BL. 
  However, over one oscillation period the vortex threads first grow in
  axial length, then detach from the sphere, whence a new double thread
  is formed.}{%
  The graphs in figure~\ref{fig-results-ref-3d-iso-lambda2-c1l}
  show visualizations of the flow field in terms of iso-surfaces of
  $\lambda_2$ for five snapshots equally distributed over one
  oscillation cycle.  
  It can be seen that the shape of the wake significantly varies over
  the oscillation period. 
  Although the vortical structure in the near-field of the sphere
  still principally
  exhibits a double-threaded character (much alike case BL), 
  over one oscillation period the vortex threads first grow in
  axial length, then detach from the sphere, whence a new double thread
  is formed.}
A more detailed picture of the cycle is provided by the sequence of
graphs of the contours of the projected relative velocity
$u_{r\parallel}$ shown in figure~\ref{fig-results-ref-contour-c1l}. 
The swaying in direction $\mathbf{e}_{p\perp}$ is confirmed, while the
simultaneous temporal evolution of the recirculation region can be
observed (figure~\ref{fig-results-ref-contour-c1l}$a$-$e$). 
In the plane given by $\mathbf{e}_{p\parallel}$ and
$\mathbf{e}_{pHz\perp}$
(figure~\ref{fig-results-ref-contour-c1l}$f$-$j$), on the other hand,  
it can be seen that the axial growth of the vortical structures in the
wake first causes a stretching and thinning of the wake, and then a
subsequent retraction and further growth in axial extent. 
In order to quantify the variation of the recirculation length $L_r$ 
over each period, we have defined an average
and a fluctuation value, henceforth denoted by $\overline{L}_r$ and
$L_r^\prime$, respectively, using the definitions given in
(\ref{equ-res-ref-def-mean}) and (\ref{equ-res-ref-def-ampli}), and
using data from the five snapshots shown in
figure~\ref{fig-results-ref-contour-c1l}.  
The values for case CL are listed in table~\ref{tab-results-ref-C}. 
We observe that the mean recirculation length $\overline{L}_r$ is
slightly larger than in case BL (same domain, but $G=178.46$). The
amplitude of the fluctuations $L_r^\prime$ measures approximately 4\%
of the mean value.  
\revision{%
  Finally, for completeness we present in
  figure~\ref{fig-results-ref-press-profiles-c1l} the surface pressure
  variation along two perpendicular great circles for the same sequence
  of snapshots. 
  While the changes over one cycle are nearly unnoticeable on the
  upstream side, the location of the local pressure maximum on the
  downstream side (figure~\ref{fig-results-ref-press-profiles-c1l}$b$,
  small negative values of $\theta_2$) considerably moves back and forth
  along the great circle (in the plane of particle motion), consistently
  with the observed lateral swaying of the wake.}{%
  Note that the surface pressure
  variation along two perpendicular great circles for the same sequence
  of snapshots is contained in the supplementary data-set (figure
  omitted). 
}
\subsubsection{Chaotic regime}
\label{sec-ref-results-chaotic}
\begin{table}
  \centering
  \setlength{\tabcolsep}{5pt}
  \begin{tabular}{*{7}{c}}
    case&
    $\langle{u}_{pV}\rangle$&
    %
    $\langle u_{pV}^{\prime\prime}u_{pV}^{\prime\prime}\rangle^{1/2}$&
    $\langle u_{pr}^{\prime\prime}u_{pr}^{\prime\prime}\rangle^{1/2}$&
    $\langle\omega_{pV}^{\prime\prime}\omega_{pV}^{\prime\prime}\rangle^{1/2}$&
    $\langle\omega_{px}^{\prime\prime}\omega_{px}^{\prime\prime}\rangle^{1/2}$
    &
    \revision{}{%
      $Re_\parallel$
    }
    \\[1ex]
    DL&
    $-1.4604$& 
    $0.0087$& 
    $0.0854$& 
    $0.0013$& 
    $0.0067$  
    &
    \revision{}{%
      $365.10$
    }
  \end{tabular}
  \caption{
    Reference results for cases DL in the chaotic flow regime
    ($G=250$), accumulated over a sampling period of $2250$ time
    units.  
    \revision{}{%
      The Reynolds number $Re_\parallel$ (defined in
      \ref{equ-res-ref-def-reynolds-par}) is computed from the average 
      velocity value. 
    }
  }
  \label{tab-results-ref-D}
\end{table}
\begin{figure}
  \figpap{
  %
  \begin{minipage}{.22\linewidth}
    \centerline{$(a)$}
    \includegraphics[width=\linewidth,clip=true,
    viewport=1050 827 1500 1750]
    {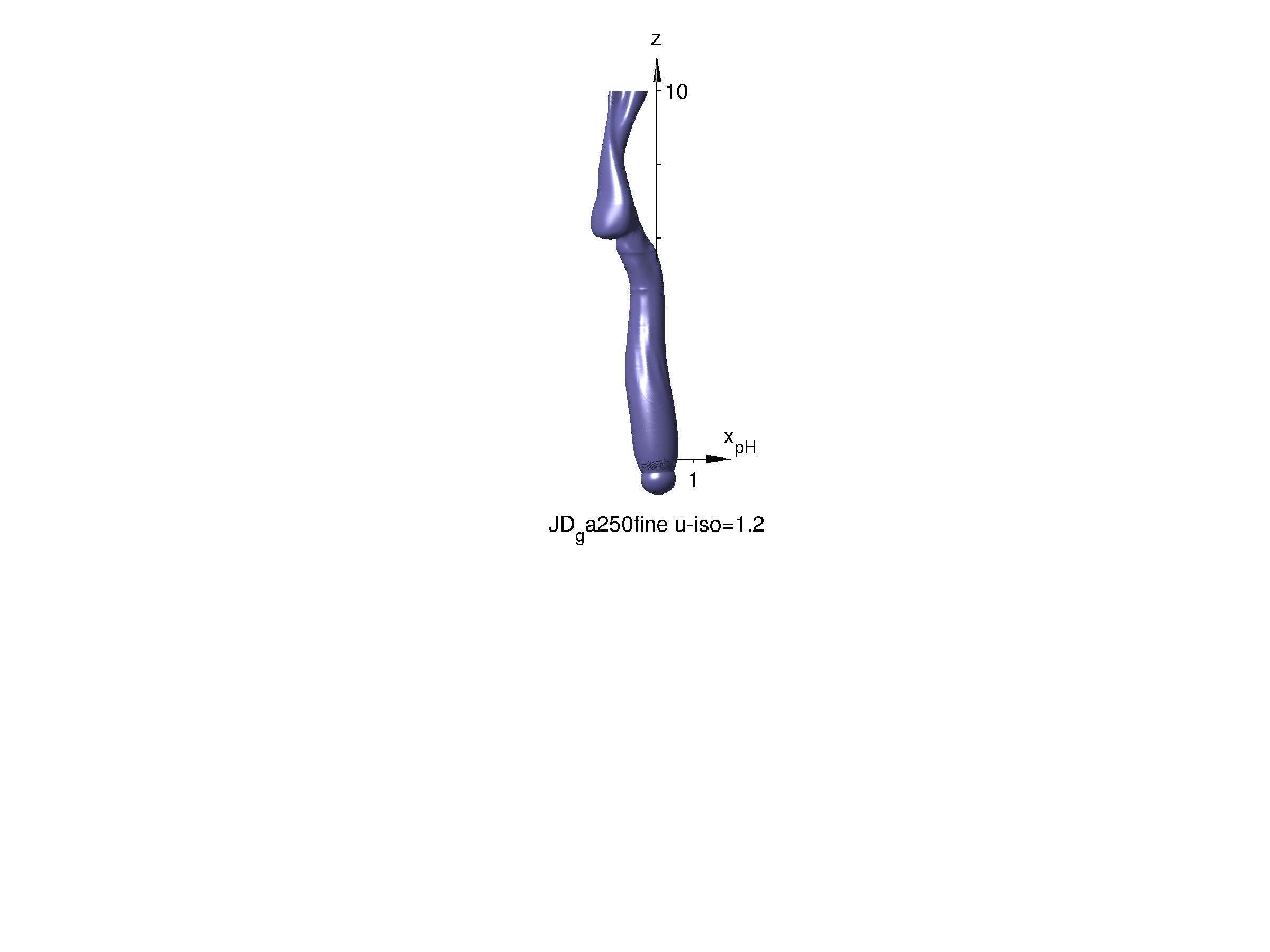}
  \end{minipage}
  \hfill
  \begin{minipage}{.22\linewidth}
    \centerline{$(b)$}
    \includegraphics[width=\linewidth,clip=true,
    viewport=1050 827 1500 1750]
    {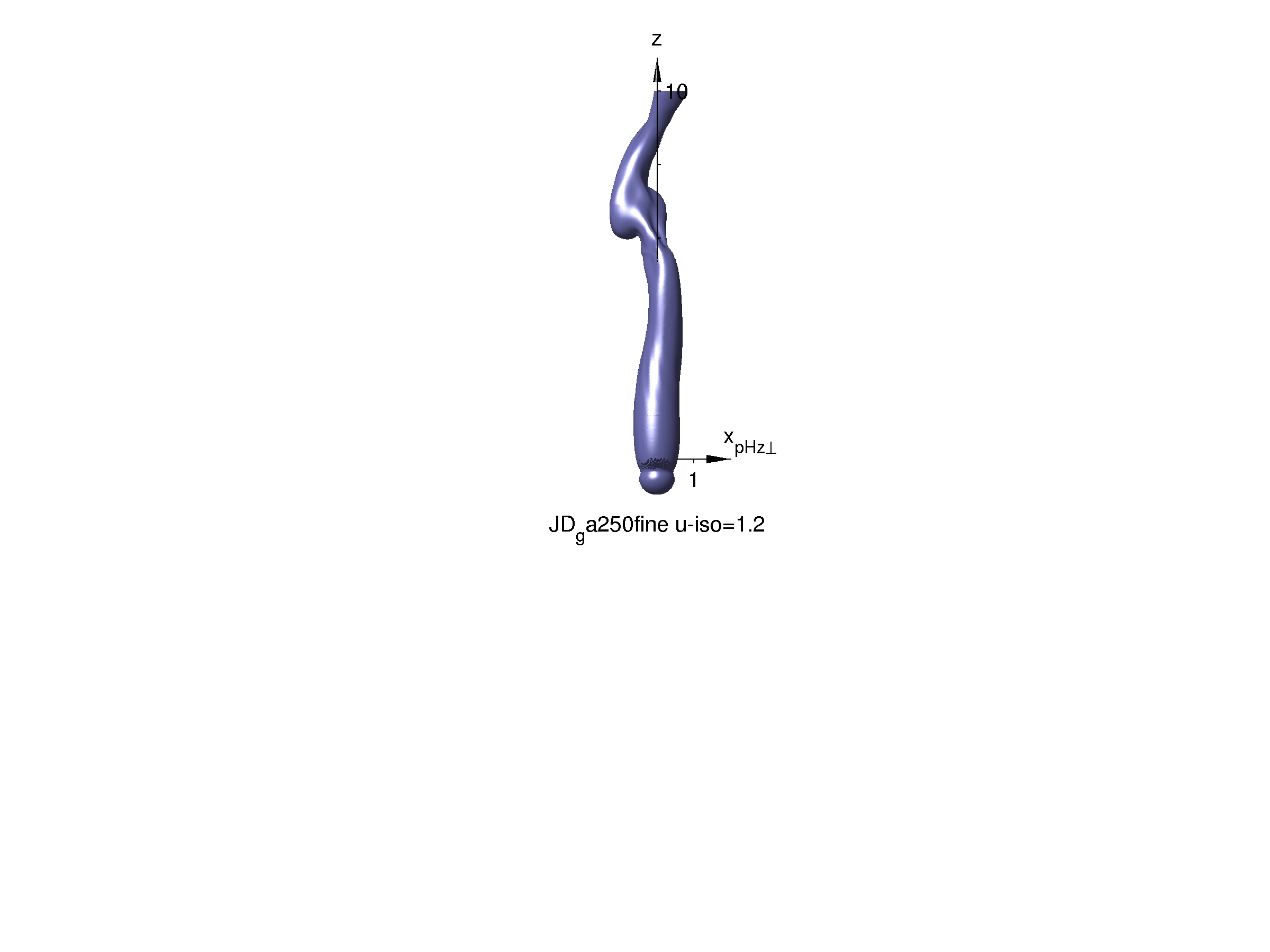}
  \end{minipage}
  \hfill
  \begin{minipage}{.22\linewidth}
    \centerline{$(c)$}
    \includegraphics[width=\linewidth,clip=true,
    viewport=1050 827 1500 1750]
    {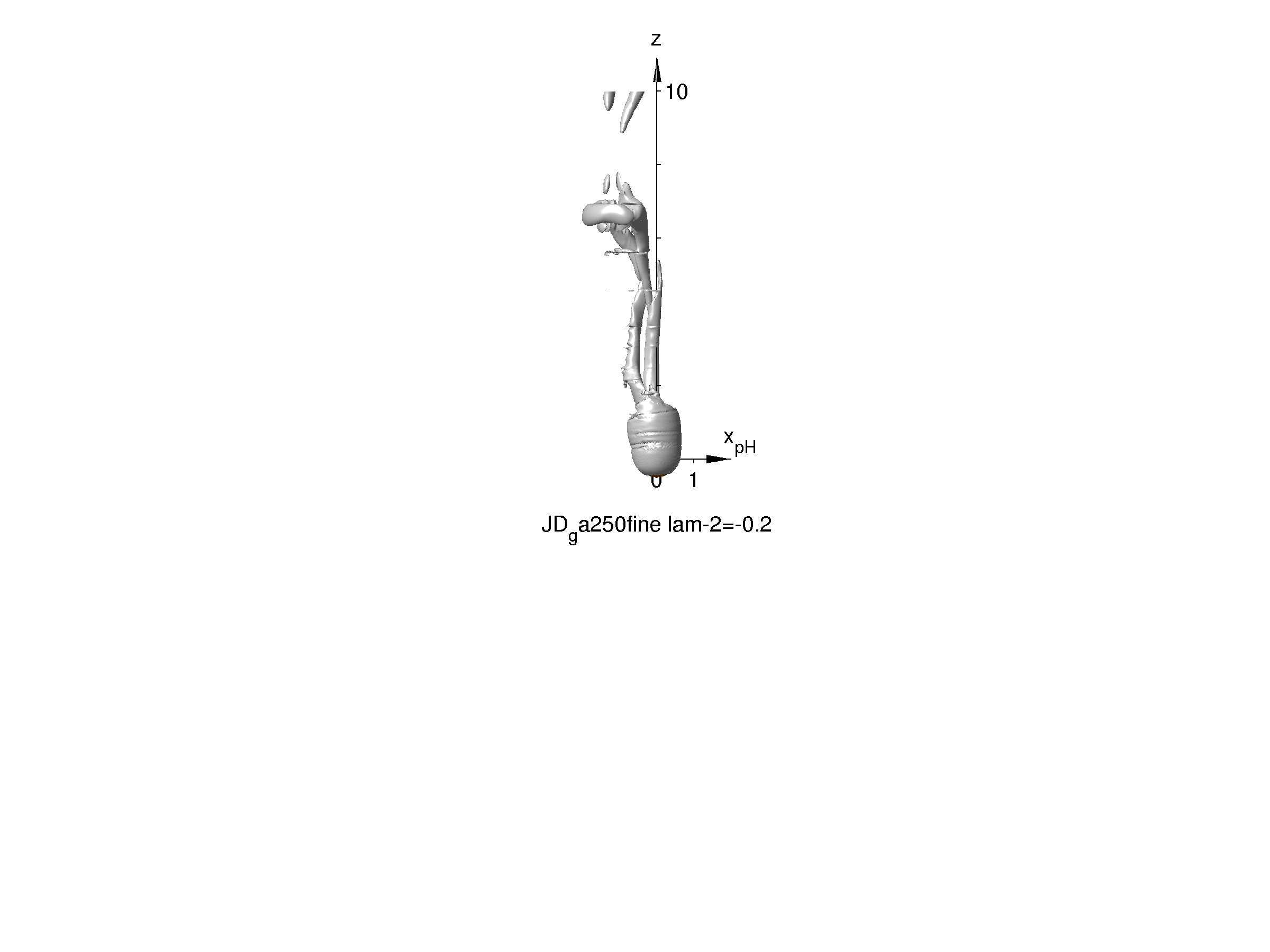}
  \end{minipage}
  \hfill
  \begin{minipage}{.22\linewidth}
    \centerline{$(d)$}
    \includegraphics[width=\linewidth,clip=true,
    viewport=1050 827 1500 1750]
    {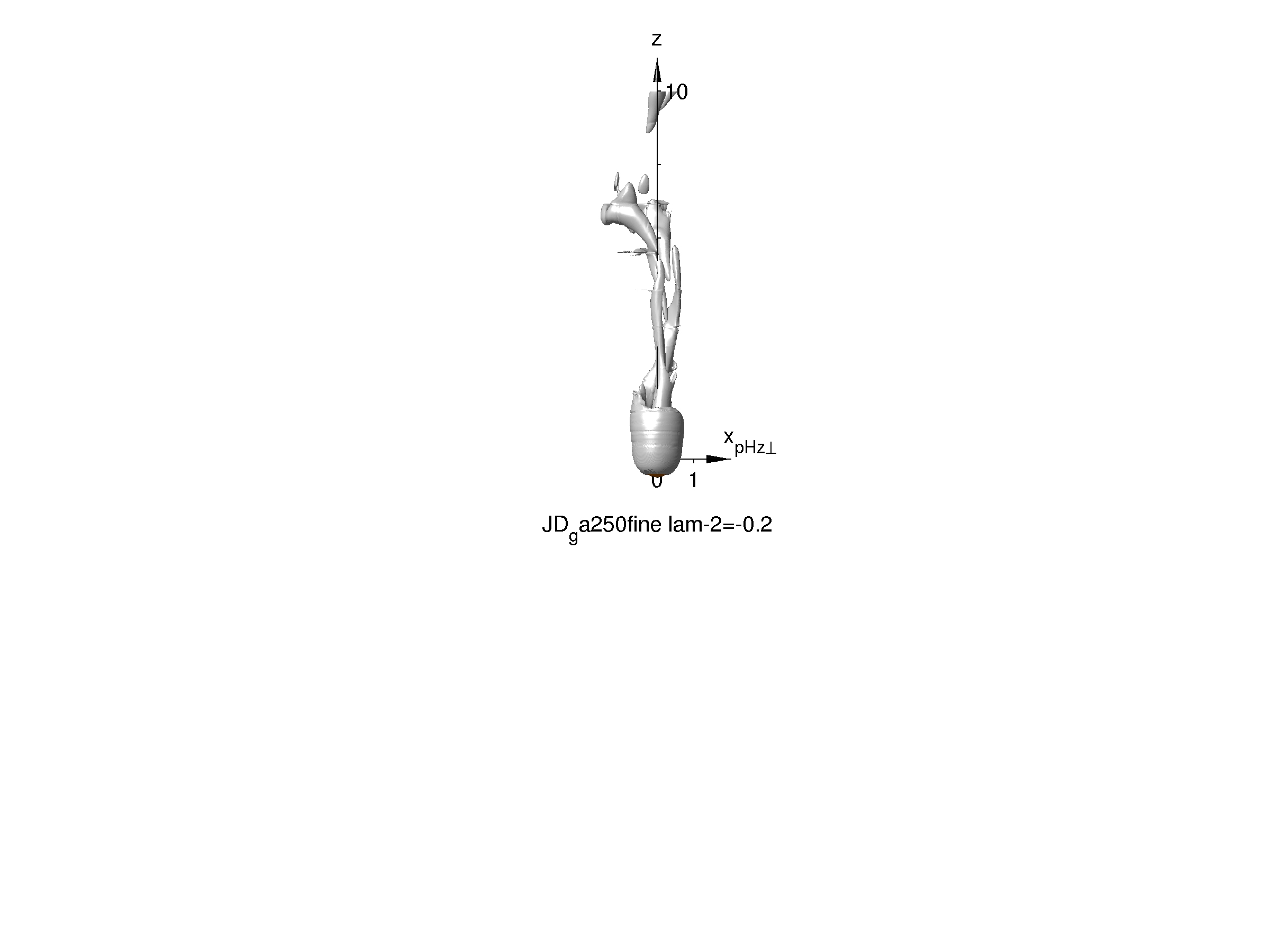}
  \end{minipage}
  }{
  \caption{%
    Reference results for case DL ($G=250$, chaotic dynamics),
    showing an instantaneous flow field. 
    Graphs $(a)$ and $(b)$ show the surface where $u_{r\parallel}=1.2$
    (corresponding to $0.82|\langle{u}_{pV}\rangle|$). 
    Graphs $(c)$ and $(d)$ show the surface where 
    $\lambda_2=-0.2$  
    (cf.\ definition in the text). 
    In $(a)$ and $(c)$ the view is directed along $\mathbf{e}_{pHz\perp}$;  
    in $(b)$ and $(d)$ it is directed along $\mathbf{e}_{pH}$.
    Note that some spectral-element boundaries are visible in $(c)$
    and $(d)$.  
    \protect\label{fig-results-ref-3d-iso-d1l}
  }
  }
\end{figure}
\begin{figure}
  \figpap{
  \centering
  \begin{minipage}{3ex}
    \rotatebox{90}{$\sigma\cdot pdf$}
  \end{minipage}
  \begin{minipage}{.45\linewidth}
    \centerline{$(a)$}
    \includegraphics[width=\linewidth]
    {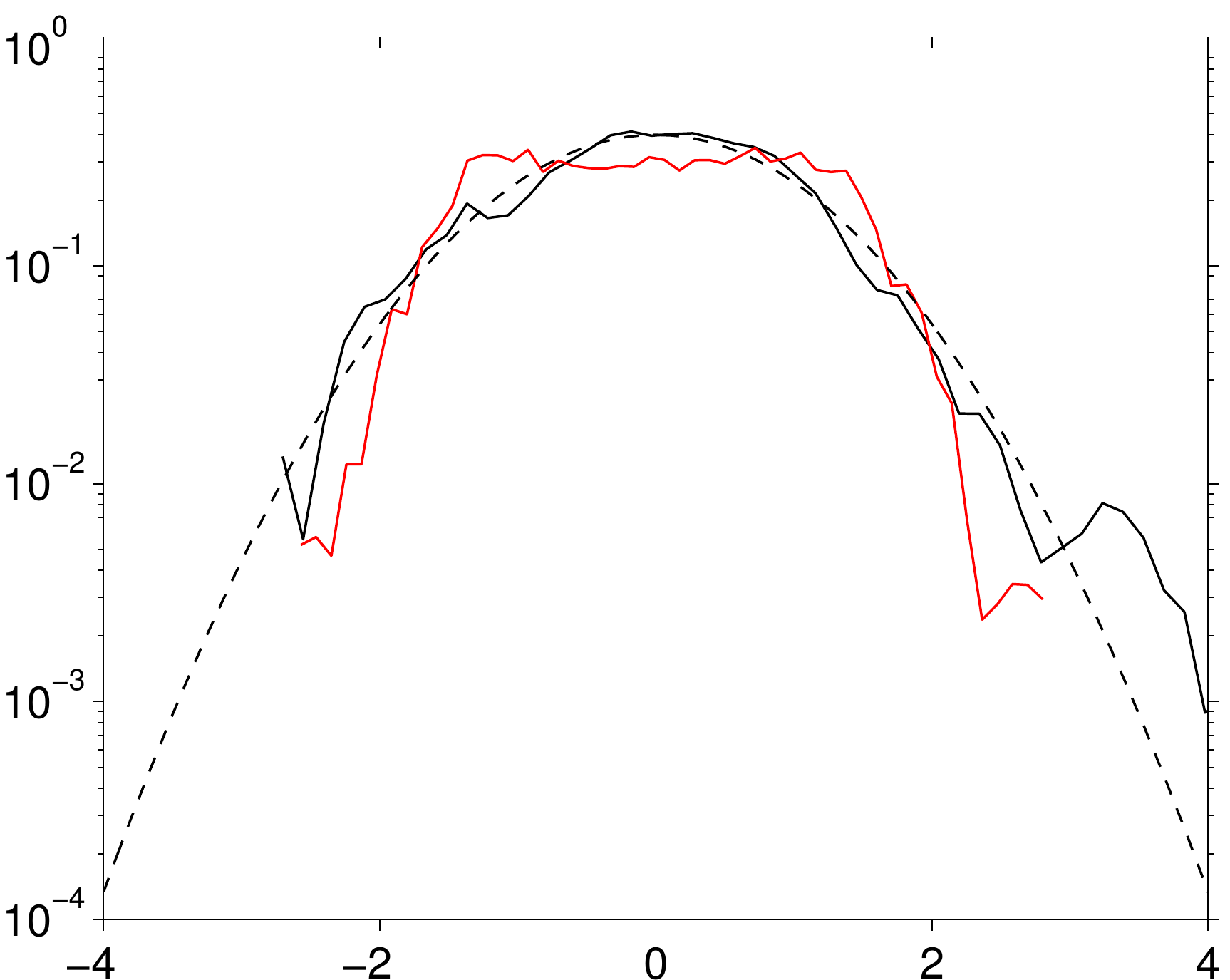}
    \\
    \centerline{$u_{p\alpha}^{\prime\prime}/\sigma$}
  \end{minipage}
  \hfill
  \begin{minipage}{3ex}
    \rotatebox{90}{$\sigma\cdot pdf$}
  \end{minipage}
  \begin{minipage}{.45\linewidth}
    \centerline{$(b)$}
    \includegraphics[width=\linewidth]
    {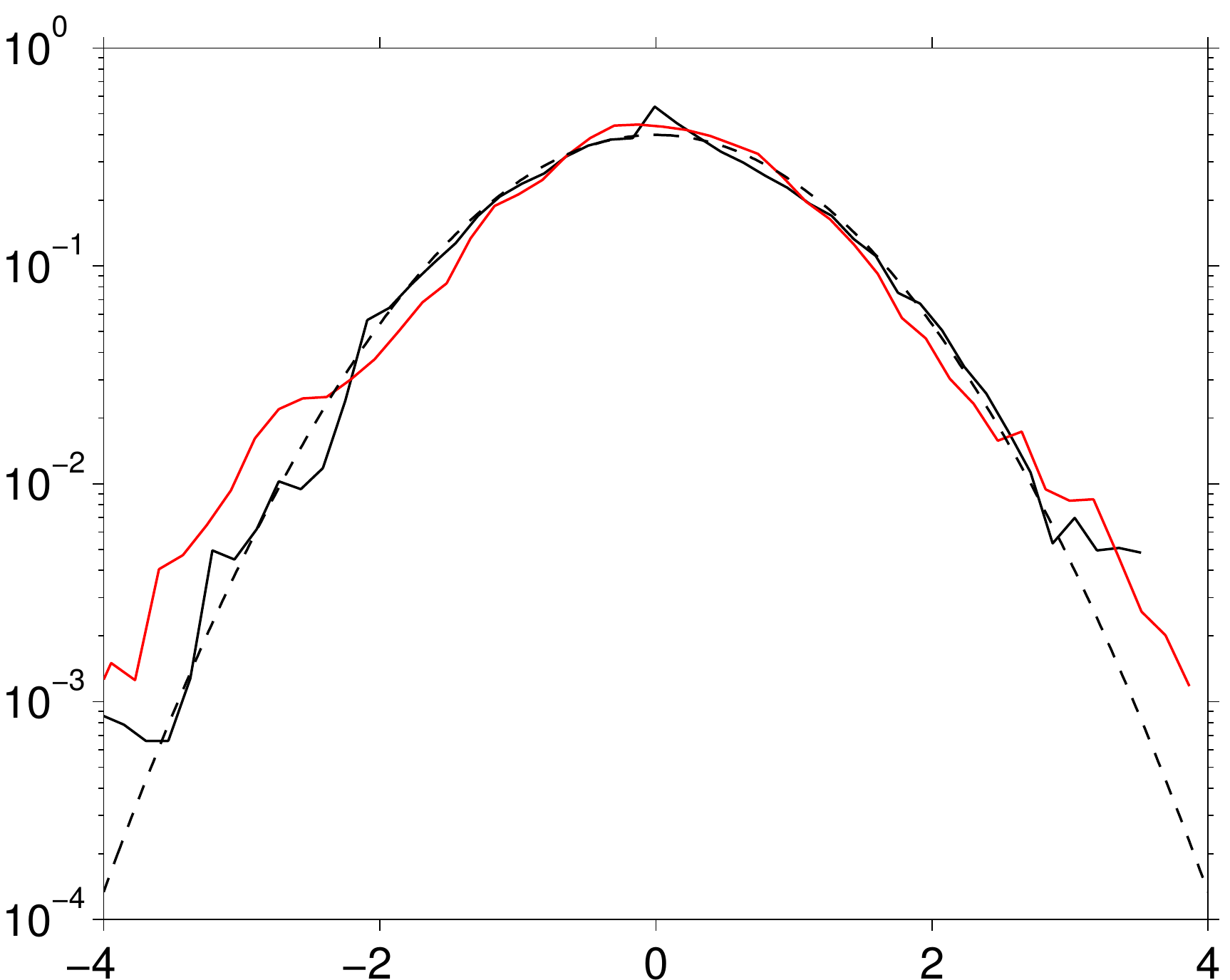}
    \\
    \centerline{$\omega_{p\alpha}^{\prime\prime}/\sigma$}
  \end{minipage}
  }{
  \caption{%
    Reference results for case DL ($G=250$). 
    Probability density functions of: 
    $(a)$ translational particle velocity; 
    $(b)$ angular particle velocity.
    The line-styles are as follows: 
    \solid, vertical component; 
    {\color{red}\solid}, horizontal component; 
    \dashed, Gaussian reference curve. 
    \protect\label{fig-results-ref-pdf-d1l}
  }
  }
\end{figure}
\begin{figure}
  \figpap{
  \centering
  \begin{minipage}{3ex}
    \rotatebox{90}{$R_{u_{p\alpha} u_{p\alpha}}$}
  \end{minipage}
  \begin{minipage}{.45\linewidth}
    \centerline{$(a)$}
    \includegraphics[width=\linewidth]
    {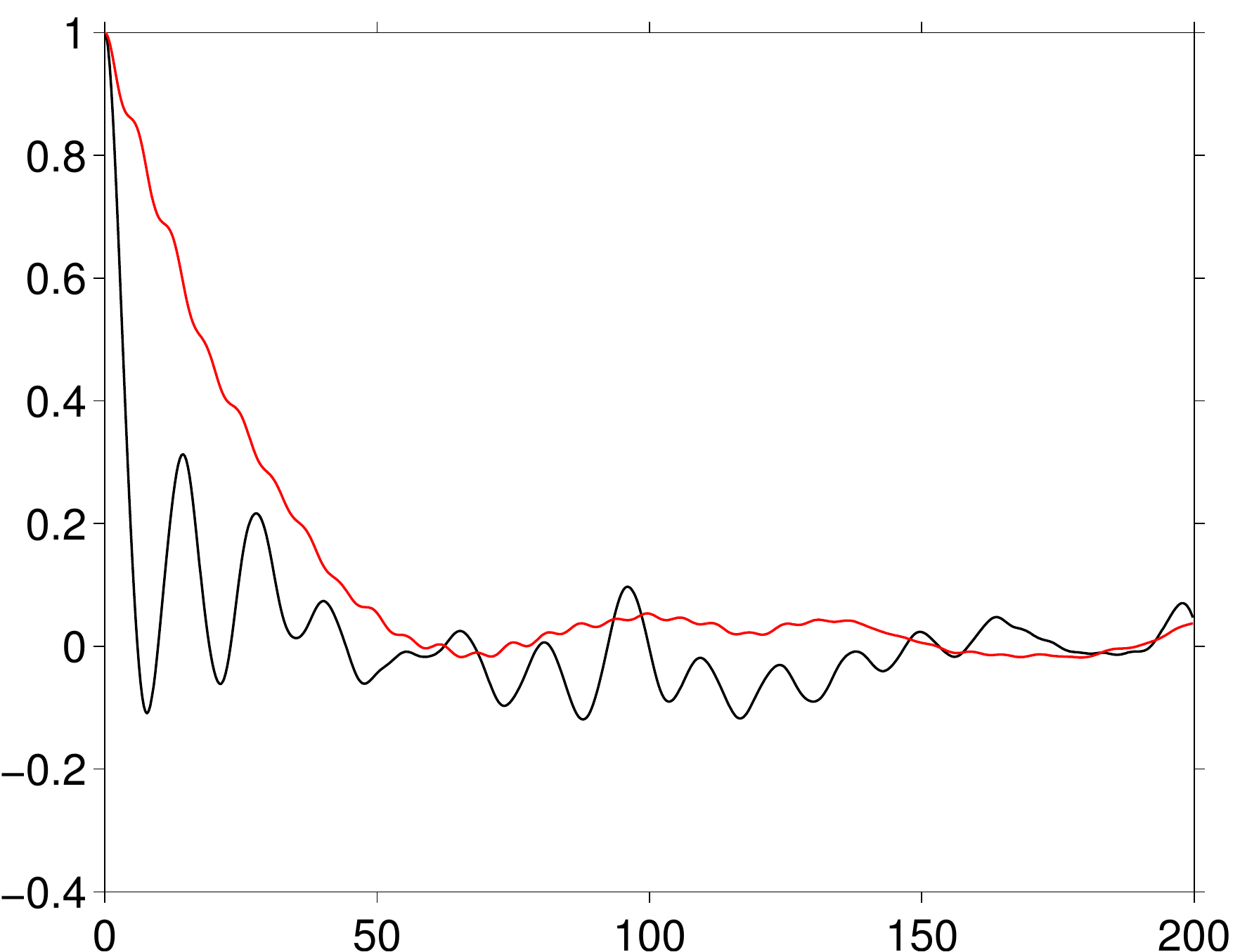}
    \\
    \centerline{$\tau_{sep}$}
  \end{minipage}
  \hfill
  \begin{minipage}{3ex}
    \rotatebox{90}{$R_{\omega_{p\alpha}\omega_{p\alpha}}$}
  \end{minipage}
  \begin{minipage}{.45\linewidth}
    \centerline{$(b)$}
    \includegraphics[width=\linewidth]
    {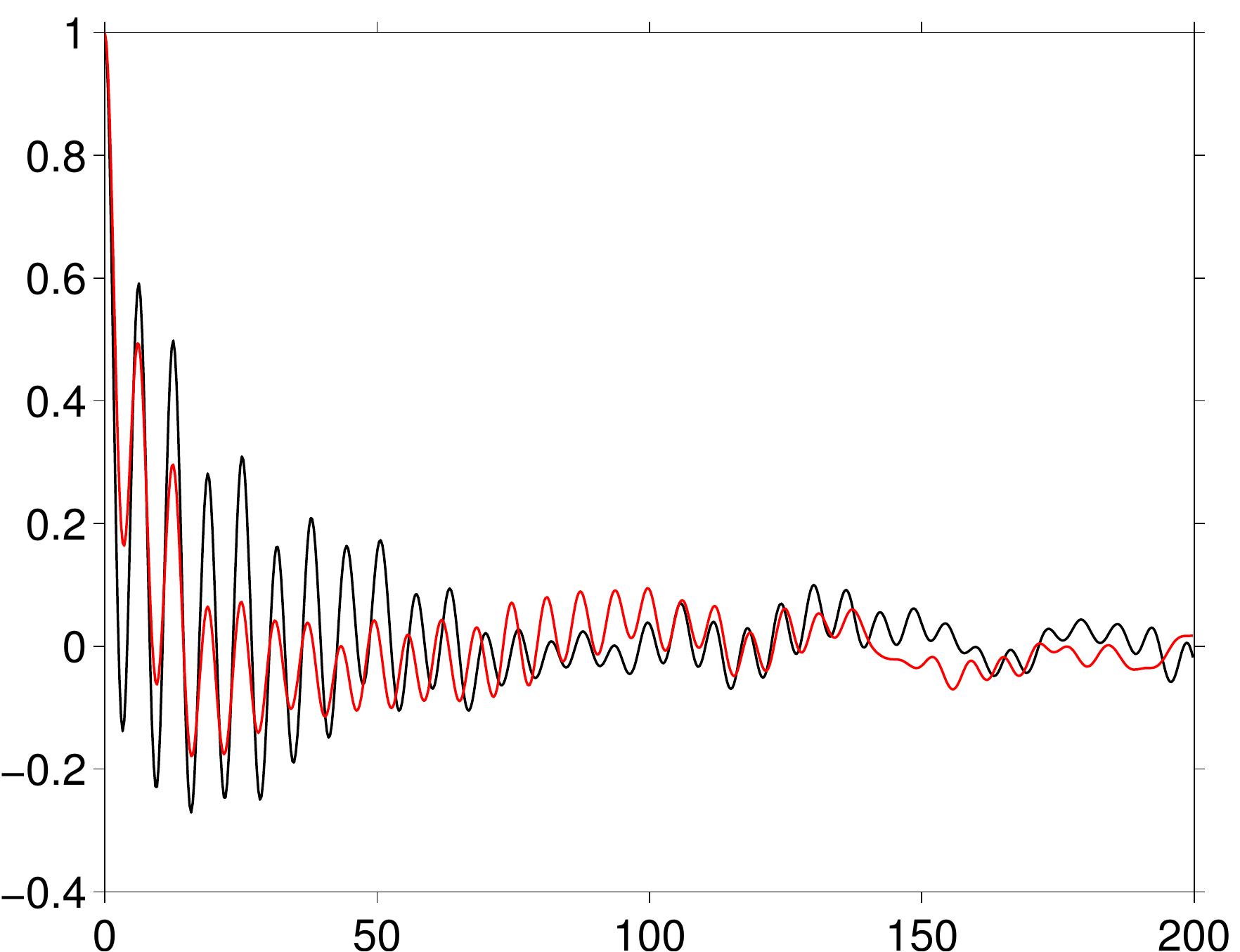}
    \\
    \centerline{$\tau_{sep}$}
  \end{minipage}
  }{
  \caption{%
    Reference results for case DL ($G=250$). 
    Temporal auto-correlations of: 
    $(a)$ translational particle velocity; 
    $(b)$ angular particle velocity.
    The line-styles are as follows: 
    \solid, vertical component; 
    {\color{red}\solid}, horizontal component.
    \label{fig-results-ref-corr-d1l}
  }
  }
\end{figure}
We have observed that the system settles into a chaotic state when the 
Galileo number is set to $G=250$ and the larger computational domain
is used (case DL). 
Contrarily, in the smaller domain (case DS) the system remains in a 
state characterized by zig-zagging motion in a vertical plane. 
%
%
It is interesting to note that the zig-zagging state is that
co-existing with the chaotic one and having the high, experimentally
evidenced, frequency $0.147$. At this Galileo number it is slightly
quasi-periodic. The chaotic and zig-zagging states do not co-exist
with the considered confinements.
We will henceforth concentrate upon case DL, which exhibits chaotic
dynamics. 

The temporal evolution of the vertical particle velocity  
component in case DL is shown in
figure~\ref{fig-results-ref-history-d1l}$(a)$, 
while figure~\ref{fig-results-ref-history-d1l}$(b)$ depicts a
phase-space diagram of the two horizontal velocity components. 
Substantial fluctuations of all degrees of freedom are recorded.  
Figure~\ref{fig-results-ref-3d-iso-d1l} shows the flow field at one
instant during the chaotic particle motion in 
\revision{case DL.}{%
  case DL (again showing iso-surfaces of the relative velocity
  $u_{r\parallel}$ and of $\lambda_2$ from two different view angles).
}
It can be seen
that in the near-field of the particle (up to approximately $4$
diameters downstream) the wake remains qualitatively similar to the above
cases at lower Galileo number, still exhibiting two principal
threads. However, further downstream the shape of the wake becomes
considerably more bent and twisted away from the direction of the
instantaneous particle motion. In particular, the vortical structure
exhibits a clear hairpin-like reconnection.  

In order to characterize the chaotic motion quantitatively, let us
define an average value of a particle-related quantity $\phi$ which
is expected to converge to the statistical average when a sufficiently
large number of samples is chosen, viz. 
\begin{equation}\label{equ-res-ref-def-stat-mean}
  \langle \phi \rangle 
  \equiv
  \frac{1}{\sum_{j=1}^{N_{runs}}N_t(j)}
  \sum_{j=1}^{N_{runs}}
  \sum_{i=1}^{N_t(j)}
  \phi(t_i^{(j)})
  \,.
\end{equation}
In (\ref{equ-res-ref-def-stat-mean}) the number of repetitions of the
``experiment'' is denoted by $N_{runs}$, the number of samples taken
in the $j$th ``experiment'' by $N_t(j)$, and $t_i^{(j)}$ is the time at
which the $i$th sample is taken in the $j$th ``experiment''.  
Note that the reference computation was only run once (i.e.\
$N_{runs}=1$), generating samples over an interval of approximately
$2250$ time units. 
In the chaotic regime the only particle velocity component which has a
non-zero mean is the vertical one; the mean angular particle velocity
is zero.  
Based upon the average defined in (\ref{equ-res-ref-def-stat-mean}),
we can define an instantaneous fluctuation around the mean value, i.e.
\begin{equation}\label{equ-res-ref-def-stat-fluct}
  \phi^{\prime\prime}(t)
  \equiv
  \phi(t)-\langle \phi\rangle
  \,.
\end{equation}
Table~\ref{tab-results-ref-D} lists the averages and fluctuation
amplitudes recorded in our simulation case DL. 
It can be seen that the fluctuations of the 
translational particle velocity in the horizontal plane are roughly a
factor of ten more intense than those of the vertical component. 
Please note that the quantity $\langle
u_{pr}^{\prime\prime}u_{pr}^{\prime\prime}\rangle^{1/2}$
measures the amplitude of the fluctuations of a velocity component
along one (fixed, but arbitrary) direction in the horizontal plane,
which is not the same as computing the rms of $u_{pH}$.  
Concerning the angular particle velocity, the ratio between a
component in the horizontal plane and the vertical component roughly
measures $5.2$.  
This result is interesting, since the large discrepancy between the
components should be measurable in laboratory experiments, where it
could equally serve for the purpose of validation.

Normalized probability density functions of the velocity components
are shown in figure~\ref{fig-results-ref-pdf-d1l}. 
It can be seen that the vertical component of the (translational) particle
velocity as well as all components of the angular particle velocity
are approximately Gaussian distributed. 
Interestingly, however, the horizontal component of the translational
particle velocity exhibits a plateau in the interval 
$u^{\prime\prime}_{pr}/\sigma(u^{\prime\prime}_{pr})\in[-1.4,1.4]$. For
values outside this interval the probability drops off sharply. 
%
%
%
This feature is due to a slow rotation of the original symmetry plane
found at these Galileo number values, with a near-helical trajectory
\citep[cf.\ ][]{jenny:04b}.

The Lagrangian auto-correlation function of a particle-related
quantity $\phi(t)$ is defined as
\begin{equation}\label{equ-res-ref-def-lag-corr}
  R_{\phi\phi}(\tau_{sep})
  \equiv
  \frac{1}{\int_{T_1}^{T_2}
  \phi(t)\,\phi(t)\,\mbox{d}t}
  \int_{T_1}^{T_2}
  \phi(t)\,\phi(t+\tau_{sep})\,\mbox{d}t
  \,.
\end{equation}
This quantity can provide the information on the temporal correlation
of the signals which has not been discussed up to this point. 
Figure~\ref{fig-results-ref-corr-d1l} shows the correlation functions
for the translational and angular velocity components. 
Again, the data for the vertical and horizontal components of the
translational particle velocity are fundamentally different. Whereas the
former 
rapidly drops to zero (first zero-crossing at $\tau_{sep}\approx6$),
the latter decays at a much slower rate  (first zero-crossing at
$\tau_{sep}\approx57$). 
Furthermore, the auto-correlation function of the vertical component
of the translational particle velocity has a marked superposed oscillation
with a period of approximately $13$ time units. This oscillating
feature is discernible for as long as $150$ time units. 
The horizontal component, on the other hand, exhibits only very weak
oscillations. 
Turning to the angular velocity signals, it can be observed from
figure~\ref{fig-results-ref-corr-d1l}$(b)$ that both vertical and
horizontal components show similar overall features, with a rapid
decay (first zero-crossing at $\tau_{sep}\approx2.5$ and $9$,
respectively) and marked superposed oscillations with a period of
approximately $6.5$ time units for both components. 
This latter oscillation period corresponds to the frequency
characterizing the ordered zig-zagging state coexisting with chaos at 
lower density ratios \cite{jenny:04b}.  
Note that the
oscillation period of the auto-correlation of the angular particle
components is half the value of the oscillation period of the
auto-correlation of the vertical component of the translational particle
velocity. 
%

%
\section{Immersed boundary computations}\label{sec-ibm}

\begin{figure}
  \figpap{
  \centering
  \begin{minipage}{.4\linewidth}
    \includegraphics[width=\linewidth,clip=true,
    viewport=90 375 400 700]
    {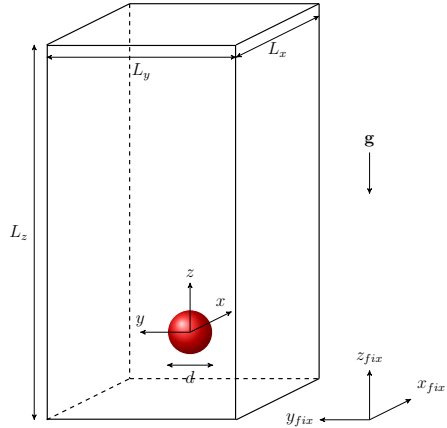}
  \end{minipage}
  }{
  \caption{%
    The computational domain as
    employed in the IBM computations described in \S~\ref{sec-ibm}. 
    The side-lengths of the cuboid were chosen as $L_x=L_y=5.34$ and
    $L_z=16$.  
    \protect\label{fig-ibm-schematic}
  }
  }
\end{figure}
\subsection{Numerical method}\label{sec-ibm-numa}
%
%
%
%
%
The numerical method employed in the current simulations is identical
to the one presented in \cite{uhlmann:04}.
The incompressible Navier-Stokes equations are solved by a fractional
step approach with implicit treatment of the 
viscous terms (Crank-Nicolson) and a three-step Runge-Kutta scheme for
the non-linear terms. The spatial discretization employs second-order
central finite-differences on a staggered mesh; the mesh is uniform
and isotropic. 
The no-slip condition at the surface of moving solid particles is
imposed by means of a specifically designed immersed boundary technique
\citep{uhlmann:04}. The motion of the particles is computed from the
Newton equations for translational and angular motion of rigid bodies, driven
by gravity and hydrodynamic forces/torque. The solid-fluid coupling
assures that the interaction forces cancel identically when
integrating over both phases. 

The numerical approach has been previously validated over a wide range
of flow configurations \citep{uhlmann:04,uhlmann:06c,uhlmann:07c}. 
%
It has been successfully employed for the simulation of various
large-scale systems involving many mobile particles 
\citep{uhlmann:08a,uhlmann:12a,villalba:12,kidanemariam:13}.   

The immersed boundary representation of particles allows for arbitrary
solid body motion with respect to the fixed computational grid. It is
this feature which makes the method suitable for the simulation of
many-body problems, where approaches such as the one employed in
\S~\ref{sec-ref} are not applicable. At the same time, the accuracy of
the representation of moving particles in the framework of
non-conforming methods, such as the IBM, needs to be carefully
established. For this purpose, we simulate the motion of a single
heavy sphere on a computational grid which is fixed in an inertial
frame. Consequently, the sphere is free to move across the
computational grid, and possible numerical perturbations due to the
sphere's translation are part of the errors to be gauged through the
validation process. 

The computational domain used in the IBM simulations is sketched in
figure~\ref{fig-ibm-schematic}. The horizontal cross-section of the
cuboid is square, with a side-length $L_x=L_y=5.34$, while the
vertical length of the domain measures $L_z=16$. This latter dimension
is slightly larger than the one used in the reference simulation of 
\S~\ref{sec-ref}. Concerning the horizontal cross-section, the
smaller domain used in the reference simulations corresponds to the 
inscribed circle of the IBM domain, while the larger domain in that
series is the circumscribed circle. 
%
%

In the IBM simulations the flow velocity at the horizontal inflow
plane (located at $z_{fix}=0$ in the inertial frame) is imposed, i.e.\
\begin{equation}\label{equ-ibm-inflow-vel}
  \mathbf{u}(x_{fix},y_{fix},z_{fix}=0,t)=\mathbf{u}_\infty=(0,0,w_\infty)
  \,.
\end{equation}
At the horizontal outflow plane ($z_{fix}=L_z$) a convective outflow
condition is employed \citep{uhlmann:04}. 
The pressure field is solved with a zero-gradient condition at the
inflow and outflow planes. 
In both horizontal directions periodicity of the flow field over the
periods $L_x$ and $L_y$, respectively, is imposed. 

It should be noted that the outer geometry and boundary conditions
used in the reference simulation and the IBM approach do not match in
the lateral directions. However, since the reference data shows that
the influence of the lateral domain size on the particle motion is
rather weak (cf.\
tables~\ref{tab-results-ref-A}-\ref{tab-results-ref-D}), it is
expected that the comparison is still conclusive. 
%

The grid width $\Delta x$ has been varied in the range $D/\Delta x=15$
to 
$48$, 
as given in tables~\ref{tab-results-ibm-steady-axi-symm} to 
\ref{tab-results-ibm-chaotic}. 
This corresponds to grid sizes from $241\times80\times80$ up to
$769\times 256\times 256$. 
The time step was adjusted such that the maximum CFL number was
approximately $0.3$, except where stated otherwise.
%
\revision{}{%
  Henceforth, the flow cases simulated with the IBM approach are
  denoted with two letters (the first indicating the flow regimes A to
  D, the second reading ``C'' as in ``Cartesian'') and two digits (for
  the number of mesh widths per sphere diameter) as in ``AC-15''.  
}

\begin{table}
  \centering
  \setlength{\tabcolsep}{5pt}
  \begin{tabular}{*{7}{c}}
    &
    $D/\Delta x$&
    $G$&
    $u_{pV}$&
    ${\cal E}(u_{pV})$&
    $L_r$&
    ${\cal E}(L_r)$
    \\[1ex]
    AC-15&
    $15$&
    $144.13$&
    $-1.2063$&
    $0.0612$&
    $1.3431$&   
    $0.0289$
    \\
    AC-18& 
    $18$&
    $144.18$&
    $-1.2131$&
    $0.0560$&
    $1.3688$&  
    $0.0102$
    \\
    AC-24& 
    $24$&
    $143.39$&
    $-1.2199$&
    $0.0507$&
    $1.3785$&  
    $0.0056$
    \\
    AC-36& 
    $36$&
    $142.91$&
    $-1.2274$&
    $0.0448$&
    $1.3835$&  
    $0.0004$
  \end{tabular}
  \caption{
    Results from IBM computations of case A (cf.\
    table~\ref{tab-parameters-ref}), where 
    $\rho_p/\rho_f=1.5$ and the nominal value of the Galileo number is 
    $G=144$. 
    The error is computed with respect to the results of the reference
    case AL (cf.\ table~\ref{tab-results-ref-A}). 
}    
  \label{tab-results-ibm-steady-axi-symm}
\end{table}
\begin{figure}[b]
  \figpap{
  \centering
  \begin{minipage}{3ex}
    $z$
  \end{minipage}
  \begin{minipage}{.2\linewidth}
    \centerline{$(a)$ AC-15}
    \includegraphics[width=\linewidth]
     {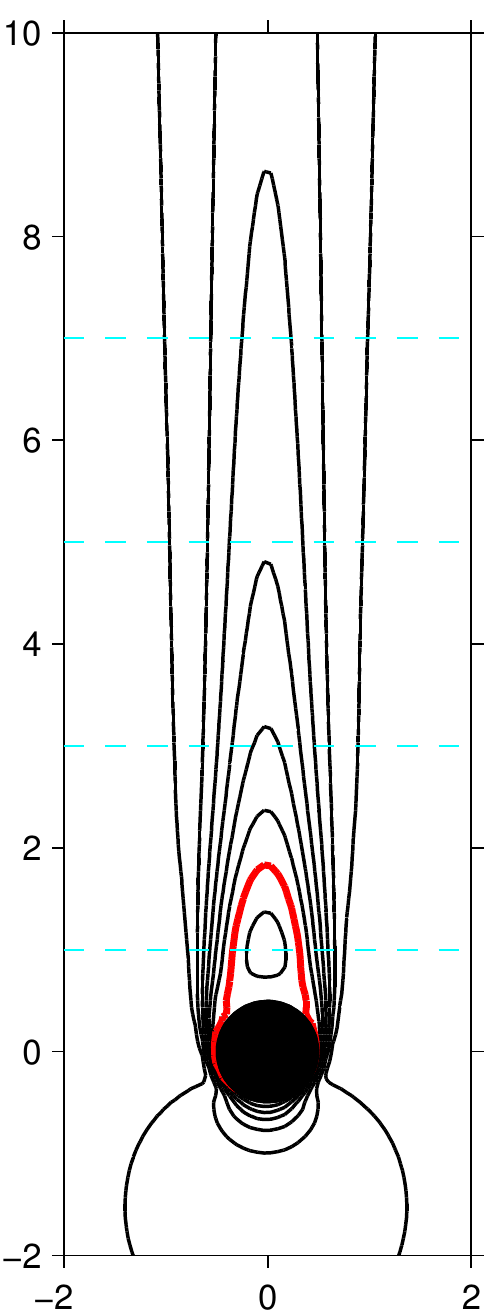}
    \centerline{$r$}
  \end{minipage}
  \begin{minipage}{.2\linewidth}
    \centerline{$(b)$ AC-18}
    \includegraphics[width=\linewidth]
    {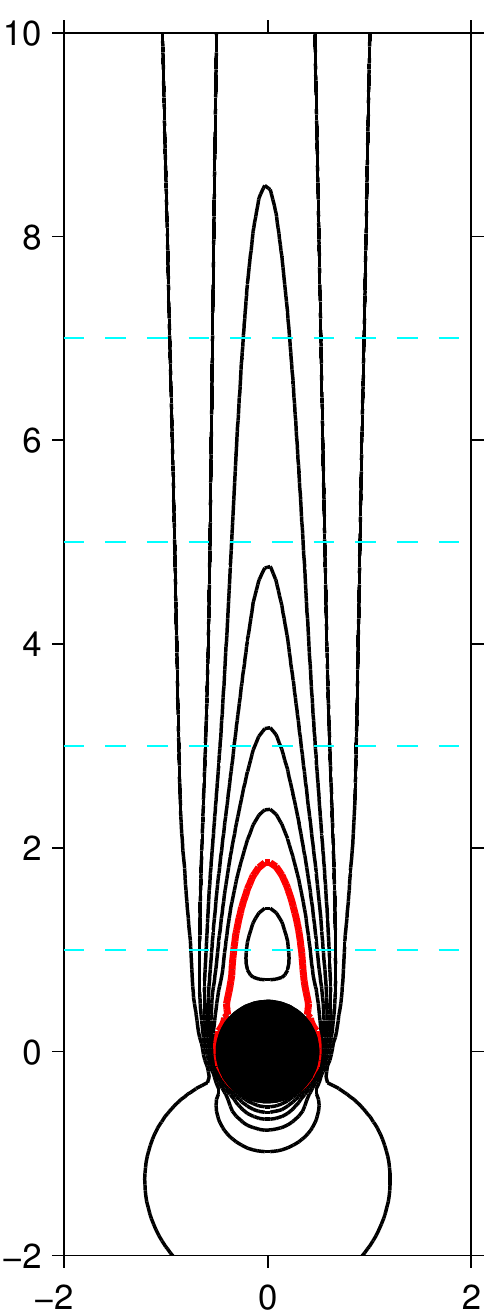}
    \centerline{$r$}
  \end{minipage}
  \begin{minipage}{.2\linewidth}
    \centerline{$(c)$ AC-24}
    \includegraphics[width=\linewidth]
    {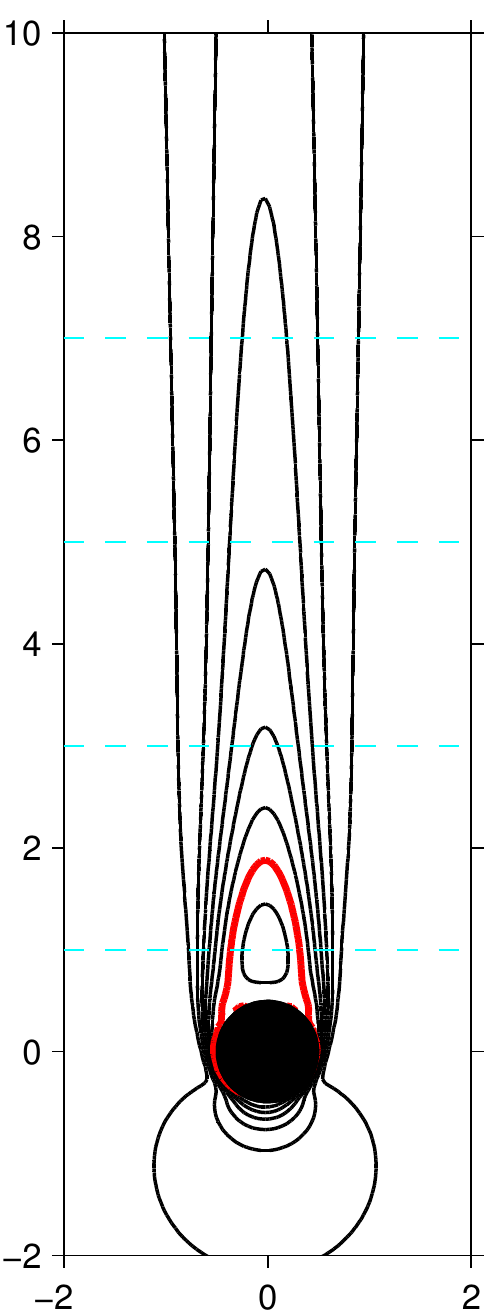}
    \centerline{$r$}
  \end{minipage}
  \begin{minipage}{.2\linewidth}
    \centerline{$(d)$ AC-36}
    \includegraphics[width=\linewidth]
    {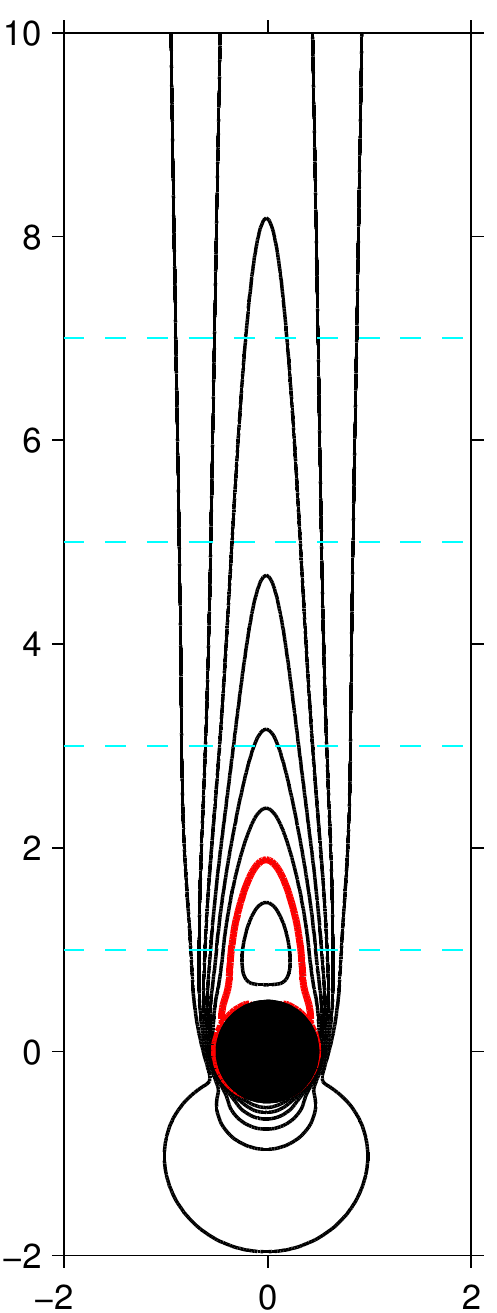}
    \centerline{$r$}
  \end{minipage}
  }{
  \caption{%
    Data from IBM computations (at different spatial resolutions) of the
    steady axisymmetric case 
    ($G=144$). 
    The graphs show contours of the vertical relative velocity 
    $u_{r\parallel}$ in the vertical/radial plane passing through the
    particle center.
    Contours are shown for values (-0.2:0.2:1.2); 
    the red line marks the extent of the recirculation region (i.e.\
    $u_{r\parallel}=0$).
    \protect\label{fig-results-ibm-contour-steady-vertical}
  }
  }
\end{figure}
The IBM simulations were initialized with a fixed particle (particle
location at a distance of $5.34$ from the inflow plane), imposing a
value for the Reynolds number $Re_\infty=|\mathbf{u}_\infty|D/\nu$ 
by adjustment of the viscosity value. After the flow around the fixed
sphere was fully established, the simulation was restarted based
upon the latest flow field, but now letting the sphere move
freely. For the mobile case two additional parameters then need to be 
prescribed, namely the density ratio (which was chosen identical to
the corresponding reference cases) and the value of the gravitational
acceleration.  
In order to allow for relatively long-time integration in a fixed
domain of limited extent, the buoyancy was chosen in order to
approximately match the magnitude of the drag force. Therefore, the
particle -- once released -- will not rapidly drift towards either the
inflow or the outflow plane, requiring a premature termination of the
simulation. In particular, the balance between drag and buoyancy
yields for the value of the gravitational acceleration 
\begin{equation}\label{equ-ibm-equi-buoy-drag}
  |\mathbf{g}|=
  \frac{|F_z^{(fix)}|}{\left(\rho_p-\rho_f\right)D^3\pi/6}
    \,,
\end{equation}
where $F_z^{(fix)}$ is the drag force acting on the particle as
obtained from the fixed-particle simulation (while the values for $D$,
$\nu$ and $\rho_f$ are kept fixed). 
With the value of the gravitational acceleration given by
(\ref{equ-ibm-equi-buoy-drag}) the Galileo number can then be computed
from its definition (\ref{equ-def-galileo}). Although the value of $G$
can be estimated, its precise magnitude is not known beforehand,
therefore requiring a certain amount of experimentation in order to
obtain a desired value. With the purpose of limiting the number of
trials, we have allowed for small deviations with respect to  
the reference value of the Galileo number. This is reflected in
tables~\ref{tab-results-ibm-steady-axi-symm} to
\ref{tab-results-ibm-chaotic} where the actual values are listed. 
It can be seen that the deviations from the nominal values are indeed
small (below 1.5\%). 

The above described procedure does not avoid vertical
drift (even in the regime where the particle motion is steady), since
mobility affects the wake characteristics and, therefore, 
leads to modified hydrodynamic forces once the sphere is released. It
simply serves the purpose of maintaining the vertical drift relatively 
low, thereby allowing for larger residence times of the particle
inside the computational domain. 
Note that in all wake regimes, except for the axisymmetric one, there
exists additionally a significant particle drift velocity in the
horizontal plane.   
\begin{figure}
  \figpap{
  \centering
  \begin{minipage}{3.5ex}
    $\frac{u_{r\,\parallel}}{u_{r\parallel\infty}}$
  \end{minipage}
  \begin{minipage}{.45\linewidth}
    \centerline{$(a)$}
    \includegraphics[width=\linewidth]
    {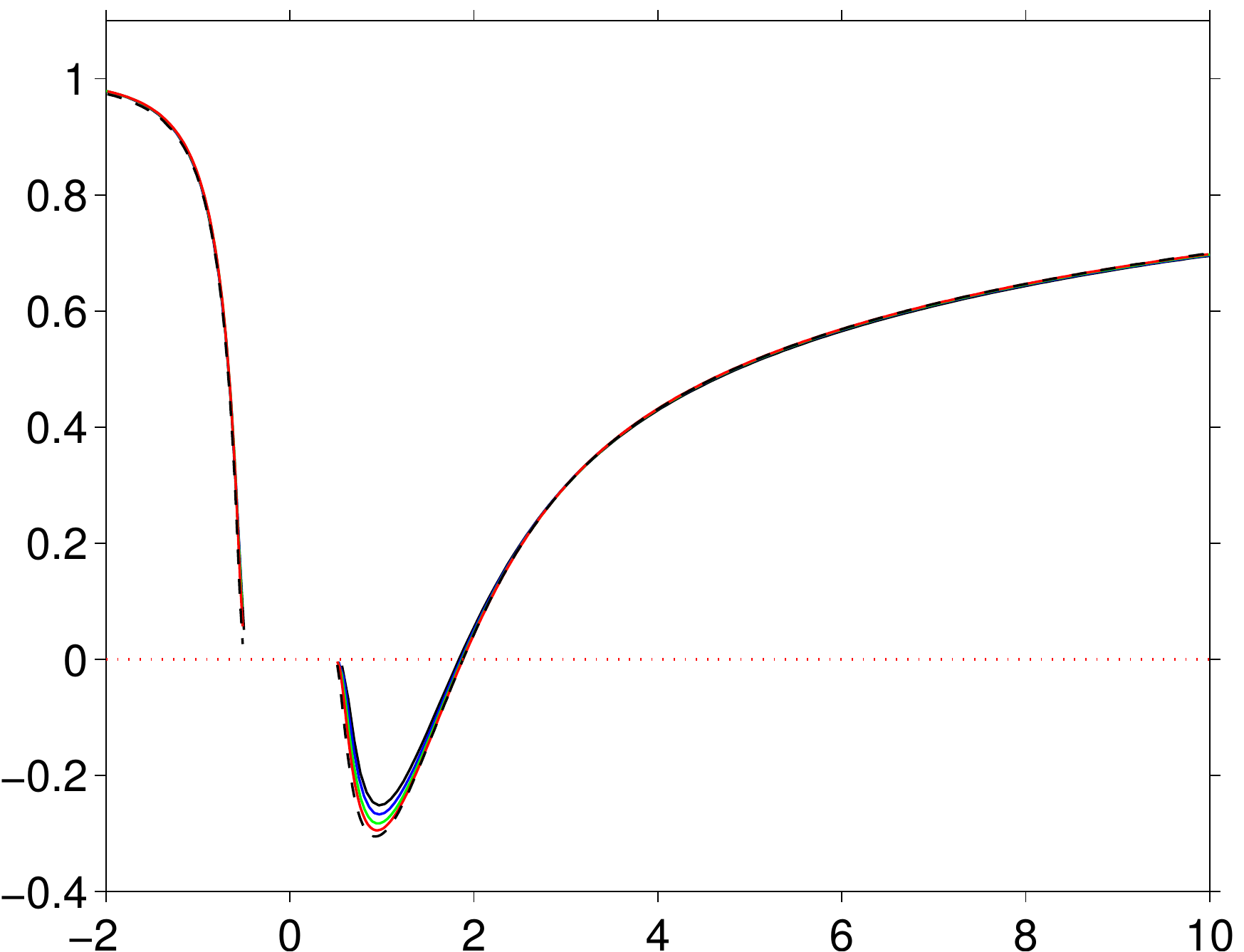}
    \centerline{$z$}
  \end{minipage}
  \begin{minipage}{.45\linewidth}
    \centerline{$(b)$}
    \includegraphics[width=\linewidth]
    {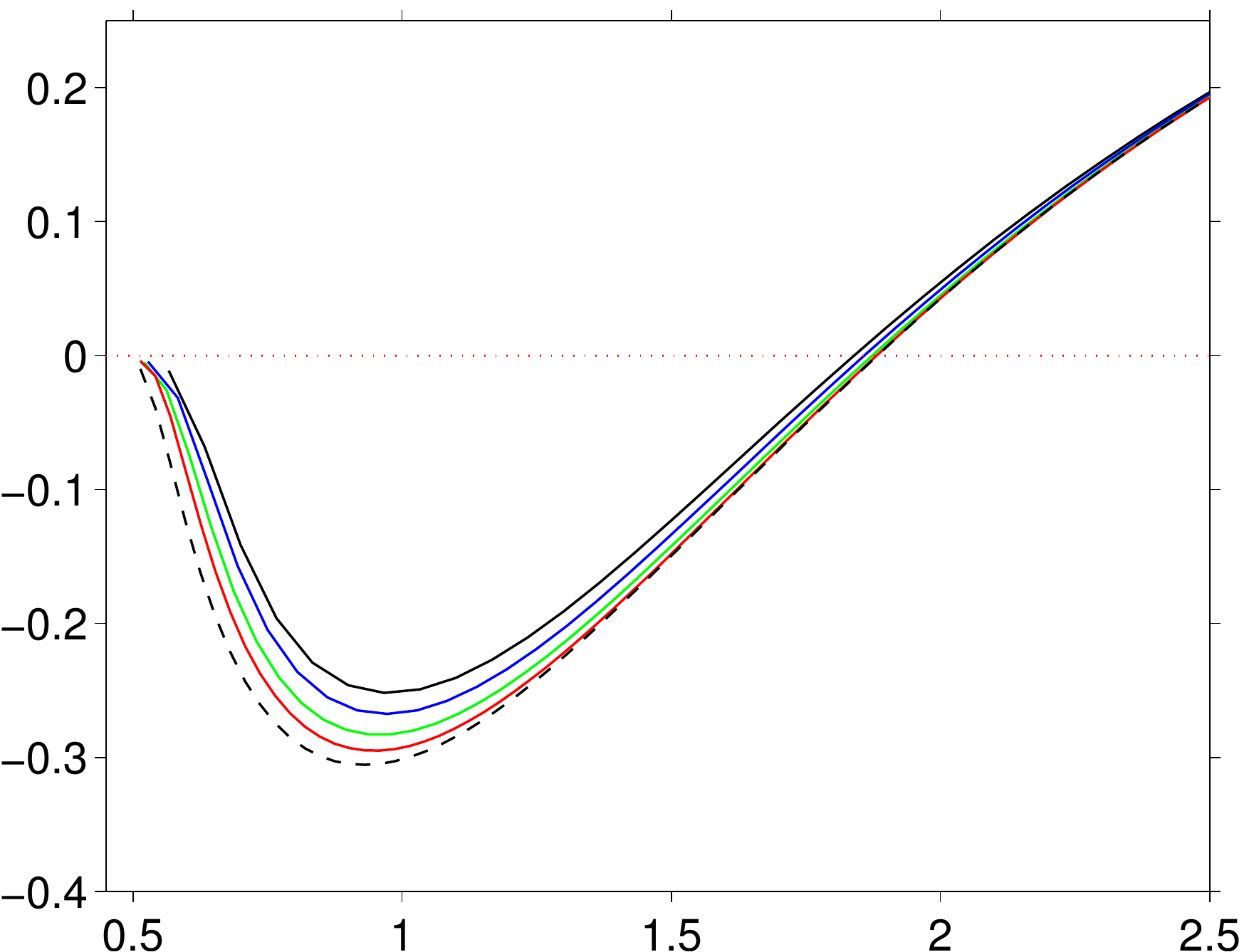}
    \centerline{$z$}
  \end{minipage}
  }{
  \caption{%
    Data from IBM computations of the steady axisymmetric case
    ($G=144$), 
    showing the quantity $u_{r\parallel}$ on the vertical axis
    through the particle center. 
    The graph in $(b)$ is a close-up of the same data in the
    recirculation region.  
    Line styles and color-coding indicate: 
    {\color{black}\solid}, case AC-15; 
    {\color{blue}\solid}, case AC-18; 
    {\color{green}\solid}, case AC-24; 
    {\color{red}\solid}, case AC-36; 
    {\color{black}\dashed}, reference case AL 
    \revision{(cf.\ figure~\ref{fig-results-ref-contour-a1l}).}{%
      (vertical cut through
      figure~\ref{fig-results-ref-contour-a1l}$c$).} 
  \protect\label{fig-results-ibm-wake-deficit-steady-vertical}
  }
  }
\end{figure}
\begin{figure}
  \figpap{
  \begin{minipage}{3ex}
    $u_{r\,\parallel}$
  \end{minipage}
  \begin{minipage}{.45\linewidth}
    \centerline{$(a)$ AC-15}
    \includegraphics[width=\linewidth]
    {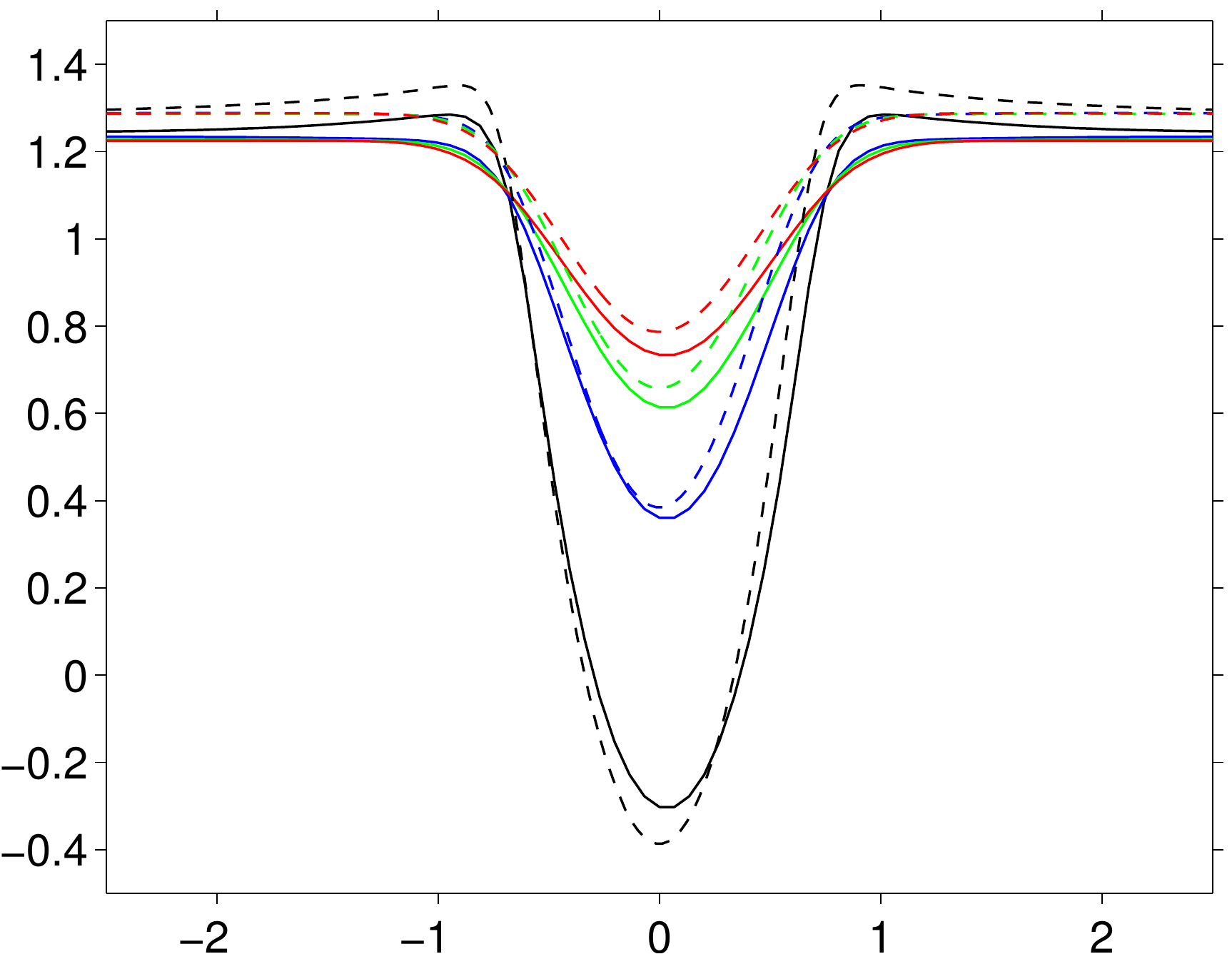}
    \centerline{$r$}
  \end{minipage}
  \hfill
  \begin{minipage}{3ex}
    $u_{rad}$
  \end{minipage}
  \begin{minipage}{.45\linewidth}
    \centerline{$(b)$ AC-15}
    \includegraphics[width=\linewidth]
    {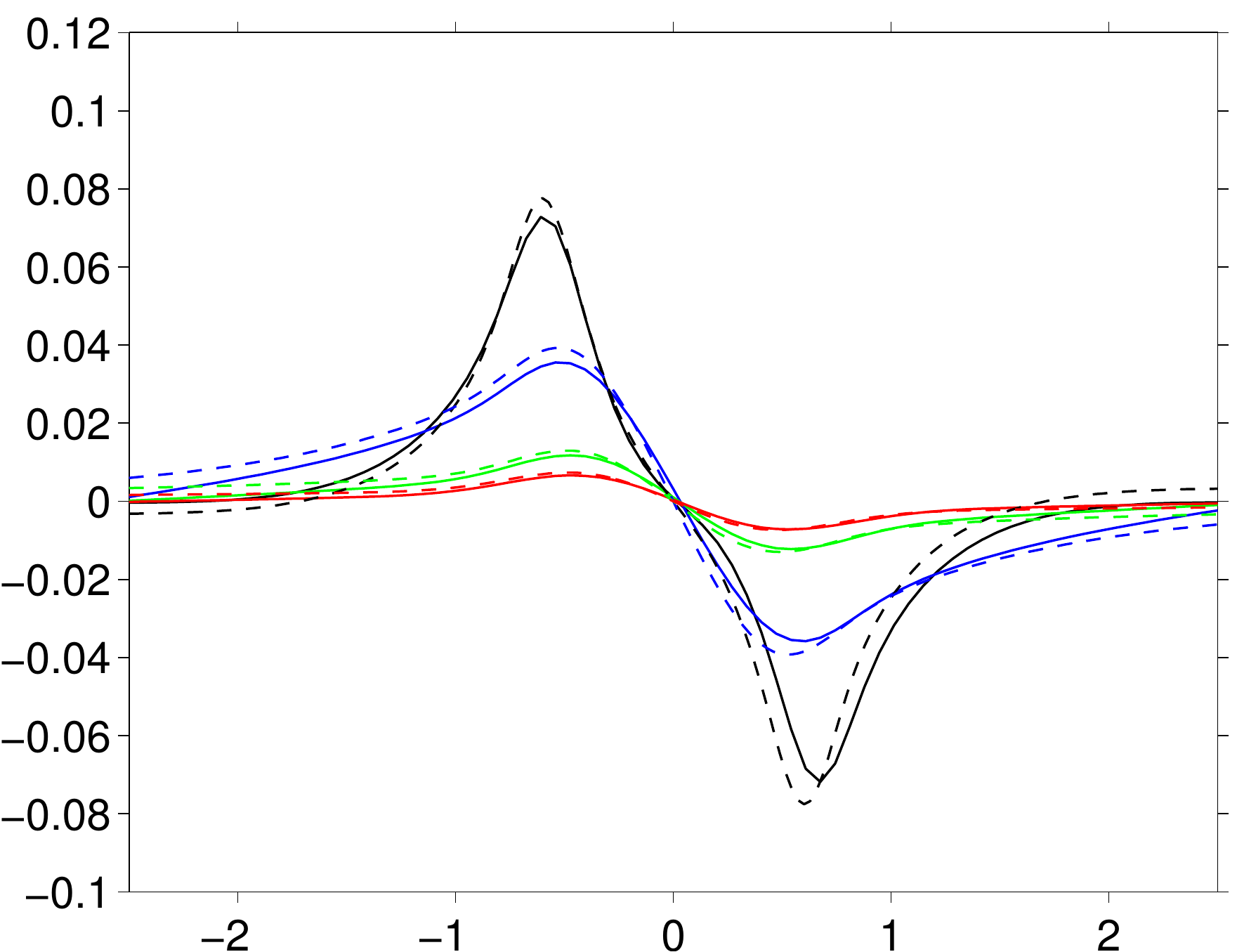}
    \centerline{$r$}
  \end{minipage}
  \\[1ex]
  \begin{minipage}{3ex}
    $u_{r\,\parallel}$
  \end{minipage}
  \begin{minipage}{.45\linewidth}
    \centerline{$(c)$ AC-36}
    \includegraphics[width=\linewidth]
    {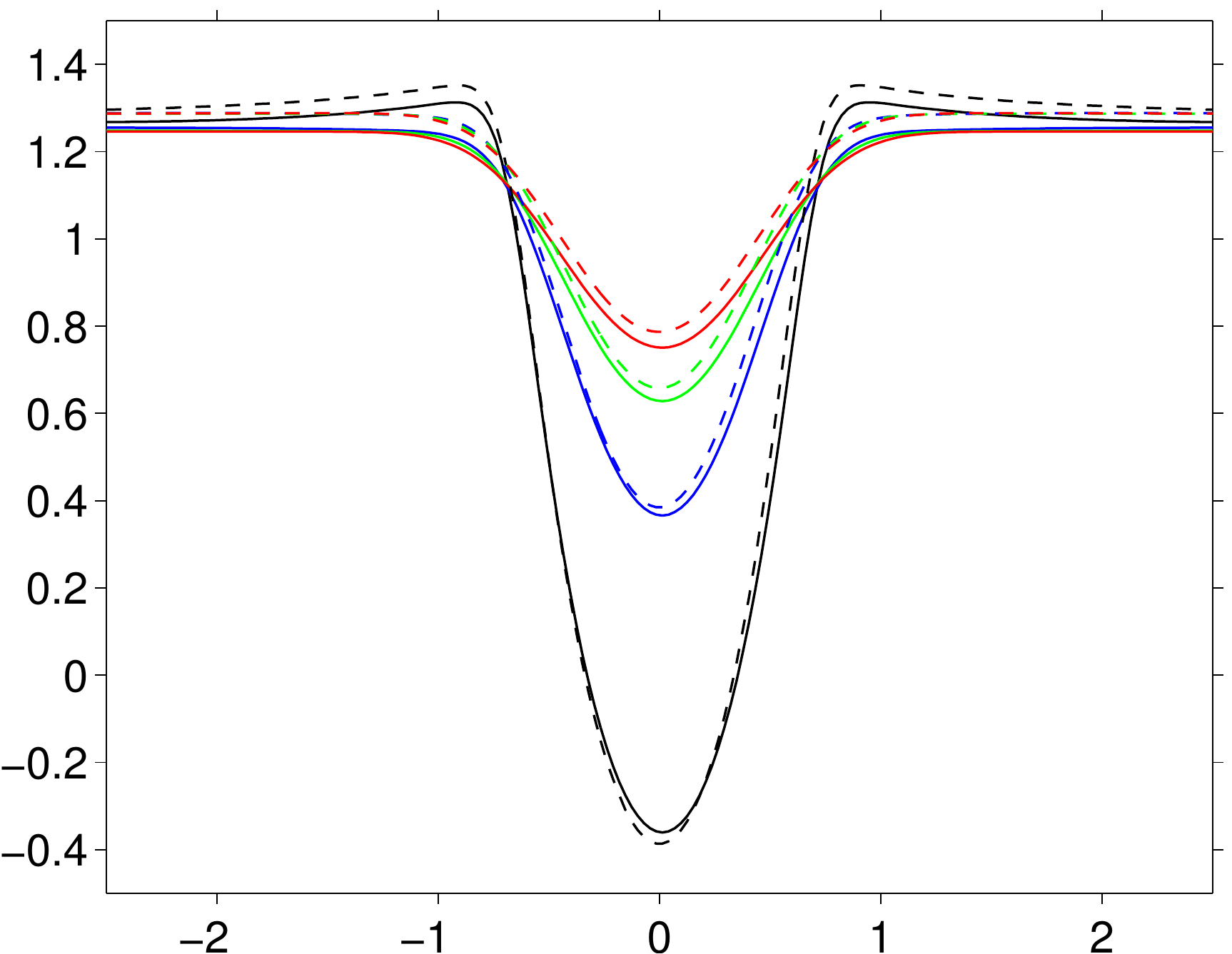}
    \centerline{$r$}
  \end{minipage}
  \hfill
  \begin{minipage}{3ex}
    $u_{rad}$
  \end{minipage}
  \begin{minipage}{.45\linewidth}
    \centerline{$(d)$ AC-36}
    \includegraphics[width=\linewidth]
    {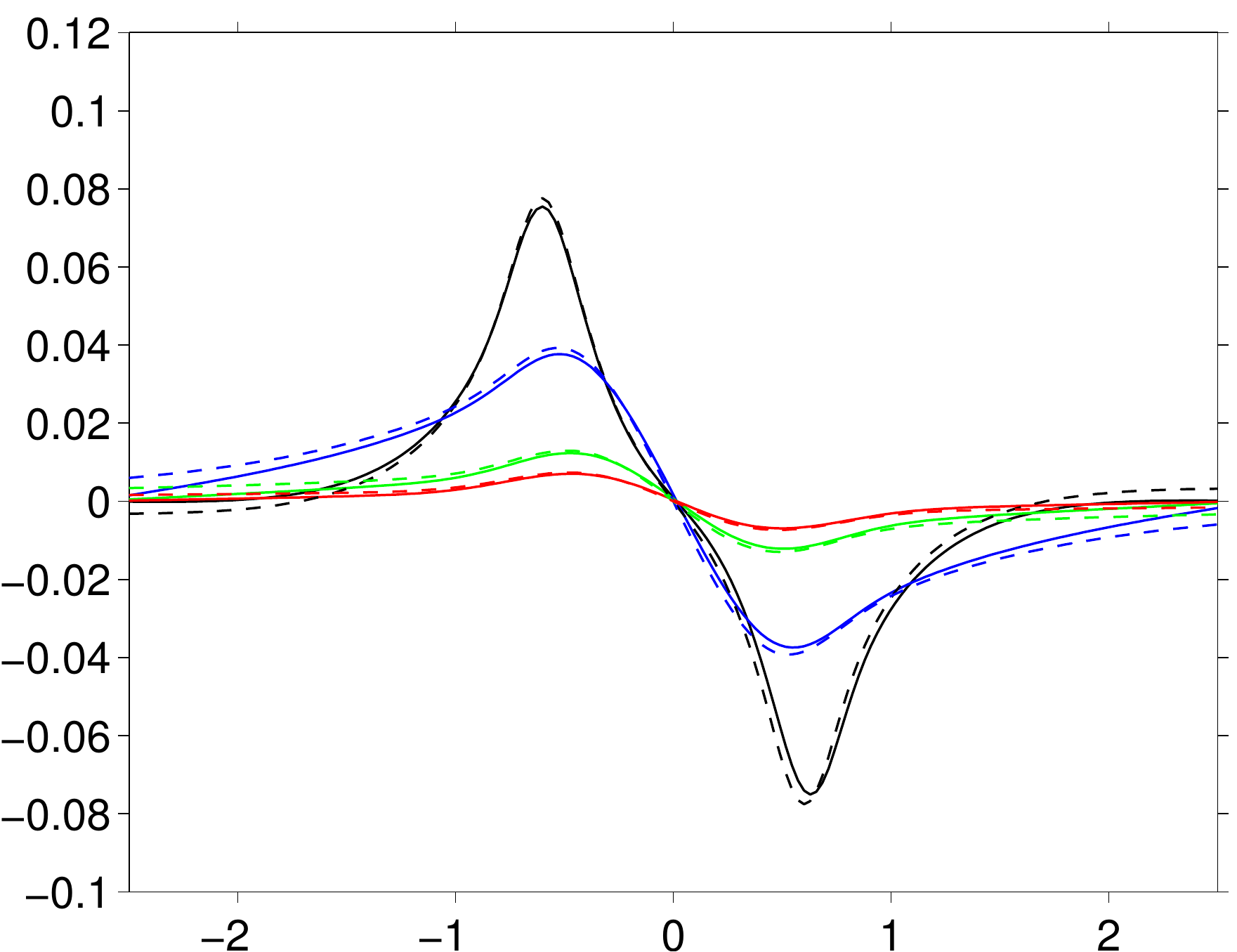}
    \centerline{$r$}
  \end{minipage}
  }{
  \caption{%
    Radial profiles of the relative flow velocity obtained with the
    IBM method in case AC ($G=144$), taken along the cyan-colored
    dashed lines in
    figure~\ref{fig-results-ibm-contour-steady-vertical}. 
    $(a,c)$ shows the axial velocity component $u_{r\,\parallel}$; 
    $(b,d)$ the radial velocity component $u_{rad}$. 
    $(a,b)$ are for case AC-15, 
    $(c,d)$ are for case AC-36.
    The color code indicates the distance downstream of the sphere: 
    {\color{black}\solid}, $x_{p\,\parallel}=-1$; 
    {\color{blue}\solid}, $x_{p\,\parallel}=-3$; 
    {\color{green}\solid}, $x_{p\,\parallel}=-5$; 
    {\color{red}\solid}, $x_{p\,\parallel}=-7$; 
    the reference data (case AL) is indicated by dashed lines%
    \revision{ (cf.\ figure~\ref{fig-results-ref-cross-profiles-a1l}).}{.}
    \protect\label{fig-results-ibm-cross-profiles-a1c}
   }
  }
\end{figure}

\subsection{Results}
\label{sec-ibm-results}
\begin{figure}
  \figpap{
  \centering
  \begin{minipage}{1.5ex}
    $c_p$
  \end{minipage}
  \begin{minipage}{.45\linewidth}
    \includegraphics[width=\linewidth]
    {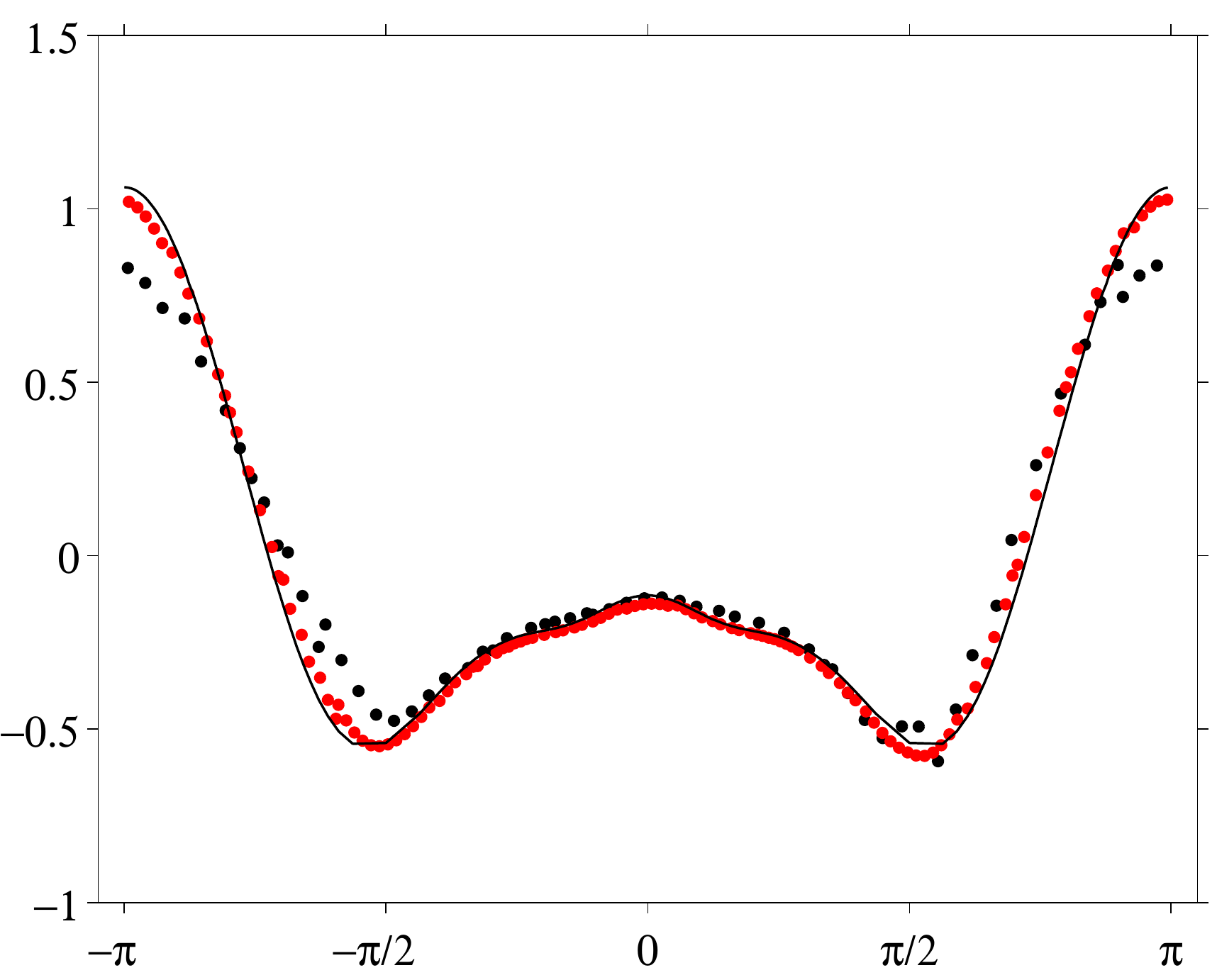}
    \centerline{$\theta$}
  \end{minipage}
  }{
  \caption{%
    Data from IBM computations of the steady axisymmetric case
    ($G=144$).
    The graph shows a profile of the pressure coefficient $c_p$ along
    a great circle on the sphere. 
    The upstream stagnation point corresponds to a value of the angle  
    $\theta=\pm\pi$. Please refer to figure~\ref{fig-ref-notation-1}
    for the definition of the angle $\theta$. 
    Line styles and symbols indicate: 
    {\color{black}$\bullet$}, case AC-15; 
    {\color{red}$\bullet$}, case AC-36; 
    {\color{black}\solid}, reference case AL 
    \revision{%
      (cf.\ figure~\ref{fig-results-ref-press-a1l}$c$).}{%
      (along blue dashed lines in
      figure~\ref{fig-results-ref-press-a1l}).} 
    \protect\label{fig-results-ibm-press-a1c}
  }
  }
\end{figure}
In the following discussion we will measure the difference between a 
particle-related quantity $\phi$ obtained from a given simulation
using the present immersed boundary method on the one hand and the
reference results of \S~\ref{sec-ref-results} on the other hand
through a relative error ${\cal E}$ defined as 
\begin{equation}\label{equ-res-ibm-def-rel-error}
  {\cal E}(\phi)=
  \frac{\left|\phi-\phi^{(ref)}\right|}{\left|\phi^{(ref)}\right|}
  \,.
\end{equation}
In the definition (\ref{equ-res-ibm-def-rel-error}) we use the
reference result $\phi^{(ref)}$ for the purpose of
normalization. 

%
At this point it should be emphasized that the present work deals with
instabilities triggered by bifurcations having definite thresholds
expressed by critical Galileo numbers which are very sensitive to
numerical accuracy. 
For this reason the numerical convergence of the threshold values 
has been systematically used to test the numerical parameters of the 
spectral/spectral-element code used for generating the reference
results \citep[cf.\ ][]{ghidersa00,jenny:04}. 
In particular, it does not make sense to normalize relative errors by
quantities becoming non-zero at instability thresholds while
investigating a parameter domain in which the instabilities set in. In
some instances it would amount to dividing by zero. For this reason
all relative errors of non-dimensional quantities are normalized
by the non-dimensional vertical velocity, i.e.\ we use the reference
value of the vertical component, $u_{pV}^{ref}$, for computing the
error of $u_{pH}$ and $\omega_{pH}$ in the denominator of
(\ref{equ-res-ibm-def-rel-error}). 

The reference data used for the present comparison is taken from the
results presented in \S~\ref{sec-ref-results}, as obtained in the
larger domain with $D_\Omega=7.54$, i.e.\ cases AL, BL, CL, DL (cf.\
tables~\ref{tab-parameters-ref}--\ref{tab-results-ref-D}). 
\subsubsection{Steady axi-symmetric regime}
\label{sec-ibm-results-axisymm}
\begin{figure}
  \figpap{
  \begin{minipage}{.19\linewidth}
    \centerline{$(a)$}
    \includegraphics[width=\linewidth,clip=true,
    viewport=1050 840 1500 1750]
    {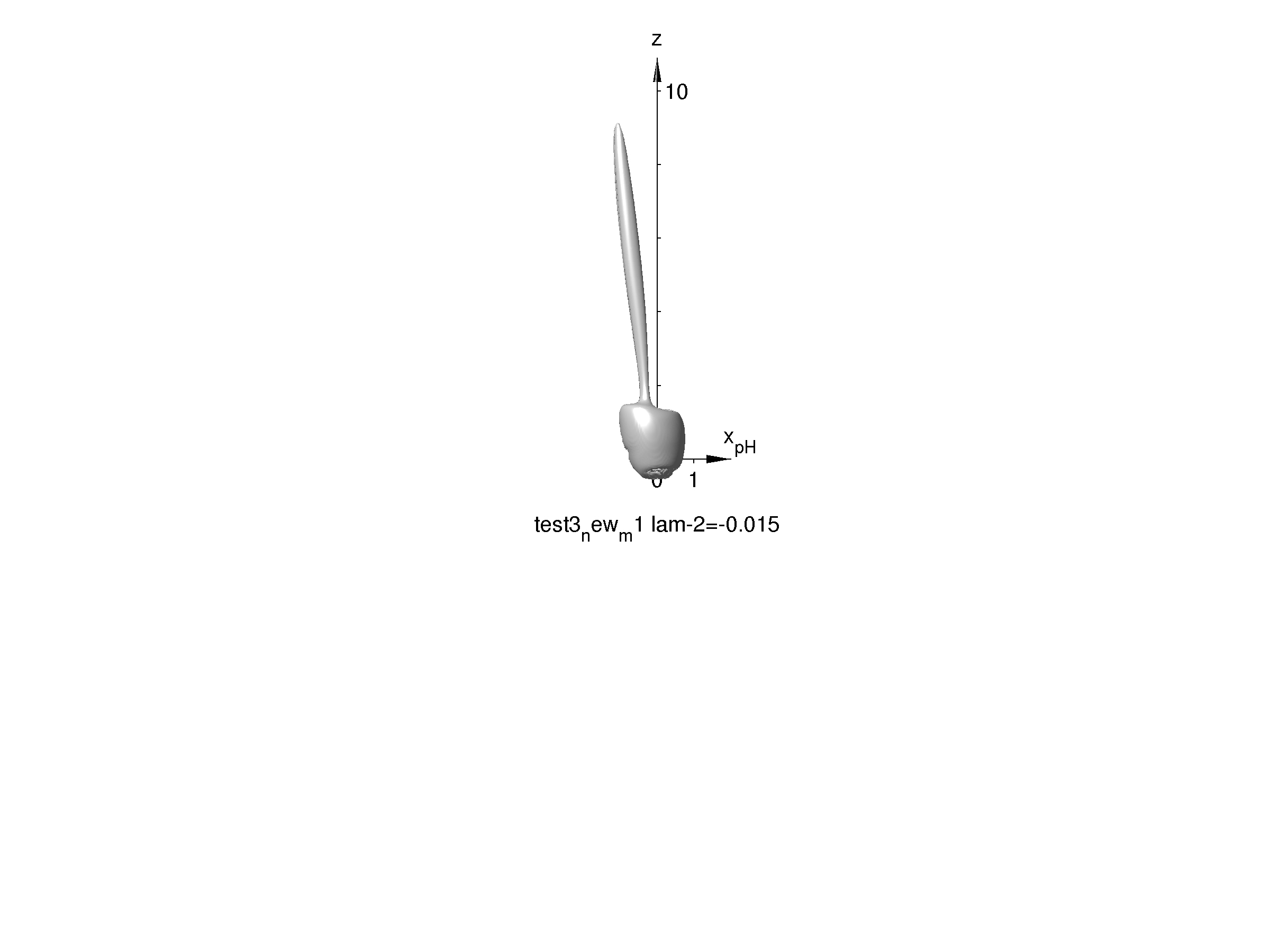}
  \end{minipage}
  \begin{minipage}{.19\linewidth}
    \centerline{$(b)$}
    \includegraphics[width=\linewidth,clip=true,
    viewport=1050 840 1500 1750]
    {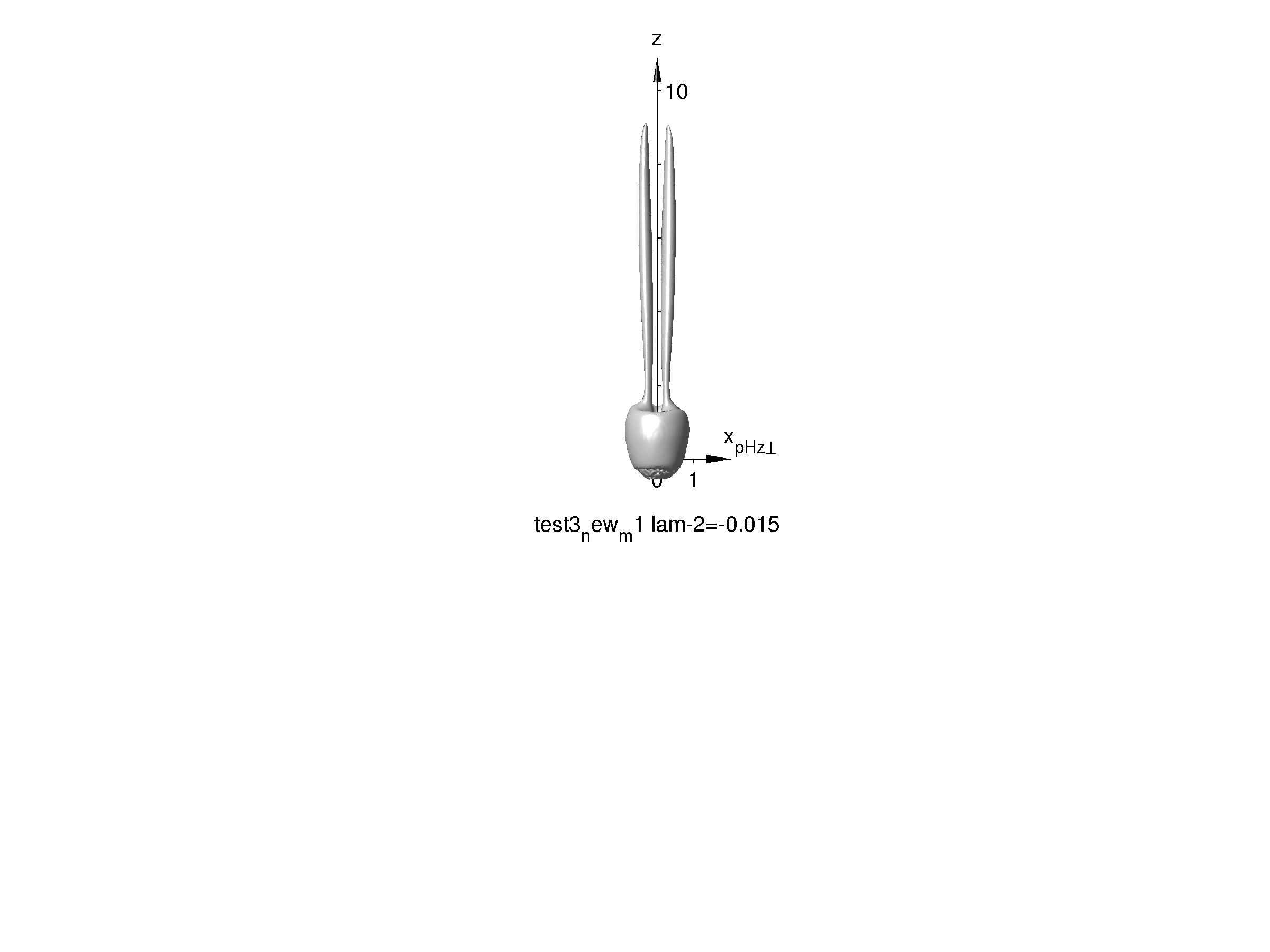}
  \end{minipage}
  \hspace*{3ex}
  \begin{minipage}{1ex}
    $z$
  \end{minipage}
  \begin{minipage}{.185\linewidth}
    \centerline{$(c)$}
    \includegraphics[width=\linewidth]
    {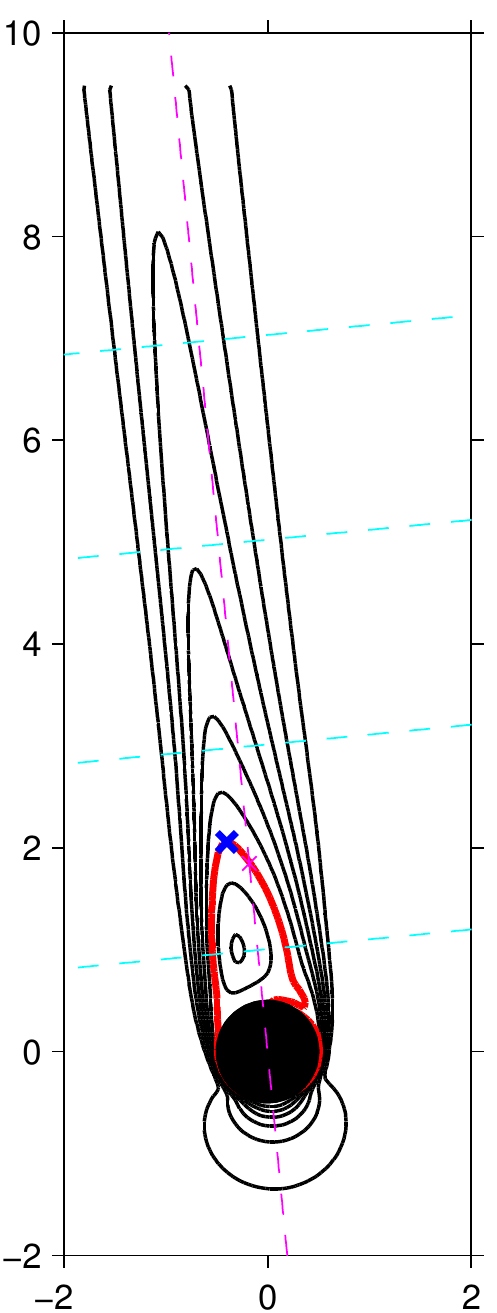}
    \centerline{${x}_{pH}$}
  \end{minipage}
  \begin{minipage}{1ex}
    $z$
  \end{minipage}
  \begin{minipage}{.185\linewidth}
    \centerline{$(d)$}
    \includegraphics[width=\linewidth]
    {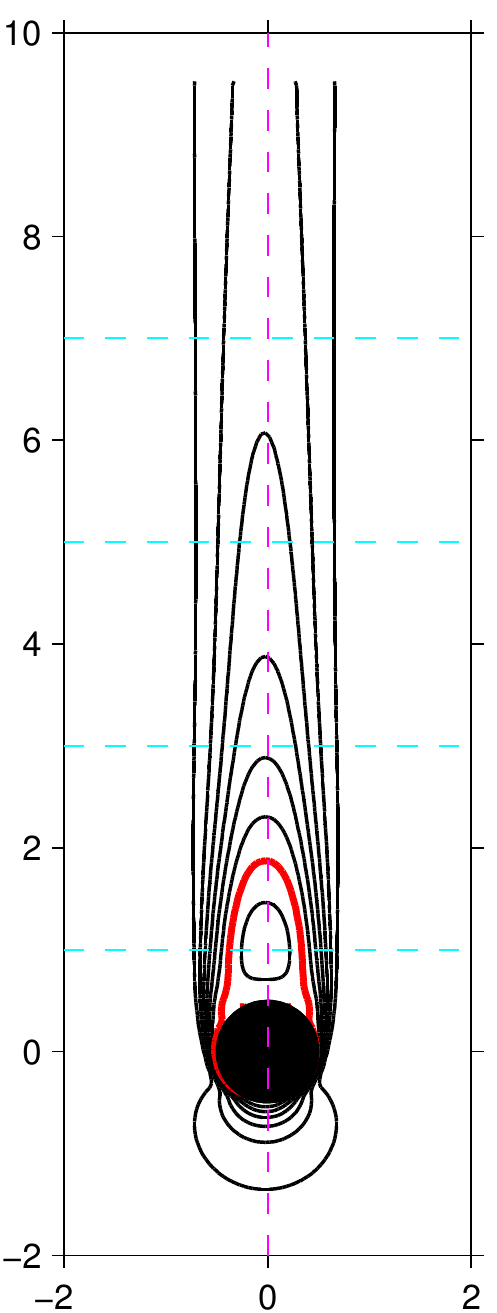}
    \centerline{${x}_{pHz\perp}$}
  \end{minipage}
  }{
  \caption{%
    Results from IBM computation in case BC-24 ($G=178.46$).
    $(a)$, $(b)$ shows the surface where $\lambda_2=-0.015$. 
    $(c)$, $(d)$ depict contours of the projected relative velocity
    $u_{r\,\parallel}$ in the plane which is spanned by the vertical
    direction $\mathbf{e}_z$ and the direction of the particle motion
    $\mathbf{e}_{p\parallel}$, passing through the sphere's center. 
    Contours are shown for values (-0.4:0.2:1.2); the red line marks
    the extent of the recirculation region (i.e.\ $u_{r\,\parallel}=0$). 
    The blue cross in $(c)$ marks the location which defines the
    recirculation length $L_r$. 
    The magenta colored dashed line indicates the direction of the particle
    motion given by the unit vector $\mathbf{e}_{p\,\parallel}$.
    In $(a)$ and $(c)$ the view is directed along $\mathbf{e}_{pHz\perp}$;  
    in $(b)$ and $(d)$ it is directed along $\mathbf{e}_{pH}$.
  \protect\label{fig-results-ibm-3d-iso-lambda2-iso-urel-b1l}
  }
  }
\end{figure}
Table~\ref{tab-results-ibm-steady-axi-symm} shows the steady-state
results for the particle motion obtained with four different spatial
resolutions, ranging from $D/\Delta x=15$ to $36$. 
It can be seen that the vertical particle velocity relative to the
ambient fluid, $u_{pV}$, is slightly over-predicted by the present
immersed boundary method, with the error decreasing from approximately
6\% at $D/\Delta x=15$ to 4.5\% at $D/\Delta x=36$.  
The particle wake obtained by the IBM simulation at these spatial
resolutions is illustrated in
figure~\ref{fig-results-ibm-contour-steady-vertical} which shows
contours of the vertical component of the relative flow velocity
$u_{r\parallel}$ at the same levels chosen in
figure~\ref{fig-results-ref-contour-a1l}$(c)$ for the reference case. 
The visual impression is that of a very good match, with the wake
spreading slightly over-predicted at the lower spatial resolutions. 
%
Figure~\ref{fig-results-ibm-wake-deficit-steady-vertical} shows
profiles of $u_{r\parallel}$ on the vertical axis passing through the
sphere's center, allowing for a direct comparison with
the reference results. Noticeable discrepancies are only found in the
recirculation region. The close-up in
figure~\ref{fig-results-ibm-wake-deficit-steady-vertical}$(b)$
illustrates the convergence towards the reference case results with
increasing spatial resolution.  
This comparison can be made quantitative by considering the prediction
of the recirculation length, $L_r$, the error of which is given in
table~\ref{tab-results-ibm-steady-axi-symm}. It is found that the
relative error decreases from approximately 3\% at $D/\Delta x=15$ to
0.04\% at $D/\Delta x=36$. 

Radial profiles of the two non-zero components of the relative flow
velocity $\mathbf{u}_r$ in this axi\-symmetric case are shown in
figure~\ref{fig-results-ibm-cross-profiles-a1c} for the two spatial
resolutions $D/\Delta x=15$ and 36. The comparison with the reference
results demonstrates the quality of the predictions and the
convergence with increasing spatial resolution. 
Note that the residual difference in $u_{r\parallel}$ at large radial
distances from the sphere directly reflects the respective difference
in the obtained settling velocity (cf.\
table~\ref{tab-results-ibm-steady-axi-symm}).  

Finally, the pressure coefficient $c_p$ along 
a great circle (as defined in figure~\ref{fig-ref-notation-1}$b$) is
shown in figure~\ref{fig-results-ibm-press-a1c}. 
Note that the IBM simulation (in the finite-difference context) yields
values of the pressure field at the nodes of the global grid which are by
definition not conforming to the spherical particle
surface. Therefore, the surface pressure is not defined without
ambiguity. In practice we have taken the approach of
\cite{uhlmann:04}, plotting the pressure at the first grid node (along
each grid line in one direction in the plane of the chosen great
circle) outside the range of the discrete delta function, i.e.\ for
which $|\mathbf{x}_{ijk}-\mathbf{x}_{sphere}|\geq D/2+3\Delta x/2$. 
The comparison in figure~\ref{fig-results-ibm-press-a1c} shows that
the general agreement is good even at a resolution of $D/\Delta x=15$,
with the largest discrepancies occurring around the upstream
stagnation point. At a spatial resolution of $D/\Delta x=36$ the match
with the reference data from the spectral element method can be
described as excellent. 
\subsubsection{Steady oblique regime}
\label{sec-ibm-results-steady-oblique}
\begin{figure}
  \figpap{
  \begin{minipage}{3.5ex}
    $\frac{u_{r\,\parallel}}{u_{r\parallel\infty}}$
  \end{minipage}
  \begin{minipage}{.45\linewidth}
    \centerline{$(a)$}
    \includegraphics[width=\linewidth]
    {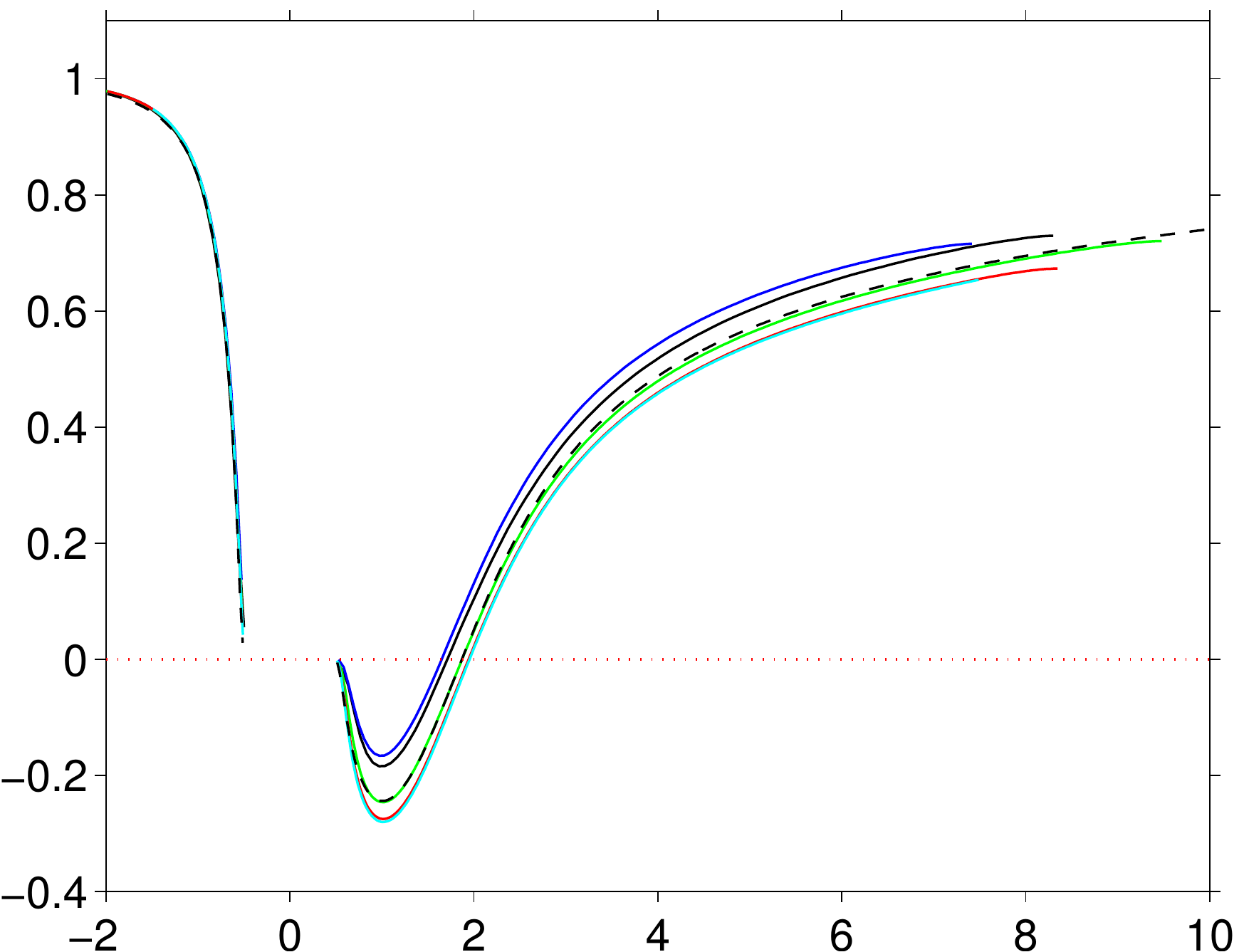}
    \centerline{$x_{p\,\parallel}$}
  \end{minipage}
  \begin{minipage}{.45\linewidth}
    \centerline{$(b)$}
    \includegraphics[width=\linewidth]
    {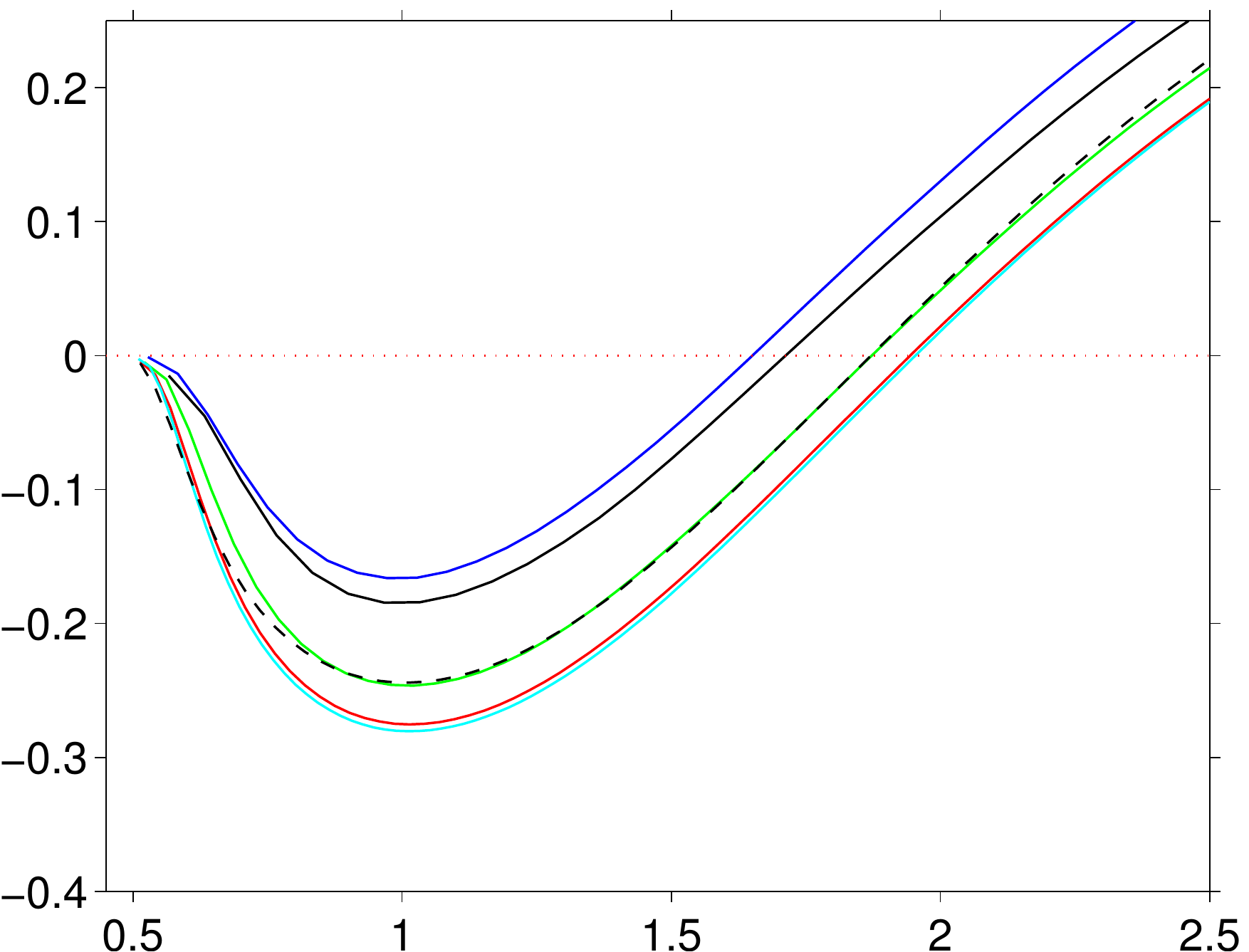}
    \centerline{$x_{p\,\parallel}$}
  \end{minipage}
  }{
  \caption{%
    Data from IBM computations (at different spatial resolutions) of the
    steady oblique case ($G=178.46$), 
    showing the quantity $u_{r\parallel}$ along the axis
    passing through the particle center and following the direction of
    the particle motion (along the magenta-colored dashed line in
    figure~\ref{fig-results-ibm-3d-iso-lambda2-iso-urel-b1l}$c,d$). 
    The graph in $(b)$ is a close-up of the same data in the
    recirculation region.  
    Line styles and color-coding indicate: 
    {\color{black}\solid}, case BC-15; 
    {\color{blue}\solid}, case BC-18; 
    {\color{green}\solid}, case BC-24; 
    {\color{red}\solid}, case BC-36; 
    {\color{cyan}\solid}, case BC-48; 
   {\color{black}\dashed}, reference case BL (cf.\ 
    figure~\ref{fig-results-ref-contour-b1l}).  
  \protect\label{fig-results-ibm-wake-deficit-steady-oblique}
  }
  }
\end{figure}
\begin{figure}[b]
  \figpap{
  \centering
  \begin{minipage}{3.5ex}
    $\frac{u_{r\,\parallel}}{u_{r\parallel\infty}}$
  \end{minipage}
  \begin{minipage}{.45\linewidth}
    \includegraphics[width=\linewidth]
    {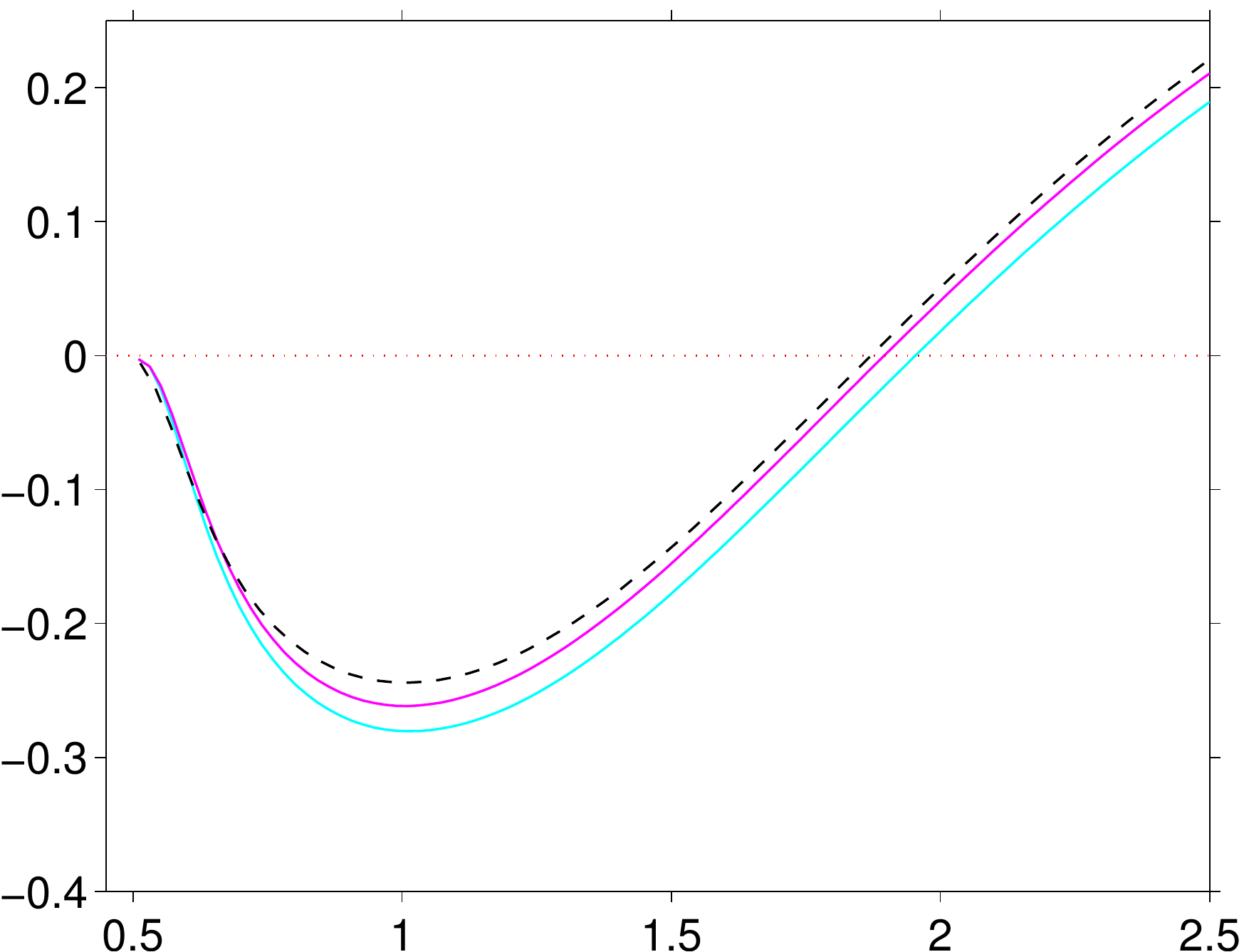}
    \centerline{$x_{p\,\parallel}$}
  \end{minipage}
  }{
  \caption{%
    As figure~\ref{fig-results-ibm-wake-deficit-steady-oblique}$(b)$,
    comparing the IBM result with spatial resolution $D/\Delta x=48$ for
    two different time steps. 
    Line styles and color-coding indicate: 
    {\color{cyan}\solid}, case BC-48 ($CFL=0.3$); 
    {\color{magenta}\solid}, case BC-48h ($CFL=0.15$); 
    {\color{black}\dashed}, reference case BL (cf.\ 
    figure~\ref{fig-results-ref-contour-b1l}).  
  \protect\label{fig-results-ibm-dthalf-wake-deficit-steady-oblique}
  }
  }
\end{figure}
The simulations with the present immersed boundary method capture the
oblique particle motion at a (nominal) Galileo number of $G=178.46$ at
all chosen grid resolutions $D/\Delta x=15$ to $48$. 
Figure~\ref{fig-results-ibm-3d-iso-lambda2-iso-urel-b1l}$(a,b)$ shows
an iso-surface of $\lambda_2$ for case B1C-24, visualizing the same
value as for the reference case in
figure~\ref{fig-results-ref-3d-iso-b1l}.  
It can be observed that the double-threaded wake structure and its 
inclination with respect to the vertical axis is faithfully
reproduced. The same observation holds for the contours of the
parallel component of the relative flow velocity, $u_{r\parallel}$, 
shown in
figure~\ref{fig-results-ibm-3d-iso-lambda2-iso-urel-b1l}$(c,d)$ which
should be compared to the reference result depicted in
figure~\ref{fig-results-ref-contour-b1l}$(a,b)$. 

Profiles of the projected relative velocity $u_{r\parallel}$ along an axis
parallel to $\mathbf{e}_{p\parallel}$ through the sphere's center are
shown in figure~\ref{fig-results-ibm-wake-deficit-steady-oblique}. 
It can be seen that the match with the reference data is good, with
some discrepancies downstream of the particle. 
At this point it should be mentioned that the profiles taken along the
chosen axis $\mathbf{e}_{p\parallel}$ are highly sensitive to small
changes in the location of the double-threaded vortices in the wake
which are attached to the particle off-center, cf.\ discussion in
\S~\ref{sec-ref-results-steady-oblique}.  
The close-up of the recirculation region provided in
figure~\ref{fig-results-ibm-wake-deficit-steady-oblique}$(b)$ suggests
that the predictions become better with refinement up to $D/\Delta
x=24$ (where an excellent match is observed), and then -- surprisingly
-- appear to converge to a profile which is slightly off the reference
result (with virtually no further change when refining from $D/\Delta
x=36$ to 48).  
We will return to this point shortly.

\begin{figure}
  \figpap{
  \centering
  \begin{minipage}{5ex}
    $u_{r\,\parallel}$
  \end{minipage}
  \begin{minipage}{.37\linewidth}
    \centerline{$(a)$}
    \includegraphics[width=\linewidth]
    {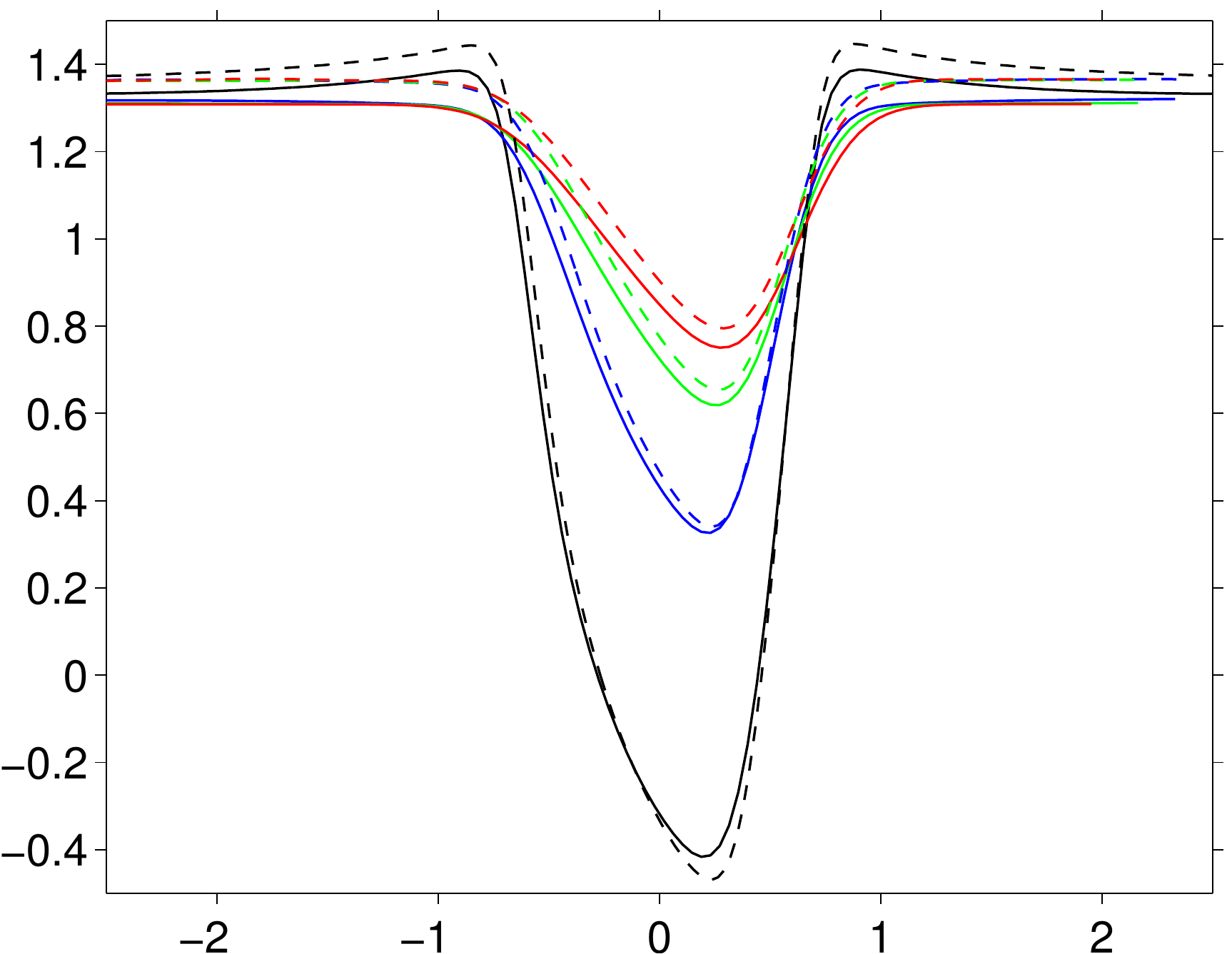}
    \centerline{$x_{p\,\perp}$}
  \end{minipage}
  \begin{minipage}{.37\linewidth}
    \centerline{$(b)$}
    \includegraphics[width=\linewidth]
    {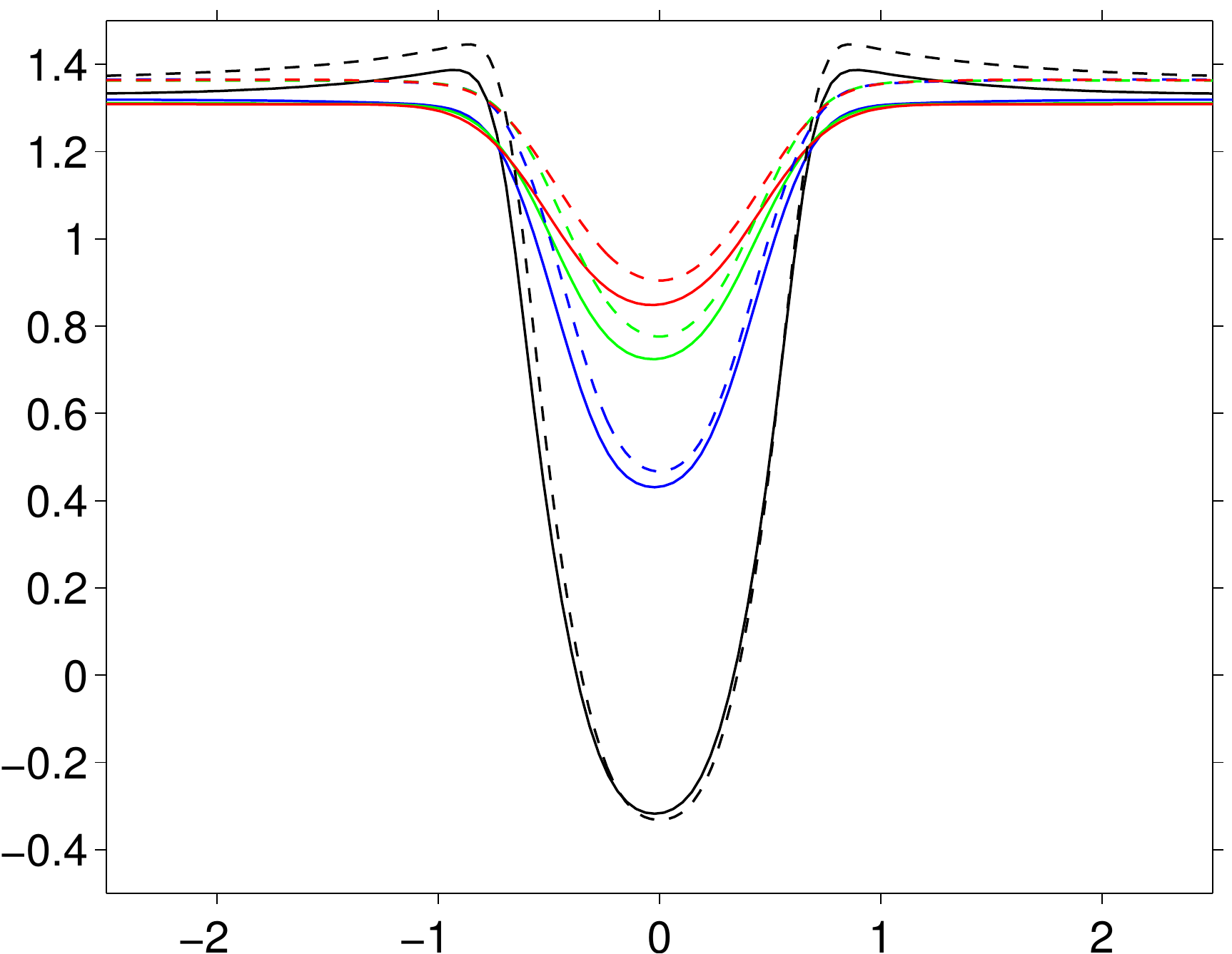}
    \centerline{$x_{pHz\perp}$}
  \end{minipage}
  \\[1ex]
  \begin{minipage}{5ex}
    $u_{r\,\perp}$
  \end{minipage}
  \begin{minipage}{.37\linewidth}
    \centerline{$(c)$}
    \includegraphics[width=\linewidth]
    {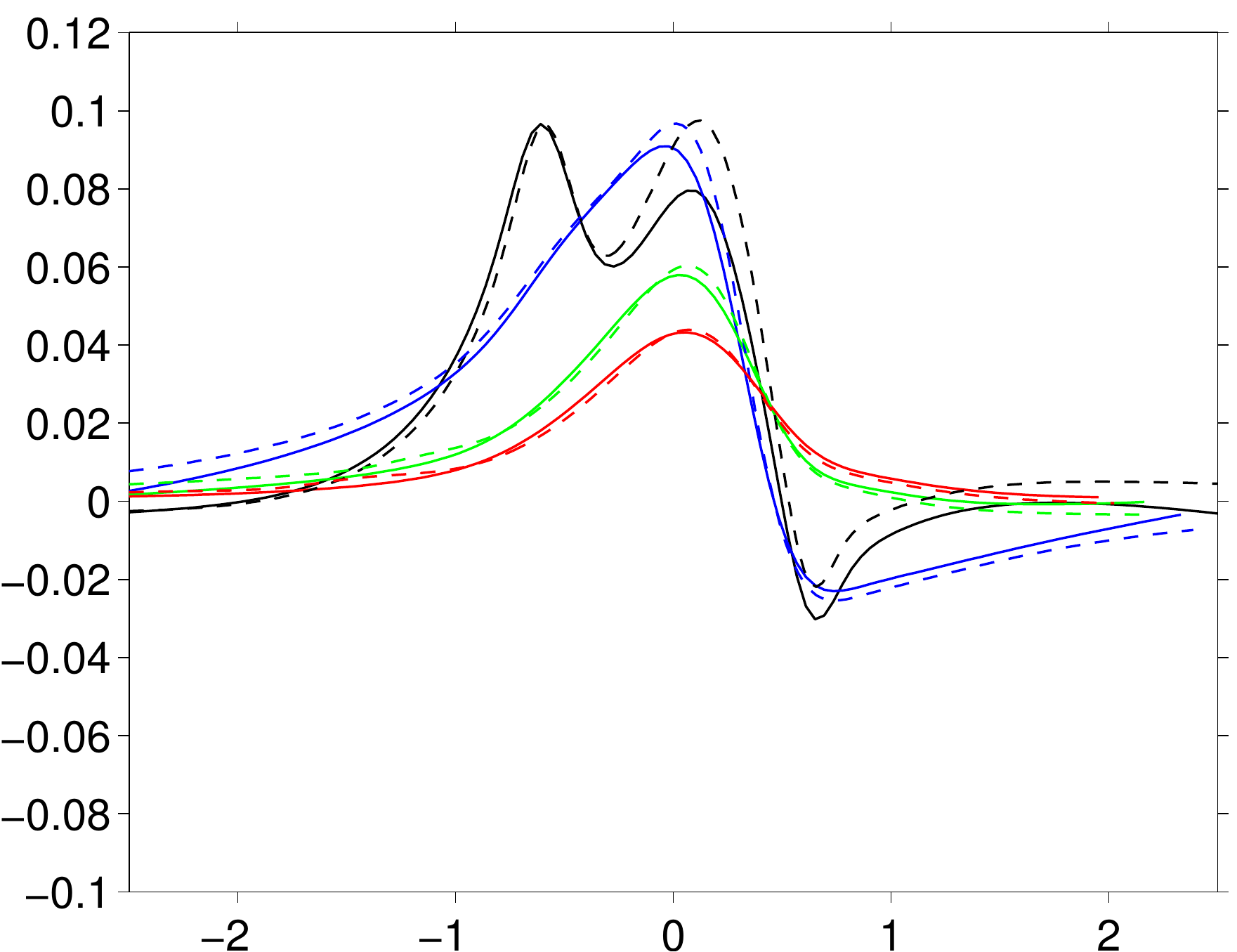}
    \centerline{$x_{p\,\perp}$}
  \end{minipage}
  \begin{minipage}{.37\linewidth}
    \centerline{$(d)$}
    \includegraphics[width=\linewidth]
    {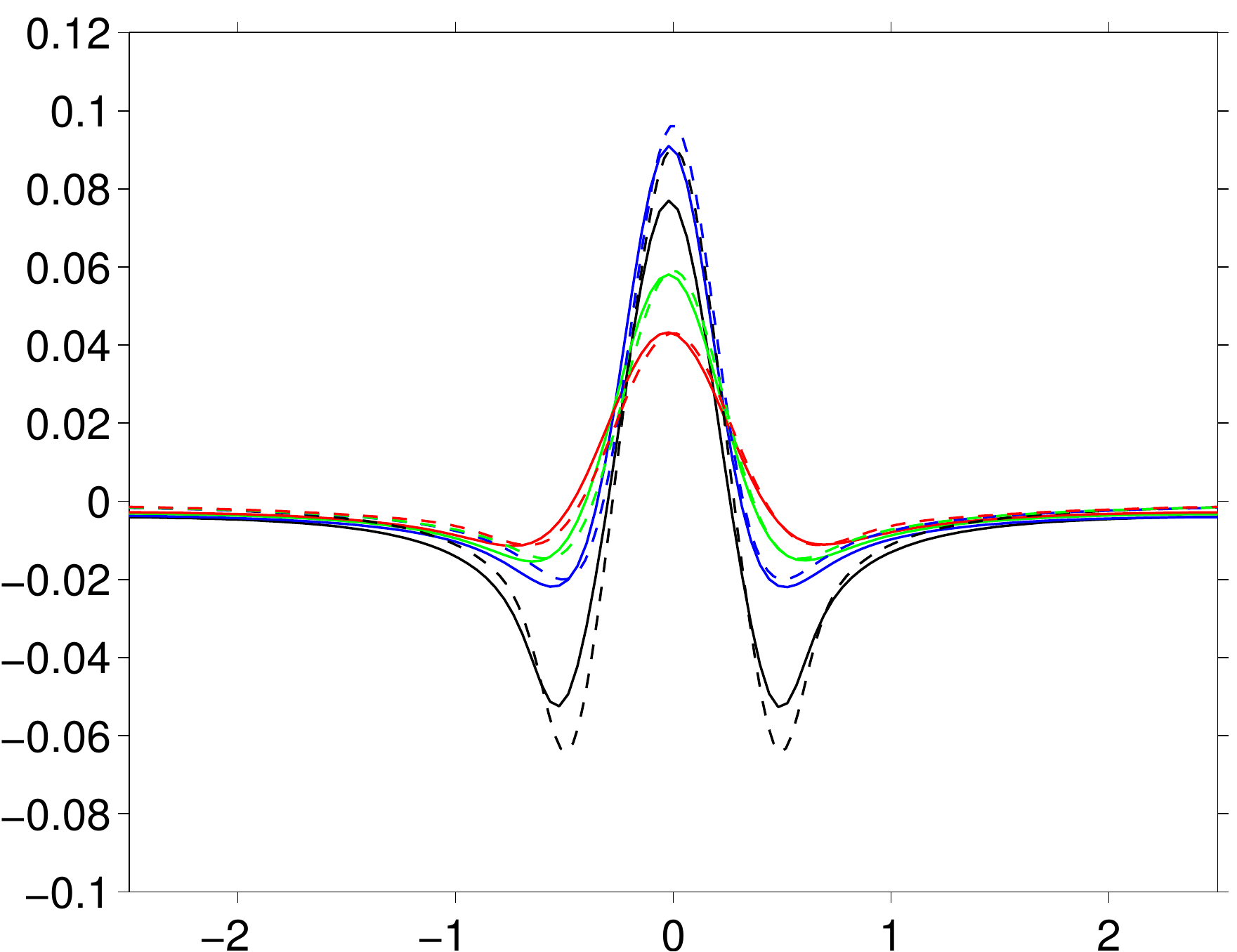}
    \centerline{$x_{pHz\perp}$}
  \end{minipage}
  \\[1ex]
  \begin{minipage}{5ex}
    $u_{rHz\perp}$
  \end{minipage}
  \begin{minipage}{.37\linewidth}
    \centerline{$(e)$}
    \includegraphics[width=\linewidth]
    {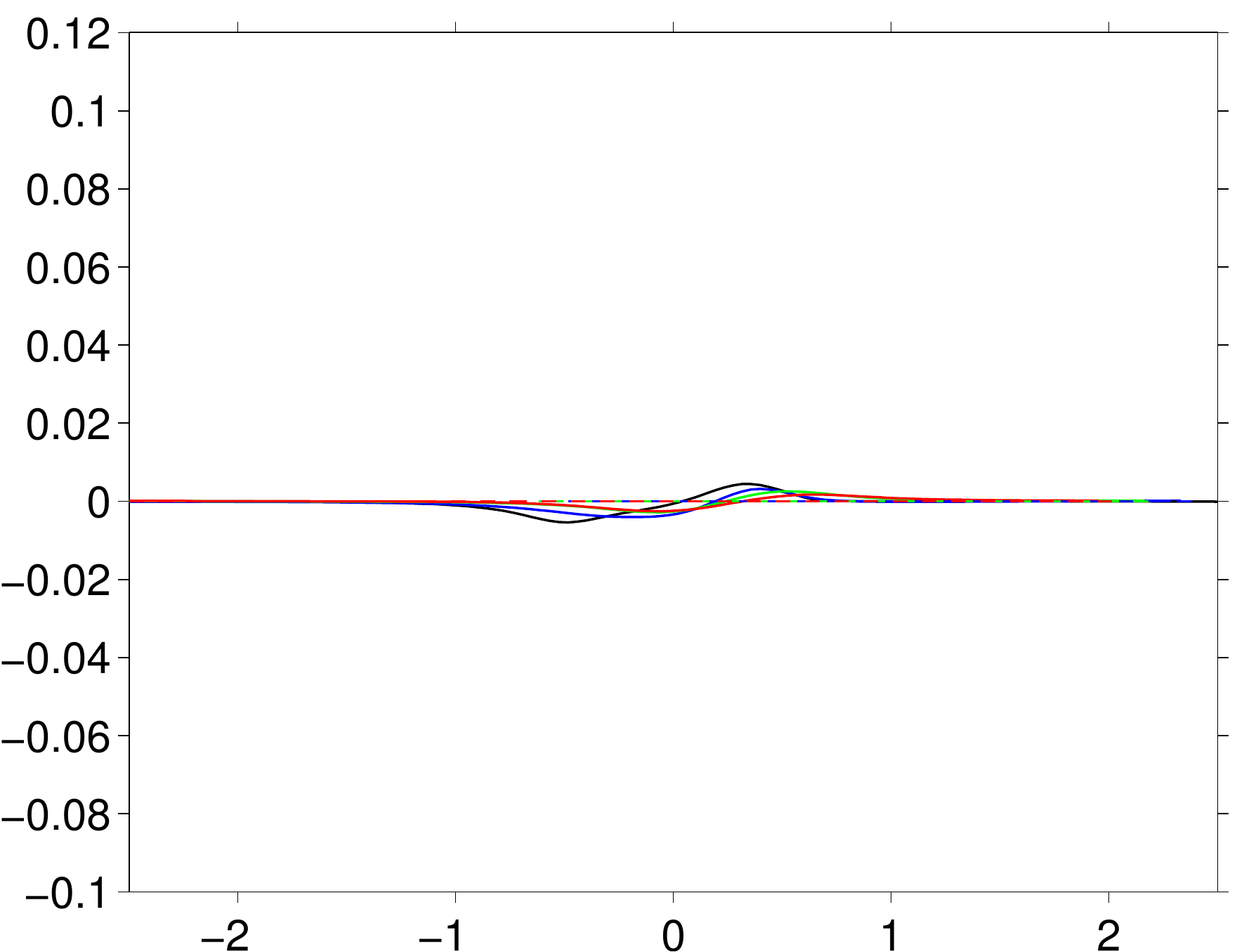}
    \centerline{$x_{p\,\perp}$}
  \end{minipage}
  \begin{minipage}{.37\linewidth}
    \centerline{$(f)$}
    \includegraphics[width=\linewidth]
    {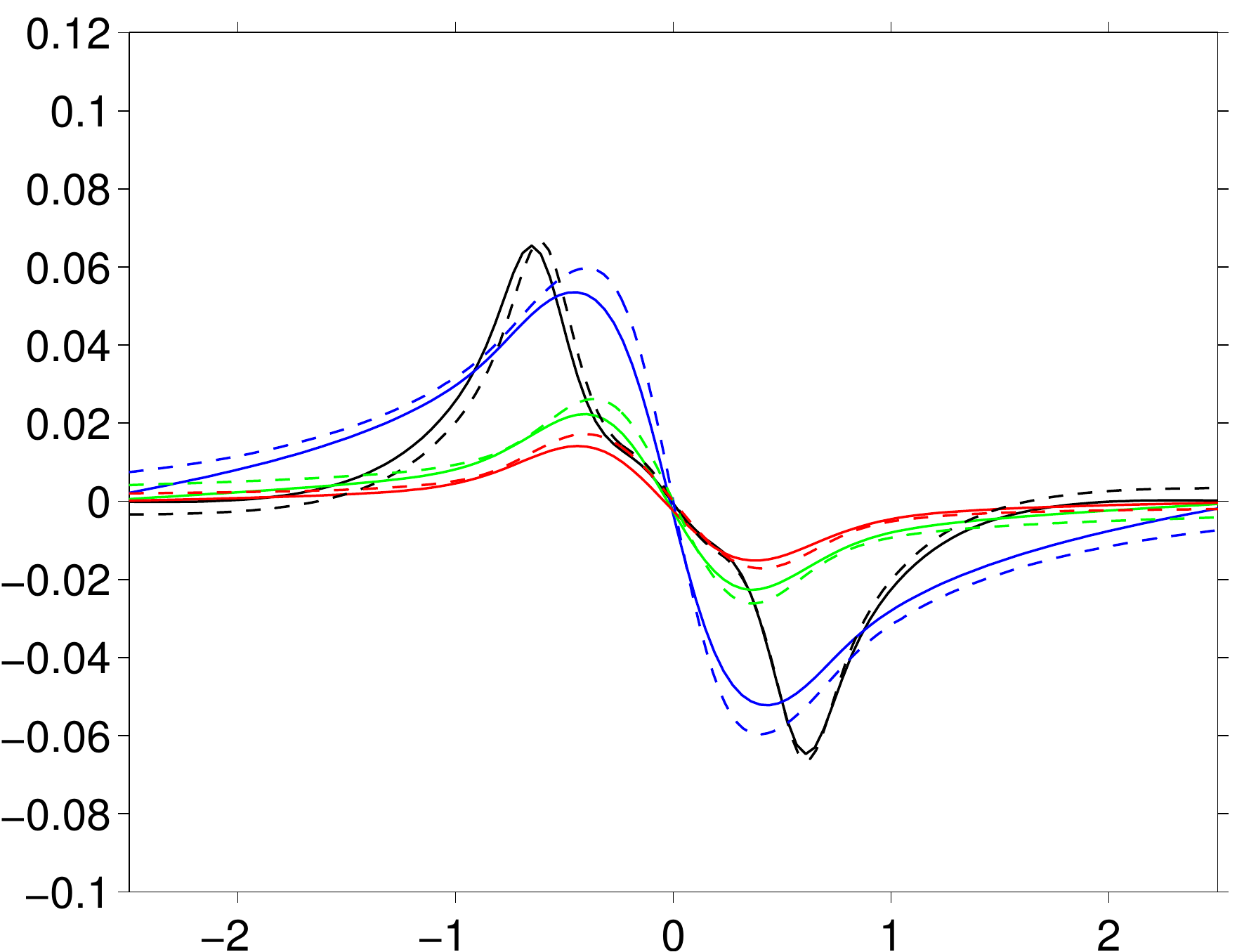}
    \centerline{$x_{pHz\perp}$}
  \end{minipage}
  \\
  }{
  \caption{%
    \revision{Cross-profiles of the relative velocity as in
    figure~\ref{fig-results-ref-cross-profiles-b1l}, but for the IBM
    computations in case BC-24. The reference data is shown as dashed
    lines.}{%
    Cross-profiles of the relative velocity components in local
    coordinates for the IBM computations in case BC-24, given along
    the lines indicated in cyan color in
    figures~\ref{fig-results-ref-contour-b1l} and
    \ref{fig-results-ibm-3d-iso-lambda2-iso-urel-b1l}$(c,d)$. 
    The color code indicates the distance downstream of the sphere: 
    {\color{black}\solid}, $x_{p\,\parallel}=-1$; 
    {\color{blue}\solid}, $x_{p\,\parallel}=-3$; 
    {\color{green}\solid}, $x_{p\,\parallel}=-5$; 
    {\color{red}\solid}, $x_{p\,\parallel}=-7$.
    The reference data is shown as dashed lines with the same color
    code. 
  }
  \protect\label{fig-results-ibm-cross-profiles-b1c-24}
  }
  }
\end{figure}
\begin{table}
  \centering
  \setlength{\tabcolsep}{5pt}
  \begin{tabular}{lc*{5}{r}l}
    &
    \multicolumn{1}{c}{$D/\Delta x$}&
    \multicolumn{1}{c}{$G$}&
    \multicolumn{1}{c}{$u_{pV}$}&
    \multicolumn{1}{c}{$u_{pH}$}&
    \multicolumn{1}{c}{$\omega_{pH}$}&
    \multicolumn{1}{c}{$L_r$}&
    \\[1ex]
    BC-15&
    $15$&  
    $177.67$&
    $-1.2514$&
    $0.1787$&
    $0.0997$&
    $1.4983$&
    \\
    ${\cal E}^{(BC-15)}$&&&
    $0.0771$&
    $0.0400$&
    $0.0634$&
    $0.0802$&
    \\[1ex]
    BC-18&
    $18$&  
    $177.42$&
    $-1.2695$&
    $0.1518$&
    $0.0700$& 
    $1.5726$&
    \\
    ${\cal E}^{(BC-18)}$&&&
    $0.0638$&
    $0.0201$&
    $0.0415$&
    $0.0346$&
    \\[1ex]
    BC-24& 
    $24$&  
    $178.46$&
    $-1.2846$&
    $0.1242$&
    $0.0376$&
    $1.5953$&
    \\
    ${\cal E}^{(BC-24)}$&&&
    $0.0527$&
    $0.0002$&
    $0.0176$&
    $0.0207$&
    \\[1ex]
    BC-36& 
    $36$&  
    $176.93$&
    $-1.2965$&
    $0.1081$&
    $0.0162$&
    $1.6215$& 
    \\
    ${\cal E}^{(BC-36)}$&&&
    $0.0439$&
    $0.0121$&
    $0.0018$&
    $0.0046$&
    \\[1ex]
    BC-48& 
    $48$&  
    $176.95$&
    $-1.3010$&
    $0.1028$&
    $0.0089$&
    $1.6291$&
    \\
    ${\cal E}^{(BC-48)}$&&&
    $0.0406$&
    $0.0160$&
    $0.0035$&
    $0.0000$&
    \\[1ex]
  \end{tabular}
  \caption{
    Results from IBM computations of case B (cf.\
    table~\ref{tab-parameters-ref}), where 
    $\rho_p/\rho_f=1.5$ and the nominal value of the Galileo number is
    $G=178.46$.
    The error is computed with respect to the results of the reference
    case BL (cf.\ table~\ref{tab-results-ref-B}).
  }    
  \label{tab-results-ibm-steady-oblique}
\end{table}
\begin{table}[b]
  \centering
  \setlength{\tabcolsep}{5pt}
  \begin{tabular}{lc*{6}{r}l}
    &
    \multicolumn{1}{c}{$D/\Delta x$}&
    \multicolumn{1}{c}{$G$}&
    \multicolumn{1}{c}{$u_{pV}$}&
    \multicolumn{1}{c}{$u_{pH}$}&
    \multicolumn{1}{c}{$\omega_{pH}$}&
    \multicolumn{1}{c}{$L_r$}&
    \\[1ex]
    BC-15h&
    $15$&  
    $177.01$& 
    $-1.2434$&
    $0.2090$&
    $0.1293$&
    $1.4479$&
    \\
    ${\cal E}^{(BC-15h)}$&&&
    $0.0830$&
    $0.0623$&
    $0.0853$&
    $0.1112$&
    \\[1ex]
    BC-18h&
    $18$&  
    $176.72$&
    $-1.2668$&
    $0.1735$&
    $0.0915$& 
    $1.5167$&
    \\
    ${\cal E}^{(BC-18h)}$&&&
    $0.0658$&
    $0.0361$&
    $0.0574$&
    $0.0689$&
    \\[1ex]
    BC-24h& 
    $24$&  
    $176.12$&
    $-1.2867$&
    $0.1375$&
    $0.0497$&
    $1.5665$&
    \\
    ${\cal E}^{(BC-24h)}$&&&
    $0.0511$&
    $0.0096$&
    $0.0265$&
    $0.0384$&
    \\[1ex]
    BC-36h& 
    $36$&  
    $176.15$&
    $-1.3005$&
    $0.1161$& 
    $0.0236$& 
    $1.5981$& 
    \\
    ${\cal E}^{(BC-36h)}$&&&
    $0.0409$&
    $0.0062$&
    $0.0073$&
    $0.0190$&
    \\[1ex]
    BC-48h& 
    $48$&  
    $176.72$&
    $-1.3067$&
    $0.1110$&
    $0.0153$&
    $1.6109$& 
    \\
    ${\cal E}^{(BC-48h)}$&&&
    $0.0364$&
    $0.0100$&
    $0.0012$&
    $0.0107$&
  \end{tabular}
  \caption{
    As table~\ref{tab-results-ibm-steady-oblique}, but computed with half the
    time step (i.e.\ $CFL\approx0.15$). 
  }    
  \label{tab-results-ibm-steady-oblique-DTHALF}
\end{table}
\begin{figure}
  \figpap{
  \centering
  \begin{minipage}{1.5ex}
    $c_p$
  \end{minipage}
  \begin{minipage}{.45\linewidth}
    \centerline{$(a)$}
    \includegraphics[width=\linewidth]
    {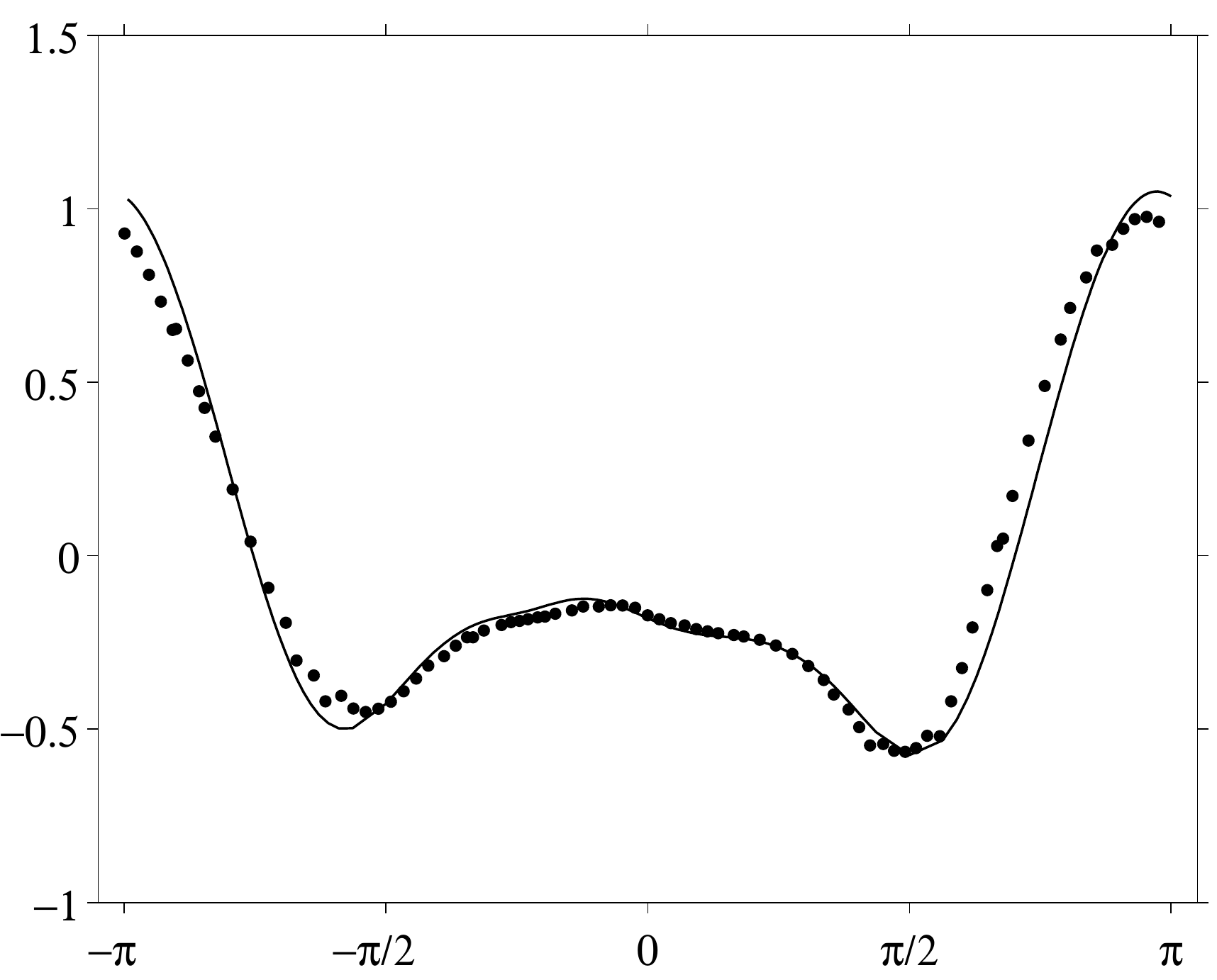}
    \centerline{$\theta_1$}
  \end{minipage}
  \hspace*{2ex}
  \begin{minipage}{.45\linewidth}
    \centerline{$(b)$}
    \includegraphics[width=\linewidth]
    {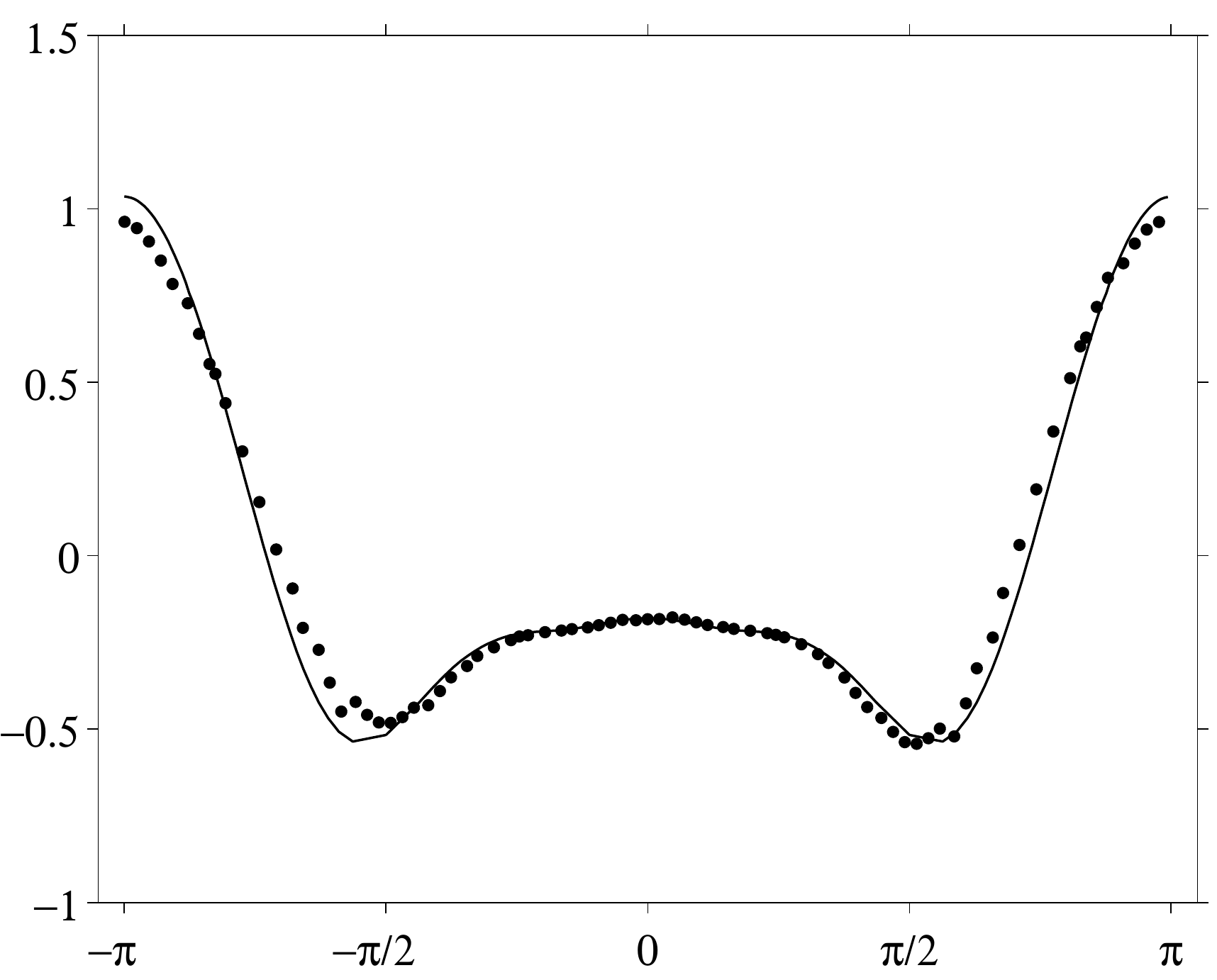}
    \centerline{$\theta_2$}
  \end{minipage}
  }{
  \caption{%
    Data from IBM computations of the steady axisymmetric case
    ($G=144$).
    The graphs show profiles of the pressure coefficient $c_p$ along
    two perpendicular great circles on the sphere surface. 
    $(a)$ in a plane given by the vertical direction $\mathbf{e}_z$
    and $\mathbf{e}_{pH}$; 
    $(b)$ in a plane given by the vertical direction $\mathbf{e}_z$
    and $\mathbf{e}_{pHz\perp}$. 
    The upstream stagnation point corresponds to a value of the angle  
    $\theta\approx\pm\pi$. Please refer to figure~\ref{fig-ref-notation-1}
    for the location of the great circles and the definition of the
    angles $\theta_1$, $\theta_2$.  
    Line styles and symbols indicate: 
    {\color{black}$\bullet$}, case BC-24; 
    {\color{black}\solid}, reference case BL 
    \revision{(cf.\ figure~\ref{fig-results-ref-press-b1l}$c$).}{%
      (taken along dashed lines in
      figure~\ref{fig-results-ref-press-b1l}). 
    }
  \protect\label{fig-results-ibm-press-b1c}
  }
  }
\end{figure}
Let us now turn to the steady-state results pertaining to the
particle motion relative to the ambient fluid, as given in
table~\ref{tab-results-ibm-steady-oblique}. 
Here it is again found that the vertical component of the relative
velocity, $u_{pV}$, is increasingly well predicted when refining in
space. The relative error amounts to 
7.7\% at $D/\Delta x=15$, 
to 
5.3\% at $D/\Delta x=24$ 
and to  
4\% at $D/\Delta x=48$. 
The horizontal component $u_{pH}$, which is non-zero at this Galileo
number value,
first appears to converge (error decreasing from 4\% at $D/\Delta
x=15$ to practically zero at $D/\Delta x=24$), but then tends towards
a value which is somewhat smaller than the reference result (error of
1.6\% at $D/\Delta x=48$). 
A similar result holds for the angular particle velocity around the
horizontal axis perpendicular to the particle motion, $\omega_{pH}$:
here the best match is obtained with $D/\Delta x=36$ (error of 0.2\%).  
Finally, it can be seen from
table~\ref{tab-results-ibm-steady-oblique} that the error in the
prediction of the length of the recirculation region, $L_r$,
monotonically decreases with spatial resolution (error insignificant
at $D/\Delta x=48$). Note that this latter observation is not in
contrast to the small discrepancy observed in the projected
recirculation length at the highest  spatial resolution in
figure~\ref{fig-results-ibm-wake-deficit-steady-oblique}$(b)$,
since $L_r$ is taken as the maximum extension of the contour 
with $u_{r\parallel}=0$ downstream of the particle (cf.\
\S~\ref{sec-ref-results-notation}).  

In order to clarify the non-monotonic behavior of the error with
spatial refinement (while keeping the CFL number fixed) observed for 
$u_{pH}$, $\omega_{pH}$ and for the profile of $u_{r\parallel}$ (taken
along the axis parallel to $\mathbf{e}_{p\parallel}$ and passing
through the sphere's center), we have repeated the above simulations
with the time step reduced by a factor of 2, i.e.\ with a maximum CFL
number of approximately 0.15. 
The results for the particle velocities at steady state, obtained with
the reduced time step and otherwise identical conditions, are given in
table~\ref{tab-results-ibm-steady-oblique-DTHALF}
and the axial velocity along the axis downstream of the particle is
shown in
figure~\ref{fig-results-ibm-dthalf-wake-deficit-steady-oblique}. 
%
Note that although the particle motion is steady, the results obtained
with the present methodology still depend upon the numerical time step
$\Delta t$ for two reasons: first, the particle still undergoes a
motion with respect to the finite-difference grid, i.e.\ the flow is
non-trivially unsteady in the fixed frame of reference; 
secondly, the use of a fractional step method introduces a ``slip
error'' on the fluid-solid interface which is of order $\Delta t$
\citep[cf.\ discussion in][]{uhlmann:05a}. 
From table~\ref{tab-results-ibm-steady-oblique-DTHALF} it can be seen
that the respective errors of all particle-related degrees of freedom 
(except for $u_{pH}$ at $D/\Delta x=48$) 
behave in a monotonic fashion at this lower value of the CFL number,
i.e.\ decreasing with decreasing grid width $\Delta x$. 
The observed convergence behavior suggests that there is one
contribution to the overall numerical error which is proportional to
the ratio $\Delta t/\Delta x$.
%
%

The profiles of the three components of the relative velocity
$\mathbf{u}_r$ in the three local coordinate directions
$\mathbf{e}_{p\perp}$, 
$\mathbf{e}_{pHz\perp}$,
$\mathbf{e}_{p\parallel}$ 
along the lines perpendicular to the axis of particle motion (as
indicated by the magenta-colored dashed lines in 
figure~\ref{fig-results-ibm-3d-iso-lambda2-iso-urel-b1l}$c,d$) are 
shown in figure~\ref{fig-results-ibm-cross-profiles-b1c-24} for one
spatial resolution ($D/\Delta x=24$). 
The graphs confirm that all aspects of the wake flow are captured with
high accuracy by the IBM simulation when a sufficient spatial
resolution is applied. 

Finally, the surface pressure on the sphere is illustrated in
figure~\ref{fig-results-ibm-press-b1c} by way of the coefficient $c_p$
taken along the previously defined two great circles (cf.\ sketch in
figure~\ref{fig-ref-notation-1}$b,c$).  
It can be seen that the simulation is able to faithfully capture the
shift of the local pressure maximum on the downstream side of the
sphere along the great circle which is parallel to
$\mathbf{e}_{pHz\perp}$ (i.e.\ towards small negative values of
$\theta_1$).  
\subsubsection{Oscillating oblique regime}
\label{sec-ibm-results-oscill-oblique}
\begin{figure}[b]
  \figpap{
  \centering
  \begin{minipage}{4ex}
    $u_{pV}$
  \end{minipage}
  \begin{minipage}{.37\linewidth}
    \centerline{$(a)$}
    \includegraphics[width=\linewidth]
    {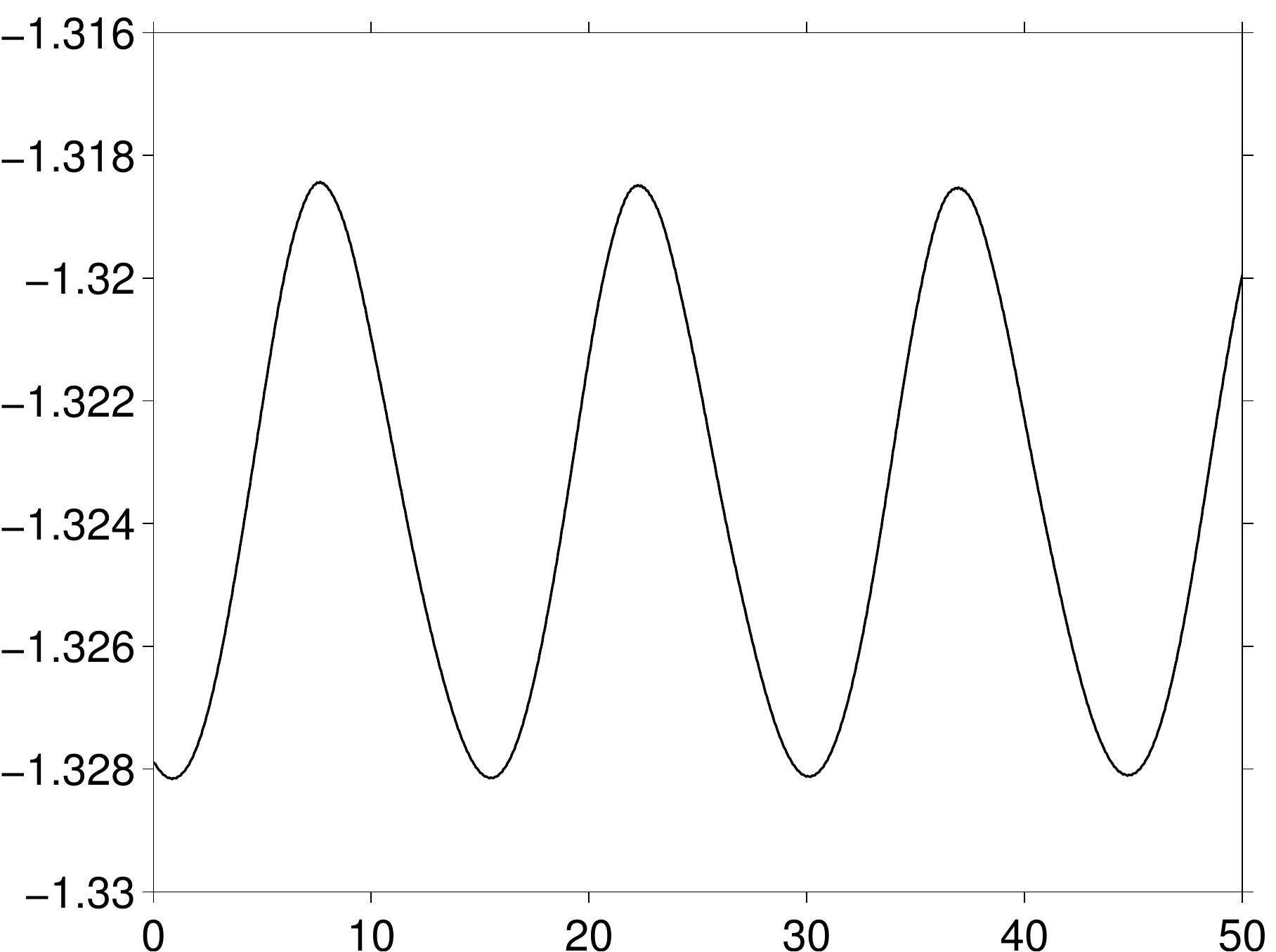}
    \\
    \centerline{$t$}
  \end{minipage}
  \revision{%
    \\[1ex]}{\hspace*{5ex}}
  \begin{minipage}{4ex}
    $u_{pH}$
  \end{minipage}
  \begin{minipage}{.37\linewidth}
    \centerline{$(b)$}
    \includegraphics[width=\linewidth]
    {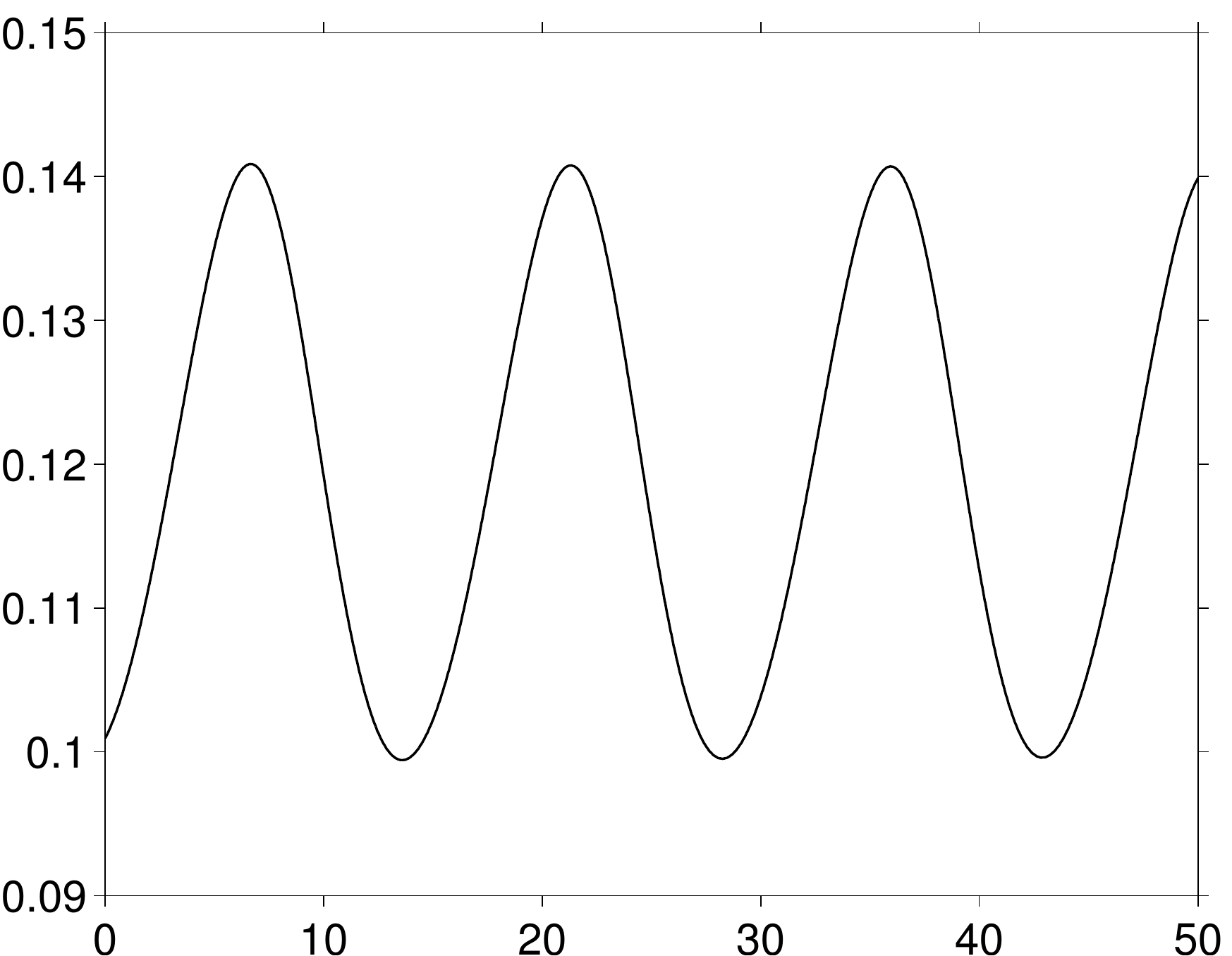}
    \\
    \centerline{$t$}
  \end{minipage}
  \revision{%
  \\[1ex]
  \begin{minipage}{4ex}
    $\omega_{pH}$
  \end{minipage}
  \begin{minipage}{.37\linewidth}
    \centerline{$(c)$}
    \includegraphics[width=\linewidth]
    {figure_31c.pdf}
    \\
    \centerline{$t$}
  \end{minipage}
  \\
  }{}
  }{
  \caption{%
    IBM results for case CC-48 ($G=190.96$, $D/\Delta x=48$), exhibiting
    time-periodic dynamics.  
    The graphs show the temporal evolution of: 
    $(a)$ the vertical particle velocity component; 
    $(b)$ the horizontal particle velocity component%
    \revision{; 
      $(c)$ the horizontal angular particle velocity component.}{.} 
    The corresponding reference data is shown in
    figure~\ref{fig-results-ref-history-c1l}. 
    \protect\label{fig-results-ibm-history-c1c}
  }
  }
\end{figure}
The extent of the regime in which the particle motion is oblique (with
respect to the vertical direction), restricted to a plane in space,
and where it is time-periodic spans a relatively narrow range of
values of the Galileo number, $185\lesssim G\lesssim215$
(according to our data from the spectral-element simulations of
\S~\ref{sec-ref-results}).  
%
%
While defining the relative error at the beginning of
\S~\ref{sec-ibm-results} 
we have mentioned the sensitivity of thresholds of bifurcation to
numerical accuracy. This is the more true the higher the order of the
bifurcation. Since the oscillating oblique regime arises as the result
of a secondary bifurcation, a small inaccuracy can induce an upward
shift of its threshold by several Galileo number units. 
Using the present immersed boundary method, simulations using spatial
resolutions of $D/\Delta x=24$ and $36$ fail to capture the secondary
instability at $G=190$.
%
These simulations yield exponentially decaying oscillations at the
correct frequencies showing that the threshold lies above $G=190$. 
%
%
In the following we will present results obtained with a
spatial resolution of $D/\Delta x=48$. 

\revision{%
  The shape of the signals of the three non-zero translational and angular
  velocity components ($u_{pV}$, $u_{pH}$, $\omega_{pH}$) is shown in
  figure~\ref{fig-results-ibm-history-c1c}.}{%
  The shape of the signals of the translational 
  velocity components ($u_{pV}$ and $u_{pH}$) is shown in
  figure~\ref{fig-results-ibm-history-c1c}.}
Comparing the time evolution
with the one of the reference signals (cf.\
figure~\ref{fig-results-ref-history-c1l}) reveals a very close match.  
%
The mean values and fluctuation amplitudes of these periodic signals,
as defined in
(\ref{equ-res-ref-def-mean}-\ref{equ-res-ref-def-ampli}), as well as
the oscillation frequency are shown in
table~\ref{tab-results-ibm-oscill-oblique}. All quantities (including
the frequency) are predicted with errors below 4\%. 
%
\begin{table}
  \centering
  \setlength{\tabcolsep}{5pt}
  \begin{tabular}{lc*{8}{r}}
    &
    \multicolumn{1}{c}{$D/\Delta x$}&
    \multicolumn{1}{c}{$G$}&
    \multicolumn{1}{c}{$\overline{u}_{pV}$}&
    \multicolumn{1}{c}{$\overline{u}_{pH}$}&
    \multicolumn{1}{c}{$\overline{\omega}_{pH}$}&
    \multicolumn{1}{c}{$u_{pV}^\prime$}&
    \multicolumn{1}{c}{$u_{pH}^\prime$}&
    \multicolumn{1}{c}{$\omega_{pH}^\prime$}&
    \multicolumn{1}{c}{$f$}
    \\[1ex]
    CC-48& 
    $48$&
    $190.96$&
    $-1.3233$&
    $0.1201$&
    $0.0061$&
    $0.0049$&
    $0.0207$&
    $0.0051$&
    $0.0683$
    \\
    ${\cal E}^{(CC-48)}$&&&
    $0.0383$&
    $0.0116$&
    $0.0043$&
    $0.0023$&
    $0.0089$&
    $0.0021$&
    $0.0380$ 
  \end{tabular}
  \caption{
    Results from IBM computations of case C (cf.\
    table~\ref{tab-parameters-ref}), where 
    $\rho_p/\rho_f=1.5$ and the nominal value of the Galileo number is
    $G=190$.
    The error is computed with respect to the results of the reference
    case CL (cf.\ table~\ref{tab-results-ref-C}).
  }    
  \label{tab-results-ibm-oscill-oblique}
\end{table}
\begin{figure}
  \figpap{
  \centering
  \begin{minipage}{4ex}
    $u_{pH}$
  \end{minipage}
  \begin{minipage}{.35\linewidth}
    \includegraphics[width=\linewidth]
    {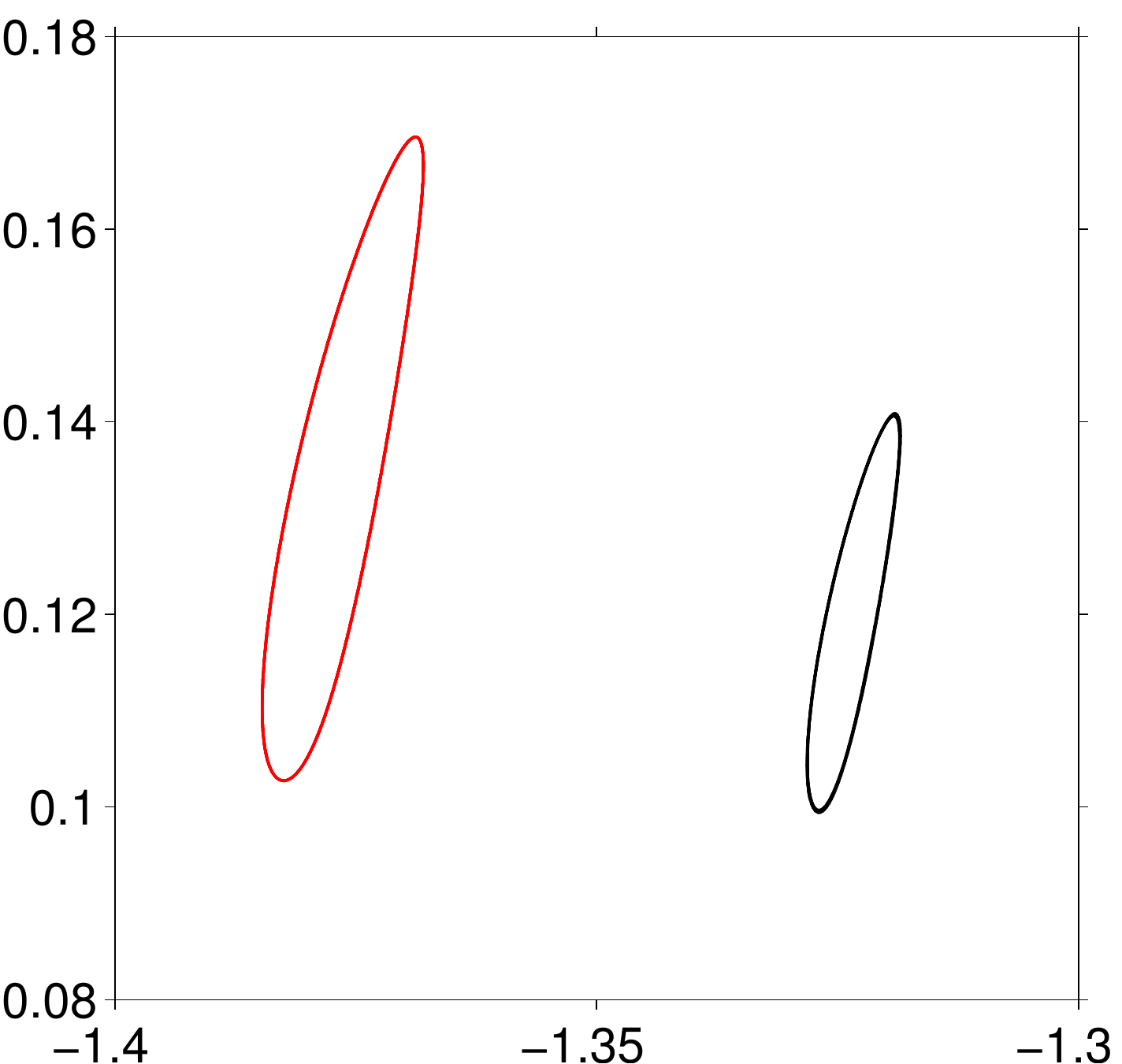}
    \centerline{$u_{pV}$}
  \end{minipage}
  }{
  \caption{%
    IBM results for case CC ($G=190$), exhibiting time-periodic
    dynamics.  
    Phase-space plot in the two-dimensional space spanned by the
    vertical and horizontal particle velocity components. 
    {\color{black}\solid}, case CC-48; 
    {\color{red}\solid}, reference data (case CL). 
    \protect\label{fig-results-ibm-phase-space-c1c}
  }
  }
\end{figure}
Finally, a phase-space plot of horizontal versus vertical particle
velocity is shown in figure~\ref{fig-results-ibm-phase-space-c1c}. 
When equal scaling of the axis is used (as in that figure), the
trajectories in phase space have a roughly elliptic shape, with a
strong vertical elongation (the fluctuations of the horizontal
component are much larger than those of the vertical one), and with a
slight inclination with respect to the vertical direction. The IBM
results reproduce the shape of the phase-space trajectory very well,
albeit at a somewhat smaller scale, i.e.\ the fluctuation amplitude is
generally under-predicted, as obvious from the results shown in
table~\ref{tab-results-ibm-oscill-oblique}. 
The smaller secondary instability amplitude is to be put, again, on
account of the upward shift of the instability threshold. 
%
\subsubsection{Chaotic regime}
\label{sec-ibm-results-chaotic}
\begin{figure}
  \figpap{
  \centering
  \begin{minipage}{3ex}
    $w_{pr}$
  \end{minipage}
  \begin{minipage}{.45\linewidth}
    \centerline{$(a)$}
    \includegraphics[width=\linewidth]
    {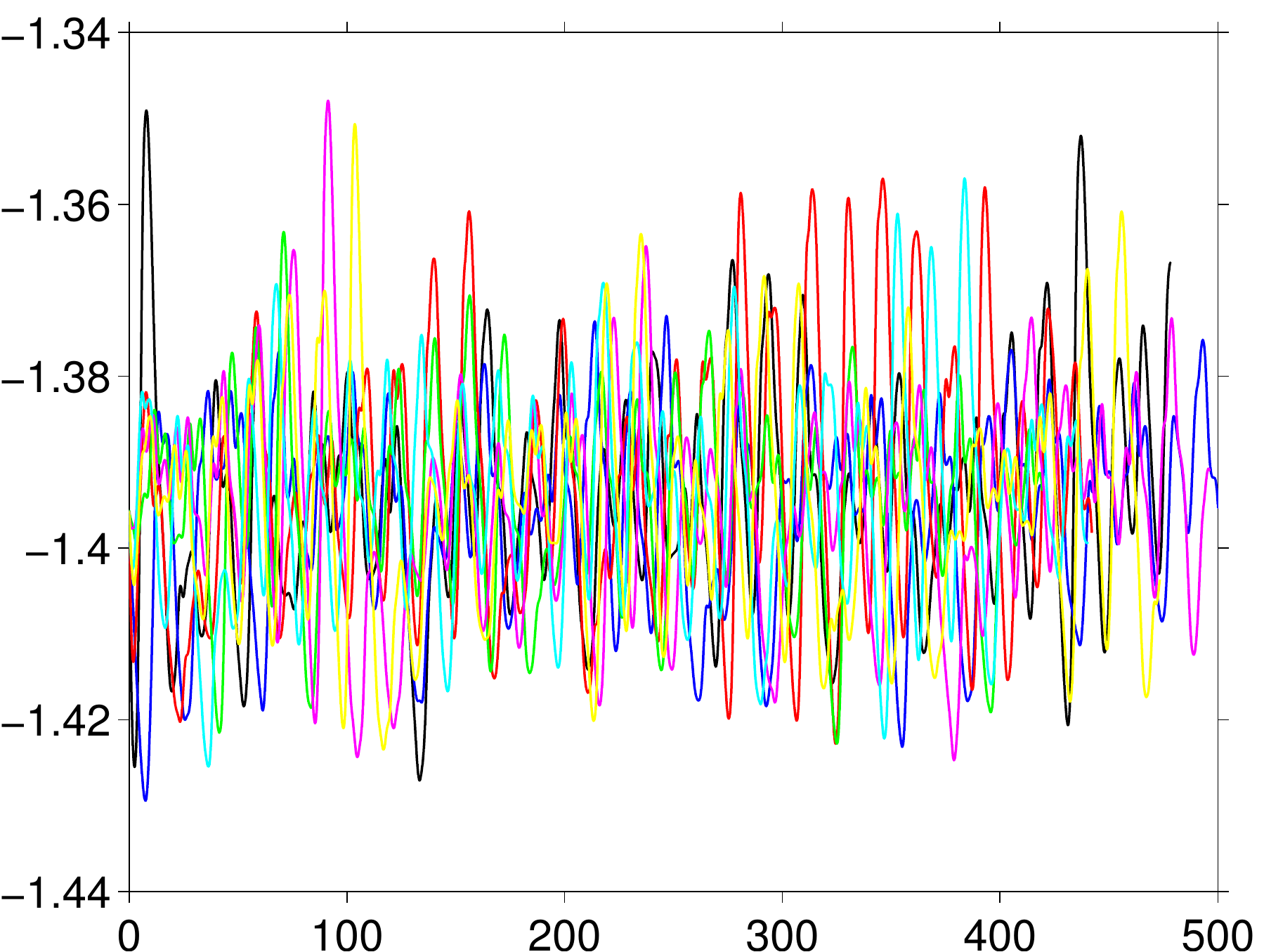}
    \\
    \centerline{$t$}
  \end{minipage}
  \hfill
  \begin{minipage}{3ex}
    $u_{pr}$
  \end{minipage}
  \begin{minipage}{.35\linewidth}
    \centerline{$(b)$}
    \includegraphics[width=\linewidth]
    {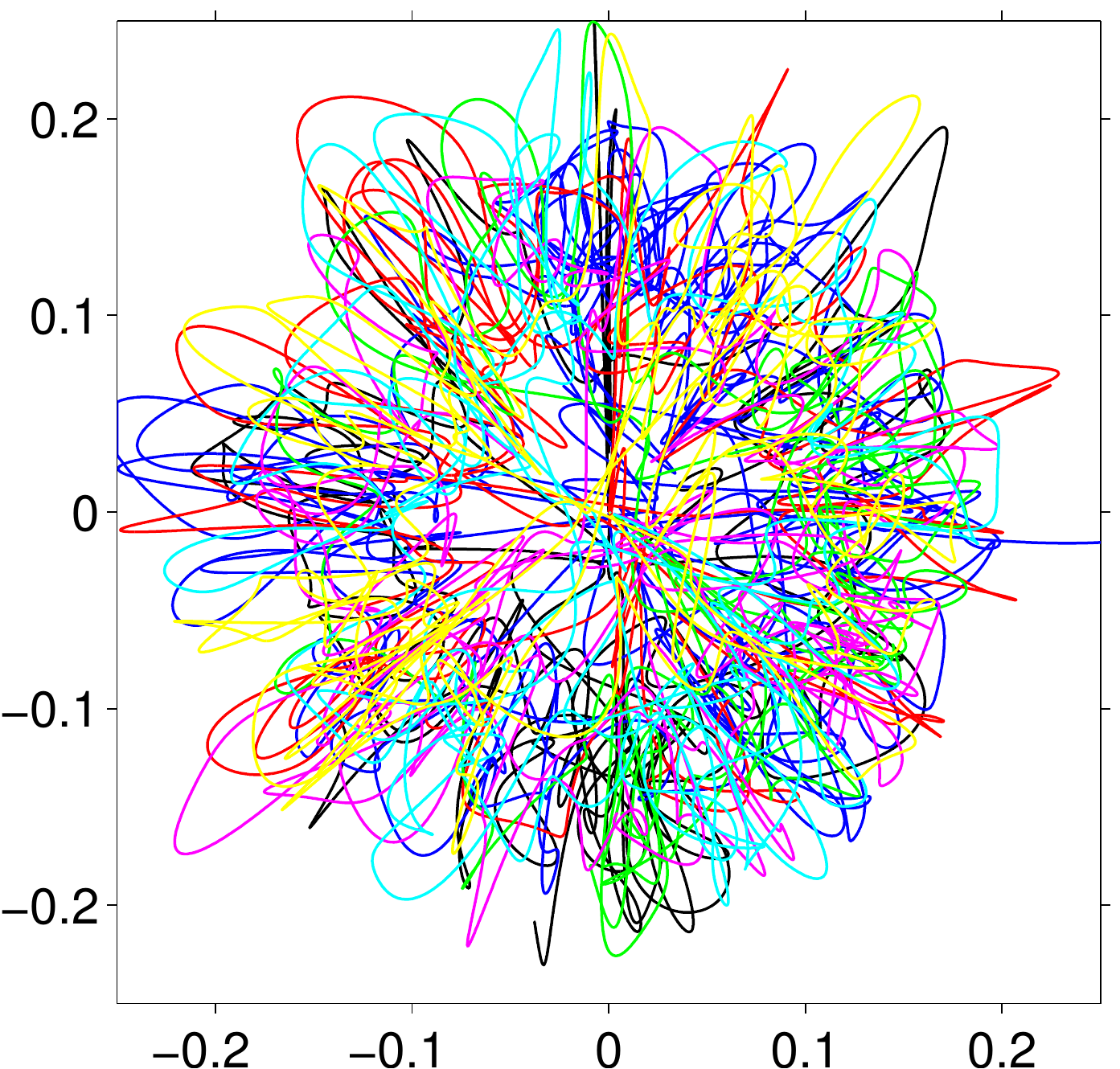}
    \\
    \centerline{$v_{pr}$}
  \end{minipage}
  }{
  \caption{%
    IBM results for case DC-36 ($G=250$), exhibiting chaotic
    dynamics. 
    $(a)$ Temporal evolution of the vertical particle velocity
    component (measured relative to the ambient fluid velocity). 
    $(b)$ Phase-space plot in the two-dimensional space spanned by the
    two horizontal particle velocity components. 
    Each color corresponds to one realization under identical physical
    and numerical conditions, but starting with different initial fields.
    \protect\label{fig-results-ibm-history-d1c}
  }
  }
\end{figure}
\begin{figure}[b]
  \figpap{
  \begin{minipage}{.22\linewidth}
    \centerline{$(a)$}
    \includegraphics[width=\linewidth,clip=true,
    viewport=1030 827 1500 1750]
    {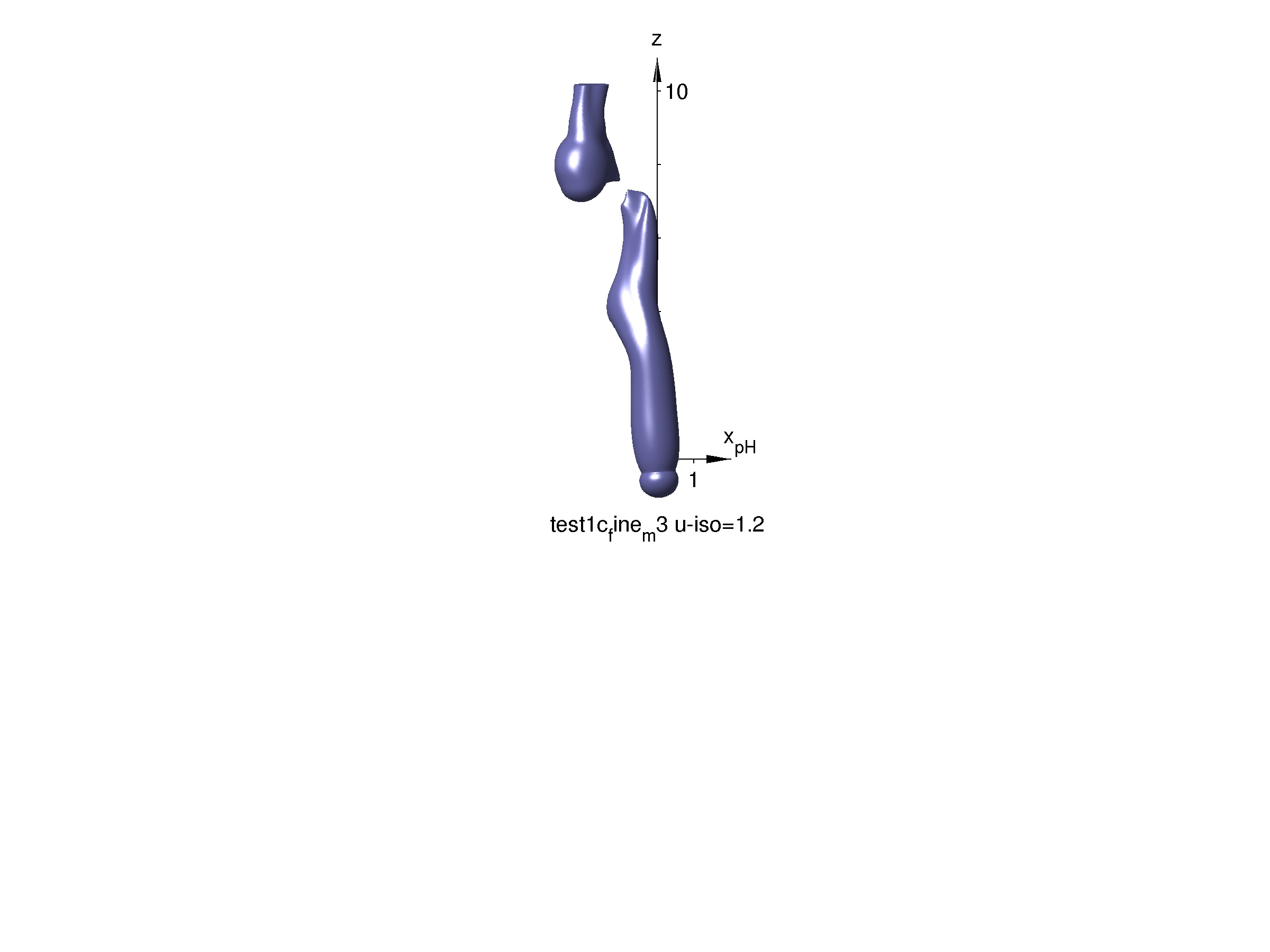}
  \end{minipage}
  \hfill
  \begin{minipage}{.22\linewidth}
    \centerline{$(b)$}
    \includegraphics[width=\linewidth,clip=true,
    viewport=1030 827 1500 1750]
    {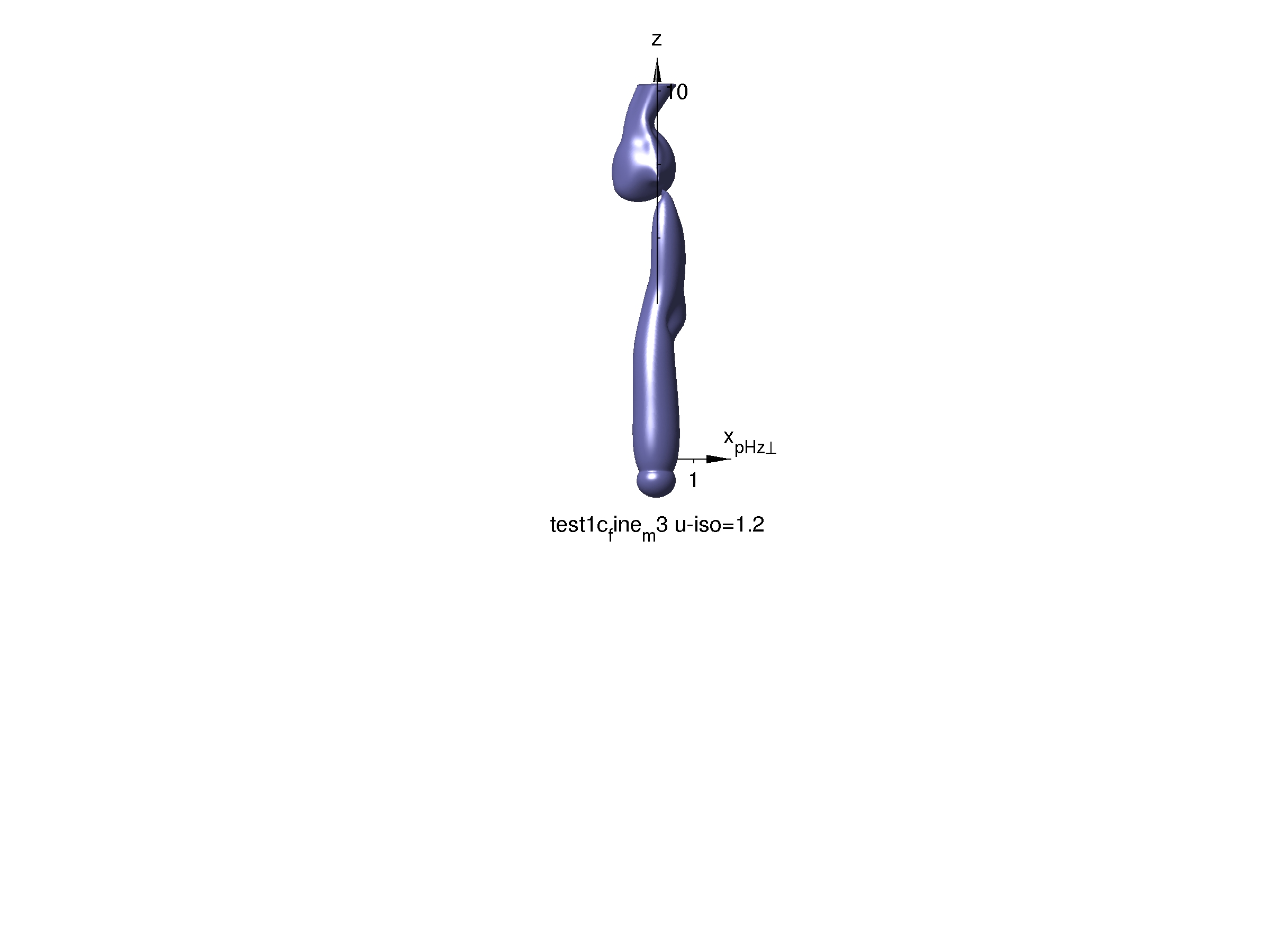}
  \end{minipage}
  \hfill
  \begin{minipage}{.22\linewidth}
    \centerline{$(c)$}
    \includegraphics[width=\linewidth,clip=true,
    viewport=1030 827 1500 1750]
    {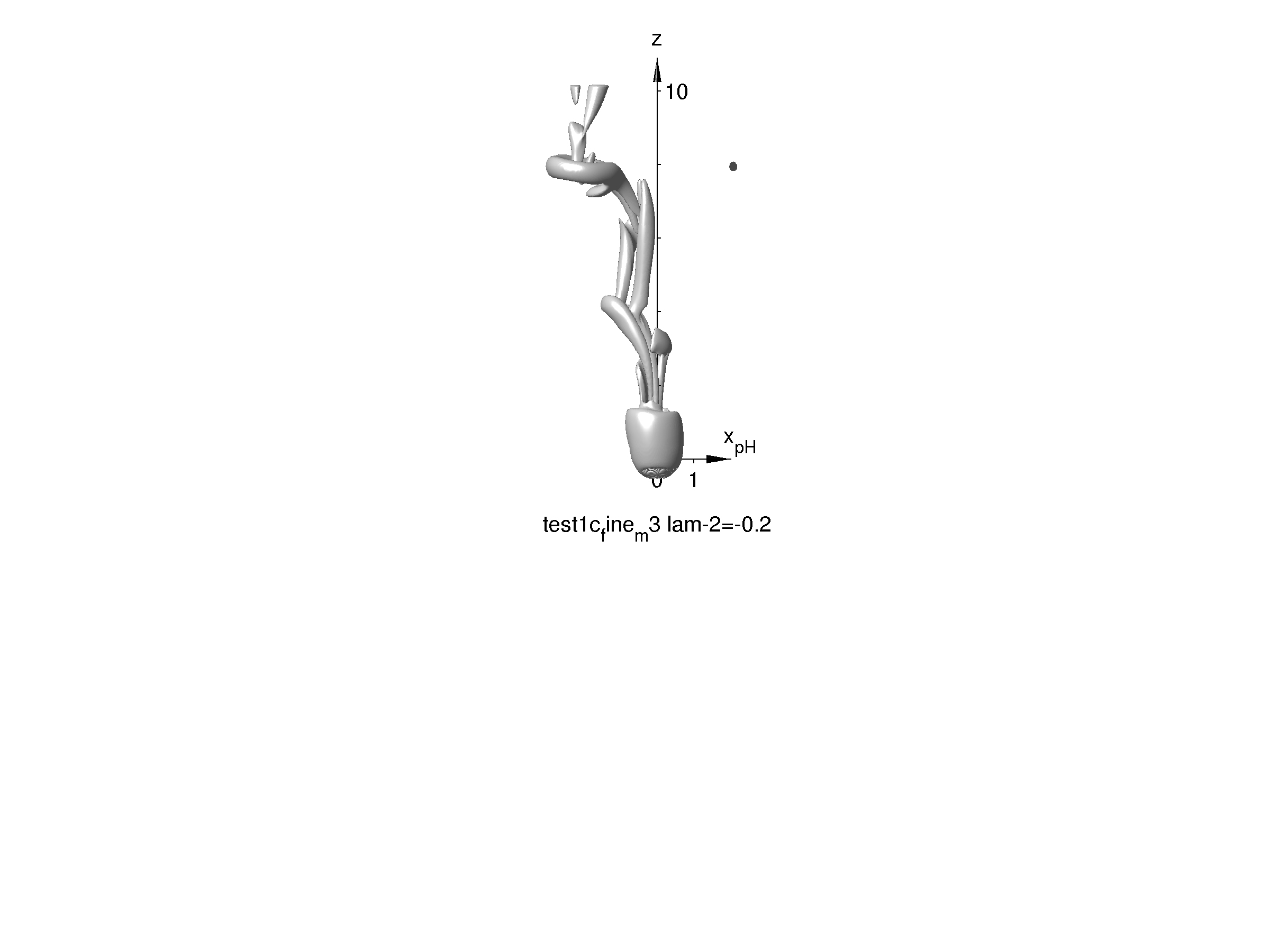}
  \end{minipage}
  \hfill
  \begin{minipage}{.22\linewidth}
    \centerline{$(d)$}
    \includegraphics[width=\linewidth,clip=true,
    viewport=1030 827 1500 1750]
    {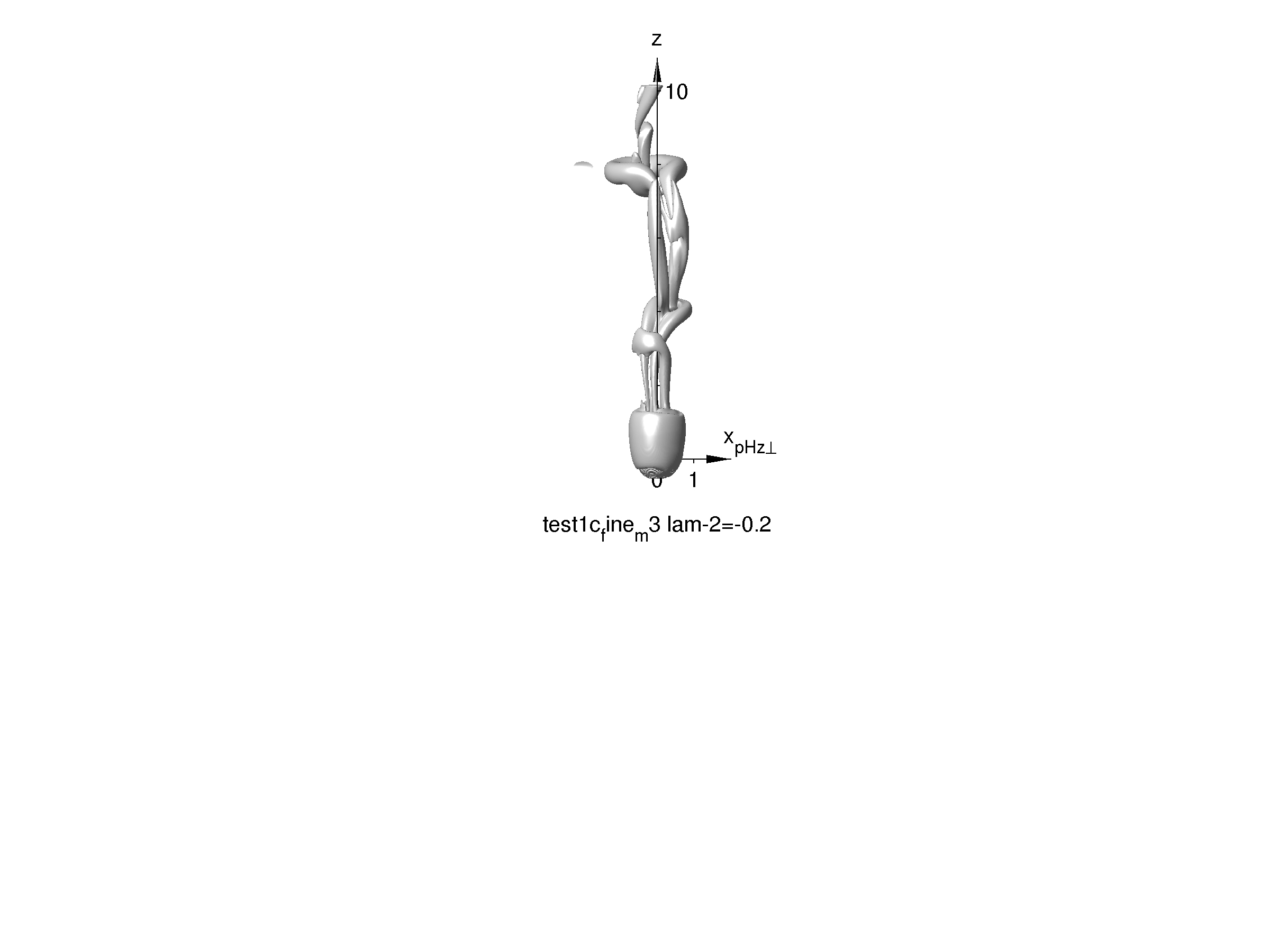}
  \end{minipage}
  }{
  \caption{%
    IBM results for case DC-36 ($G=250$),
    showing an instantaneous flow field. 
    Graphs $(a)$ and $(b)$ show the surface where
    $u_{r\parallel}=1.2$.  
    Graphs $(c)$ and $(d)$ show the surface where 
    $\lambda_2=-0.2$. 
    In $(a)$ and $(c)$ the view is directed along $\mathbf{e}_{pHz\perp}$;  
    in $(b)$ and $(d)$ it is directed along $\mathbf{e}_{pH}$.
    \protect\label{fig-results-ibm-3d-iso-d1c}
  }
  }
\end{figure}
In order to capture the chaotic particle motion observed at $G=250$
(cf.\ \S~\ref{sec-ref-results-chaotic}) it was found that a spatial
resolution of $D/\Delta x\leq24$ is not sufficient when employing the
present immersed boundary technique. Therefore, we have computed this
case with $D/\Delta x=36$. 

A number of $N_{runs}=7$ independent realizations has been simulated,
each at identical physical and numerical conditions, but starting from 
a different initial field. The total time simulated amounts to 3570
units. 
%
Figure~\ref{fig-results-ibm-history-d1c}$(a)$ shows the time history
of the vertical component of the particle velocity relative to the
ambient fluid, $w_{pr}$, over the various simulations; 
figure~\ref{fig-results-ibm-history-d1c}$(b)$ gives an impression of
the corresponding trajectories in phase space spanned by the two
horizontal components $u_{pr}$ and $v_{pr}$. Both graphs have a
similar appearance as the counterparts obtained with the
spectral-element method (cf.\
figure~\ref{fig-results-ref-history-d1l}).  
Additionally, the flow field for one snapshot is visualized in
figure~\ref{fig-results-ibm-3d-iso-d1c}, where iso-surfaces of the
relative velocity projected upon the instantaneous particle motion,
$u_{r\parallel}$, as well as of $\lambda_2$ are shown. Clearly, a
similar wake as in the reference case (cf.\
figure~\ref{fig-results-ref-3d-iso-d1l}) is obtained.  

\begin{table}
  \centering
  \setlength{\tabcolsep}{5pt}
  \begin{tabular}{lccr*{4}{c}}
    &
    \multicolumn{1}{c}{$D/\Delta x$}&
    \multicolumn{1}{c}{$G$}&
    \multicolumn{1}{c}{$\langle u_{pV}\rangle$}&
    \multicolumn{1}{c}{$\langle u_{pV}^{\prime\prime}u_{pV}^{\prime\prime}\rangle^{1/2}$}&
    \multicolumn{1}{c}{$\langle u_{pr}^{\prime\prime}u_{pr}^{\prime\prime}\rangle^{1/2}$}&
    \multicolumn{1}{c}{$\langle\omega_{pV}^{\prime\prime}\omega_{pV}^{\prime\prime}\rangle^{1/2}$}&
    \multicolumn{1}{c}{$\langle\omega_{px}^{\prime\prime}\omega_{px}^{\prime\prime}\rangle^{1/2}$}
    \\[1ex]
    DC-36& 
    $36$&
    $250.00$&
    $-1.3946$&
    $0.0129$&
    $0.0959$&
    $0.0012$&
    $0.0136$
    \\
    ${\cal E}^{(DC-36)}$&&&
    $0.0451$&
    $0.0029$&
    $0.0072$&
    $0.0000$&
    $0.0047$ 
    %
  \end{tabular}
  \caption{
    %
    Results from IBM computations of case D (cf.\
    table~\ref{tab-parameters-ref}), where 
    $\rho_p/\rho_f=1.5$ and the nominal value of the Galileo number is
    $G=250$.
    Note that an ensemble of $N_{runs}=7$ particle paths was analyzed with a
    total averging interval of $3570$ non-dimensional
    time units.  
    The error is computed with respect to the results of the reference
    case DL (cf.\ table~\ref{tab-results-ref-D}).
  }
  \label{tab-results-ibm-chaotic}
\end{table}
The mean value of the vertical particle velocity component (relative
to the ambient fluid) as well as the rms values of all translational and
angular velocity components computed according to the definitions
(\ref{equ-res-ref-def-stat-mean}-\ref{equ-res-ref-def-stat-fluct}) are
shown in table~\ref{tab-results-ibm-chaotic}. 
It is found that the agreement with the reference data is very good. 
The relative error associated to the mean settling velocity measures
4.5\%, while the remaining components are all predicted with errors
below one percent. 
In particular, the rms values are systematically higher in the IBM
simulation
(except for
$\langle\omega_{pV}^{\prime\prime}\omega_{pV}^{\prime\prime}\rangle^{1/2}$). 
This can be explained by a less chaotic behavior as
compared to the reference simulation due to an upward shift of the
onset of chaos. 

\begin{figure}[b]
  \figpap{
  \centering
  \begin{minipage}{3ex}
    \rotatebox{90}{$\sigma\cdot pdf$}
  \end{minipage}
  \begin{minipage}{.45\linewidth}
    \centerline{$(a)$}
    \includegraphics[width=\linewidth]
    {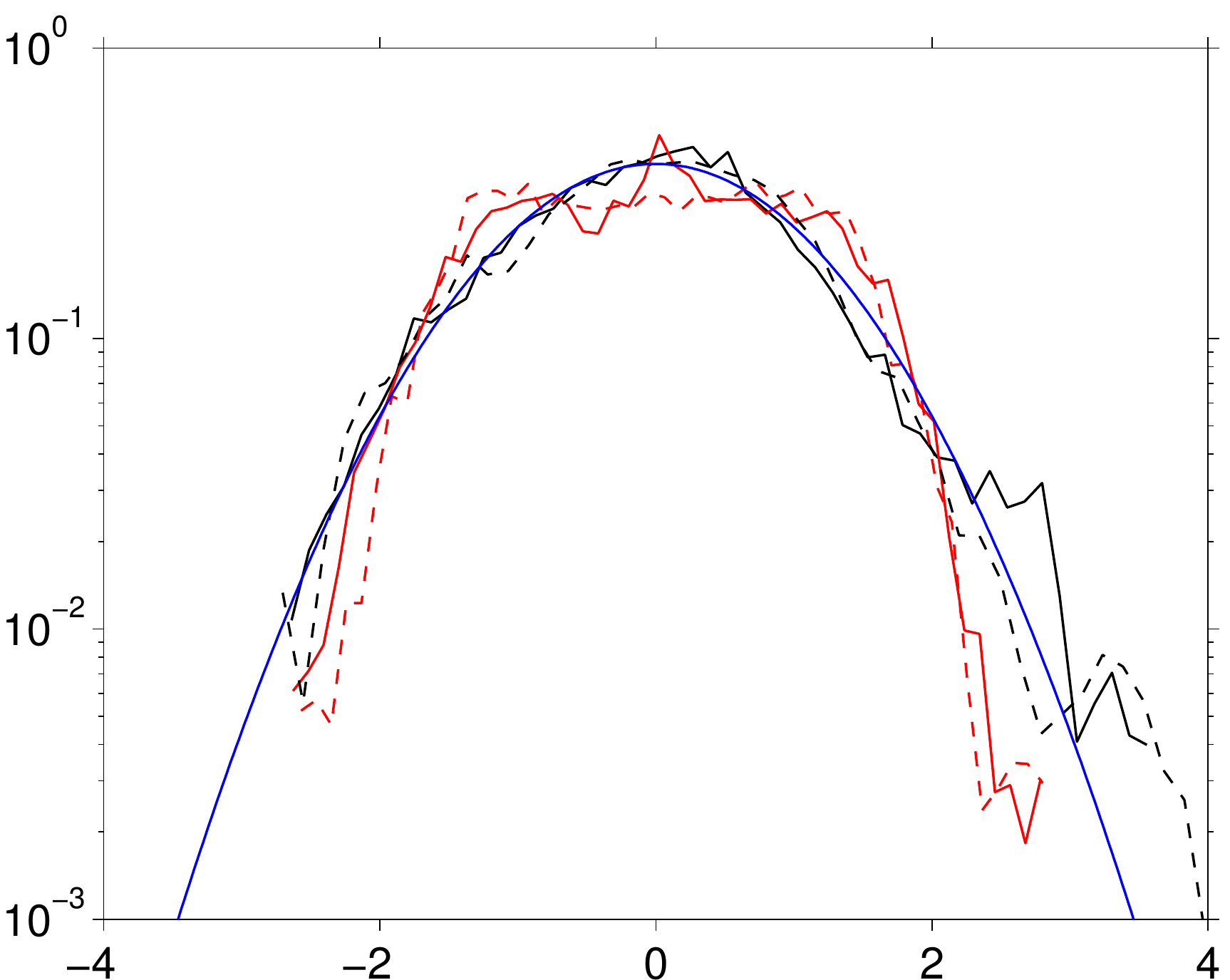}
    \\
    \centerline{$u_{p\alpha}^{\prime\prime}/\sigma$}
  \end{minipage}
  \hfill
  \begin{minipage}{3ex}
    \rotatebox{90}{$\sigma\cdot pdf$}
  \end{minipage}
  \begin{minipage}{.45\linewidth}
    \centerline{$(b)$}
    \includegraphics[width=\linewidth]
    {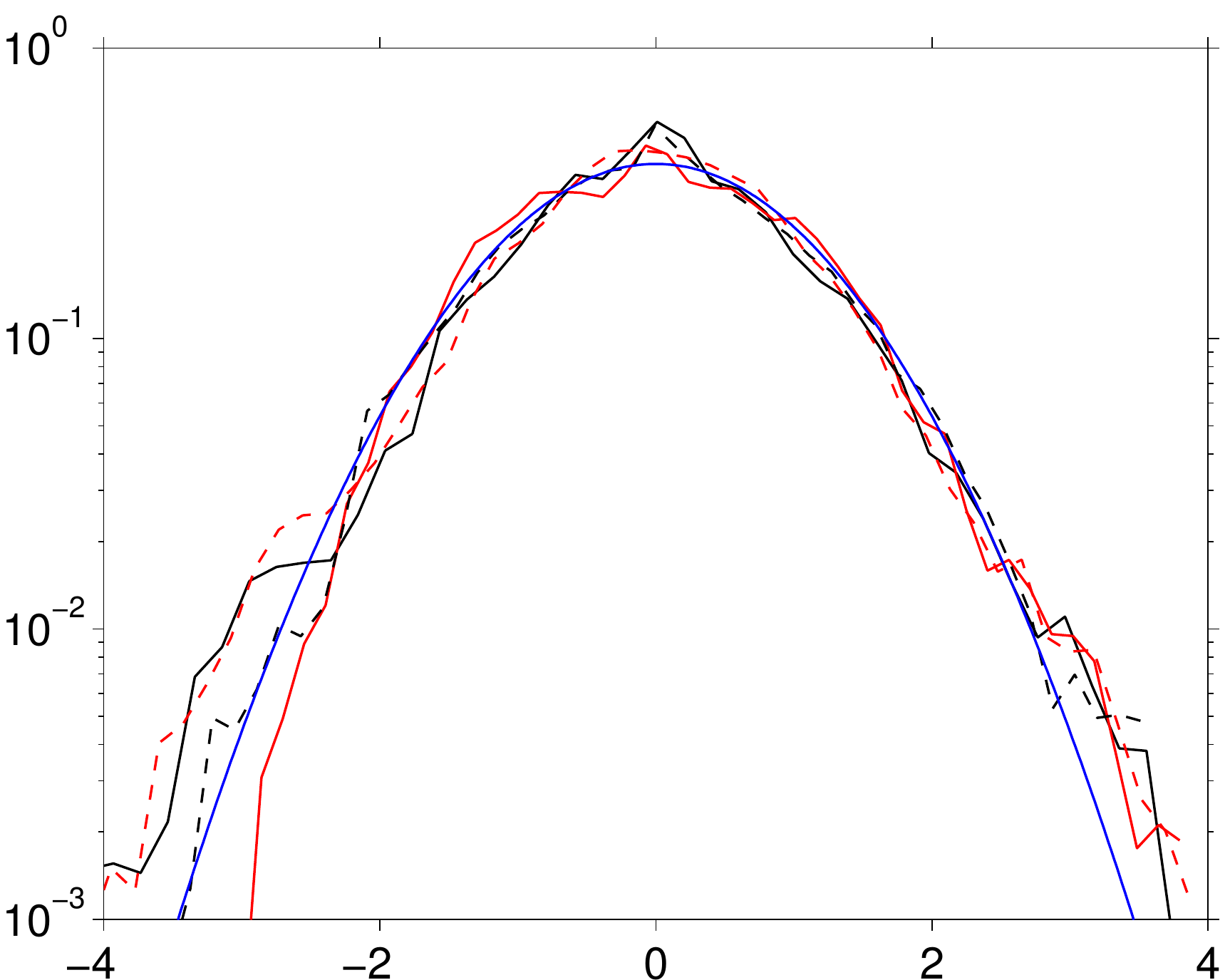}
    \\
    \centerline{$\omega_{p\alpha}^{\prime\prime}/\sigma$}
  \end{minipage}
  }{
  \caption{%
    IBM results for case DC-36 ($G=250$). 
    Probability density functions of: 
    $(a)$ translational particle velocity; 
    $(b)$ angular particle velocity.
    The line-styles are as follows: 
    \solid, vertical component; 
    {\color{red}\solid}, horizontal component; 
    {\color{blue}\solid}, Gaussian reference curve. 
    The reference data is shown as dashed lines. 
    \protect\label{fig-results-ibm-pdf-d1c}
  }
  }
\end{figure}
The normalized probability density functions corresponding to the
translational and angular velocity components are shown in
figure~\ref{fig-results-ibm-pdf-d1c} alongside the reference data
which is included in order to facilitate a direct comparison. 
It can be seen that -- up to the statistical uncertainty inherent in
both data-sets -- all significant features found in the reference
data are reproduced faithfully by the IBM simulation using a spatial
resolution of $D/\Delta x=36$. In particular, the plateau-like shape
of the pdf of $u_{pH}$ and its sharp drop-off around approximately
twice the standard deviation is captured; the same is true for the
roughly Gaussian-shaped pdf of $u_{pV}$ and the mild peak around the
mean value in $\omega_{pV}$. 
On the other hand, it can be observed that the horizontal components
of the translational and angular particle velocity, $u_{pH}$ and
$\omega_{pH}$, exhibit mild peaks around their mean values which are
not present in the reference data.  

\begin{figure}
  \figpap{
  \centering
  \begin{minipage}{3ex}
    \rotatebox{90}{$R_{u_{p\alpha} u_{p\alpha}}$}
  \end{minipage}
  \begin{minipage}{.45\linewidth}
    \centerline{$(a)$}
    \includegraphics[width=\linewidth]
    {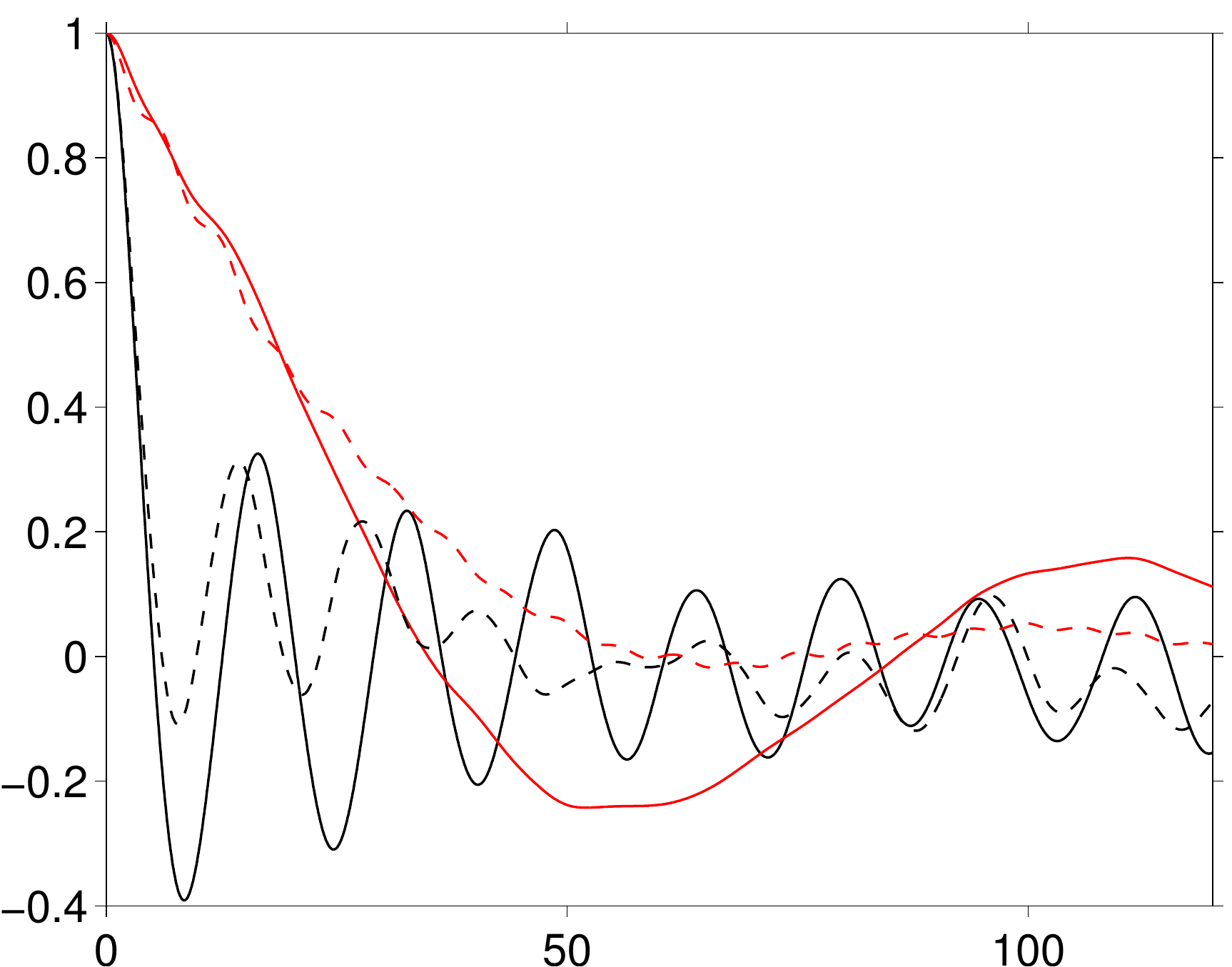}
    \\
    \centerline{$\tau_{sep}$}
  \end{minipage}
  \hfill
  \begin{minipage}{3ex}
    \rotatebox{90}{$R_{\omega_{p\alpha}\omega_{p\alpha}}$}
  \end{minipage}
  \begin{minipage}{.45\linewidth}
    \centerline{$(b)$}
    \includegraphics[width=\linewidth]
    {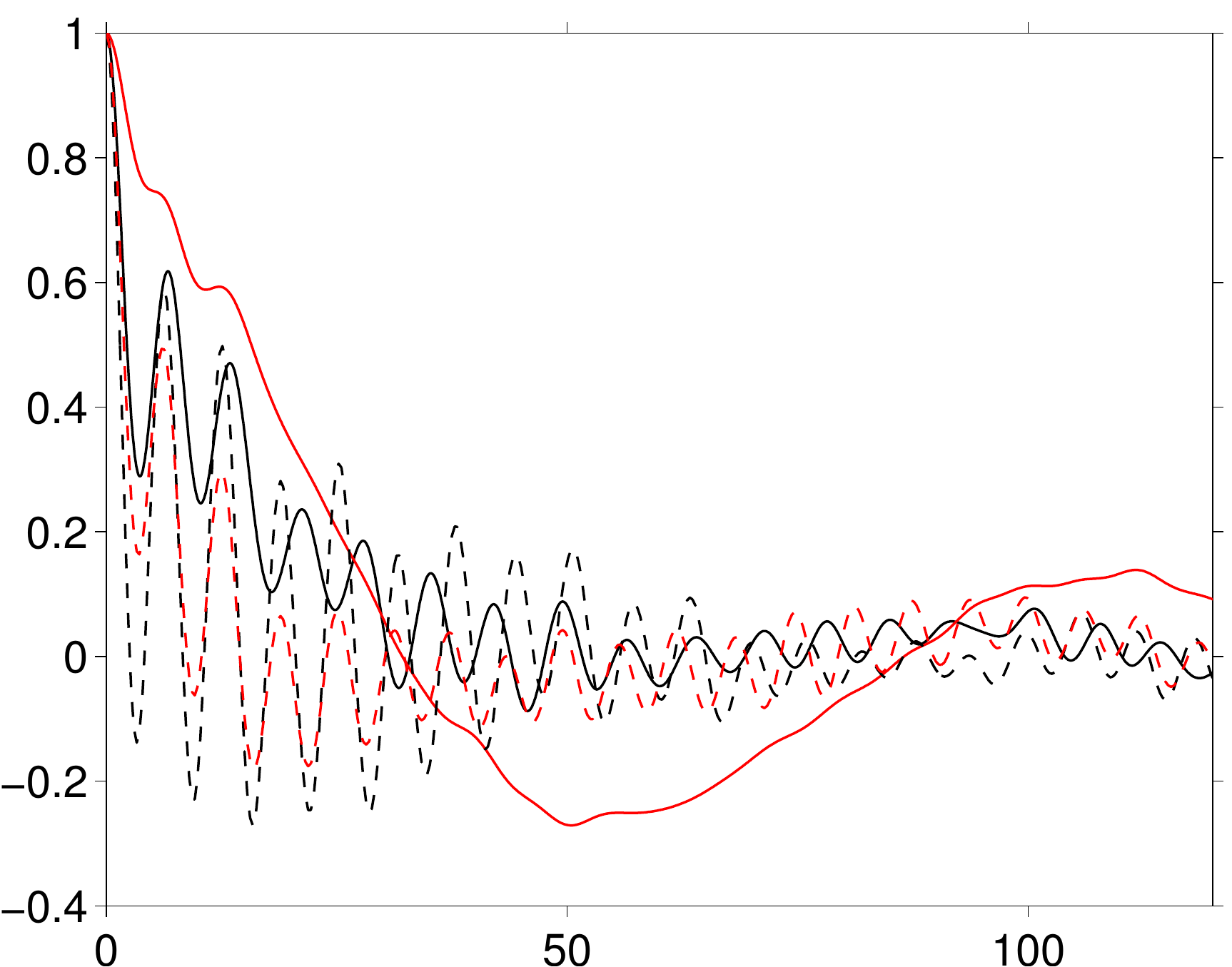}
    \\
    \centerline{$\tau_{sep}$}
  \end{minipage}
  }{
  \caption{%
    IBM results for case DC-36 ($G=250$). 
    Temporal auto-correlations of: 
    $(a)$ translational particle velocity; 
    $(b)$ angular particle velocity.
    The line-styles are as follows: 
    \solid, vertical component; 
    {\color{red}\solid}, horizontal component.
    The reference data is shown as dashed lines. 
    \label{fig-results-ibm-corr-d1c}
  }
  }
\end{figure}
Finally, let us turn to the auto-correlations (defined in eqn.\ 
\ref{equ-res-ref-def-lag-corr}) of the different particle velocity
components shown in figure~\ref{fig-results-ibm-corr-d1c}.  
The auto-correlations of the IBM simulation are characterized by a
slower decay as compared to the reference case. This confirms the
conjecture that, in the IBM simulation, the threshold of chaos lies
closer. Nevertheless, qualitatively the characteristic features of the
auto-correlation functions are well reproduced except for that of the
horizontal component of the angular velocity.  
%
%
First, the decay rate for short times is very well predicted
for both (horizontal and vertical) components. 
Second, the frequencies of the dominant oscillations are relatively well
predicted; in particular, this is true for the fact that the frequency
of the oscillation of the vertical component is approximately 
eight
times larger than the one of the horizontal component. 
It can be observed that the amplitude of the oscillations of
the auto-correlation for both translational velocity components is
overestimated in the simulations with the immersed boundary method,
which is related to a less disordered behavior.   
%

%
\section{Conclusion}\label{sec-conclusion}
We have presented data for the motion of a single solid sphere
settling in ambient fluid. The solid to fluid density ratio has been
set to $\rho_p/\rho_f=1.5$, while the Galileo number was varied from
144 to 250 such as to cover all four regimes of sphere motion. 
The data was generated by means of high-fidelity numerical simulation
employing a Fourier/spectral-element method applied to the problem
formulated in a coordinate system attached to the sphere, thereby
avoiding remeshing \citep{jenny:04b}. A moderately-sized computational
domain was chosen in order to keep the computational cost of   
grid convergence studies tractable. 
%

The data-set provided includes the sphere's degrees of freedom as well
as extracts of the flow field: the recirculation length, the relative
velocity along the axis of particle motion and along various
cross-profiles, the pressure on the sphere's surface. 
In the case of time-periodic motion, the time-evolution of the
particle motion as well as that of the flow field in its wake is
analyzed in detail.  
In the case of chaotic particle motion, a statistical analysis of the
translational and angular particle velocity is performed, presenting
moments, probability density functions and Lagrangian auto-correlation
data. 

In the second part of this contribution we have presented results of
simulations of the above solid-fluid system performed with an immersed
boundary method \citep{uhlmann:04} and using various spatial
resolutions varying from 15 points per diameter up to 48. 
The errors with respect to the reference solution (obtained with the
Fourier/spectral-element method) were presented. It was found that a
spatial resolution of $D/\Delta x=15$ is capable of reproducing the
particle motion in the steady axi-symmetric regime (at $G=144$) with
the dominant error measuring approximately 6\%.  
In the steady oblique regime ($G=178.46$) the particular IBM requires
a higher resolution of 24 points per diameter in order to produce
results of comparable quality. 
In this case we have also observed that for the immersed boundary
method applied in a Navier-Stokes fractional step context the choice
of the time step is important in order to achieve the desired accuracy
at high spatial resolution. 
In the regime where the particle motion is still restricted to a plane
in space, but varying periodically in time (at $G=190$), a higher 
spatial accuracy of $D/\Delta x=48$ was necessary in order to capture
the state correctly. 
Finally, it was observed that the chaotic particle motion at $G=250$
can be simulated with good accuracy when using $D/\Delta x=36$. 
This includes errors committed on the first and second moments of the
particle velocity components which are bounded by 4.5\%; 
it also yields a good representation of the probability density
functions of particle velocities as well as reasonable
auto-correlation functions of translational particle velocity. 

The present study provides benchmark data which is expected to be
useful for the validation of numerical approaches to finite-size
particulate flow. As we have shown, it also provides a basis for
determining the required spatial and temporal resolution corresponding
to a particular numerical method as a function of the parameter
range. As such it can be instrumental in preliminary studies towards
simulations of large-scale multi-particle systems where it is often
important to determine the minimum numerical requirements for a
desired accuracy. 

\revision{}{%
  An additional aspect of fluid--particle interaction processes is the 
  presence of turbulent background flow. In that case, direct
  numerical simulation methods should be able to faithfully take into
  account the time-dependent multi-scale ``forcing'' exerted by the
  turbulent fluid motion upon each particle. Since rigorous benchmark
  cases for this situation are not available, further efforts should
  be made in the future to fill this gap.}
%

%
\revision{%
  The data presented herein will be available under the following URL:\\ 
  \href{http://www.ifh.kit.edu/dns_data}
  {\tt http://www.ifh.kit.edu/dns\_data}}{%
  The data presented herein is available as supplementary material
  from the journal website. 
  It is also accessible under the following URL:\\
  \href{http://www.ifh.kit.edu/dns_data/particles/single_sphere_sedimentation}
  {\tt www.ifh.kit.edu/dns\_data/particles/single\_sphere\_sedimentation}}
%
\section*{Acknowledgments}
Fruitful discussions with Todor Doychev and Aman G.\ Kidanemariam
throughout this work are gratefully acknowledged. 
This work was supported by the German Research Foundation (DFG) under
projects UH 242/1-1 and UH 242/1-2. 
%

%
\bibliographystyle{model2-names}
\addcontentsline{toc}{section}{Bibliography}
%

%
\end{document}